\documentclass[ppnp-LaTeX,notitlepage,preprint,tightenlines,eqsecnum,nofootinbib,superscriptaddress,floatfix]{elsarticle}

\usepackage{graphicx} 
\usepackage{bm}
\usepackage{amssymb}  
\usepackage[pdftex]{color}
\usepackage{slashed}
\usepackage{booktabs}
\usepackage{multirow}
\usepackage{amsmath}
\usepackage{stackrel}
\usepackage{rotating}
\usepackage{rotfloat}
\usepackage{hyperref}
\usepackage{amsmath}
\usepackage[utf8]{inputenc}
\allowdisplaybreaks
\usepackage{enumitem}

\def\b0{\boldsymbol{0}}

\newcommand{\be}{\begin{equation}}
\newcommand{\ee}{\end{equation}}
\newcommand{\bea}{\begin{eqnarray}}
\newcommand{\eea}{\end{eqnarray}}
\newcommand{\nn}{\nonumber}

\def\barr{\left(\begin{array}{c}}
\def\earr{\end{array}\right)}
\def\bmat{\left(\begin{array}{cc}}
\def\emat{\end{array}\right)}

\def\Figref#1{Fig.~\ref{fig:#1}}

\def\pd{\partial}
\def\a {\alpha}
\def\m {\mu}
\def\n {\nu}
\def\b {\beta}
\def\r {\rho}
\def\l {\lambda}
\def\si{\sigma} 

\begin{document}

\begin{frontmatter}

\title{The hadronic light-by-light contribution to the muon's anomalous magnetic moment}
\author[lab1,lab2]{Igor Danilkin}
\author[lab1,lab2]{Christoph Florian Redmer}
\author[lab1,lab2]{Marc Vanderhaeghen}

\address[lab1]{Institut f\"ur Kernphysik, Johannes Gutenberg-Universit\"at,  55099 Mainz, Germany}
\address[lab2]{PRISMA$^+$ Cluster of Excellence, Johannes Gutenberg-Universit\"at, 55099 Mainz, Germany}

\begin{abstract}

In view of the current 3 - 4~$\sigma$ deviation between theoretical and experimental values for the muon's anomalous magnetic moment, we review the ongoing efforts in constraining the hadronic light-by-light contribution to $a_\mu$ by using dispersive techniques combined with a dedicated experimental program to obtain the required hadronic input. 

\end{abstract}

\begin{keyword}
muon's anomalous magnetic moment, meson transition form factors, two-photon fusion processes, dispersion theory
\end{keyword}

\end{frontmatter}


%
\tableofcontents

\section{Introduction}
\label{sec1}
The anomalous magnetic moment of the muon $a_\mu = (g - 2)_\mu  /2$ has since long been studied as a test of the Standard Model (SM) of particle physics, and for its high potential of probing new, beyond the Standard Model, physics. Its experimental value is dominated by the measurements made at Brookhaven National Laboratory~\cite{Bennett:2002jb,Bennett:2004pv,Bennett:2006fi}, 
resulting in a world average of \cite{Tanabashi:2018oca} 
\bea 
a_\mu^{exp} = (11659209.1 \pm 6.3) \times 10^{-10}.
\eea
The SM prediction to $a_\mu$ was recently updated in Refs.~\cite{Jegerlehner:2017gek,Keshavarzi:2018mgv}, resulting in the value~\cite{Keshavarzi:2018mgv}:
\bea 
a_\mu^{SM} = (11659182.04 \pm 3.56) \times 10^{-10}.
\label{amusm}
\eea
The presently observed $3 - 4~\sigma$ deviation between theory and experiment in this observable has indicated that with the obtained precision, one may be tantalizingly close to probe new physics. On the experimental side, this discrepancy has triggered new $a_\mu$ measurements both at FERMILAB (E989) \cite{LeeRoberts:2011zz,Grange:2015fou} as well as at J-PARC ~\cite{Iinuma:2011zz} within the next few years with the aim to reduce the experimental error on $a_\mu$ by a factor of four over the present value.  However, the interpretation of $a_\mu$ critically depends on the knowledge of the strong-interaction contributions, which at present totally dominate the Standard Model uncertainty. This has motivated an intense activity also on the theoretical side to reliably estimate contributions of hadrons to $a_\mu$. 

The hadronic uncertainties mainly originate from the hadronic vacuum polarization (HVP) and the hadronic light-by-light (HLbL) processes. Forthcoming data from high-luminosity $e^+ e^-$ colliders, particularly from the BESIII and Belle~II experiments, will further reduce the uncertainty in the HVP over the next years to make it commensurate with the experimental precision on $a_\mu$.

Unlike the HVP contribution, in most of the existing estimates of the HLbL contribution, the description of the non-perturbative light-by-light matrix element is based on hadronic models rather than being determined from data. 
Unfortunately, a reliable estimate based on such models is possible only within certain kinematic regimes, resulting in a large, mostly uncontrolled uncertainty of $a_\mu$. 
The SM value of Eq.~(\ref{amusm}) corresponds with the chosen estimate for the HLbL contribution~\cite{Keshavarzi:2018mgv}:
\bea
a_\mu^{HLbL} = (9.8 \pm 2.6) \times 10^{-10}, 
\eea
which at present dominates the uncertainty of the SM value for $a_\mu$. 
To reduce the model dependence entering the present estimates for $a_\mu^{HLbL}$ implies to resort to ab initio approaches such as lattice QCD or data-driven dispersive approaches. The lattice efforts in constraining the HVP and HLbL contributions to $a_\mu$ have been reviewed recently in Ref.~\cite{Meyer:2018til}, to which we refer for details and references. The present work aims to review the progress in our understanding of the empirical constraints on the HLbL contribution to $a_\mu$. For this purpose, we will discuss both the progress in the experimental information available on two-photon to hadron processes, as well as in the theoretical understanding of such two-photon to hadron processes,  and the resulting data-driven estimates for the HLbL contribution to $a_\mu$. After a general introduction to the observables for the HLbL process and related sum rules in Section \ref{sec2}, the subsequent sections aim to provide a connection between these observables and existing theoretical approaches. The organizing  principle is according to the produced hadronic state.

The dominant contributions to HLbL are given by single pseudoscalar states, which are discussed in Section~\ref{sec3}. The coupling of $\pi^0, \eta$ and $\eta^\prime$ to two photons is described by transition form factors (TFFs), which can be determined by experiments. Data for arbitrary virtualities of the photons are required as input to the data-driven approaches and for the validation of hadronic models. 
We will review the experimental status of these TFFs, in particular new measurements of the single-virtual TFF of $\pi^0$ from the BESIII Collaboration, and of the double-virtual TFF of $\eta^\prime$ from the BaBar Collaboration. We will also review the theoretical understanding of these TFFs, in particular the model-independent approaches based on dispersion theory, lattice QCD, and the short-distance constraints from perturbative QCD. For the purpose of a new estimate of the $\pi^0, \eta, \eta^\prime$ pole contributions to $a_\mu$, we will provide a new parameterization for the double-virtual TFFs satisfying the perturbative QCD constraints, and provide a new fit to the world data for $\pi^0, \eta, \eta^\prime$. We will compare the resulting estimate for $a_\mu$ with existing predictions.  

The next important contribution to $a_\mu$ comes from states of two pseudoscalar mesons, which are considered in Section \ref{sec4}. The main ingredients needed for this channel are double-virtual photon fusion processes $\gamma^*\gamma^* \to MM$, where $M=\pi,K,\eta$. Taking these channels into account allows one to cover not only the pion (kaon) loop contributions, but also scalar and tensor meson contributions. While the $f_0(500)$ resonance can be considered in the elastic approximation, a contribution from $f_0(980)$ and $a_0(980)$ resonances requires a coupled channel treatment. The contributions beyond two mesons are in general much more complicated and typically considered in the narrow resonance approximation. Section \ref{sec5} will briefly cover that.

In Section \ref{sec6} a summary and outlook will be provided, together with a new estimate for $a_\mu^{HLbL}$. Furthermore, we will emphasize the open challenges (both experimental and theoretical) needed to be addressed to further constrain the HLbL input to $a_\mu$.

\section{Observables for hadronic light-by-light processes and sum rules}
\label{sec2}
In this section we will detail the observables for the coupling of two photons to hadrons. Furthermore, we will show how causality relates these observables to the forward light-by-light scattering process. This leads to relations, such as superconvergence sum rules, which provide model independent constraints on the general light-by-light scattering amplitude.  

\subsection{Observables for photon-photon-hadron processes}
\label{sec2.1}

Experimentally, the two-photon fusion process $\gamma^\ast \gamma^\ast \to \mathrm{X}$ for two spacelike virtual photons into a general hadronic system $\mathrm{X}$, 
which is either a single meson state with $C$-parity $C = +1$, or more generally a multi-meson state such as $\pi \pi$ or $\pi^0 \eta$, is accessed at $e^+e^-$ colliders through the 
$e^+ + e^- \to e^+ + e^- + \mathrm{X}$ reaction. 

Following Refs.~\cite{Bonneau:1973kg,Budnev:1974de}, the kinematics of the process 
$e (p_1) + e (p_2) \to e (p^\prime_1) + e (p^\prime_2) + \mathrm{X}$, 
in the lepton c.m. system, i.e. the c.m. system of the colliding beams (denoted by  {\it ee c.m.}) is characterized by the four-vectors of the incoming leptons~:
\begin{eqnarray}
p_1 (E, \vec p_1), \quad \quad \quad  p_2 (E, - \vec p_1), 
\end{eqnarray}
with beam energy $E = \sqrt{s_{ee}} / 2$, and $s_{ee} = (p_1 + p_2)^2$. 
\\
The kinematics of the outgoing leptons can be related to the virtual photon four-momenta as~:
\begin{eqnarray}
q_1 = p_1 - p^\prime_1, \quad \quad  \quad  q_2 = p_2 - p^\prime_2 . 
\end{eqnarray}
The kinematics of the outgoing leptons then determines five kinematical quantities: 
\begin{itemize}
\item
the energies of both virtual photons:
\begin{eqnarray}
\omega_1 \equiv q_1^0 = E - E^\prime_1, \quad \quad \quad  \omega_2 = q_2^0 \equiv E - E^\prime_2,
\end{eqnarray}
with $E^\prime_1$ and $E^\prime_2$ the energies of both outgoing leptons;

\item
the spacelike virtualities of both virtual photons ($q_1^2 < 0$ and $q_2^2 < 0$):
\begin{eqnarray}
Q_1^2 \equiv - q_1^2 &=& 4 E E^\prime_1 \sin^2 \theta_1 / 2 + Q_{1, \, min}^2 \; ,  \nonumber \\
Q_2^2 \equiv - q_2^2 &=& 4 E E^\prime_2 \sin^2 \theta_2 / 2 + Q_{2, \, min}^2 \; , 
\end{eqnarray}
where $\theta_1$ and $\theta_2$ are the (polar) angles of the scattered electrons relative to the 
respective beam directions, and where the minimal values of the virtualities are given by (in the limit where $E^\prime_1\gg m_e$ and $E^\prime_2 \gg m_e$, with $m_e$ the electron mass)~:
\begin{eqnarray}
Q_{1, \, min}^2 \simeq m_e^2 \frac{\omega_1^2}{E E^\prime_1},
\quad \quad  \quad
Q_{2, \, min}^2 \simeq m_e^2 \frac{\omega_2^2}{E E^\prime_2};
\end{eqnarray}

\item 
the azimuthal angle $\phi$ between both lepton planes, which in the lepton c.m. frame  
can be obtained as~:
\begin{eqnarray}
\bigl( \cos \phi \bigr)_{ee~c.m.} \equiv - \frac{p^\prime_{1 \perp} \cdot p^\prime_{2 \perp}}{\left[ 
(p^\prime_{1 \perp})^2 \; (p^\prime_{2 \perp})^2 \right]^{1/2}},
\end{eqnarray} 
where $p^\prime_{1 \perp}$ and $p^\prime_{2 \perp}$ denote the components of the outgoing lepton four-vectors which are perpendicular to the respective beam directions, and are defined 
in the lepton c.m. frame  as~:
\begin{eqnarray}
\left ( p^\prime_{1 \perp} \right)^\mu = - R^{\mu \nu} (p_1, p_2) \, \left ( p^\prime_1 \right)_\nu , 
\quad \quad  \quad
\left( p^\prime_{2 \perp} \right)^\mu = - R^{\mu \nu} (p_1, p_2) \, \left ( p^\prime_2 \right)_\nu , 
\end{eqnarray}
with projector
\begin{eqnarray}
R^{\mu \nu} (p_1, p_2) = - g^{\mu \nu} +  
\frac{\left[ (p_1 \cdot p_2) \left( p_1^\mu \, p_2^\nu + p_2^\mu \, p_1^\nu \right)
- m_e^2 \left( p_1^\mu \, p_1^\nu + p_2^\mu \, p_2^\nu \right)
\right]}{\left[ (p_1 \cdot p_2)^2 - m_e^4 \right]} . 
\end{eqnarray}

\end{itemize}

In the following it will also turn out to be useful to determine kinematical quantities in the c.m. system of the virtual photons (denoted by $\gamma \gamma$ c.m.). 
In particular, the azimuthal angle between both lepton planes, in the $\gamma \gamma$ c.m. frame,  which we denote by $\tilde \phi$ is given by~:
\begin{eqnarray}
\bigl(  \cos \tilde \phi \bigr)_{\gamma \gamma~c.m.}  \equiv - \frac{\tilde p_{1 \perp} \cdot \tilde p_{2 \perp}}{\left[ 
(\tilde p_{1 \perp})^2 \; (\tilde p_{2 \perp})^2 \right]^{1/2}},
\label{eq:costilde}
\end{eqnarray} 
where $\tilde p_{1 \perp}$ and $\tilde p_{2 \perp}$ denote the transverse components of the incoming lepton four-vectors in the $\gamma \gamma$  c.m. frame  and are defined in a covariant way as~:
\begin{eqnarray}
\left ( \tilde p_{1 \perp} \right)^\mu = - R^{\mu \nu} (q_1, q_2) \, \left ( p_1 \right)_\nu , 
\quad \quad  \quad
\left(\tilde  p_{2 \perp} \right)^\mu = - R^{\mu \nu} (q_1, q_2) \, \left ( p_2 \right)_\nu , 
\end{eqnarray}
with
\begin{eqnarray}
R^{\mu \nu} (q_1, q_2) = - g^{\mu \nu} + 
\frac{\left[ (q_1 \cdot q_2) \left( q_1^\mu \, q_2^\nu + q_2^\mu \, q_1^\nu \right)
- q_1^2 \, q_2^\mu \, q_2^\nu  - q_2^2 \, q_1^\mu \, q_1^\nu 
\right]}{\left[ (q_1 \cdot q_2)^2 - q_1^2 q_2^2 \right]}. 
\end{eqnarray}
As the {\it rhs} of Eq.~(\ref{eq:costilde}) is expressed in a Lorentz invariant way, one can 
then evaluate all four-momenta in the lepton c.m. frame, 
to obtain the expression of $\cos \tilde \phi$ in terms of the lepton c.m. kinematics.

The cross section for the process 
$e (p_1) + e (p_2) \to e (p^\prime_1) + e (p^\prime_2) + \mathrm{X}$, with $\mathrm{X}$ the produced hadronic system, 
can then be expressed in terms of eight response functions for the $\gamma^\ast \gamma^\ast \to \mathrm{X}$ 
process as~\cite{Budnev:1974de}:  
\begin{eqnarray}
d \sigma &=& \frac{\alpha^2}{16\, \pi^4 \, Q_1^2 \, Q_2^2} \, \frac{2 \sqrt{X}}{s_{ee}\, (1 - 4\,m_e^2 / s_{ee})^{1/2}}  
\cdot \frac{d^3 \vec p_1^{\, \prime}}{E_1^{\prime}} 
\cdot \frac{d^3 \vec p_2^{\, \prime}}{E_2^\prime} \nonumber \\
&\times& \left\{ 
4\,  \rho_1^{++} \, \rho_2^{++} \, \sigma_{TT} 
+  \rho_1^{00} \, \rho_2^{00} \, \sigma_{LL} 
+ 2\,  \rho_1^{++} \, \rho_2^{00} \, \sigma_{TL} 
+ 2\,  \rho_1^{00} \, \rho_2^{++}  \, \sigma_{LT} 
\right. \nonumber \\
 && + 2\,  \left( \rho_1^{++} - 1 \right)  \left( \rho_2^{++} - 1 \right)  \cos (2 \tilde \phi)  \tau_{TT}  \nonumber \\
&& + 8  \left[ \frac{\left( \rho_1^{00} + 1 \right)  \left( \rho_2^{00} + 1 \right)}{\left( \rho_1^{++} - 1 \right)  \left( \rho_2^{++} - 1 \right)}\right]^{1/2}  \cos \tilde \phi \, \tilde  \tau_{TL} 
\nonumber \\ 
&& + h_1 h_2 \, 4 \, \left[ \left( \rho_1^{00} + 1 \right)  \left( \rho_2^{00} + 1 \right) \right]^{1/2} \,  \tilde \tau_{TT} \nonumber \\
&& +  \left.
h_1 h_2\, 8 \, \left[ \left( \rho_1^{++} - 1 \right)  \left( \rho_2^{++} - 1 \right) \right]^{1/2} \cos \tilde \phi  \,   \tau_{TL}
\right\}, 
\label{eq:gagacross}
\end{eqnarray}
where $h_1 = \pm 1$ and $h_2 = \pm 1$ are both lepton beam helicities, and where we defined  
 \begin{equation}
X \equiv (q_1 \cdot q_2)^2 - q_1^2\, q_2^2, 
\label{eq:defX}
 \end{equation}
as well as the virtual photon density matrix elements~:
\begin{eqnarray}
\rho_1^{++} &=& \frac{1}{2} \left\{ 1 - \frac{4\, m_e^2}{Q_1^2} + \frac{1}{X} \left( 2 \, p_1 \cdot q_2 - q_1 \cdot q_2  \right)^2 \right\}\, , \nonumber \\
\rho_2^{++} &=& \frac{1}{2} \left\{ 1 - \frac{4\, m_e^2}{Q_2^2} + \frac{1}{X} \left( 2 \, p_2 \cdot q_1 - q_1 \cdot q_2  \right)^2 \right\}\, , \nonumber \\
\rho_1^{00} &=& \frac{1}{X} \left( 2 \, p_1 \cdot q_2 - q_1 \cdot q_2 \right)^2 - 1 \, , \nonumber \\
\rho_2^{00} &=& \frac{1}{X} \left( 2 \, p_2 \cdot q_1 - q_1 \cdot q_2  \right)^2 - 1\, . 
\label{eq:kincoeff}
\end{eqnarray}
The $\gamma^\ast \gamma^\ast \to \mathrm{X}$ process is described by eight response functions: four positive definite cross sections, $\sigma_{TT},  \sigma_{LL},  \sigma_{TL},  \sigma_{LT}$; as well as four responses which can have either sign, $\tau_{TT},  \tau_{TL}, \tilde \tau_{TT},  \tilde \tau_{TL}$, , with $T (L)$ indicating a transverse (longitudinal) photon respectively. We will give the definitions of these response functions in terms of the forward light-by-light amplitudes in the next section. These eight response functions depend upon three kinematical variables: $s = (q_1 + q_2)^2, Q_1^2$, and $Q_2^2$. 

In the following we will also need the cross section for the unpolarized single-tagged process 
$e (p_1) + e (p_2) \to e (p^\prime_1) + e (p^\prime_2) + \mathrm{X}$, where the lepton momentum $p^\prime_1$ is detected, whereas the second lepton momentum $p^\prime_2$ goes undetected. This corresponds with the kinematical situation where the  
photon $q_1$ has a finite virtuality $Q_1^2$, whereas the second photon $q_2$ is quasi-real, i.e. $Q_2^2 \simeq 0$. 
By measuring the outgoing lepton $p^\prime_1$ (energy and angle), as well as the invariant mass $s$ of the hadronic system, the 
energy $\omega_2$ of the quasi-real photon is fixed as:
\begin{eqnarray}
\omega_2 = E \left( \frac{s + Q_1^2}{4E\,\omega_1 + Q_1^2} \right),
\end{eqnarray}
The resulting single-tagged cross section is differential in $Q_1^2$, $\omega_1$, and $s$ and is given by (for $E \gg m_e$):
\begin{eqnarray}
\frac{d \sigma}{d \omega_1\,  dQ_1^2\,  d s} &=& \frac{1}{\left(\omega_1 + \frac{Q_1^2}{4E} \right) \, Q_1^2 \, (s + Q_1^2)} 
\label{eq:cross1tag2} \\
&\times&
\bigg\{ F_1^{++} \, \sigma_{TT}(s, Q_1^2, Q_2^2 = 0) + F_1^{00} \,  \sigma_{LT}(s, Q_1^2, Q_2^2 = 0)  
\bigg\}, \nonumber 
\end{eqnarray}
where $F_1^{++}$ and $F_1^{00}$ are correspondingly defined virtual photon flux factors, which can be obtained from 
the general expression of Eq.~(\ref{eq:gagacross}).

As the $\gamma^\ast \gamma$ cross sections in Eq.~(\ref{eq:cross1tag2}) do not depend on $\omega_1$, one can integrate Eq.~(\ref{eq:cross1tag2}) over the experimentally accepted range of $\omega_1$ values, i.e. 
$\omega_{1, min}^{exp} < \omega_1 < \omega_{1, max}^{exp}$ yielding:
\begin{eqnarray}
\frac{d \sigma}{dQ_1^2 \, d s} &=& \frac{\tilde F_1^{++} }{Q_1^2 \, (s + Q_1^2)}  
\bigg\{ \sigma_{TT}(s, Q_1^2, Q_2^2 = 0) + \varepsilon  \, \sigma_{LT}(s, Q_1^2, Q_2^2 = 0)  
\bigg\}, \nonumber \\
\label{eq:cross1tag3}
\end{eqnarray}
with (dimensionless) integrated transverse virtual photon flux factor $\tilde F_1^{++} $ given by
\begin{eqnarray}
\tilde F_1^{++} &=& \int_{\omega_{1, min}^{exp}}^{\omega_{1, \max}^{exp}} \frac{d \omega_1}{\left(\omega_1 + Q_1^2/ (4E) \right)}\, F_1^{++}, 
\end{eqnarray}
and longitudinal photon polarization parameter $\varepsilon$, which depends on $Q_1^2$ and $W^2$, given by
\begin{eqnarray}
\varepsilon &=& \frac{1}{\tilde F_1^{++} } \int_{\omega_{1, min}^{exp}}^{\omega_{1, \max}^{exp}} \frac{d \omega_1}{\left(\omega_1 +  Q_1^2 / (4E) \right)}\, F_1^{00}.
\end{eqnarray}

\subsection{Forward light-by-light scattering}
\label{sec2.2}

The response functions, which appear in the general expression for the $e^+ + e^- \to e^+ + e^- + \mathrm{X}$ cross section, 
Eq.~(\ref{eq:gagacross}), can be related through unitarity with the amplitudes for the forward scattering of virtual photons on virtual photons: 
\begin{equation}
\gamma^\ast(\lambda_1, q_1) + \gamma^\ast(\lambda_2, q_2) \to 
\gamma^\ast(\lambda^\prime_1, q_1) + \gamma^\ast(\lambda^\prime_2, q_2), 
\label{eq:process1}
\end{equation} 
where $q_1$, $q_2$ are photon four-momenta, and 
$\lambda_1, \lambda_2$ ($\lambda^\prime_1, \lambda^\prime_2$) are the helicities 
of the initial (final) virtual photons, which can take on the values  $\pm 1$ (transverse polarizations) and zero (longitudinal). The total helicity in the $\gamma^\ast \gamma^\ast$ c.m. system is given by 
$\Lambda = \lambda_1 - \lambda_2 = \lambda^\prime_1 - \lambda^\prime_2$. To define the kinematics, besides the photon virtualities $Q_1^2 = - q_1^2$, $Q_2^2 = -q_2^2$, we also use the Mandelstam invariants: 
$s = (q_1 + q_2)^2$, $u = (q_1 - q_2)^2$, and the following crossing-symmetric variable: 
\begin{eqnarray}
\nu \equiv \mbox{$\frac14$}\,(s - u) = q_1 \cdot q_2.
\end{eqnarray}

The $\gamma^\ast \gamma^\ast \to \gamma^\ast \gamma^\ast$ forward scattering amplitudes, denoted as $M_{\lambda^\prime_1 \lambda^\prime_2, \lambda_1 \lambda_2}$,  are functions of $\nu$, $Q_1^2$, $Q_2^2$.  
Parity invariance ($P$) and time-reversal invariance ($T$) imply the following relations among the matrix elements with different helicities~: 
\begin{eqnarray}
P: \quad M_{\lambda^\prime_1 \lambda^\prime_2, \lambda_1 \lambda_2} &=&
M_{- \lambda^\prime_1 - \lambda^\prime_2, - \lambda_1 - \lambda_2}, 
\\
T: \quad M_{\lambda^\prime_1 \lambda^\prime_2, \lambda_1 \lambda_2} &=&
M_{\lambda_1 \lambda_2, \lambda^\prime_1 \lambda^\prime_2},
\end{eqnarray}
which leaves out only eight independent amplitudes~\cite{Budnev:1971sz}
\footnote{Note that the phase of the mixed longitudinal-transverse amplitudes depends on the phase convention for the longitudinal polarization vectors. 
We are using a purely real phase for the longitudinal polarization vector in this work, whereas a purely imaginary phase was chosen in Ref.~\cite{Budnev:1974de}.}:
\begin{equation}
M_{++,++}, \, M_{+-,+-}, \, M_{++,--}, \,  
M_{00,00}, \, M_{+0,+0}, \, M_{0+,0+}, \,  M_{++,00}, \, M_{0+,-0}.
\end{equation}
 
We next look at the constraint imposed by crossing symmetry, which requires that the amplitudes for the process (\ref{eq:process1}) equal the amplitudes for the process where the photons with e.g.\ label 2 are crossed:
\begin{equation}
\gamma^\ast(\lambda_1, q_1) + \gamma^\ast(- \lambda^\prime_2, - q_2) \to 
\gamma^\ast(\lambda^\prime_1, q_1) + \gamma^\ast(- \lambda_2, - q_2).  
\label{eq:process2}
\end{equation} 
As under photon crossing $\nu \to -\nu$, one obtains 
\begin{eqnarray}
M_{\lambda^\prime_1 \lambda^\prime_2, \lambda_1 \lambda_2}(\nu, Q_1^2, Q_2^2) &=&
M_{\lambda^\prime_1 - \lambda_2, \lambda_1 - \lambda^\prime_2}(-\nu, Q_1^2, Q_2^2),
\end{eqnarray}
it becomes convenient to introduce amplitudes which are either even or odd in $\nu$ (at fixed $Q_1^2$ and $Q_2^2$). 
One easily verifies that  the following six amplitudes are {\it even} in $\nu$~:
\begin{eqnarray}
&& \left( M_{++,++} + M_{+-,+-}  \right), \quad 
M_{++,--},  \quad
M_{00,00}, \nonumber \\
&& M_{+0,+0}, \quad
M_{0+,0+}, \quad
 \left( M_{++,00} + M_{0+,-0}  \right), 
\label{eq:even}
\end{eqnarray}
whereas the following two amplitudes are {\it odd} in $\nu$~:
\begin{eqnarray}
 \left( M_{++,++} - M_{+-,+-}  \right), \quad \quad
\left( M_{++,00} - M_{0+,-0}  \right). 
\label{eq:odd}
\end{eqnarray}

The optical theorem allows one to relate the absorptive part of the $\gamma^\ast \gamma^\ast \to \gamma^\ast \gamma^\ast$ forward scattering amplitudes to the response functions for the process $\gamma^\ast \gamma^\ast \to \mathrm{X}$, which appear in Eq.~(\ref{eq:gagacross}).  Denoting the absorptive part as 
\begin{eqnarray}
W_{\lambda^\prime_1 \lambda^\prime_2, \lambda_1 \lambda_2} \equiv  \mathrm{Abs}\,M_{\lambda^\prime_1 \lambda^\prime_2, \lambda_1 \lambda_2}, 
\end{eqnarray} 
the optical theorem yields:
\begin{eqnarray}
&& W_{\lambda^\prime_1 \lambda^\prime_2, \lambda_1 \lambda_2} = 
\frac{1}{2} \int d \Gamma_\mathrm{X}\, (2 \pi)^4\, \delta^4(q_1 + q_2 - p_\mathrm{X})  \nonumber \\
&& \hspace{3.5cm} \times {\cal M}_{\lambda_1 \lambda_2} (q_1, q_2; p_\mathrm{X}) \,
{\cal M}^\ast_{\lambda^\prime_1 \lambda^\prime_2} (q_1, q_2; p_\mathrm{X}), 
\label{eq:abs}
\end{eqnarray}
where ${\cal M}_{\lambda_1 \lambda_2} (q_1, q_2; p_\mathrm{X})$ denotes the invariant amplitude 
for the process
\begin{equation}
\gamma^\ast(\lambda_1, q_1) + \gamma^\ast(\lambda_2, q_2) \to \mathrm{X}(p_\mathrm{X}).
\end{equation}
As a result,  the absorptive parts are expressed in terms of eight independent 
$\gamma^\ast \gamma^\ast \to \mathrm{X}$ response functions:
\begin{eqnarray}
W_{++,++} + W_{+-,+-}  &\equiv& 2 \sqrt{X} \, \left(\sigma_0 + \sigma_2 \right) = 2 \sqrt{X} \,  \left(\sigma_\parallel + \sigma_\perp \right)
\equiv 4 \sqrt{X} \, \sigma_{TT},   \nonumber \\
W_{++,++} - W_{+-,+-}   &\equiv& 2 \sqrt{X} \, \left(\sigma_0 - \sigma_2 \right) \equiv 4 \sqrt{X} \, \tilde \tau_{TT} ,  \nonumber \\
W_{++,--} &\equiv& 2 \sqrt{X} \,  \left(\sigma_\parallel - \sigma_\perp \right) \equiv 2 \sqrt{X} \, \tau_{TT} ,   \nonumber \\
W_{00,00} &\equiv& 2 \sqrt{X} \, \sigma_{LL},  \nonumber \\
W_{+0,+0} &\equiv& 2 \sqrt{X} \, \sigma_{TL},  \nonumber \\
W_{0+,0+} &\equiv& 2 \sqrt{X} \, \sigma_{LT},  \nonumber \\
W_{++,00} + W_{0+,-0}  &\equiv& 4 \sqrt{X} \, \tau_{TL},   \nonumber\\
W_{++,00} - W_{0+,-0}  &\equiv& 4 \sqrt{X} \, \tilde \tau_{TL}, 
\label{eq:vcross}
\end{eqnarray}
where the virtual photon flux factor $X$ defined as in Eq.~(\ref{eq:defX}).
In Eq.~(\ref{eq:vcross}), 
 $\sigma_0 (\sigma_2)$ are the $\gamma^\ast \gamma^\ast \to \mathrm{X}$ cross sections for total helicity 0 (2) respectively, 
 and $\sigma_\parallel (\sigma_\perp)$ are the  cross sections for linear photon polarizations with both photon polarization 
 directions parallel (perpendicular) to each other respectively. 
 The remaining cross sections (positive definite quantities $\sigma$) 
 involve either one transverse ($T$) and one longitudinal ($L$) photon polarization, or two longitudinal photon polarizations, with $\sigma_{LT}$ and $\sigma_{TL}$ related as~:
 \begin{equation}
\sigma_{LT}(\nu, Q_1^2, Q_2^2) = \sigma_{TL}(\nu, Q_2^2, Q_1^2). 
 \end{equation}  
The quantities $\tau_{TT}, \tilde \tau_{TT}, \tau_{TL}, \tilde \tau_{TL}$ denote interference cross sections (which are not sign-definite) 
with either both photons transverse ($TT$), or for one transverse and one longitudinal photon ($TL$). 

\subsection{Light-by-light scattering sum rules}
\label{sec2.3}

The principle of (micro-)causality allows to make exact statements
about analytic properties of the forward light-by-light scattering amplitudes in the complex energy plane, and to derive several sum rules, as was done in  
Refs.~\cite{Pascalutsa:2010sj,Pascalutsa:2012pr}, see also the recent pedagogical review of Ref.~\cite{Pascalutsa:2018ced}. 
The causality principle translates into the statement of analyticity
of the forward $\gamma^\ast \gamma^\ast$ scattering amplitude in the
entire $\nu$ plane, except for the real axis where the branch cuts associated with
particle production are located. Assuming that the threshold for particle production is $\nu_0 > 0$, one can write down the usual dispersion relations, 
in which the amplitude is given by integrals over the non-analyticities, which in
this case are branch cuts extending from $\pm \nu_0$ to $\pm \infty$.
Finally, for amplitudes that are even or odd in $\nu$ we can write (for any fixed values of $Q_1^2, Q_2^2 > 0$):
\begin{subequations}
\begin{eqnarray}
f_{even}(\nu)  & = & \frac{2}{\pi} \int_{\nu_0}^\infty \! d \nu^\prime \frac{\nu^\prime}{\nu^{\prime \, 2} - \nu^2-i0^+} \mathrm{Abs} \, f_{even}(\nu^\prime),\\
\label{eq:dreven}
f_{odd}(\nu) & = & \frac{2 \nu}{\pi} \int_{\nu_0}^\infty \! d \nu^\prime \frac{ 1}{\nu^{\prime \, 2} - \nu^2 - i0^+} \mathrm{Abs}\, f_{odd}(\nu^\prime), 
\label{eq:drodd}
\end{eqnarray}
\end{subequations}
where $0^+$ is an infinitesimal positive number.

These dispersion relations are derived with the provision that the integrals converge.
If they do not, subtractions must be made; e.g., the once-subtracted
dispersion relation for the even amplitudes reads:
\begin{eqnarray}
f_{even}(\nu)  & = & f_{even}(0) + \frac{2\,\nu^2}{\pi} \int_{\nu_0}^\infty \! d \nu^\prime \frac{1}{\nu^\prime (\nu^{\prime \, 2} - \nu^2-i0^+)} \mathrm{Abs} \, f_{even}(\nu^\prime).
\label{eq:drsubt}
\end{eqnarray}
Using a Regge pole model assumption for the high-energy asymptotics of the 
light-by-light forward amplitudes. Refs.~\cite{Pascalutsa:2010sj,Pascalutsa:2012pr} have derived 
the following subtracted dispersion relations for the case of one real and one virtual photon 
(when the virtual photon flux factor becomes $X = \nu^2$):
\begin{eqnarray}
M_{++,++} (\nu)+ M_{+-,+-}(\nu)   &=& 
\frac{4\,\nu^2}{\pi} \int_{\nu_0}^\infty \!\! d \nu^\prime \,  \frac{  \sigma_\parallel(\nu^\prime) + \sigma_\perp(\nu^\prime) }{\nu^{\prime \, 2} - \nu^2-i0^+}    , 
\label{eq:sr4}   \\
  M_{++,--} (\nu) &=& 
\frac{4\,\nu^2}{\pi} \int_{\nu_0}^\infty \!\!d \nu^\prime  \, \frac{ \sigma_\parallel(\nu^\prime) - \sigma_\perp(\nu^\prime)  }{\nu^{\prime \, 2} - \nu^2-i0^+}   , 
\label{eq:sr5}  \\
  M_{0+,0+} (\nu) &=& 
\frac{4\,\nu^2}{\pi} \int_{\nu_0}^\infty \!\! d \nu^\prime \,  \frac{ \sigma_{LT}(\nu^\prime) }{\nu^{\prime \, 2} - \nu^2-i0^+} , \\
  M_{+0,+0} (\nu) &=& 
\frac{4\,\nu^2}{\pi} \int_{\nu_0}^\infty \!\! d \nu^\prime \,  \frac{ \sigma_{TL}(\nu^\prime) }{\nu^{\prime \, 2} - \nu^2-i0^+}. 
\label{eq:sr6} 
\end{eqnarray}
For all the remaining amplitudes 
the asymptotic behavior justifies the use of unsubtracted dispersion relations which, upon substituting Eq.~(\ref{eq:vcross}), lead to the following sum rules, valid for both photons virtual: 
\begin{eqnarray}
 M_{++,++}(\nu) - M_{+-,+-} (\nu) &=& 
\frac{4\,\nu}{\pi} \int_{\nu_0}^\infty \!\!d \nu^\prime  \,\frac{\sqrt{X^\prime}\,\big[ \sigma_0(\nu^\prime) - \sigma_2(\nu^\prime) \big]  }{\nu^{\prime \, 2} - \nu^2-i0^+}   , 
\label{eq:sr1} \\
 M_{++,00}(\nu) - M_{0+,-0}(\nu)  &=& 
\frac{8\,\nu}{\pi} \int_{\nu_0}^\infty \!\!d \nu^\prime  \,\frac{\sqrt{X^\prime} \, \tilde \tau_{TL} (\nu^\prime)}{\nu^{\prime \, 2} - \nu^2-i0^+}   , 
\label{eq:sr1b}\\
M_{++,00} (\nu)+ M_{0+,-0} (\nu)   &=& 
\frac{8}{\pi} \int_{\nu_0}^\infty \!\!d \nu^\prime \, \frac{\nu^\prime \sqrt{X^\prime}\, \tau_{TL} (\nu^\prime) }{\nu^{\prime \, 2} - \nu^2-i0^+} ,
\label{eq:sr2} 
\end{eqnarray}
where the dependence on virtualities $Q_1^2$, $Q_2^2$ is tacitly  assumed.
\label{eq:sumrules} 

The above sum rules, relating all the forward $\gamma^\ast \gamma^\ast$ elastic scattering
amplitudes to the energy integrals of  the $\gamma^\ast \gamma^\ast$ fusion cross sections,
should hold for any space-like photon virtualities in the unsubtracted cases, and for
one of the virtualities equal to zero in the subtracted cases. To obtain more specific relations, the low-energy (small $\nu$) 
behavior of the $\gamma^\ast \gamma^\ast \to \gamma^\ast \gamma^\ast$ forward scattering amplitudes $M$ has to be considered.  
At lowest order in the energy, the self-interactions of the electromagnetic field  are described by an effective Lagrangian (of fourth order in the photon energy and/or momentum, and fourth order in the electromagnetic field):
\begin{equation}
\mathcal{L}^{(8)} = c_1 \,(F_{\mu\nu}F^{\mu\nu})^2 + c_2\, (F_{\mu\nu}\tilde F^{\mu\nu})^2,
\label{EHL} 
\end{equation}
where $F_{\mu\nu} = \partial_\mu A_\nu - \partial_\nu A_\mu$, $\tilde F^{\mu\nu} = 
\varepsilon^{\mu\nu\alpha\beta} \partial_\alpha A_\beta$, 
and where $c_1, c_2$ are two low-energy constants (LECs) which contain the structure dependent information. 
It is often referred to as Euler-Heisenberg Lagrangian due to 
the seminal work~\cite{Heisenberg:1935qt}. 

At the next order in energy, one considers the terms involving two derivatives on the field tensors, corresponding with the sixth order in the photon energy and/or momentum. Writing down all such dimension-ten operators and reducing their number using the antisymmetry of the field tensors, the Bianchi identities, as well as adding or removing total derivative terms, 
there are 6 independent terms at that order, which were chosen in Ref.~\cite{Pascalutsa:2012pr} as~:
\begin{eqnarray}
\mathcal{L}^{(10)}&=&
c_{3}\, (\pd_{\a} F_{\m\n}) (\pd^{\a} F^{\l\n}) F_{\l\r}F^{\m\r}+
c_4\, (\pd_{\a} F_{\m\n}) (\pd^{\a} F^{\m\n}) F_{\l\r}F^{\l\r}
\nonumber \\
&+ & c_5\, (\pd^{\a} F_{{\a}\n}) (\pd_\b F^{\b\n}) F_{\l\r}F^{\l\r} 
+c_{6}\, (\pd_{\a}\pd^{\a} F_{\m\n}) F^{\l\n}F_{\l\r}F^{\m\r}\nonumber \\
&+&
c_7\, ( \pd_{\a}\pd^{\a} F_{\m\n}) F^{\m\n}F_{\l\r}F^{\l\r}+
c_{8}\, (\pd^{\a}  F_{\a\m}) (\pd_{\b} F^{\b\l})  F_{\r\l}F^{\r\m},
\end{eqnarray}
where $c_3, \ldots, c_8$ are the new LECs arising at this order. 
Only $c_3$ and $c_4$ appear in the case of real photons. 

This specifies the low-energy limit of the light-by-light scattering
amplitudes in terms of the LECs describing the low-energy self-interactions of
the electromagnetic field. As a result, 
the following set of super-convergence relations, valid for at least one real photon (e.g., $Q_1\geq 0$,  $Q_2^2 = 0$) 
were derived in Refs.~\cite{Pascalutsa:2010sj,Pascalutsa:2012pr}: 
\begin{eqnarray}
0 &=& \int\limits_{s_0}^\infty d s \, \frac{ 1 }{(s + Q_1^2)} \, \tilde \tau_{TT} (s, Q_1^2, 0), 
\label{SR1} \\
0 &=& \int\limits_{s_0}^\infty d s  \, \frac{1}{(s + Q_1^2)^2} 
\left[ \sigma_\parallel + \sigma_{LT} + \frac{(s + Q_1^2)}{Q_1 Q_2} \tilde \tau_{TL} 
\right]_{Q_2^2 = 0}, 
\label{SR2} \\
0 &=& \int\limits_{s_0}^\infty d s  \, \left[ \frac{\tau_{TL} (s, Q_1^2, Q_2^2) }{Q_1 Q_2} 
\right]_{Q_2^2 = 0}. 
\label{SR3}
\end{eqnarray}
and the following set of sum rules for the LECs of the dimension-8 (Euler-Heisenberg)
Lagrangian, in terms of cross sections where both  photons are real:
\begin{eqnarray}
 c_1 & = & \frac{1}{8 \pi }\int\limits_{s_0}^{\infty} \frac{d s}{s^2}
 \,\sigma_\parallel (s,0,0),
 \label{s1rule1} \\ 
 c_2 & = & \frac{1}{8 \pi }\int\limits_{s_0}^{\infty} \frac{d s}{s^2}\,  \sigma_\perp(s,0,0),
 \label{s0rule2}
\end{eqnarray}
where $s_0 = 2\, \nu_0 - Q_1^2 - Q_2^2$. 
There are as well the sum rules for the LECs of the dimension-10 Lagrangian, for which we refer to 
Ref.~\cite{Pascalutsa:2012pr}.

All of the above relations were verified exactly in perturbation theory  
at leading order in scalar and spinor QED~\cite{Pascalutsa:2012pr}, 
and a proof to all orders in perturbation theory was given within the context of the $\phi^4$ quantum field theory~\cite{Pauk:2013hxa}. 
These super-convergence relations were subsequently applied to the 
$\gamma^\ast \gamma$ -production of mesons, and it was shown 
quantitatively that they lead to relations between the 
$\gamma^\ast \gamma$ transition form factors (TFFs)  for scalar, pseudo-scalar, axial-vector, and tensor mesons. 
In particular, it was shown in ~\cite{Pascalutsa:2012pr} that for isospin $I = 0$ states, the helicity-2 minus helicity-0 difference sum rule for transverse photons, Eq.~(\ref{SR1}), was saturated by the pseudo-scalar mesons $\eta, \eta^\prime$, and the tensor meson $f_2(1270)$. Furthermore for the second sum rule of Eq.~(\ref{SR2}), which involves both transverse and longitudinal photons it was found that it was saturated by the axial-vector mesons $f_1(1285)$, $f_1(1420)$, and the tensor meson $f_2(1270)$. This has allowed Ref.~\cite{Pascalutsa:2012pr} to provide  empirical estimates for the dominant helicity $\Lambda = 2$ TFF for $f_2(1270)$, which was found to be in very good agreement with recent Belle data~\cite{Masuda:2015yoh}. 

The analysis of \cite{Pascalutsa:2012pr} has been improved in Ref.~\cite{Danilkin:2016hnh} after the release of the Belle data~\cite{Masuda:2015yoh} by including contributions beyond the $\Lambda = 2$ TFF for the tensor meson $f_2(1270)$, as well as including contributions of higher mesons. 
For the sum rule of Eq.~(\ref{SR1}), the previous findings were confirmed that the $\eta, \eta^\prime$ and $\Lambda = 2$ production of $f_2(1270)$ saturate this sum rule within the experimental uncertainty up to around 1 GeV$^2$. For larger values of $Q^2$, a clear signal was found for additional $\Lambda = 2$ strength. Adding the second lowest tensor meson, $f_2(1565)$, allows to saturate the helicity sum rule up to $Q^2 \simeq 5$~GeV$^2$, corresponding with the whole range of the Belle data. This has allowed Ref.~\cite{Danilkin:2016hnh} to make a prediction for the $\Lambda = 2$ TFF for the tensor meson $f_2(1565)$ over the whole range in $Q^2$, which can be tested by future data at lower $Q^2$.
Furthermore, Ref.~\cite{Danilkin:2016hnh} also analyzed the sum rules of Eqs.~(\ref{SR2},\ref{SR3}), which both involve the TFFs for longitudinal and transverse photons. By accounting for the contributions of the 
$f_1(1285)$, $f_1(1420)$, $f_0(980)$, $f_2(1270)$, and $f_2(1565)$ mesons to the sum rule of Eq.~(\ref{SR2}), and 
the $f_1(1285)$, $f_1(1420)$, and $f_2(1270)$ mesons to sum rule of Eq.~(\ref{SR3}), 
It was demonstrated that both sum rules can be well satisfied up to around  $Q^2 \simeq 1$~GeV$^2$ 
within the experimental uncertainty. This has for the first time allowed to extract the $\Lambda = 1$ and $\Lambda = (0,L)$ TFF for the $f_2(1270)$ meson in the low $Q^2$ region, up to around 1 GeV$^2$. A very sizable value for the longitudinal, i.e. $\Lambda = (0,L)$, TFF of the tensor meson $f_2(1270)$ was found. A direct measurement of this longitudinal TFF may be very worthwhile and may be possible in the near future at BESIII in a double-tagged experiment. 

The estimates for the different meson TFF were then used in \cite{Danilkin:2016hnh} to provide updates for the scalar, axial-vector and tensor meson  light-by-light contributions to the muon's $a_\mu$.
We will discuss these estimates in 
Sections~\ref{sec:2mesonamu} and \ref{sec5}.

The eight forward light-by-light amplitudes of Eqs.~(\ref{eq:even},\ref{eq:odd}) have also directly been studied for two spacelike photons in $N_f = 2$ lattice QCD~\cite{Green:2015sra,Gerardin:2017ryf}. 
Via dispersive sum rules, as given by Eqs.~(\ref{eq:sr1},\ref{eq:sr1b},\ref{eq:sr2}) the real parts for the forward light-by-light amplitude were then be compared with the integral over the absorptive parts. The latter were estimated in Refs.~\cite{Green:2015sra,Gerardin:2017ryf} using an empirical model for the TFFs of the dominant pseudoscalar, scalar, axial, and tensor mesons. 
It was found that the monopole and dipole masses parametrizing the TFFs compare reasonably well in magnitude with phenomenological determinations for $I = 0$ states, with the notable exception of the subdominant TFFs for the tensor mesons with helicity $\Lambda = 1$ and $\Lambda = (0,T)$, where the TFFs were found to fall off more slowly. The pioneering work of Ref.~\cite{Green:2015sra,Gerardin:2017ryf} paves the way for future lattice calculations to directly test, in a broad kinematic regime and by a completely independent method, the resonance-exchange model widely used in calculating $a_\mu$.

\section{Pseudoscalar meson transition form factors and $a_\mu$ contribution}
\label{sec3}
\subsection{Overview}
The interaction of two virtual photons with a pseudoscalar meson is described by the S-matrix element
\begin{equation*}
    {\mathcal M}_{\lambda_1 \lambda_2} = -i e^2 \varepsilon_{\mu\nu\alpha\beta}\, \varepsilon^\mu(q_1,\lambda_1)\, \varepsilon^\nu(q_2,\lambda_2)\, q_1^\alpha \,q_2^\beta \, F_M(q_1^2,q_2^2),
\end{equation*}
where $\varepsilon_{\mu\nu\alpha\beta}$ is the fully antisymmetric tensor, $\varepsilon^\alpha(q_i,\lambda_i)$ are the polarization vectors of the photons depending on the helicity $\lambda$ and the four-momentum $q$ of the photons, and $F_M(q_1^2,q_2^2)$ is the meson ($M = \pi^0, \eta, \eta^\prime$) transition form factor (TFF), which is a function of the squared four-momenta, the virtualities of the photons. The TFF comprises the structural information of the hadron.

\begin{figure}
    \centering
    \includegraphics[width=0.3\textwidth]{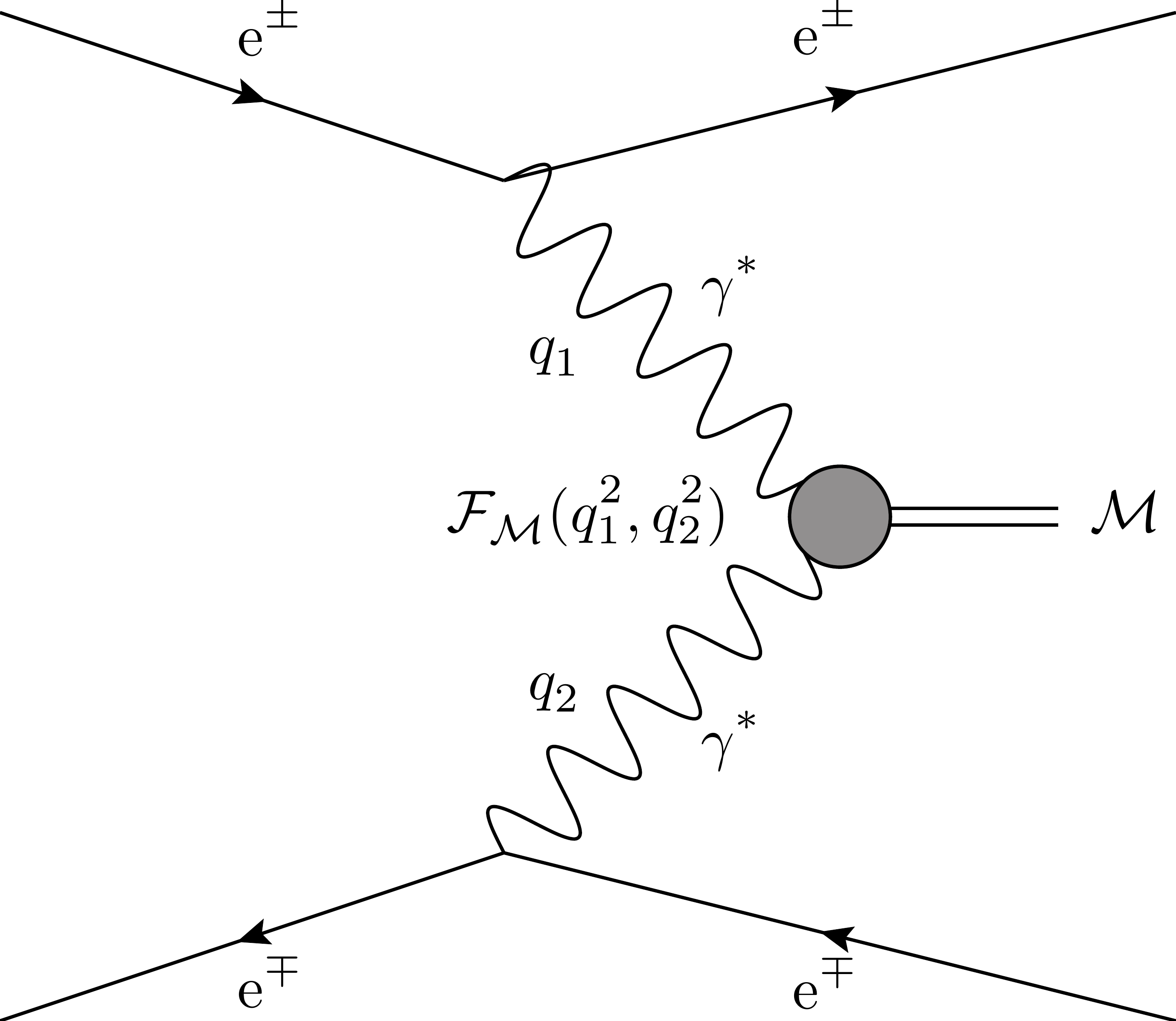}\hspace*{2em}%
    \raisebox{-1.19em}{\includegraphics[width=0.3\textwidth]{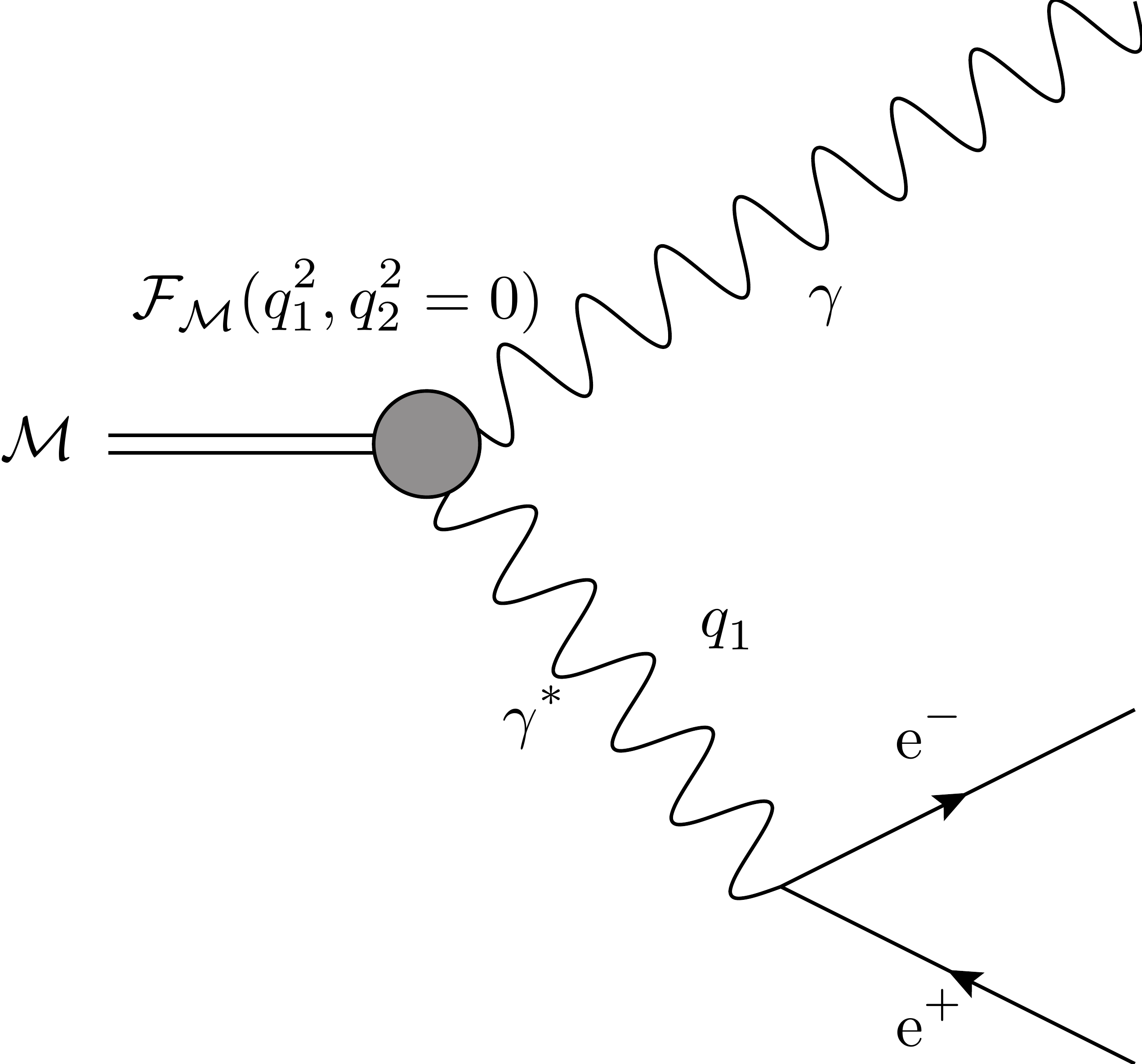}}\hspace*{2em}%
    \raisebox{1.25em}{\includegraphics[width=0.3\textwidth]{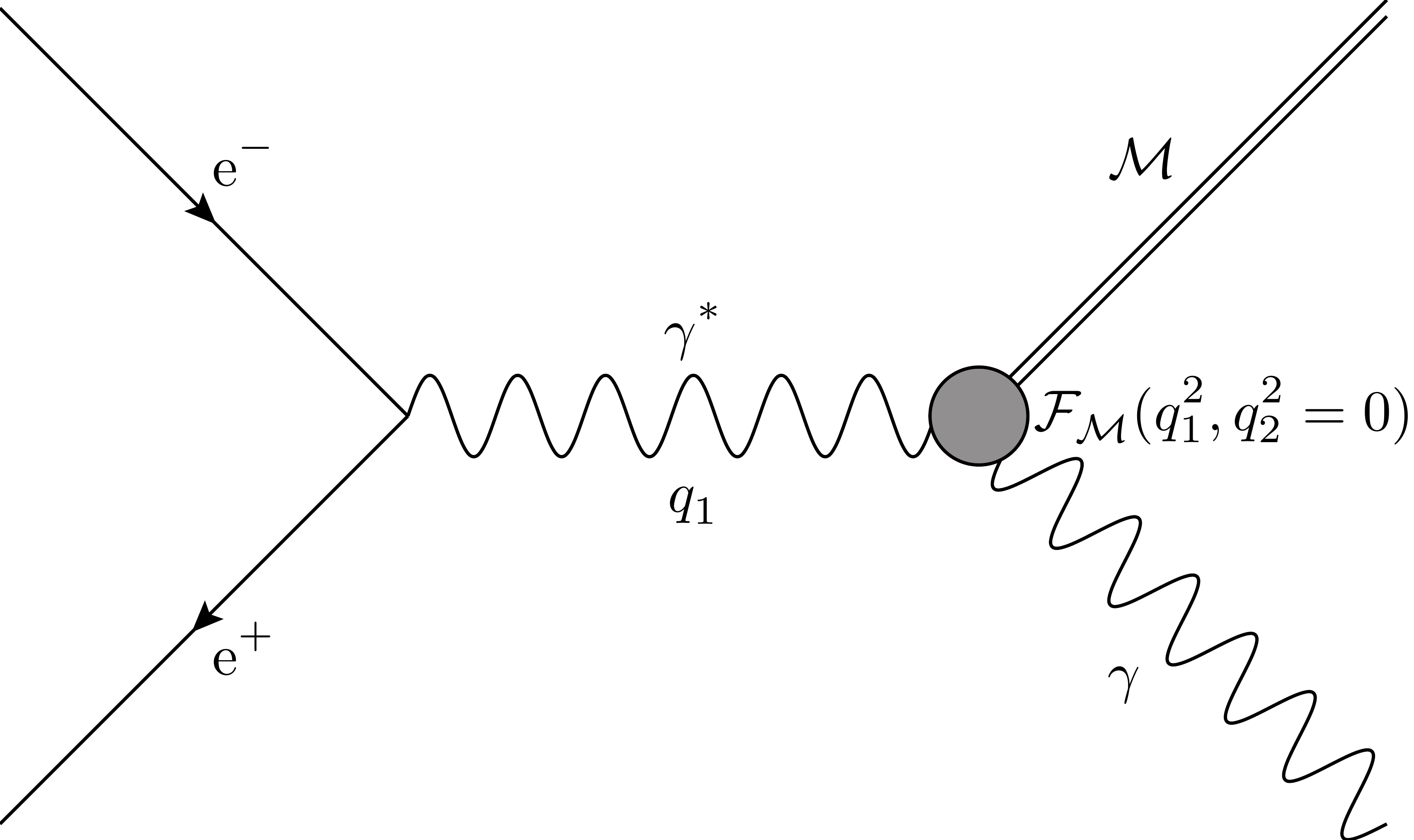}}
    \caption{\label{fig:TFF_exp_access} Feynman diagrams of the three processes, which allow to measure pseudoscalar TFFs $F_M(q_1^2,q_2^2)$ in experiments. \emph{Left:} Two-photon scattering at lepton colliders with spacelike virtualities $q_1^2 < 0,q_2^2 < 0$; \emph{Center:} Dalitz decays with timelike virtualities $q^2=m_{ee}^2$; \emph{Right:} Radiative production of mesons at $e^+e^-$ colliders with timelike virtuality $q^2=s$.}
\end{figure}

Experimentally, the pseudoscalar TFFs can be investigated in different approaches, giving access to different virtualities. The spacelike regime ($q_i^2 < 0$) can be studies at lepton colliders in the process of two-photon scattering. As depicted in the left panel of \Figref{TFF_exp_access}, both leptons exchange a virtual photon. Their virtuality is defined by the momentum transfer of the leptons. The fusion of both photons to form a meson is described by the TFF.

The timelike regime ($q_i^2 > 0$) can be studied in two different ways, covering two separate kinematic regions. In Dalitz decays, as illustrated in the central panel of \Figref{TFF_exp_access}, pseudoscalar mesons decay into a real and a virtual photon. The latter converts in a lepton pair, allowing to measure its four-momentum. The virtuality is constrained by the square of the meson mass. Double Dalitz decays, \emph{i.e.} decays of pseudoscalars into two lepton pairs allow to investigate the timelike TFF as a function of two virtualities. Above the boundary of the meson mass, the timelike TFF can still be investigated through the radiative production of pseudoscalar mesons at $e^+e^-$ colliders. As can be seen from the right panel in \Figref{TFF_exp_access}, the virtuality is fixed to the center of mass energy $s$ of the collider. When the radiative production proceed with a virtual photon in the final state, which decays into a lepton pair, also here the timelike TFF can be studied as a function of two virtualities.

In the following, a summary of the available experimental information on pseudoscalar TFFs is provided, starting with the spacelike TFF measurements in two-photon scattering reactions, followed by the studies of the timelike TFFs in the radiative meson production at $e^+e^-$ machines, as well as mesons decays.

The overview of the experimental status is followed by a review of the theoretical approaches to understand the pseudoscalar meson TFFs.

\subsection{Experimental situation on spacelike $\pi^0, \eta, \eta^\prime$ TFFs}
Spacelike TFFs $F_M(-Q_1^2,-Q_2^2)$ of pseudoscalar mesons are experimentally accessible through the investigation of the two-photon scattering process at $e^+e^-$ machines. If in the scattering each of the beam particles emits a photon, these can fuse to form states of quantum numbers $J^{PC} = 0^{-+}, 0^{++},2^{++},...$, which are not directly accessible in the dominating scattering and annihilation processes involving the exchange of a single photon. The mass of mesons produced in two-photon scattering is typically well below the center-of-mass energies at which the lepton colliders are operated. Thus, the lightest pseudoscalar mesons $\pi^0, \eta$ and $\eta^\prime$ are predominantly produced. Their production cross section is directly proportional to the square of their TFF.

The momentum dependence of the TFF is studied by measuring the momentum transfer $q^2=-Q^2$ of the scattered leptons. Since information at small momentum transfers is most relevant for the estimate of the HLbL contribution to $a_\mu$, the outgoing leptons in the events of interest are predominantly emitted at small angles. The typical layout of particle detectors at $e^+e^-$ machines limits the geometric acceptance in this range. The beam optics of the accelerator, required to bend and focus the beam towards the collision region, do usually not allow to place active detectors close to zero degree scattering angles. At the same time, this region also suffers high rates of beam background and small angle Bhabha scattering, making the readout and identification of the signal process challenging. A few facilities, \textit{e.g.} CELLO~\cite{Behrend:1981uv}, KEDR~\cite{Anashin:2013twa}, and KLOE-2~\cite{Babusci:2015osa}, have installed special tagging detectors, covering small fractions of the solid angle at small scattering angles to aid the measurement of two-photon scattering. 

The conventional approach to measure the momentum dependence of the TFF is to register the scattered leptons and the produced meson in the main detectors. This puts a lower limit on the accessible range of momentum transfer, which, neglecting the lepton mass, is defined as $Q^2=4EE^\prime\sin^2\frac{\theta}{2}$, with the beam energy $E$ and the energy $E^\prime$ of a lepton emitted at the scattering angle $\theta$. As the cross section of the process $e^+e^-\to e^+e^- M$  (with $M = \pi^0, \eta, \eta^\prime$)  drops rapidly with increasing momentum transfers, large data sets are required. Only recently the BaBar Collaboration succeeded to measure the momentum dependence of the TFF of $F_{\eta^\prime}(-Q_1^2,-Q_2^2)$ as a function of both virtualities~\cite{BaBar:2018zpn}.

A viable approach to the investigation of the momentum dependence of TFF, which has been applied frequently over the past 30 years, is referred to as single-tagged technique. Instead of an exclusive reconstruction of the final state, besides the produced mesons, only one of the outgoing leptons is required to be registered in the detector. The second lepton is assumed to be scattered along the beam axis with momentum transfer $Q^2\approx0\,\textrm{GeV}^2$, having exchanged a quasi-real photon. Using energy and momentum conservation, the scattering angle of the unmeasured lepton can be reconstructed from the known center-of-mass energy and the measured particles. By requiring the reconstructed angle of the untagged lepton to be close to $\cos\theta\approx1$, the two-photon process can be selected efficiently. The TFF is determined as $F_M(-Q_1^2,-Q_2^2=0) \equiv F_M(Q_1^2)$, depending only on a single virtuality. In the following sections an overview of the experimental results on TFFs of single pseudoscalar mesons is given. The TFFs are presented as the product $Q^2\cdot F_M(Q^2)$, which takes into account the momentum dependence of the pQCD limit $\propto Q^{-2}$. Additionally, the TFFs are normalized to their value at $Q^2=0\,\textrm{GeV}^2$, defined as
\bea
F_M(0,0) = \sqrt{\frac{4\,\Gamma_M^{\gamma\gamma}}{\pi\,\alpha^2\,m_M^3}},
\eea
with $\Gamma_M^{\gamma\gamma}$ and $m_M$ the radiative width and rest mass of the meson, and $\alpha$ being the fine structure constant. The current average values of the PDG~\cite{Tanabashi:2018oca} are considered to obtain the numerical values \begin{align}
 F_\pi(0,0) &= (0.2725\pm0.0029)\,\textrm{GeV}^{-1}, \\
 F_\eta(0,0) &= (0.2736\pm0.0048)\,\textrm{GeV}^{-1}, \\ 
 F_{\eta^\prime}(0,0) &= (0.3412\pm0.0076)\,\textrm{GeV}^{-1}.
\end{align}

An essential tool for the experimental determination of pseudoscalar TFFs are Monte Carlo generators. Using the simulation of the process $e^+e^-\to e^+e^- M$ with a known description of the TFF $F^{\rm MC}_M(- Q^2_1,- Q^2_2)$ allows to obtain the TFF from the measured cross section $\sigma^{\rm exp}$ according to
\begin{equation*}
    |F^{\rm exp}_M(- Q^2_1, -Q^2_2)|^2 = \frac{\frac{\textrm{d}\sigma^{\rm exp}}{\textrm{d}Q^2}}{\frac{\textrm{d}\sigma^{\rm MC}}{\textrm{d}Q^2}} |F^{\rm MC}_M(-Q^2_1, -Q^2_2)|^2.
\end{equation*}
Various generator codes have been developed in the past~\cite{Uehara:2013dna}. Most of the codes make use of the equivalent photon approximation, which puts limitations on the accuracy of the simulation especially for large virtualities. Thus, most recent generator developments use exact equations for the matrix element~\cite{Czyz:2010sp,Druzhinin:2014sba}.

The \textsc{GGResRC} code developed and used by the BaBar Collaboration also takes into account radiative effects of QED, albeit only for leptons with large momentum transfer. The more recently developed \textsc{Ekhara~3.0} generator includes the full calculation of radiative effects, where the terms neglected in \textsc{GGResRC} turn out to have a sizable effect~\cite{Czyz:2018jpp}.

\subsubsection{Results for $\gamma^\ast \gamma \to \pi^0$}
\label{sec:slpi0}

\begin{figure}
    \centering
    \includegraphics[width=\textwidth]{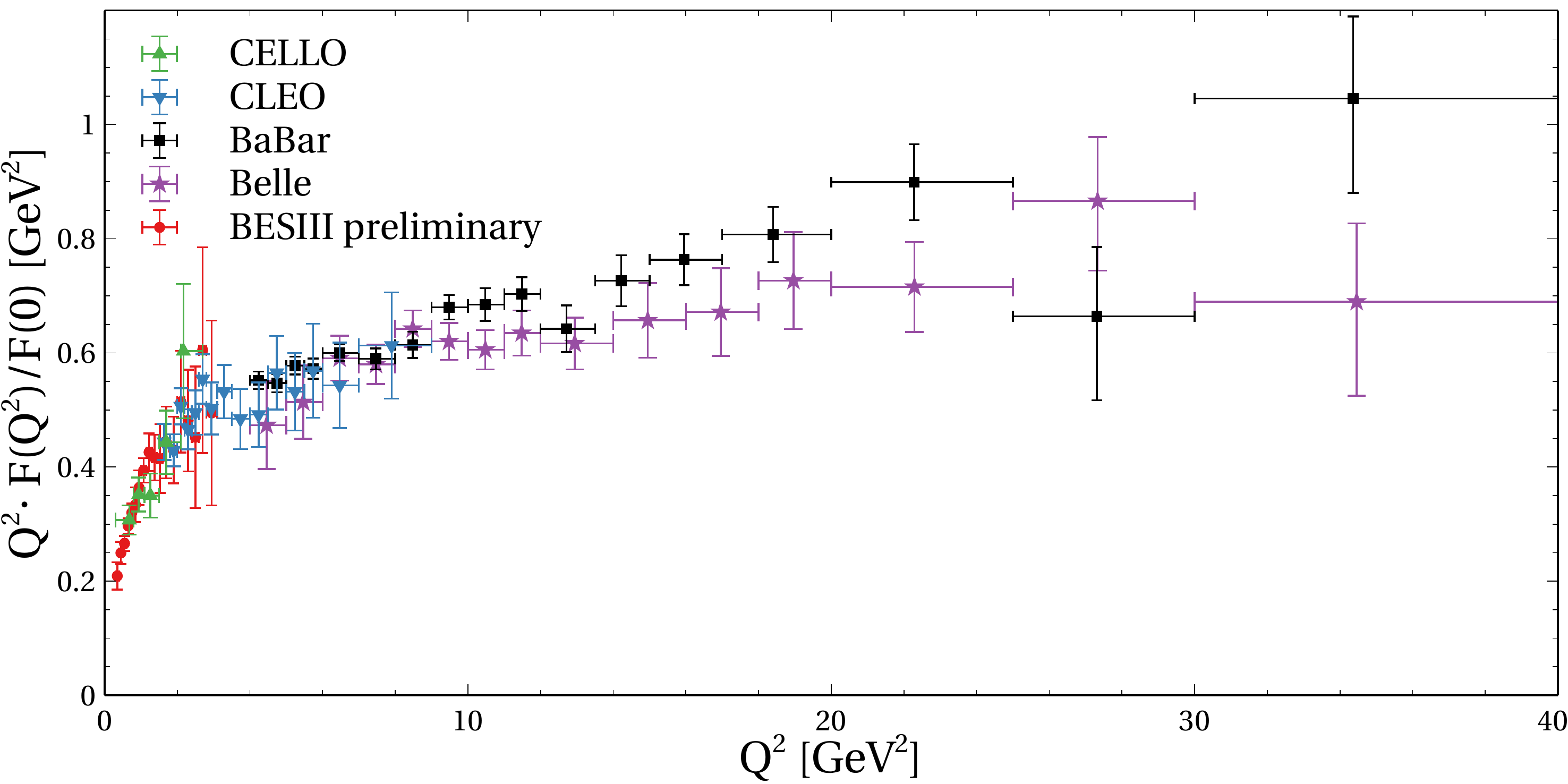}
    \caption{\label{fig:q2fpi0}(Color online) Momentum dependence of the spacelike TFF of $\pi^0$. The TFF is normalized to its value at $F_{\pi^0}(0,0)$ and multiplied with $Q^2$. Data are shown from CELLO~\cite{Behrend:1990sr} (green triangles (up), CLEO~\cite{Gronberg:1997fj} (blue triangles (down)), BaBar~\cite{Aubert:2009mc} (black squares), Belle~\cite{Uehara:2012ag} (purple stars), and preliminary data from BESIII (red circles). Error bars indicate the total uncertainties.}
\end{figure}

\begin{figure}
    \centering
    \includegraphics[width=0.5\textwidth]{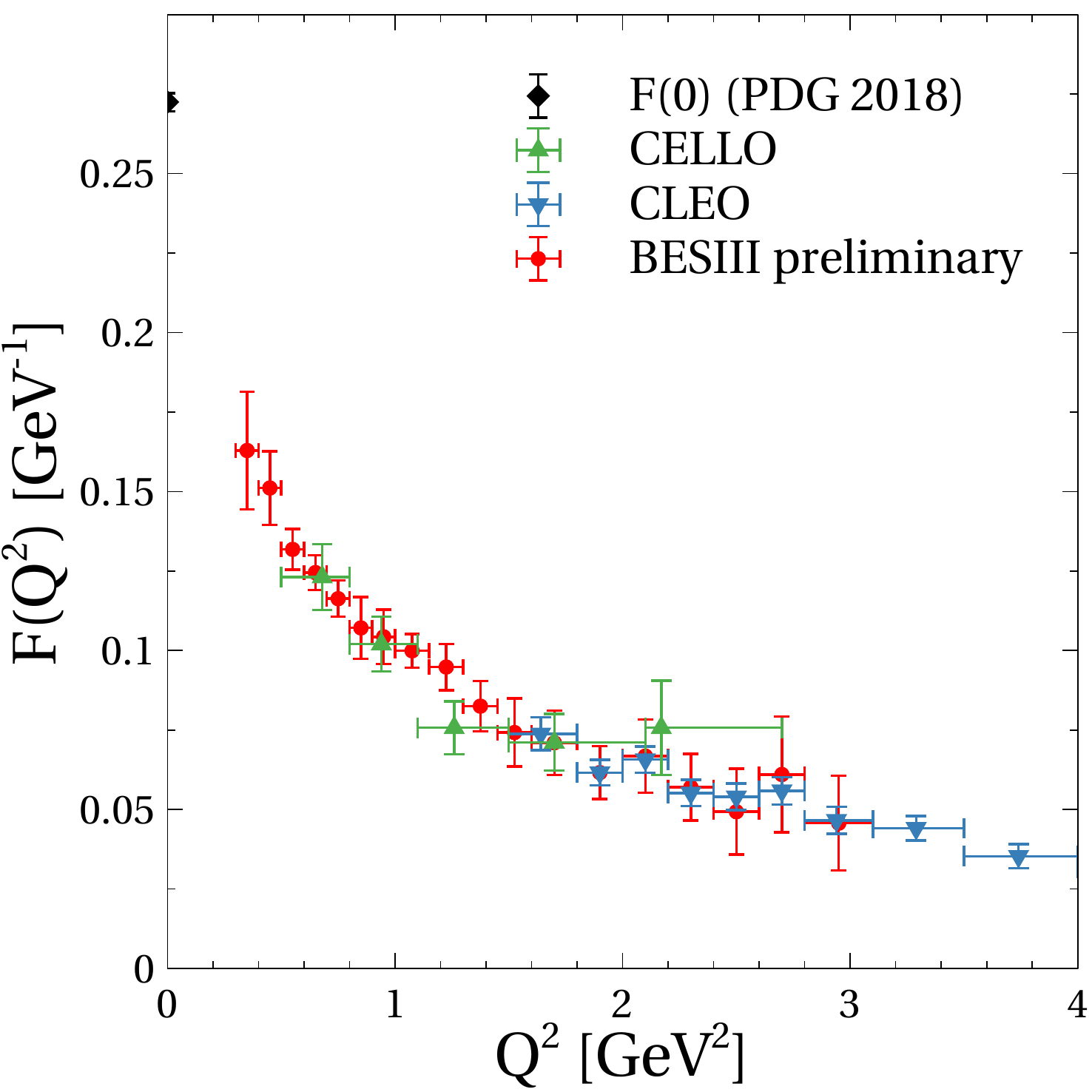}
    \caption{\label{fig:fpi0}(Color online) Momentum dependence of the spacelike TFF of $\pi^0$ for $Q^2\leq4\,\textrm{GeV}^2$. Data from CELLO~\cite{Behrend:1990sr} (green triangles (up)), CLEO~\cite{Gronberg:1997fj} (blue triangles (down)), and preliminary data from BESIII (red circles).}
\end{figure}

The measurement published in 1990 by the CELLO Collaboration, obtained from the experiment operated at the PETRA storage ring at Deutsches Elek\-tro\-nen-Syn\-chro\-tron, Hamburg, Germany~\cite{Behrend:1990sr}, is considered the first measurement of the momentum dependence of the $\pi^0$ TFF. A data sample of $86\,\textrm{pb}^{-1}$ collected at $\sqrt{s}=35\,\textrm{GeV}$ was evaluated. $137\pm12\pm16$ events of the type $e^+e^-\to e^+e^-\pi^0$ were reconstructed requiring the tag of one lepton in the forward calorimeters $(40-100\,\textrm{mrad})$ or in the end caps of the central calorimeter $(100-400\,\textrm{mrad})$, allowing to study the TFF at momentum transfers $0.5\leq Q^2\,[\textrm{GeV}^2]\leq2.7$. The results are presented in Fig.~\ref{fig:q2fpi0} and \Figref{fpi0} with solid green triangles. The extrapolation of this spacelike TFF result to zero, based on a VMD-like model, dominated the PDG's average value of the TFF slope until the recent high statistics measurements of the $\pi^0$ Dalitz decays became available. These are discussed in Sec.~\ref{sec:dalitz}.

The CLEO Collaboration published a measurement of the spacelike $\pi^0$ TFF in 1997, which covers a wider range of momentum transfers, from $1.5\,\textrm{GeV}^2$ up to $9.0\,\textrm{GeV}^2$~\cite{Gronberg:1997fj}. It is obtained from a data set of $2.88\pm0.03\,\textrm{fb}^{-1}$, which was acquired with the CLEO~II detector at the Cornell Electron Storage Ring. Two-thirds of the data were taken at $\sqrt{s}=10.58\,\textrm{GeV}$ and one third at $\sqrt{s}=10.52\,\textrm{GeV}$. As illustrated by the solid blue triangles in \Figref{q2fpi0} and \Figref{fpi0}, the significantly larger statistics allowed for a more precise determination of the TFF compared to the CELLO results, as well as for the investigation of the momentum dependence at values of $Q^2\leq 9\,\textrm{GeV}^2$.

A new era for two-photon physics started with the appearance of the B-factories. The high intensity beams allowed for detailed studies in a wide range of momentum transfer. However, the asymmetric beam energies of the storage rings, optimized to study time-dependent $CP$-violation effects, restrict the possibilities to study the momentum dependence of the TFF to $Q^2 \geq 4\,\textrm{GeV}^2$ due to the detector acceptance. Both, BaBar at PEP-II (SLAC) and Belle at KEK-B (Tsukuba) extended the range of investigation up to $Q^2=40\,\textrm{GeV}^2$ based on $485\,\textrm{fb}^{-1}$ at and close to the $\Upsilon(4S)$ resonance (BaBar, black squares in \Figref{q2fpi0}), and $759\,\textrm{fb}^{-1}$ taken at and close to the $\Upsilon(3S)$, $\Upsilon(4S)$, and $\Upsilon(5S)$ resonances (Belle, purple stars in \Figref{q2fpi0}), respectively. In the overlap region with the CLEO measurement of $4 \leq Q^2[\textrm{GeV}^2] \leq 9$, the accuracy of the TFF measurement is significantly improved. It should be noted that the BaBar Collaboration is the first to explicitly consider radiative effects of QED in their Monte Carlo simulations to determine efficiency corrections assuming that only the tagged leptons are affected. Recently, it was pointed out that the neglected terms might actually have a significant effect on the result~\cite{Czyz:2018jpp}.

The results of the B factories are predominantly suited to study the asymptotic behavior of the TFF towards high energies, as predicted by pQCD. A puzzling feature of the B-factory measurements, also referred to as BaBar-Belle-puzzle, is found in the behavior at large momenta. The BaBar result exceeds the Brodsky-Lepage-limit, which indicates the asymptotic behavior expected by pQCD as discussed in Section~\ref{sec:pqcd}, while the Belle result is well compatible with it.

Recently, the BESIII Collaboration started to investigate the momentum dependence of pseudoscalar TFFs. Based on $2.93\,\textrm{fb}^{-1}$ of data collected at $\sqrt{s}=3.773\,\textrm{GeV}$ with the BESIII detector at the Beijing Electron Positron Collider-II a preliminary result for the $\pi^0$ TFF is obtained~\cite{Redmer:2018uew}. As illustrated with red circles in \Figref{q2fpi0} and \Figref{fpi0}, the momentum dependence is studied from $0.3\,\textrm{GeV}^2$ up to $3.1\,\textrm{GeV}^2$. The preliminary BESIII result extends the CELLO measurement towards lower values of $Q^2$, which is important for the hadronic light-by-light scattering calculations for $a_\mu$, and it exceeds its accuracy. In the overlap region with the CLEO measurement at $Q^2\geq1.5\,\textrm{GeV}^2$ both results show good agreement.

Even though the BESIII measurement uses similar means to suppress radiative effects of QED on the determination of momentum transfer as the BaBar Collaboration, the preliminary result does not yet take into account radiative effects in the efficiency corrections. This will be part of the final result, performed based on the full calculations included in the \textsc{Ekhara~3.0} Monte Carlo generator.

\subsubsection{Results for $\gamma^\ast \gamma \to \eta, \eta^\prime$}
\begin{figure}
    \centering
    \includegraphics[width=\textwidth]{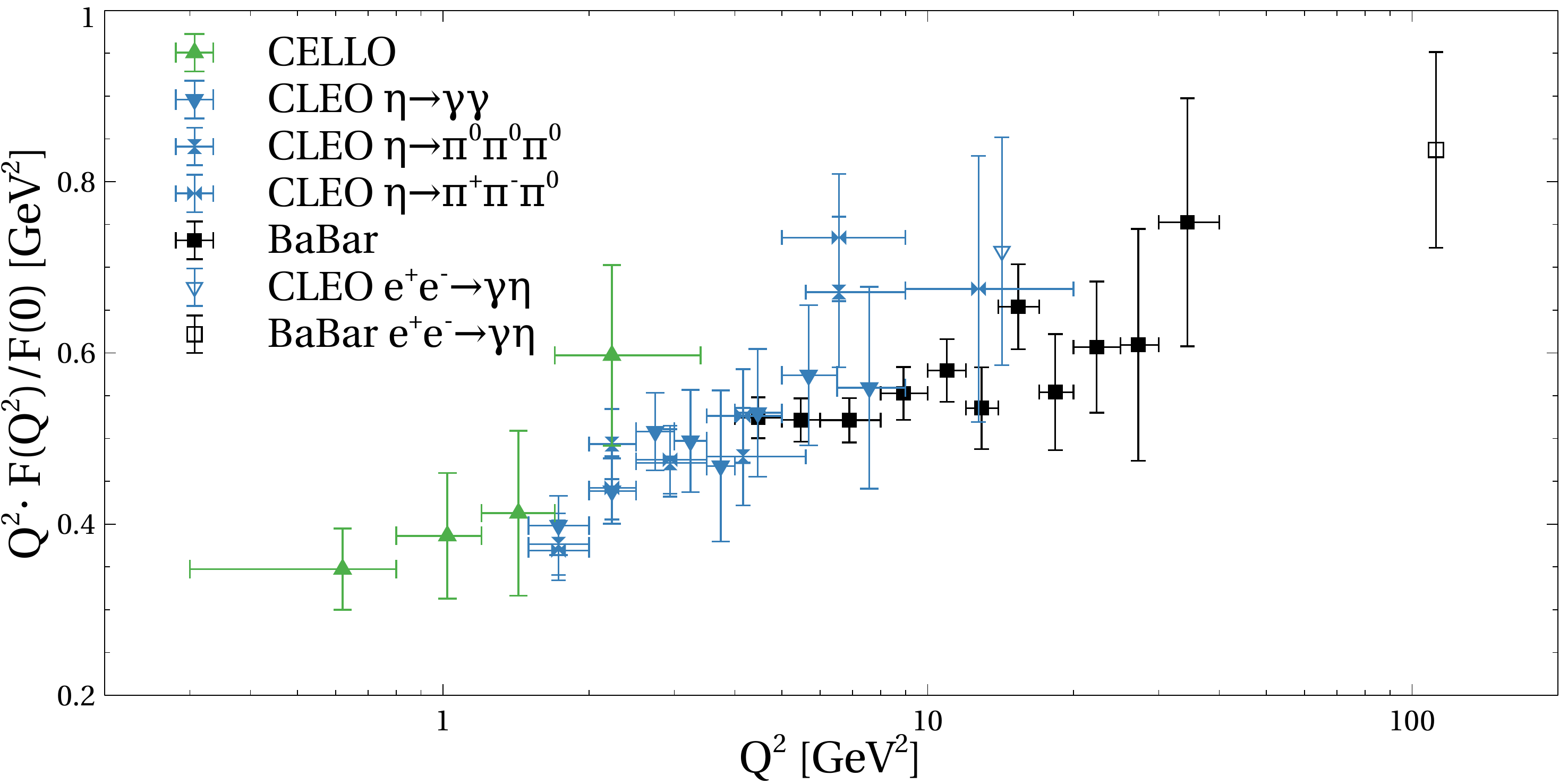}
    \caption{\label{fig:q2feta}(Color online) Momentum dependence of the spacelike TFF of $\eta$. The TFF is normalized to its value at $F_\eta(0,0)$ and multiplied with $Q^2$. Data from CELLO~\cite{Behrend:1990sr} (red), CLEO~\cite{Gronberg:1997fj} (blue), and BaBar~\cite{BABAR:2011ad} (orange) are shown with error bars corresponding to the total uncertainty. The open symbols show timelike data of CLEO~\cite{Pedlar:2009aa} and BaBar~\cite{Aubert:2006cy}. The error bars indicate the total uncertainties.}
\end{figure}

\begin{figure}
 \begin{minipage}{0.48\textwidth}
  \centerline{\includegraphics[width=\textwidth]{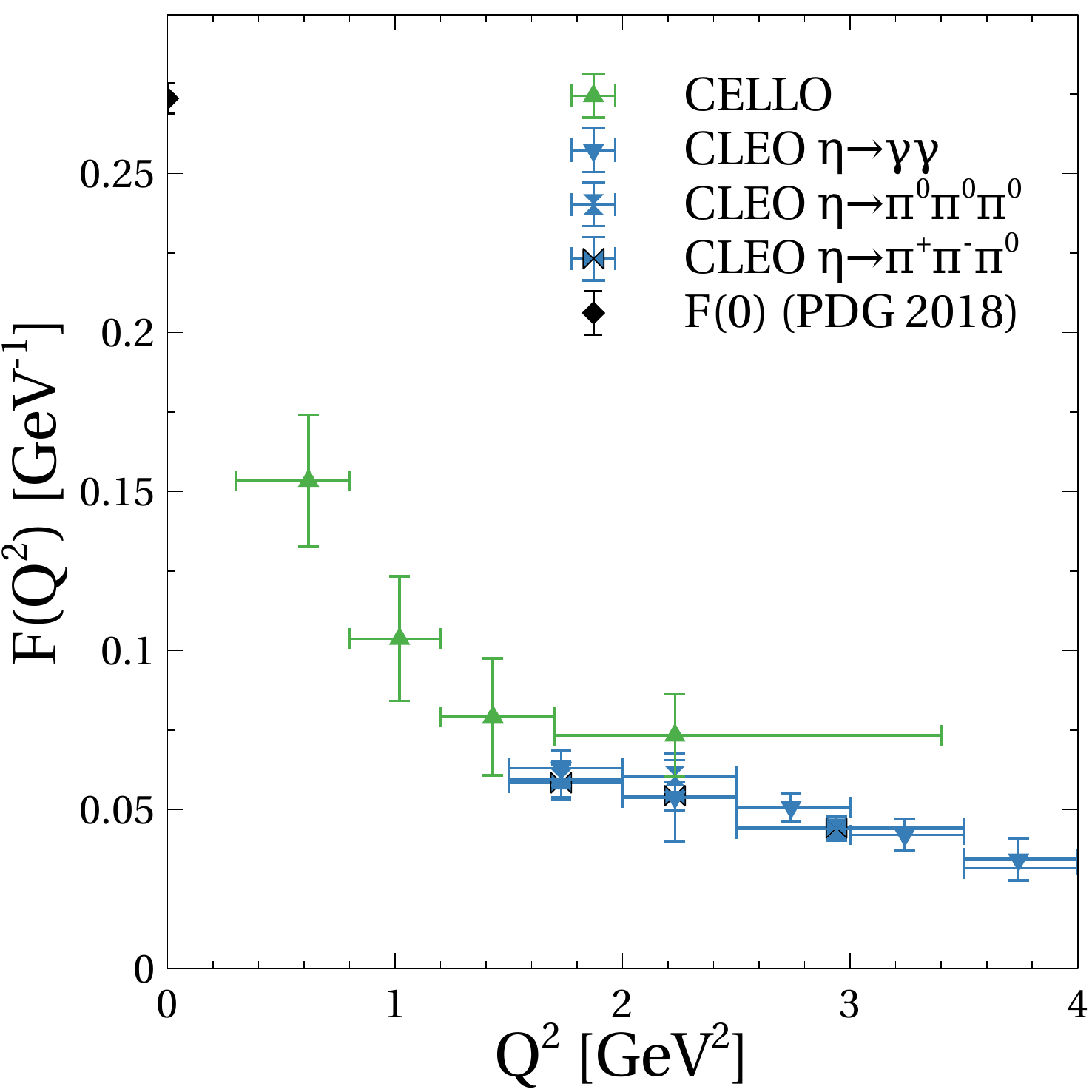}}
  \caption{Momentum dependence of the spacelike TFF of $\eta$. Data from CELLO~\cite{Behrend:1990sr} (green) and CLEO~\cite{Gronberg:1997fj} (blue) are shown. The value $F_{\eta}(0,0)$ is taken from PDG~\cite{Tanabashi:2018oca}. Error bars indicate total uncertainties.}
  \label{fig:slfeta}  
 \end{minipage}\hfill%
 \begin{minipage}{0.48\textwidth}
  \centerline{\includegraphics[width=\textwidth]{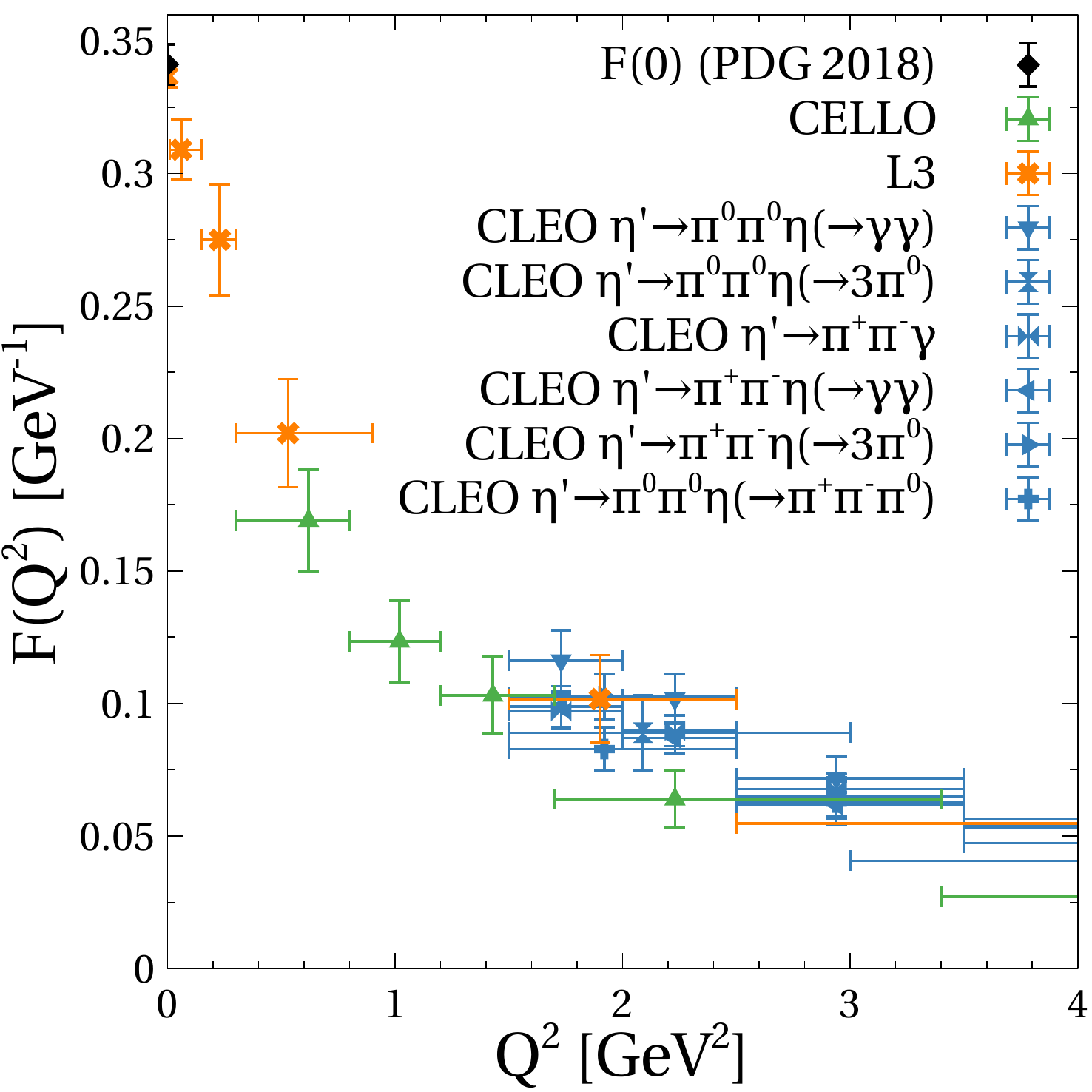}}
  \caption{Momentum dependence of the spacelike TFF of $\eta^\prime$. Data from CELLO~\cite{Behrend:1990sr} (green), L3~\cite{Acciarri:1997yx} (orange), and  CLEO~\cite{Gronberg:1997fj} (blue) are shown. The value $F_{\eta^\prime}(0,0)$ is taken from PDG~\cite{Tanabashi:2018oca}. Error bars indicate total uncertainties.}
  \label{fig:slfetap}
 \end{minipage}
\end{figure}

The TFFs of $\eta$ and $\eta^\prime$ have been studied by the CELLO and CLEO Collaborations in the single-tag technique using the same data discussed in Sec.\ref{sec:slpi0}. In contrast to the investigations of the $\pi^0$ TFF several decay modes were considered to tag the meson production. The CELLO Collaboration provides information on the momentum dependence of the $\eta$ TFF for $0.3\leq Q^2\,[\textrm{GeV}^2]\leq3.4$, by combining the three decay modes $\eta\to\gamma\gamma, \eta\to\pi^+\pi^-\pi^0$ and $\eta\to\pi^+\pi^-\gamma$~\cite{Behrend:1990sr}. The combined results are shown with solid green triangles in \Figref{q2feta} and \Figref{slfeta}. In contrast, the CLEO Collaboration published the momentum dependence of the TFF of $\eta$ separately for each decay channel. Instead of the radiative decay $\eta\to\pi^+\pi^-\gamma$ the more abundant decay into three neutral pions was considered~\cite{Gronberg:1997fj}. The intervals of momentum transfer differ between the individual decay modes. All results are illustrated with triangular blue solid symbols in \Figref{q2feta} and \Figref{slfeta}, covering momentum transfer from $1.5\,\textrm{GeV}^2$ to $9.0\,\textrm{GeV}^2$, while for $\eta\to\pi^+\pi^-\pi^0$ information is 
provided even up to $20\,\textrm{GeV}^2$.

The BaBar Collaboration considered only the charged decay into three pions, which was studied based on $469\,\textrm{fb}^{-1}$ taken at and close to the $\Upsilon(4S)$ resonance~\cite{BABAR:2011ad}.The results, shown with black solid squares in \Figref{q2feta} cover the momentum range $3\leq Q^2\,[\textrm{GeV}^2]\leq40$. In contrast to the measurement of the $\pi^0$ TFF, a deviation from the expected asymptotic behavior was not observed.

\begin{figure}
    \centering
    \includegraphics[width=\textwidth]{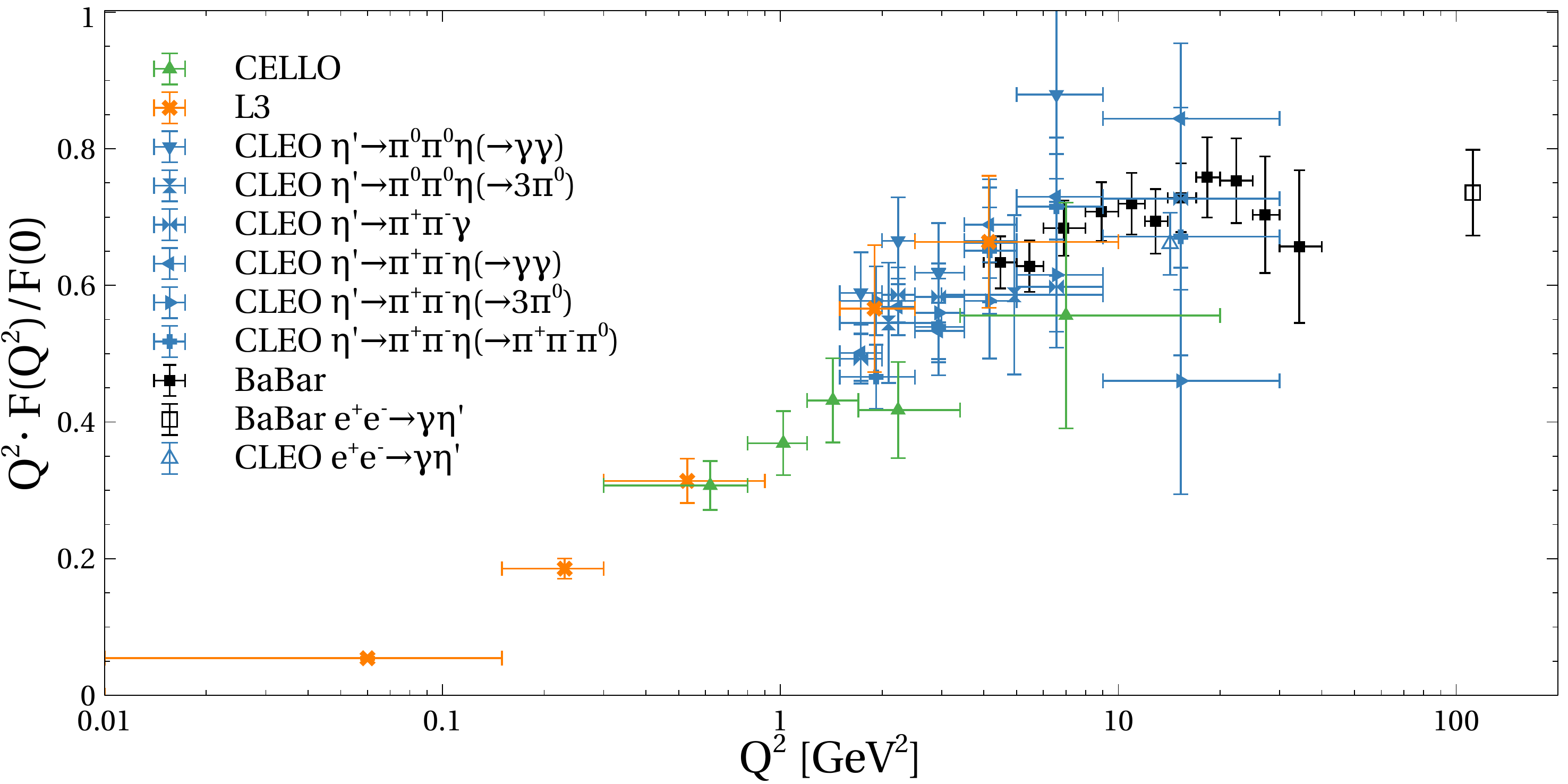}
    \caption{\label{fig:q2fetap}(Color online) Momentum dependence of the spacelike TFF of $\eta^\prime$. The TFF is normalized to its value at $F_{\eta^\prime}(0,0)$ and multiplied with $Q^2$. Data are from CELLO~\cite{Behrend:1990sr} (green), L3~\cite{Acciarri:1997yx} (orange), CLEO~\cite{Gronberg:1997fj} (blue), and BaBar~\cite{BABAR:2011ad} (black). Open symbols show timelike data of CLEO~\cite{Pedlar:2009aa} and BaBar~\cite{Aubert:2006cy}. The error bars indicate the total uncertainties.}
\end{figure}

Analyzing the same data samples, these three collaborations also provided information on the $Q^2$ dependence of the TFF of $\eta^\prime$. The CELLO Collaboration combined the results of the two most abundant decay modes $\eta^\prime\to\pi^+\pi^-\gamma$ and $\eta^\prime\to\pi^+\pi^-\eta$, where the same subsequent decay modes of the $\eta$ meson were considered as in the study of the $\eta$ TFF mentioned above~\cite{Gronberg:1997fj}. In this way, information on the TFF is provided for momentum transfers from $Q^2\leq0.3\,\textrm{GeV}^2$ to $Q^2\geq20\,\textrm{GeV}^2$, as illustrated with green triangles in \Figref{q2fetap} and \Figref{slfetap}. The CLEO Collaboration also published their results on the momentum dependence from $Q^2 \geq 1.5\,\textrm{GeV}^2$ to $Q^2 \leq 30\,\textrm{GeV}^2$. The data are provided separately for each of the six investigated final states in six intervals of momentum transfer, as illustrated with solid blue triangular symbols in \Figref{q2fetap} and \Figref{slfetap}. The BaBar Collaboration investigated the $\eta^\prime$ TFF only in the decay mode $\eta^\prime\to\pi^+\pi^-\eta$ with the subsequent decay $\eta\to\gamma\gamma$. The results, shown as solid black squares in \Figref{q2fetap} cover momentum transfers up to $Q^2\leq40\,\textrm{GeV}^2$ and provide an increased accuracy compared to previous measurements.

An additional measurement of the spacelike $\eta^\prime$ TFF was provided by the L3 Collaboration at the Large Electron Positron collider at CERN~\cite{Acciarri:1997yx}. Using $100\,\textrm{pb}^{-1}$ taken at $\sqrt{s}\simeq91\,\textrm{GeV}$ $F_{\eta^\prime}$ is determined at momentum transfers up to $Q^2=10\,\textrm{GeV}^2$. The results are shown in \Figref{q2fetap} and \Figref{slfetap} with orange crosses. In addition to the regular single-tag technique, another analysis strategy is applied to determine the TFF at the lowest values of $Q^2$. Exploiting the direct proportionality of the squared transverse momentum of the produced meson $p_t^2(\eta^\prime)$ and the squared total momentum transfer $Q^2_{\rm tot} = Q_1^2+Q_2^2$ in an event, as suggested by Monte Carlo simulations, another $129\,\textrm{pb}^{-1}$ of data containing events with untagged leptons could be included in the investigation. In this way the TFF of $\eta^\prime$ was investigated down to $Q^2\geq0.01$ using the most prominent decay mode $\eta^\prime\to\pi^+\pi^-\gamma$.

\subsubsection{Information on double virtual TFFs $\gamma^\ast \gamma^\ast \to \pi^0, \eta, \eta^\prime$}

The experimental determination of pseudoscalar TFFs in two-photon scattering at $e^+e^-$ colliders for arbitrary momentum transfers $F_M(-Q_1^2,-Q_2^2)$ is challenging. The rapid decrease of the differential cross section $\frac{\textrm{d}\sigma_{ee}}{\textrm{d}Q_1^2\textrm{d}Q_2^2}$ calls on the one hand for special detectors and on the other hand for data samples with large integrated luminosities.

Recently, the BaBar Collaboration published the first measurement of the double virtual TFF $F_{\eta^\prime}(-Q_1^2,-Q_2^2)$~\cite{BaBar:2018zpn}. Using a total integrated luminosity of $468.6\,\textrm{fb}^{-1}$, which corresponds to the complete BaBar data set taken at the peak of the $\Upsilon(4S)$ resonance, including a $43.9\,\textrm{fb}^{-1}$ sample taken $40\,\textrm{MeV}$ below the peak. The meson production was investigated through the decay chain $\eta^\prime\to\pi^+\pi^-\eta$ with the subsequent decay $\eta\to\gamma\gamma$, which is also exploited in the single tagged measurements of the $\eta^\prime$ TFF of the BaBar Collaboration. Applying conditions based on energy and momentum conservation and exploiting specific correlations due to the reaction kinematics $46.2^{+8.3}_{-7.0}$ signal events are reported. The TFF is determined in seven intervals of $(Q_1^2,Q_2^2)$. Three intervals are placed along the diagonal $Q_1^2=Q_2^2$: $2<Q_{1,2}^2<10$, $10<Q_{1,2}^2<30$, and $30<Q_{1,2}^2<60$. Another four intervals are placed for asymmetrical kinematic conditions between both virtualities at $10<Q_{1,2}^2<30; 2<Q_{2,1}^2<10$ and $30<Q_{1,2}^2<60; 2<Q_{2,1}^2<30$. In these intervals the TFF is reported to be in agreement with LO and NLO pQCD predictions, as will be discussed in more detail in Section~\ref{sec:phenoparam}. However, the prediction of the VMD model significantly underestimates the data.

A similar measurement is currently performed by the BESIII Collaboration~\cite{Redmer:2018gah}. Based on more than $10\,\textrm{fb}^{-1}$ of collision data collected at and above the peak of the $\Psi(3770)$ resonance, the double virtual TFF of $\pi^0, \eta$, and $\eta^\prime$ are under investigation. It is expected to provide information at $Q_1^2 \approx Q_2^2 \simeq \mathcal{O}(1\,\textrm{GeV}^2)$, which is of special interest for the calculations of the hadronic light-by-light contribution for $a_\mu$.

\subsection{Experimental situation on timelike $\pi^0, \eta, \eta^\prime$ TFFs}
Timelike TFFs of pseudoscalar mesons $M$ are investigated at $e^+e^-$ collider experiments by measuring the cross section of their radiative production in $e^+e^-\to M\gamma$. The cross section of the annihilation reaction is directly proportional to the squared, singly-virtual TFF: $\sigma_{M\gamma}(q^2) = \frac{2\pi^2\alpha^3}{3}|F_{M\gamma\gamma^*}(q^2)|^2$. The virtuality is defined by $\sqrt{s}$ of the collider. Thus, investigations of the momentum dependence call for energy scan experiments. In principle, all momentum transfers $q^2>m_M^2$ are accessible in this approach.

\begin{figure}
 \begin{minipage}{0.48\textwidth}
  \centerline{\includegraphics[width=\textwidth]{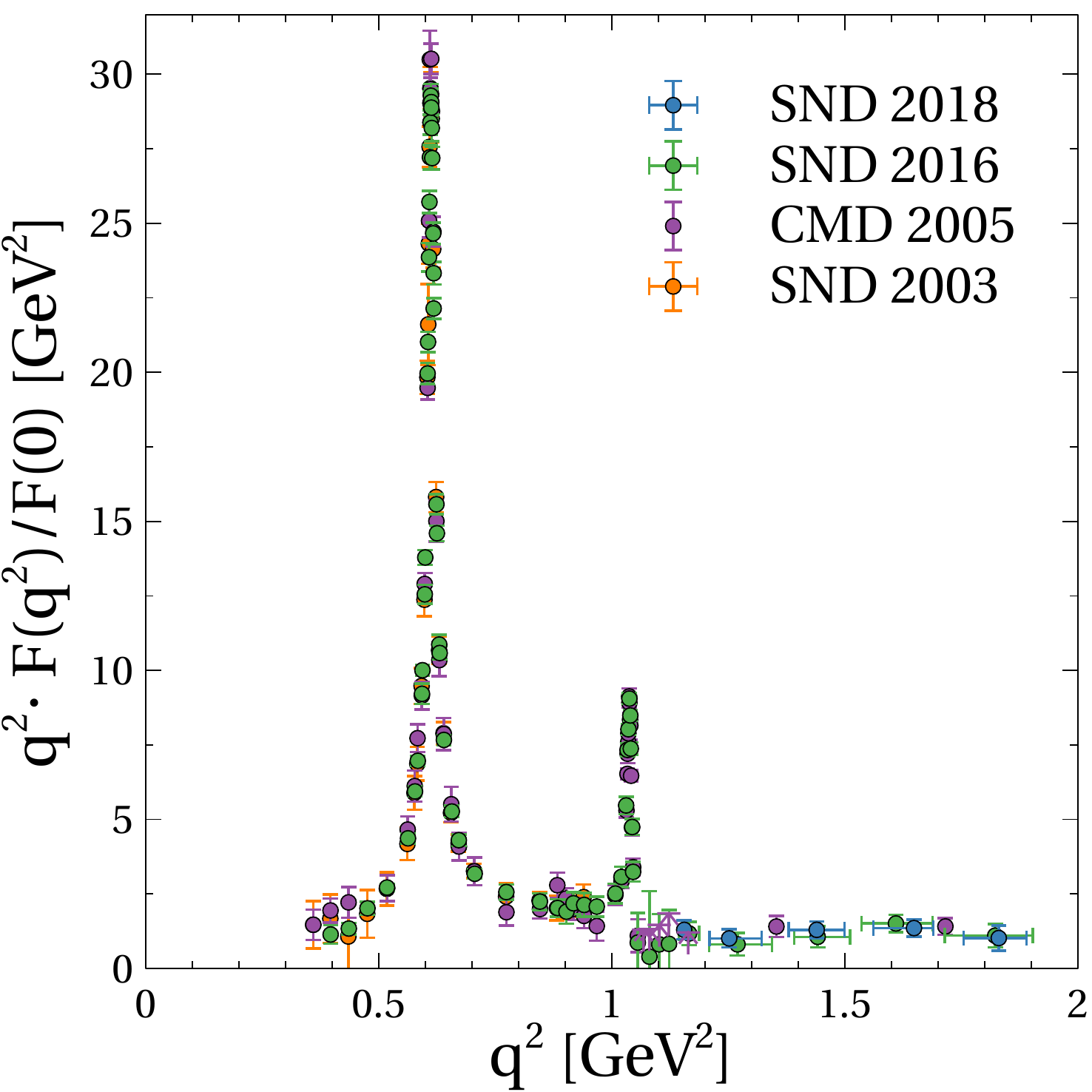}}
  \caption{Momentum dependence of the timelike TFF of $\pi^0$. Data from CMD-2~\cite{Akhmetshin:2004gw} and SND~\cite{Achasov:2003ed,Achasov:2016bfr,Achasov:2018ujw}. Arrows indicate upper limits.}
  \label{fig:tlq2fpi0}  
 \end{minipage}\hfill%
 \begin{minipage}{0.48\textwidth}
  \centerline{\includegraphics[width=\textwidth]{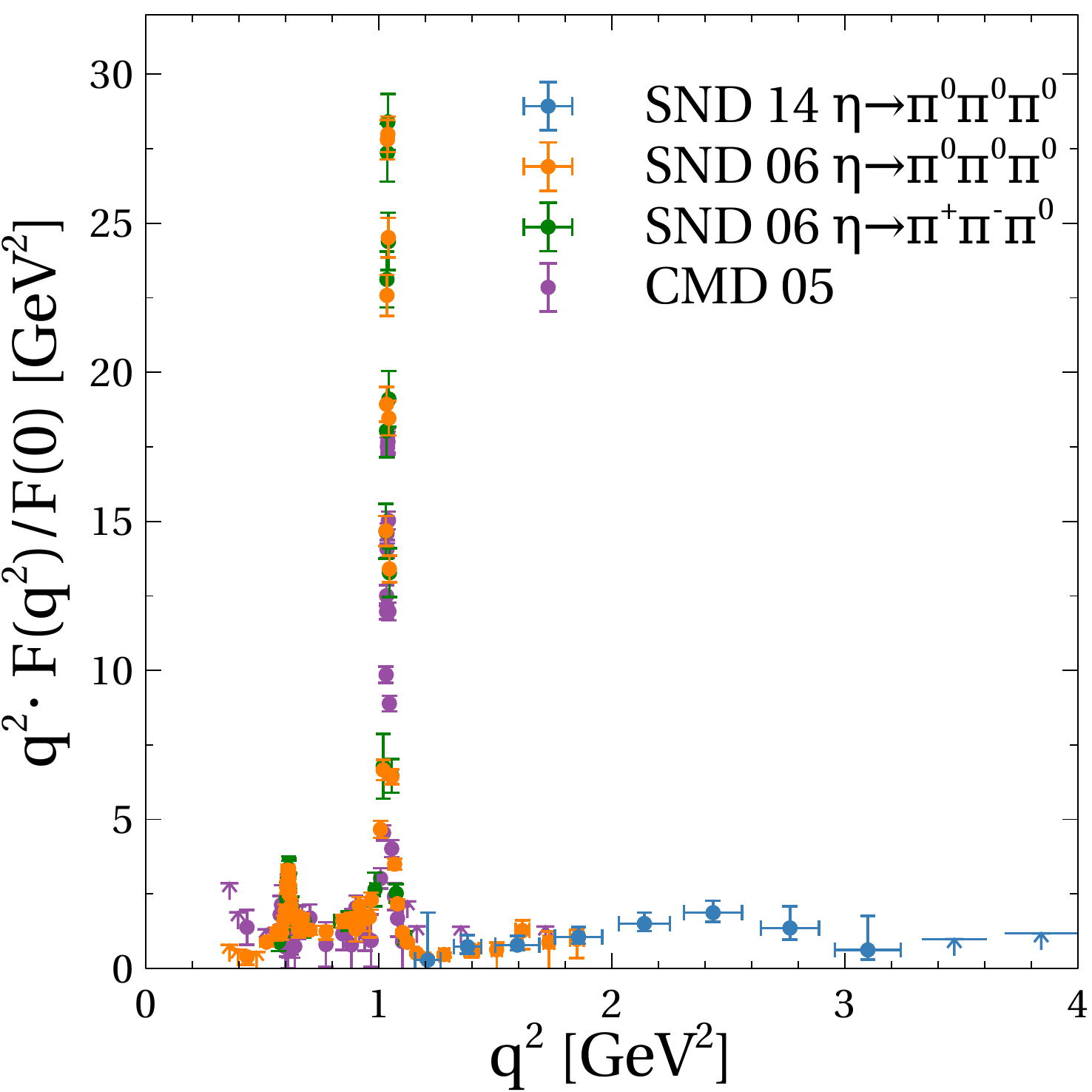}}
  \caption{Momentum dependence of the timelike TFF of $\eta$. Data from 
SND~\cite{Achasov:2006dv, Achasov:2013eli} and CMD-2~\cite{Akhmetshin:2004gw}.}
  \label{fig:tlq2feta}
 \end{minipage}
\end{figure}

The measurement of the radiative production of $\pi^0$ is experimentally challenging. Detecting three photons in the final state has irreducible background from the pure QED annihilation process $e^+e^-\to\gamma\gamma(\gamma_{\rm ISR})$. The only results published so far come from the CMD and SND experiments in Novosibirsk in Russia. The experiments have a long tradition in performing energy scans with $\sqrt{s}$ ranging from the hadron production threshold up to $2\,\textrm{GeV}$. Measurements have been performed over recent years exploiting the various stages of expansion of the storage rings as well as the detectors. In \Figref{tlq2fpi0} the most recent data are presented, taken with the CMD-2 detector at VEPP-2M for $0.6\leq\sqrt{s}\,[\textrm{GeV}]\leq1.38$~\cite{Akhmetshin:2004gw} and the SND detector at VEPP-2M for $0.6\leq\sqrt{s}\,[\textrm{GeV}]\leq1.4$~\cite{Achasov:2003ed,Achasov:2016bfr}. More recently 
the investigation were extended up to $\sqrt{s}=2\,\textrm{GeV}$ using the VEPP-2000 rings, however, signal events were not observed for $\sqrt{s}>1.4\,\textrm{GeV}$~\cite{Achasov:2018ujw}. The narrow peaks of the $\omega$ and $\phi$ resonances dominate the distribution of the TFF.

Data above $\sqrt{s}=2\,\textrm{GeV}$ has not been published, yet. The BESIII Collaboration started to investigate the timelike TFF of $\pi^0$ at $3.773\leq\sqrt{s}\,[\textrm{GeV}]\leq4.36$. Combining data set with at total integrated luminosity of $9\,\textrm{fb}^{-1}$, a result of $F_{\pi^0\gamma\gamma^*}(q^2)$ can be expected at $q^2=16.49_{-2.25}^{+2.51}\,\textrm{GeV}^2$ with a statistical precision in the order of 15\%~\cite{MScThLenz18,Redmer:2018nxj}.

Also the radiative production of $\eta$ mesons has been studied by the CMD and SND experiments in Novosibirsk. At the CMD experiment at VEPP-2M, the investigation has been performed in parallel to the measurement of the  $\pi^0$ discussed above exploiting the decay into two photons~\cite{Akhmetshin:2004gw}. Meson production is separated from QED background using the Dalitz plot of the three photons. The resulting TFF is shown in \Figref{tlq2feta} with purple symbols, covering momentum transfers of $0.6\leq\sqrt{s}\,[\textrm{GeV}]\leq1.38$. Towards the edges of this interval, only upper limits of the cross section are determined. In the same energy region the radiative production of $\eta$ mesons was also studied by the SND experiment at VEPP-2M. Using $27.8\,\textrm{pb}^{-1}$, the mesons are tagged by their decay into three pions, both charged and neutral. The resulting cross sections are illustrated in \Figref{tlq2feta} by orange and green symbols. The energy range was extended up to $q^2=4\,\text{GeV}^2$ by the more recent measurement of SND at the VEPP-2000 machine, which allowed to measure the production of $\eta$ mesons in the neutral three pion decay mode. 

At higher values of momentum transfer, data is available from the CLEO and the BaBar Collaborations. In the investigation of radiative transitions of $\psi(3770)$, the CLEO Collaboration also estimated the cross section of the continuum contribution $e^+e^- \to \gamma \eta^{(\prime)}$~\cite{Pedlar:2009aa}. In a data set of $814\,\textrm{pb}^{-1}$ acquired at $\sqrt{s}=3.773\,\textrm{GeV}$, $\eta$ mesons are reconstructed from both, charged and neutral three pion decay modes, while the $\eta^\prime$ is studied in $\eta^\prime\to\pi^+\pi^-\eta$ with the subsequent decay of $\eta$ to two photons or three pions. Extrapolating the $q^2$ behavior from the timelike TFF  measurement of BaBar, and taking into account the spacelike CLEO result for the TFF, the cross sections $\sigma_{\eta\gamma} = 0.19\pm0.07\,\textrm{pb}$ and $\sigma_{\eta^\prime\gamma} = 0.25\pm0.05\,\textrm{pb}$ are determined. The corresponding timelike TFFs are shown with open blue triangles in \Figref{q2feta} and \Figref{q2fetap}.

The measurement of the BaBar Collaboration has been performed on the $\Upsilon(4S)$ peak, corresponding to a momentum transfer of $q^2=112\,\textrm{GeV}^2$~\cite{Aubert:2006cy}. Analyzing a data set of $232\,\textrm{fb}^{-1}$, the common final state $\pi^+\pi^-\gamma\gamma$ is exploited to determine the cross section for radiative production of $\eta$ and $\eta^\prime$. After carefully evaluating potential background contributions, $20^{+5}_{-6}\,\eta\gamma$ and $50^{+8}_{-7}\,\eta^\prime\gamma$ events are reconstructed, corresponding to the cross sections $\sigma_{\eta\gamma} = (4.5^{+1.2}_{-1.1}\pm0.3)\,\textrm{fb}$ and $\sigma_{\eta^\prime\gamma} = 5.4\pm0.8\pm0.3\,\textrm{fb}$. The timelike TFFs are calculated from the dressed cross section and displayed with open black squares in \Figref{q2feta} and \Figref{q2fetap}.

The large momentum transfer of the data from CLEO and BaBar allows to test the pQCD prediction of timelike and spacelike  TFF running towards the same values, the Brodsky-Lepage-limit. In \Figref{q2feta} and \Figref{q2fetap} the CLEO results are illustrated with open blue triangles and the BaBar results are shown as open black squares. For both, the $\eta$ as well as the $\eta^\prime$ mesons, the results of the timelike TFF either agree well with the spacelike measurement or are compatible with the extrapolation of the spacelike data to larger values of momentum transfer.

Measurements similar to those of CLEO and BaBar are currently performed at BESIII. Radiative transitions of charmonium resonances have been studied and the continuum contributions are found to be negligible~\cite{Ablikim:2017vhb}. First studies of the continuum region are being performed. The performance of the BEPCII accelerator allows to study the timelike TFF form momentum transfers $4\,\textrm{GeV}^2\leq q^2 \leq 21.16\,\textrm{GeV}^2$ at BESIII~\cite{Redmer:2018nxj}. A region, which is of interest for testing the validity of pQCD prediction, and to shed more light on the BaBar-Belle puzzle.

\subsection{Dalitz decays of $\pi^0, \eta, \eta^\prime$}\label{sec:dalitz}
Decays of pseudoscalar mesons $M$ into one real and one virtual photon, which in turn decays into a lepton pair $l^+l^-$, are referred to as Dalitz decays. Since the mesons are not pointlike objects, the decay rate depends on the squared timelike TFF. The mass of the lepton pair $m_{l^+l^-}$ corresponds to the momentum transfer of the virtual photon. Thus, Dalitz decays provide a unique tool to study the TFF $F_M(q_1^2,q_2^2)$ in the momentum range $m^2_{l^+l^-} < q_1^2 < m_{M}^2, q_2^2=0$. Depending on the mass $m_{M}$ of the decaying meson, the leptons $l$ can be either $e^\pm$ or $\mu^\pm$. The experimental results are usually presented in the form of a slope parameter $\Lambda^2$ at $q^2 = 0$, extracted by a fit of the mass distribution with a single-pole approximation. In the VMD picture, the slope parameter corresponds to the effective mass of a virtual vector meson mediating the interaction.

\begin{figure}[t]
 \begin{minipage}{0.49\linewidth}
  \includegraphics[width=\textwidth]{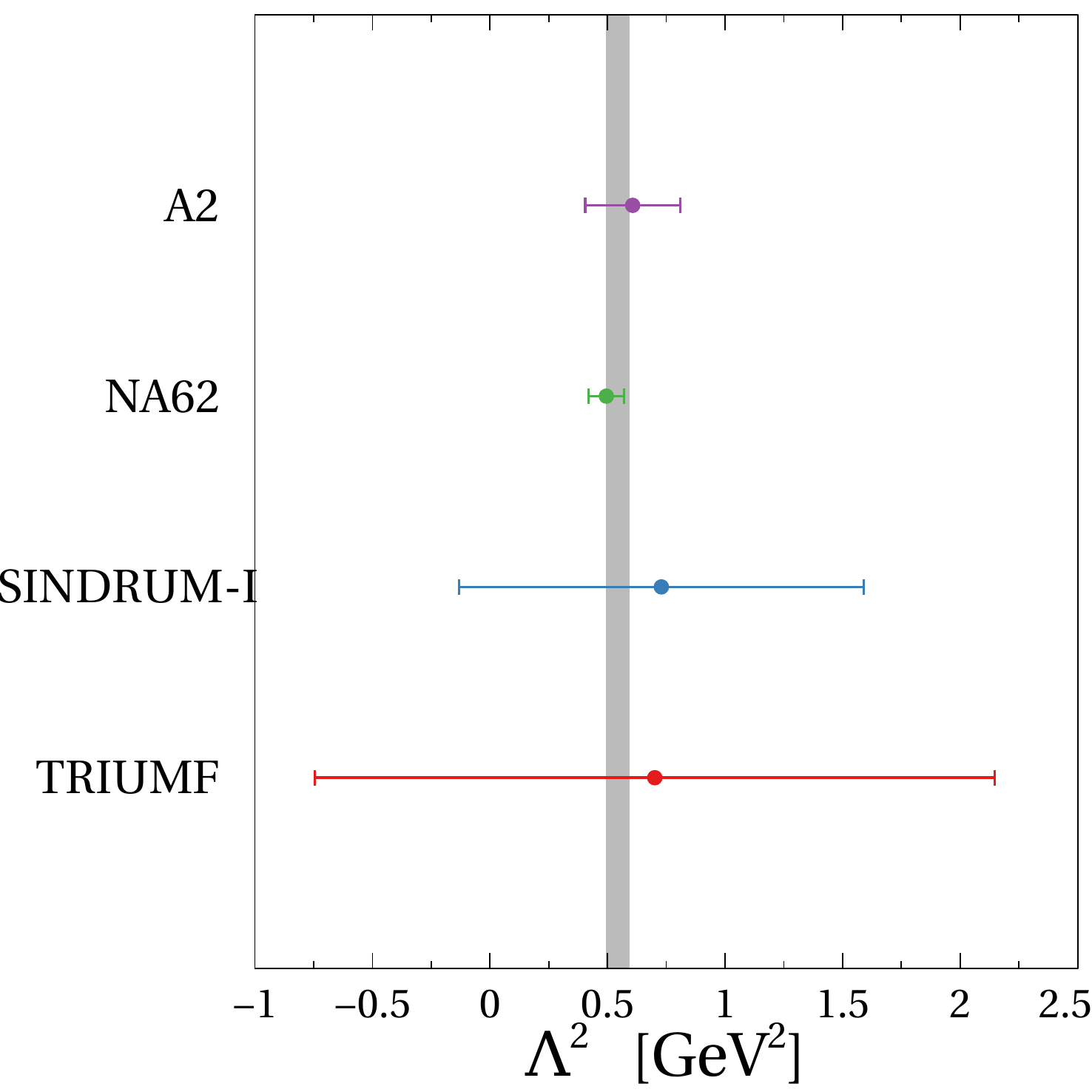}
  \caption{\label{fig:pi0slope}Slope parameter of the timelike $\pi^0$ TFF from Dalitz  decays~\cite{Farzanpay:1992pz,MeijerDrees:1992qb,Adlarson:2016ykr,TheNA62:2016fhr}. The gray band shows the current average value and its uncertainty listed by the PDG~\cite{Tanabashi:2018oca}.}
 \end{minipage}\hfill%
 \begin{minipage}{0.49\linewidth}
  \includegraphics[width=\textwidth]{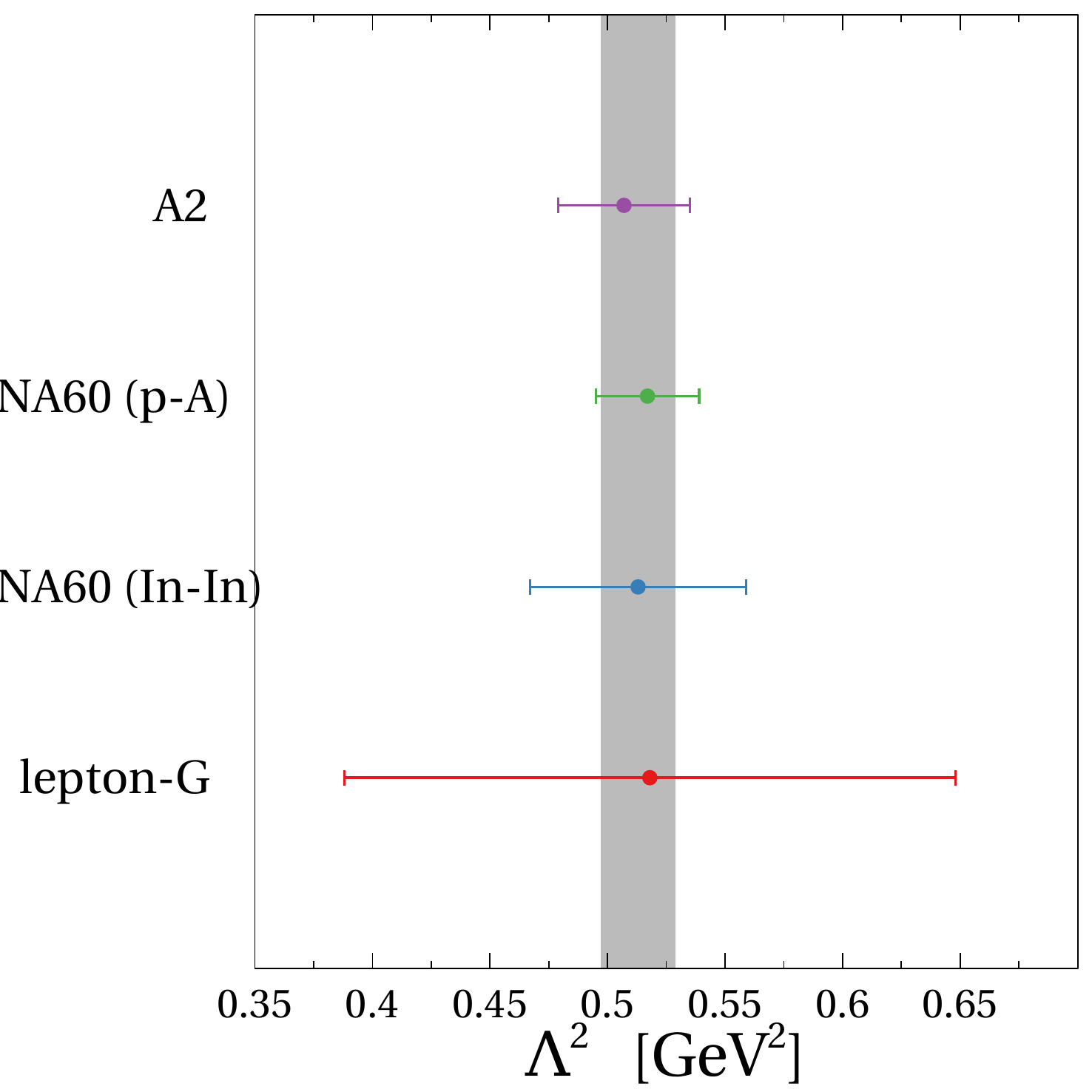}
  \caption{\label{fig:etaslope}Slope parameter of the timelike $\eta$ transition form factor from Dalitz decays~\cite{Dzhelyadin:1980kh,Arnaldi:2009aa,Arnaldi:2016pzu,Adlarson:2016hpp}. The gray band shows the current average value and its uncertainty listed by the PDG~\cite{Tanabashi:2018oca}.}
 \end{minipage}
\end{figure}

Due to its small rest mass, the Dalitz decay of the $\pi^0$ can only proceed through $e^+e^-$ pairs. For more than twenty years, the most precise measurements of the slope parameter were provided by the pion induced experiments at TRIUMF~\cite{Farzanpay:1992pz} and the SINDRUM-I Collaboration at PSI~\cite{MeijerDrees:1992qb}. The former used a non-magnetic setup of three telescopes consisting of tracking detectors and calorimeters, the latter performed their measurement with a magnetic spectrometer in cylindrical geometry composed of tracking chambers and plastic scintillators. As illustrated in \Figref{pi0slope}, both measurements yield rather large uncertainties. The PDG included in their average also the slope parameter determined from the measurement of the spacelike TFF by the CELLO Collaboration~\cite{Behrend:1990sr} discussed in Sec.~\ref{sec:slpi0}. It should be noted that this slope parameter is obtained by an extrapolation of the measurement at $0.3\leq Q^2\,[\textrm{GeV}^2]\leq2.7$ to $Q^2=0\,\textrm{GeV}^2$ based on a VMD model. The larger values of momentum transfer, compared to the Dalitz decay, from which the slope parameter was extrapolated gives room for model dependent effects.

Recently, two high precision measurements of the $\pi^0$ Dalitz decay and the slope parameter of the TFF have been published. The A2 Collaboration used the non-magnetic Crystal Ball and TAPS spectrometer setup at the tagged photon beam facility of the Mainz Microtron MAMI to produce pions off protons~\cite{Adlarson:2016ykr}. Photo-production offers a large cross section close to the $\Delta(1232)$ without any background from other physics processes. The published value of the slope of the TFF of $\Lambda_{\pi^0}^2 = 0.61\pm0.20\,\textrm{GeV}^2$, where the uncertainty is the total uncertainty, has been determined from $4\cdot10^5$ reconstructed Dalitz decay events. A new measurement has already been announced by the A2 Collaboration, aiming at doubling the statistical accuracy of the recent result of the NA62 Collaboration~\cite{TheNA62:2016fhr}.

The NA62 result has been obtained from $1.11\cdot10^6$ reconstructed Dalitz decays of the $\pi^0$, yielding a slope parameter value of $\Lambda^2_{\pi^0}=0.495\pm0.076\,\textrm{GeV}^2$. The pions were produced in the Kaon decays $K^\pm\to\pi^\pm\pi^0$, which were observed from secondary beams with a central momentum of $74\,\textrm{GeV}/c$ at the modified NA48 beam line at CERN. The momenta of the charged decay products were measured in a magnetic dipole spectrometer using drift chambers, while a LKr calorimeter was used to determine the energies of photons.

Both, the result of the A2 Collaboration as well as of the NA62 Collaboration consider radiative corrections according to Ref.~\cite{Husek:2015sma}. In contrast to previous calculations, the one-photon irreducible contribution at one-loop level, and the virtual muon loop contribution are included. Also, the terms of order higher than $\mathcal{O}(m^2)$ are taken into account.

The Dalitz decay of the $\eta$ meson can proceed through $e^+e^-$ as well as $\mu^+\mu^-$ pairs. One of the first measurements of the TFF was carried out with the \mbox{lepton-G} setup at the Institute for High Energy Physics in Serpukhov, Russia~\cite{Dzhelyadin:1980kh}. A $33\,\textrm{GeV}/c$ secondary pion beam impinging on a liquid hydrogen target was used to produce the mesons. Charged decay products are measured in a magnetic spectrometer, while neutral particles are registered in a lead glass calorimeter. The slope parameter of the $\eta$ TFF is determined as $\Lambda^2_\eta = (0.52\pm0.13)\,\textrm{GeV}^2$ based on 600 reconstructed Dalitz decays $\eta\to\mu^+\mu^-$.

More recently, the NA60 Collaboration also determined the TFF slope parameter using $\eta$ Dalitz decay with muon pairs. A first measurement was performed in peripheral In-In collisions at 158A\,GeV~\cite{Arnaldi:2009aa}. The muons are detected in a spectrometer consisting of tracking stations in a toroidal magnetic field, which is placed behind a hadron absorber. After subtracting combinatorial background as well as the contribution of narrow resonances decaying directly into muon pairs, the inclusive low-mass dimuon spectrum is fitted with the sum of the expected contributions. The yield and the slope parameter of the $\eta$ Dalitz decay are floated in the fit. 9000 signal event candidates are identified, resulting in the value of the slope parameter $\Lambda^2_\eta = (0.513\pm0.047_{\rm tot})\,\textrm{GeV}^{2}$.

The measurement has been repeated using a high statistics data set of $p-A$ collisions, where a 400\,GeV proton beam from the CERN SPS is impinging on an arrangement of Be, Cu, In, W, Pb, and U targets~\cite{Arnaldi:2016pzu}. From an approximately ten times larger data sample compared to the peripheral In-In collisions, the slope parameter of the $\eta$ TFF is extracted as $\Lambda^2_\eta = (0.517\pm0.022_{\rm tot})\,\textrm{GeV}^{2}$ by fitting the background subtracted, inclusive dimuon mass spectrum. As illustrated in \Figref{etaslope} it is the most precise measurement of the slope parameter published so far.

In contrast to the previously mentioned investigations, the A2 Collaboration measured the $\eta$ Dalitz decay into $e^+e^-$ pairs. Thus, the TFF can also be measured in the momentum range $m_{ee}^2\leq q^2\leq m_{\mu\mu}^2$, which puts additional constraints to fits determining the slope parameter at $q^2=0$. The same setup is used by the A2 Collaboration as discussed for the $\pi^0$ Dalitz decay, which allows for the exclusive reconstruction of the final state. The meson are produced in the photon induced reaction $\gamma p\to\eta p$. Several measurements have been performed determining the TFF slope parameter with increasing accuracy~\cite{Berghauser:2011zz,Aguar-Bartolome:2013vpw}. The latest measurement is based on two rounds of data taking, from which a kinematic fit based analysis determines $5.4\cdot10^4$ Dalitz decay events~\cite{Adlarson:2016hpp}. Studying the background contributions as well as systematic uncertainties for different intervals of dilepton masses allows to present the TFF in 34 bins of $m_{ee}$ from $35\pm5\,\textrm{MeV}$ to $475\pm15\,\textrm{MeV}$. The slope of the TFF is determined by a fit of the distribution as $\Lambda^2_\eta = (0.507\pm0.028_{\rm tot})\,\textrm{GeV}^{2}$.

The first measurement of the Dalitz decay of the $\eta^\prime$ meson in $e^+e^-$ pairs has been reported recently by the BESIII Collaboration~\cite{Ablikim:2015wnx}. The mesons are produced in the radiative decay of the $J/\psi$ resonance. A data set of $1.31\cdot10^9$ inclusive $J/\psi$ decays has been analyzed and $864\pm36$ events of $\eta^\prime\to\gamma e^+e^-$ are observed. Since the pole of the $\rho$ meson resides within the mass range covered by the lepton pairs, the single pole approximation of the TFF takes the form $|F(q^2)|^2 = \frac{\Lambda^2(\Lambda^2+\gamma^2)} {(\Lambda^2-q^2)^2 -\Lambda^2\gamma^2}$, where $\gamma$ can be considered the width of the effective vector meson mediating the interaction. The respective values are determined as $\Lambda_{\eta^\prime}=0.79\pm0.04_{\rm stat}\pm0.02_{\rm syst}$ and $\gamma_{\eta^\prime}=0.13\pm0.06_{\rm stat}\pm0.03_{\rm syst}$, which improves on the values previously determined in the Dalitz decay into muon pairs~\cite{Landsberg:1986fd}. Further measurements of Dalitz decays of the $\eta^\prime$ meson have been announced by the A2~\cite{Steffen:2017jem} and CLAS~\cite{Kunkel:2016upp} Collaborations.

\subsection{Meson decay processes involving two virtual photons}
Pseudoscalar mesons decaying into two virtual photons are referred to as double Dalitz decays. They allow to determine the TFF $F_M(q_1^2,q_2^2)$ as function of both virtualities. The range of virtualities covered is limited by the mass $m_M$ of the decaying meson according to $m_M^2 = (q_1+q_2)^2$, with $q_i^2=m_{ee}^2$. The double Dalitz decay was recently measured for the first time by the KLOE-2 Collaboration using data of the KLOE experiment~\cite{KLOE2:2011aa}. A data set of $1.7\,\textrm{fb}^{-1}$ taken at $\sqrt{s}\simeq1.02\,\textrm{GeV}$ has been analyzed tagging the $\eta$ meson production with the monochromatic photon of $\phi\to\gamma\eta$. Background from external photon conversion is rejected based on the correlation of decay vertex and mass of each lepton pair, resulting in $362\pm29$ signal events, which allow to determine the branching ratio.

Another decay mode of pseudoscalar mesons involving two virtual photons in a loop are decays into lepton pairs. These are very rare processes and good knowledge of the TFF is required for the prediction of the branching ratios. The latest measurement of $\pi^0 \to e^+e^-$ was performed by the KTeV-E779 experiment at Fermilab~\cite{Abouzaid:2006kk}. The pions are tagged in $K_L\to3\pi^0$, where two of the three pions are reconstructed in the two-photon decay mode. The signal mode is separated from the Dalitz decay, which is used for normalization, by requiring $(m_{ee}/m_{\pi^0})^2>0.95$. From the complete data set of E779-II 792 candidate events are obtained with an expected background of $52.7\pm11.2$ events. The resulting branching ratio exceeds the unitary bound by seven standard deviation.

Also Dalitz decay of vector mesons into pseudoscalars and a lepton pair can be related to timelike double virtual TFFs of pseudoscalar mesons. One virtuality is fixed to the mass of the vector meson, which is \emph{e. g.} produced from the virtual photon in an $e^+e^-$ annihilation. The second virtuality is determined by the mass of the lepton pair.

Recently, some attention has been drawn to the Dalitz decay of the $\omega$ meson. The NA60 Collaboration provided a high statistics analysis of $\omega\to\pi^0\mu^+\mu^-$ using the same data set and analysis methods as discussed for the $\eta$ Dalitz decay in section~\ref{sec:dalitz}~\cite{Arnaldi:2009aa}. A, compared to VMD models, unexpectedly steep increase of the TFF towards large values of $q^2$ is observed. The NA60 Collaboration confirmed the result in a second measurement~\cite{Arnaldi:2016pzu}. The A2 Collaboration studied the TFF in $\omega\pi^0 e^+ e^-$~\cite{Adlarson:2016hpp}. They observed a less pronounced increase at large $q^2$, due to limited statistics their result is compatible with the NA60 data within errors.

\begin{figure}[t]
\includegraphics[width=0.5\textwidth]{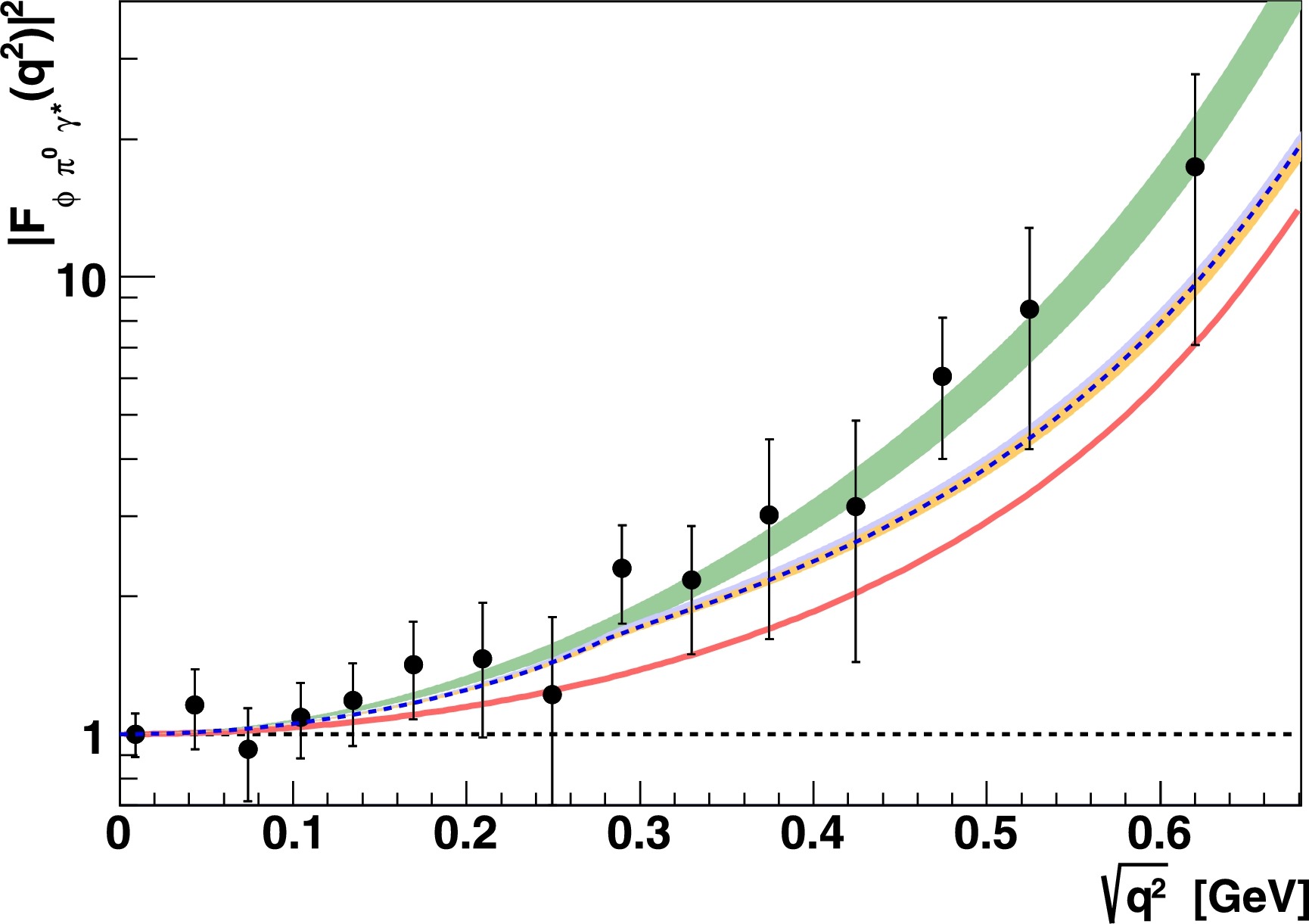}
 \includegraphics[width=0.5\textwidth]{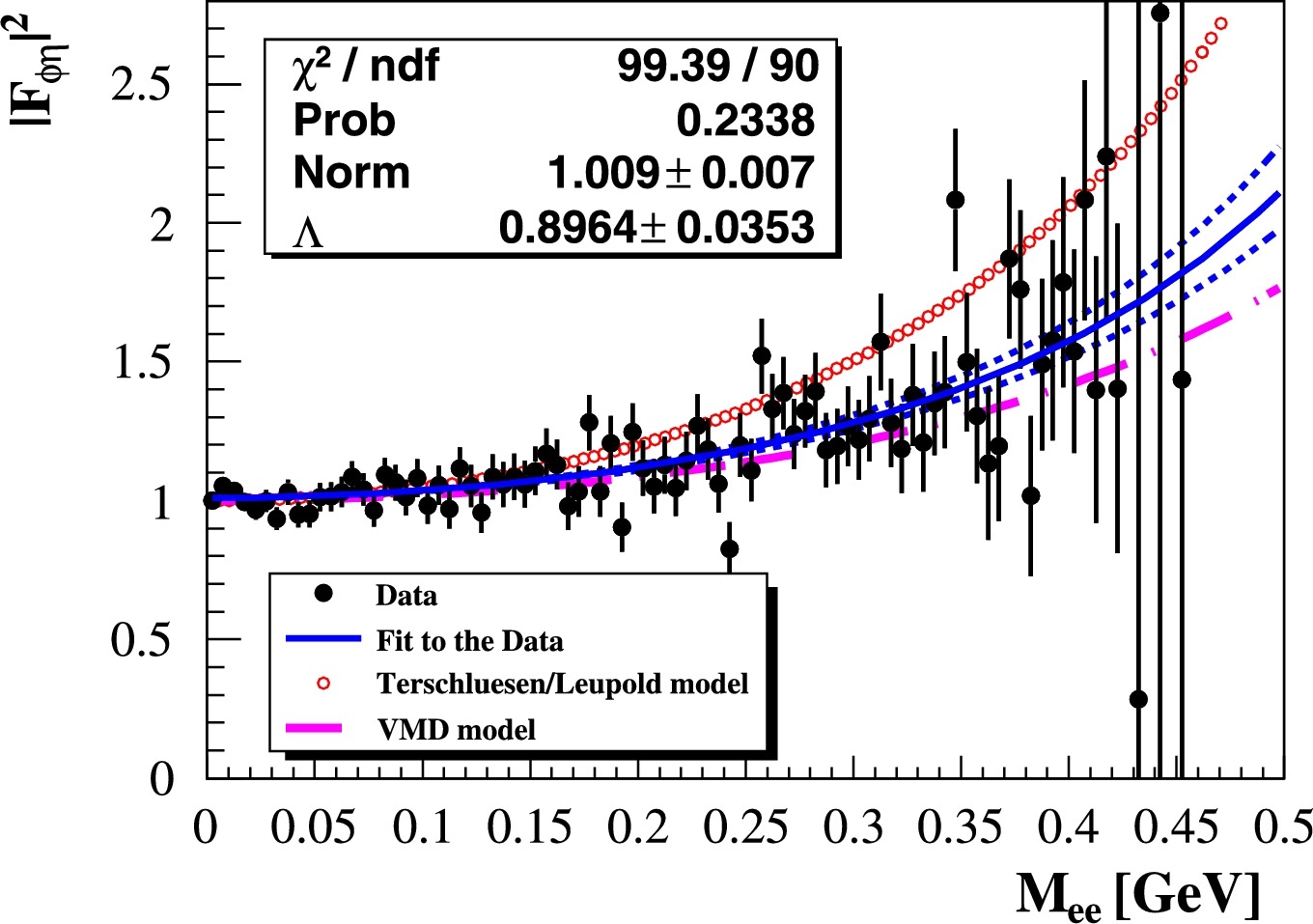}
 \caption{\textbf{Left:} Distribution of $|F_{\phi\pi^0}(m_{e^+e^-})|^2$ measured by KLOE-2. The data are confronted with the dispersive analysis of Ref.~\cite{Schneider:2012ez} (orange lower band and cyan middle band) and Ref.~\cite{Danilkin:2014cra} (blue dashed line), the chiral theory approach of Ref.~\cite{Ivashyn:2011hb} (upper green band), and the one-pole VMD model (solid red line) of Ref.~\cite{Sakurai1969}. The figure is taken from Ref.~\cite{Anastasi:2016qga} (Fig.4).
 \textbf{Right:} Distribution of $|F_{\phi\eta}(m_{e^+e^-})|^2$ measured by KLOE-2. A fit to the data is shown with the solid blue line. The blue dashed lines indicate the change of the result by varying $\Lambda_{\phi\eta}$ within $\pm 1\sigma$. The expectations according to VMD is shown in the pink dash-dotted line, and for Ref.~\cite{Terschluesen:2010ik} with open red circles. The figure is taken from Ref.~\cite{Babusci:2014ldz} (Fig.6).}
 \label{fig:KLOEdalitz}
\end{figure}

The KLOE-2 Collaboration studied the Dalitz decays $\phi\to M e^+e^-$, with $M=\eta$~\cite{Babusci:2014ldz} and more recently also $M=\pi^0$~\cite{Anastasi:2016qga} using the $1.7\,\textrm{fb}^{-1}$ data set of the KLOE experiment at $\sqrt{s}\simeq1.02\,\textrm{GeV}$. Reconstructing the $\eta$ meson in $\eta\to3\pi^0$ yields $29625\pm178$ Dalitz events after background subtraction. The right panel of \Figref{KLOEdalitz} shows the TFF as function of the lepton mass together with a fit yielding a slope $\Lambda^2 = 0.781\pm0,061^{+0.055}_{-0,048}\,\textrm{GeV}^2$, and the prediction of~\cite{Terschluesen:2010ik}, which aimed at providing a better description for the enhancement observed in the TFF of $\omega\to\pi^0 l^+l^-$. The analysis of $\phi\to\pi^0 e^+e^-$ yields about 9500 signal events after subtracting the background contributions from external photon conversion in $\phi\to\pi^0\gamma$ and Bhabha scattering. The left panel of \Figref{KLOEdalitz} shows the distributions of the TFF as function of the dilepton mass. The data are confronted with different predictions, where the best agreement is achieved by an Unconstrained Resonant Chiral Theory approach~\cite{Ivashyn:2011hb}, illustrated with the green band, the parameters of which were fitted to the $\omega\to\pi^0 l^+l^-$ data of the NA60 experiment, discussed above.

\subsection{Theoretical situation on pseudoscalar meson TFFs}
\subsubsection{Chiral anomaly}

It is well known, that closed-loop triangle graphs result in the divergence of the axial vector current, even for massless quarks. This effect is known as Adler-Bell-Jackiw anomaly \cite{Adler:1969gk,Bell:1969ts,Bardeen:1969md} and allows to couple $\pi^0$ to two vector currents. On the level of hadrons, the chiral anomaly can be expressed in the form of the parity-odd Wess-Zumino-Witten (WZW) effective Lagrangian \cite{Wess:1971yu,Witten:1983tw}. At tree level it fixes the normalization of the $\pi^0$ TFF as:
\bea
F_\pi^0(0,0) = \frac{N_c}{12\, \pi^2 f_\pi},
\label{eq:fm0}
\eea
where $f_\pi = 0.0924$~GeV is the pion decay constant and $N_c$ is a number of colors. For $N_c=3$, the obtained value of $F_M(0,0)\simeq 0.274$ GeV$^{-1}$ can be used to calculate the neutral pion two photon decay width. Its value agrees very well with the recent Primakov measurement \cite{Larin:2010kq}.

\subsubsection{Model approaches}
To calculate the pseudoscalar pole contribution to $a_\mu$ requires to account for the internal structure of the mesons through their TFFs. 
The simplest model is to assume vector-meson dominance (VMD) in a factorized form 
\bea
F^{\rm{VMD}}_M(- Q_1^2, - Q_2^2) = F_M(0,0)\, \frac{M_V^4}{(Q_1^2 + M_V^2)(Q_2^2 + M_V^2)},
\label{eq:vmd}
\eea
with $F_M(0,0)$ taken according to Eq.~(\ref{eq:fm0}). We quote the values of the parameters $f_M$ and $M_V$ used in Refs.~\cite{Jegerlehner:2009ry,Nyffeler:2016gnb}, which are used below as one model estimate of the pseudoscalar pole contribution to $a_\mu$:
\bea
\pi^0&:& \quad \quad f_{\pi^0} = 0.0924~\mathrm{GeV}, \quad M_V = 0.7755~\mathrm{GeV}, \nonumber \\
\eta&:& \quad \quad f_{\eta} = 0.093~\mathrm{GeV}, \quad \quad M_V = 0.774~\mathrm{GeV}, \nonumber \\
\eta^\prime&:& \quad \quad f_{\eta^\prime} = 0.074~\mathrm{GeV}, \quad \quad M_V = 0.859~\mathrm{GeV}, 
\label{eq:vmdparam}
\eea
The VMD TFF has the drawback that it falls too fast for $Q_1^2=Q_2^2=Q^2\gg 0$, namely as $\sim 1/Q^4$ whereas  $\sim 1/Q^2$ is expected from the operator product expansion (OPE) \cite{Nesterenko:1982dn,Novikov:1983jt}. 

Another model for the TFF, which was intensively used in the literature \cite{Jegerlehner:2009ry,Nyffeler:2016gnb,Nyffeler:2009tw} is the so called LMD+V (lowest meson dominance + vector) parametrization \cite{Knecht:2001xc}. It incorporates certain short-distance constraints from the operator product expansion and has the following form
\bea
F^{\rm{LMD+V}}_{\pi^0}(- Q_1^2, - Q_2^2) =  \frac{f_\pi}{3} \frac{Q_1^2 Q_2^2 (Q_1^2 + Q_2^2) - h_2 Q_1^2 Q_2^2 + h_5 (Q_1^2 + Q_2^2) - h_7}{(Q_1^2 + M_{V_1}^2)(Q_1^2 + M_{V_2}^2)(Q_2^2 + M_{V_1}^2)(Q_2^2 + M_{V_2}^2)}\,.
\label{eq:lmdv}
\eea
The values of the parameters used for the $\pi^0$ in Refs.~\cite{Jegerlehner:2009ry,Nyffeler:2016gnb} are given by:
\bea
M_{V_1} &=& M_\rho = 0.7755~\mathrm{GeV},   \nonumber \\
M_{V_2} &=& M_{\rho^\prime} = 1.465~\mathrm{GeV},   \nonumber \\
h_2 &=& - 10.63~\mathrm{GeV}^2, \nonumber \\
h_5 &=& (6.93 \pm 0.26)~\mathrm{GeV}^4, \nonumber \\
h_7 &=& - \frac{3 M_{V_1}^4 M_{V_2}^4}{4 \pi^2 f_\pi^2}= - 14.83~\mathrm{GeV}^6, 
\label{eq:lmdvparam}
\eea
Although this form improves on the VMD parameterization to implement the perturbative QCD limit for the double virtual symmetric case 
$Q_1^2=Q_2^2 \gg 0$, for the double virtual asymmetric case ($Q_1^2 \neq Q_2^2$) it does introduce wiggles in the parameterization at large $Q^2$ due to the numerator, as will be shown further on. 

A generalization of the LMD+V form for the TFF was proposed in \cite{Masjuan:2012wy,Escribano:2013kba,Masjuan:2017tvw}. It is based on Canterbury approximants, which was first applied for the singly virtual process \cite{Masjuan:2012wy,Escribano:2013kba} (in the form of  Pad\'e approximants) and later on in \cite{Masjuan:2017tvw} generalized for the doubly virtual process. For the singly virtual process the Pad\'e approximant is defined as
\begin{equation}
F^{\text{Pad\'e}}_{M}(-Q^2)=\frac{Q_N(Q^2)}{R_{N'}(Q^2)}=  F_M(0,0)\left(1-b_M\,Q^2+...+{\cal O}(Q^2)^{N+N'+1}\right),
\end{equation}
where $Q_N(Q^2)$, $R_{N'}(Q^2)$ are polynomials of degree $N$ and $N'$, respectively. From one side, the low energy parameters of the Taylor expansion were determined by fitting to the experimental data. From the other side, the final TFF is reconstructed via the use of Pad\'e approximant, which is constructed in such a way that it has the same Taylor expansion up to order ${\cal O}(Q^2)^{N+N'+1}$ and incorporates the correct high-energy behavior. The convergence is guaranteed by an assumption the TFF is a meromorphic function of Stieltjes type. A comparison between two consecutive elements in the sequence serves as an estimate of the systematic error. The results for TFFs have been obtained in \cite{Masjuan:2012wy,Escribano:2013kba}. The generalization to the doubly virtual process is expressed as
\begin{align}
F^{\text{Pad\'e}}_{M}(-Q_1^2,-Q_2^2)&=\frac{Q_N(Q_1^2,Q_2^2)}{R_{N'}(Q_1^2,Q_2^2)}\\
&=  F_M(0,0)\left(1-b_M\,(Q_1^2+Q_2^2)+a_{M;1,1}\,Q_1^2\,Q_2^2+...\right)\,,
\end{align}
where $Q_N(Q_1^2,Q_2^2)$ and $R_{N'}(Q_1^2,Q_2^2)$ are bi-variate symmetric polynomials. The lowest two approximant reads as
\begin{align}
C_{1}^0(Q_1^2,Q_2^2)&=\frac{F_M(0,0)}{1+b_M\,(Q_1^2+Q_2^2)}\,,\\
C_{2}^1(Q_1^2,Q_2^2)&=\frac{F_M(0,0)(1+\alpha_1(Q_1^2+Q_2^2)+\alpha_{1,1}Q_1^2\,Q_2^2)}{1+\beta_1(Q_1^2+Q_2^2)+\beta_2(Q_1^4+Q_2^4)+\beta_{1,1}Q_1^2Q_2^2+\beta_{2,1}Q_1^2Q_2^2(Q_1^2+Q_2^2)}\,.\nonumber
\end{align}
Since there is no data yet for the double-virtual TFF, not all the parameters could be fixed and one has to assume the generous band for $a_{M;1,1} \in [a_{M;1,1}^{min},\,a_{M;1,1}^{max}] $ which add additional uncertainty in the $a_\mu$ calculation. The range of $a_{M;1,1}$ is taken to be the most physically accessible which does not spoil the TFFs analytic properties. New BaBar data for the double virtual $\eta^\prime$ TFF, as discussed in Section~\ref{sec:phenoparam}, may provide a test for such parameterization.  

In addition, TFFs of the pseudoscalar mesons had been also analyzed using the framework of Dyson-Schwinger \cite{Goecke:2010if,Eichmann:2017wil,Weil:2017knt} and various effective Lagrangian based models \cite{Bijnens:1995xf,deRafael:1993za,Knecht:2001qg,Czyz:2017veo,Guevara:2018rhj}.

\subsubsection{Dispersion theory}

Recently in \cite{Hoferichter:2018dmo,Hoferichter:2018kwz} an updated dispersive framework has been presented that now incorporates the asymptotic behavior expected from perturbative QCD, as discussed in Section~\ref{sec:pqcd}. The framework based on the existing data for $e^+e^-\to 3\pi$ \cite{Achasov:2002ud,Achasov:2003ir,Aubert:2004kj}, $e^+e^-\to e^+e^-\pi^0$ \cite{Achasov:2003ed,Akhmetshin:2004gw,Achasov:2016bfr}, the $\pi^0\to\gamma\gamma$ decay width \cite{Larin:2010kq} and fundamental principles of the quantum field theory, namely unitarity, analyticity and crossing symmetry. At low energies, isospin quantum numbers are key to see which hadronic intermediate states contribute. Isovector and isoscalar photons couples predominantly to two and three pions, respectively. The latter can be approximated by the narrow vector resonances $\omega/\phi$, whose contributions can be related to their transition form factors. The unsubtracted double-spectral representation reads
\begin{align}
\label{low_energy}
&F_{\pi^0}^\text{disp}(-Q_1^2,-Q_2^2)=
\frac{1}{\pi^2} \int_{4M_\pi^2}^{\infty} d x \int_{s_{thr}}^{\infty}d y   \frac{\rho^\text{disp}(x,y)}{\big(x+Q_1^2\big)\big(y+Q_2^2\big)}+(Q_1\leftrightarrow Q_2),\,\nonumber\\
&\rho^\text{disp}(x,y)=\frac{q_\pi^3(x)}{12\pi\sqrt{x}} \text{Im}\Big[\big(F_\pi^{V}(x)\big)^*f_1(x,y)\Big]
,\quad q_{\pi}(s)=\sqrt{\frac{s}{4}-m_\pi^2},
\end{align}
where $F_\pi^{V}(x)$ is the electromagnetic form factor of the pion and $f_1(x,q^2)$ is the partial-wave amplitude for the $\gamma^*(q)\pi\to\pi\pi$ process. The $F_\pi^{V}(x)$ can be well described by an Omn\`es representation~\cite{Omnes:1958hv, Schneider:2012ez} with the input from the $\pi\pi$ $P$-wave phase shift \cite{Colangelo:2001df,Caprini:2011ky,GarciaMartin:2011cn} and the fit to the data \cite{Fujikawa:2008ma} of the unknown inelastic contributions from $\rho'$ and $\rho''$. On the other side, the amplitude $f_1(x,q^2)$ is more complicated and requires the inclusion of the left-hand cuts by solving a set of Khuri-Treiman equations \cite{Khuri:1960zz,Hoferichter:2014vra}. For $q^2=M_\omega^2,\,M_\phi^2$ it probes the corresponding Dalitz plot distributions \cite{Niecknig:2012sj,Schneider:2012ez,Danilkin:2014cra}. In the initial version of (\ref{low_energy}) given in \cite{Hoferichter:2014vra} at least one subtraction was employed in order to reproduce low energy theorem which relates the normalization to the $\pi^0\to 2\gamma$ decay. In this way, the asymptotic behavior was spoiled and formally speaking the TFF was unfavored for an estimate of $a_\mu$. Progress was made by switching to the unsubtracted representation with a cutoff in the integral and restoring the sum rule by adding an effective pole. The part above the cutoff was approximated by the perturbative QCD constraints. 
As a result, the full contribution can be written in the double-spectral representation as

\begin{figure}[t]
\centering
\includegraphics[width=0.47\textwidth]{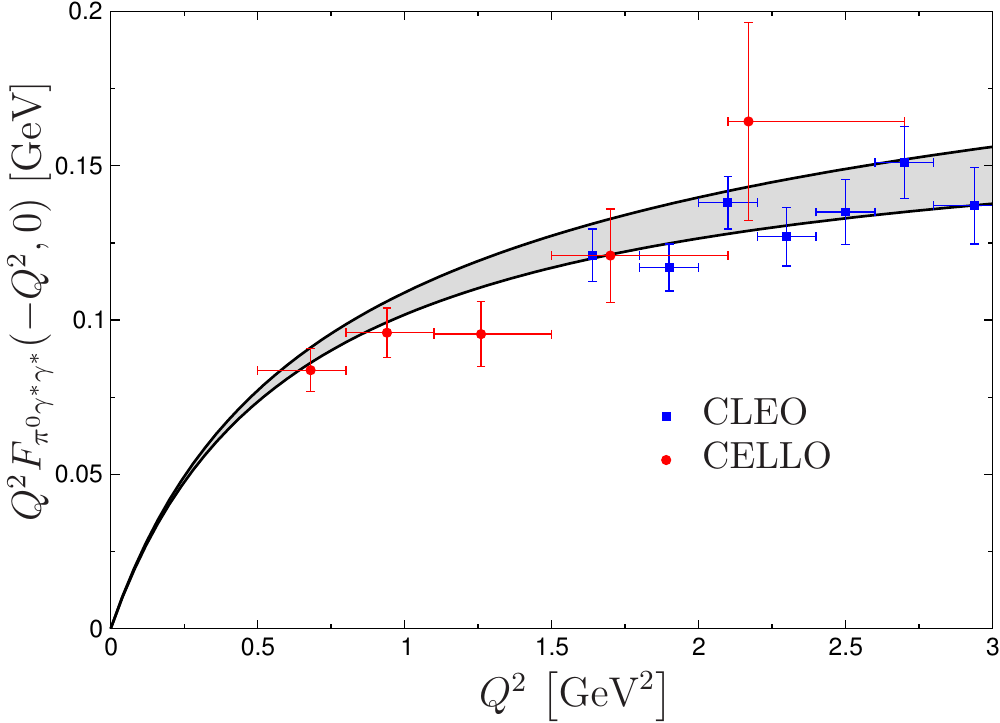}\quad 
\includegraphics[width=0.47\textwidth]{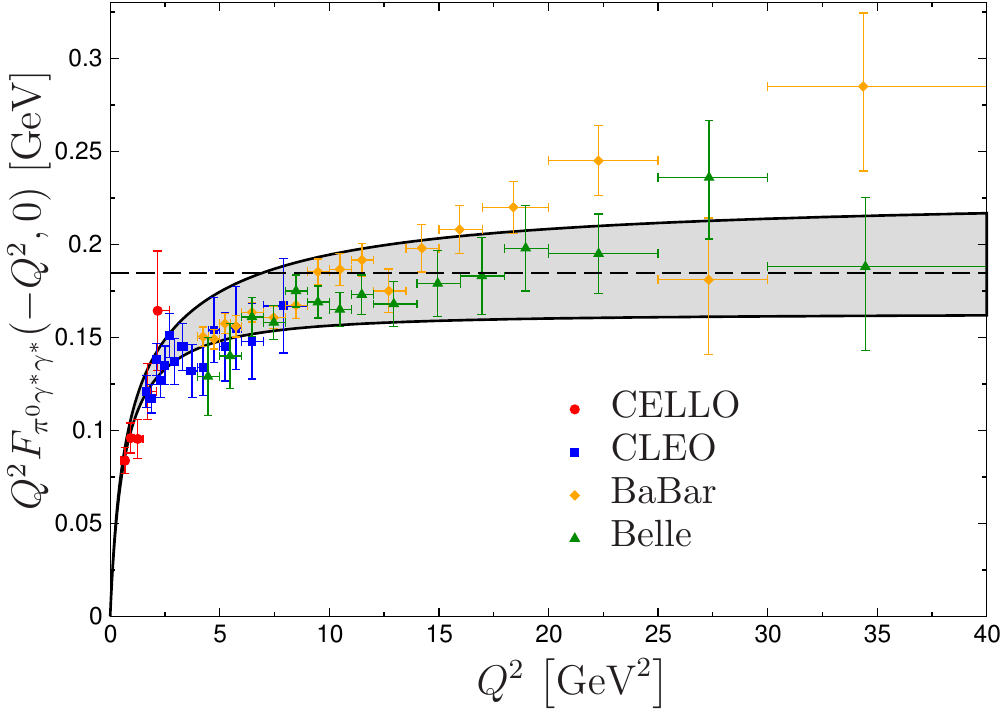}\quad 
\caption{Dispersive description of the $\pi^0$ spacelike single-virtual TFF for the low and high energy regions. The dashed line refers to the Brodsky--Lepage limit $2 f_\pi$. Figure from Refs.~\cite{Hoferichter:2018dmo,Hoferichter:2018kwz}.}
 \label{fig:TFF_spacelike}
\end{figure}
\begin{align}\label{fullTFF}
F_{\pi^0}(-Q_1^2,-Q_2^2)&=
\frac{1}{\pi^2} \int_{0}^{s_m} d x \int_{0}^{s_m}d y   \frac{\rho^\text{disp}(x,y)}{\big(x+Q_1^2\big)\big(y+Q_2^2\big)}\\
&+\frac{1}{\pi^2} \int_{s_m}^{\infty} d x \int_{s_m}^{\infty} d y   \frac{\rho^\text{asym}(x,y)}{\big(x+Q_1^2\big)\big(y+Q_2^2\big)}\nonumber\\
&+\frac{g_\text{eff}}{4\pi^2 f_{\pi}}\frac{M_\text{eff}^4}{(M_\text{eff}^2-q_1^2)(M_\text{eff}^2-q_2^2)}\, , \nonumber
\end{align}
with asymptotic part 
\begin{equation}
\rho^\text{asym}(x,y)=-2\,\pi^2 f_{\pi}\, x\, y\, \delta''(x-y),
\end{equation}
and where $s_m$ is the matching scale between the dispersive and asymptotic parts. 
The last VMD type term in (\ref{fullTFF}) was added in \cite{Hoferichter:2018dmo,Hoferichter:2018kwz} and adjusted to restore the sum rule related to $\pi^0\to 2 \gamma$ decay width and the asymptotic value in the singly-virtual direction without affecting the doubly-virtual behavior at ${\cal O}(1/Q^2)$. The representation without adding an effective pole saturates the low energy theorem at the level of 90\% and Brodsky--Lepage (BL) limit~\cite{Lepage:1979zb,Lepage:1980fj,Brodsky:1981rp} $\lim_{Q^2\to\infty}Q^2F_{\pi^0}(-Q^2,0)=2\,f_\pi$ at the level of $55\%$. We note that  in Eq.~(\ref{fullTFF})  the integrals with the mixed regions were discarded assuming that the effective pole term is sufficient to absorb them. The resulting parameters $g_\text{eff}\simeq 0.1$ and $M_\text{eff}=1.5-2$ GeV consistent with the assumption that it effectively takes into account higher intermediate states and high-energy contributions. In Fig. \ref{fig:TFF_spacelike} we show the result for the $\pi^0$ space-like single-virtual TFF with the uncertainty band \cite{Hoferichter:2018dmo,Hoferichter:2018kwz}.
The prediction for the 
pion-pole contribution to HLbL scattering reads
\bea
\label{result_final}
 a_\mu^{\pi^0\text{-pole}}&=&6.26(0.17)_{F_{\pi\gamma\gamma}}(0.11)_\text{disp}(^{0.22}_{0.14})_\text{BL}(0.05)_\text{asym}\times 10^{-10} \nonumber \\
 &=& 6.26^{+0.30}_{-0.25}\times 10^{-10}.
\eea
The uncertainties come from $\pi^0\to 2 \gamma$ decay width, the dispersive part (which includes variation of the cutoffs between 1.8 and 2.5 GeV, difference between several $\pi\pi$ phase shift representations, error from $F_\pi^V$ and $e^+e^- \to 3\pi$ fits), the uncertainty related to Brodsky--Lepage (BL) limit (which is roughly $^{+20}_{-10}\%$ around the leading order value) and finally the uncertainty from approximating the asymptotic piece with $s_m = 1.7(3)$ GeV$^2$ (it ensures a smooth matching for $q_1^2 = q_2^2 = -Q^2$). The first steps towards a similar dispersive analysis of the double-virtual TFF for $\eta/\eta'$ have been shown in \cite{Hanhart:2013vba,Xiao:2015uva}. However, the full analysis of matching to perturbative QCD has not been performed so far.

\subsubsection{Lattice QCD}
\label{sec:lattice}

\begin{figure}[t]
\includegraphics[width=0.5\textwidth]{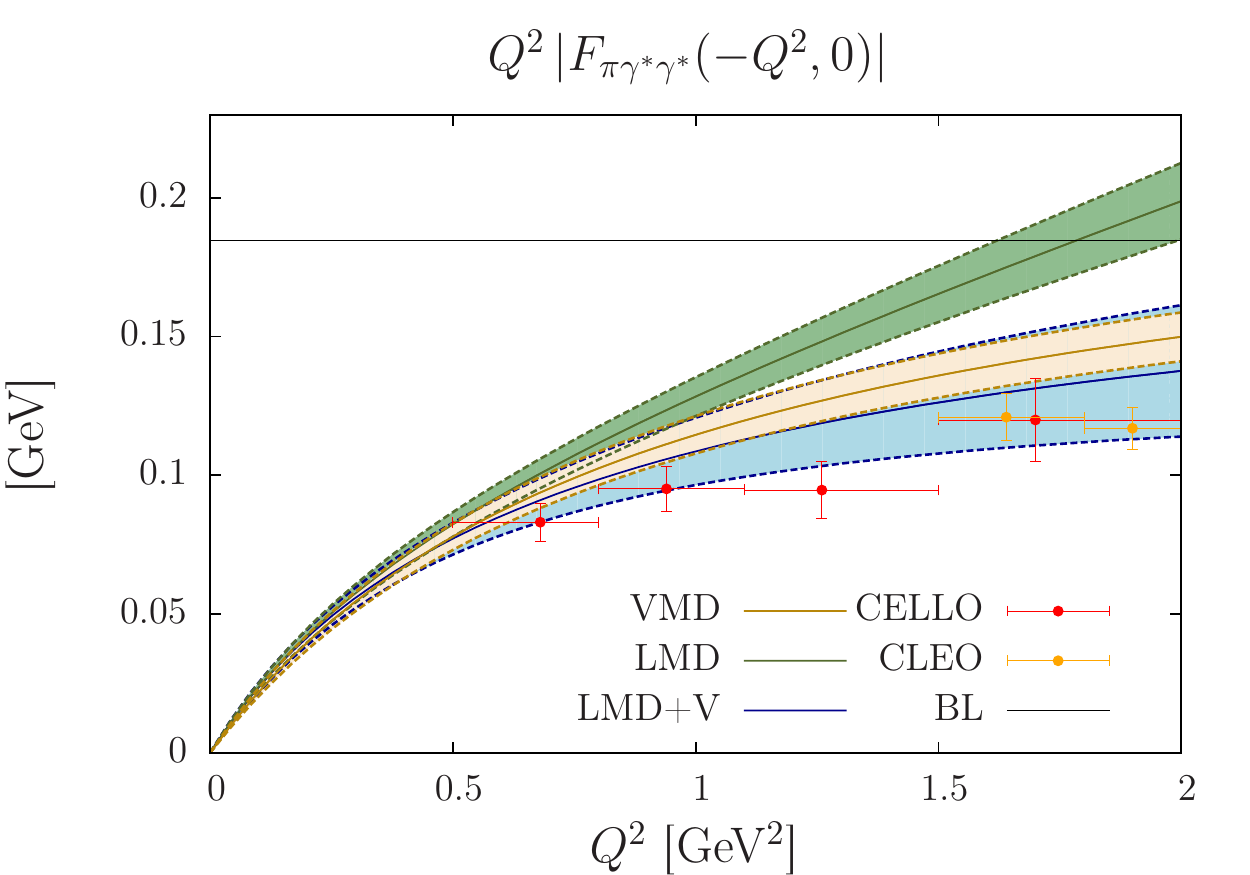}
\includegraphics[width=0.5\textwidth]{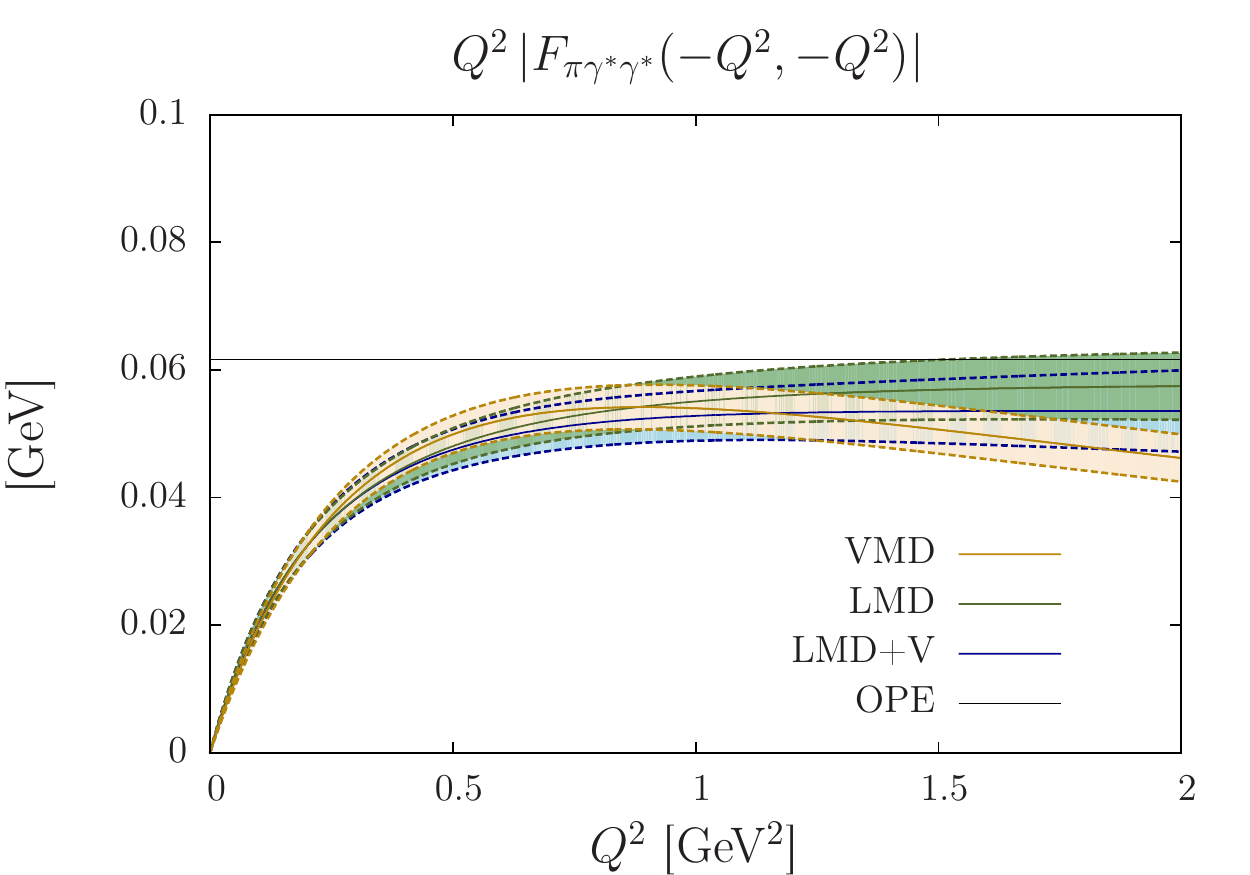}
\caption{Lattice calculation and extrapolations for the VMD, LMD and LMD+V $\pi^0$ TFF models~\cite{Gerardin:2016cqj}. Left: Single-virtual form factor. The horizontal black line corresponds to the prediction from the Brodsky-Lepage limit of Eq.~(\ref{eq:bl}). Right: Double-virtual form factor at $Q_1^2=Q_2^2$, the horizontal black line corresponds to the OPE prediction of Eq.~(\ref{eq:dvsymm}). Figure from Ref.~\cite{Gerardin:2016cqj}.}
\label{fig:lattice}
\end{figure}

In Ref.~\cite{Gerardin:2016cqj}, a calculation of the double-virtual $\pi^0$ TFF was performed in lattice QCD with two flavors of quarks, as is shown in Fig.~\ref{fig:lattice} The lattice data were found to be described by a three-parameter fit, either using the LMD model or the LMD+V model. In both cases, the overall normalization of the form factor comes out consistent with the prediction of the chiral anomaly, with a statistical accuracy of $8-9\%$. In the case of LMD+V, the functional form contains a sufficient number of parameters to be consistent with the theoretically predicted leading behavior at large $Q^2$, both in the single-virtual and the double-virtual case. Being unable to fit all the parameters, some of these parameters have been set to their phenomenological or to their ``preferred'' theory values. In particular, the parameter determining the $F_{\pi^0}(-Q^2,-Q^2)$ behavior at large $Q^2$ has been set to the OPE prediction, Eq.~(\ref{eq:dvsymm}).  The parameter determining the large $Q^2$ behavior in the single-virtual case comes out consistent, albeit with large uncertainties, with the Brodsky-Lepage expectation as given by Eq.~(\ref{eq:bl}) and the value for $h_5$ in the LMD+V fit of Eq.~(\ref{eq:lmdv}) is consistent with a fit to the CLEO data. Furthermore, the parameter $h_2$ which only enters the double-virtual and not the single-virtual form factor, comes out as expected from theoretical expectations from higher-twist corrections in the OPE, although with rather large uncertainty. 

On the other hand, the popular VMD form factor model yields a bad fit to the lattice data. The extracted normalization is not consistent with the chiral anomaly and the VMD form factor fails to reproduce the double-virtual lattice data for increasing spacelike momenta.

Ref.~\cite{Gerardin:2016cqj} has also presented a  value for the $\pi^0$-pole HLbL contribution to $a_\mu$, using the LMD+V fit, as:
\bea
\label{result_lattice}
a_\mu^{\pi^0\text{-pole}}&=& (6.50 \pm 0.83) \times 10^{-10}.
\eea

Very recently, Ref.~\cite{Gerardin:2019vio} reported an update 
of this result, including $N_f = 2+1$ ensembles with $m_\pi \simeq 200$~MeV. This new calculation improved on the error of the previous result by more than a factor of two, resulting in a value:  
\bea
\label{result_lattice2}
a_\mu^{\pi^0\text{-pole}}&=& (5.97 \pm 0.36) \times 10^{-10}.
\eea
The dominant sources of uncertainty in this latest lattice estimate are the statistical precision and the extrapolation to the physical point, followed by the error on the disconnected contribution.

As an outlook for the future, a further improvement of the lattice result calls for additional calculations with 
a larger volume at the physical pion mass. 

\subsubsection{Perturbative QCD}
\label{sec:pqcd}

At sufficiently large momentum transfer, the $\pi^0 \gamma^\ast \gamma^\ast$ TFF is a paradigm for the application of perturbative QCD (pQCD) 
techniques to exclusive processes~\cite{Lepage:1980fj,Brodsky:1981rp,Efremov:1979qk}. In such limit, perturbative QCD predicts that the $\pi^0 \gamma^\ast \gamma^\ast$ TFF factorizes into a perturbatively calculable hard-scattering kernel and a non-perturbative meson distribution amplitude.  The latter encodes the amplitude to find the meson in a collinear $q \bar q$ Fock state.  In the collinear factorization scheme (choosing the factorization scale proportional to the large virtuality), the $\pi^0$ TFF with two non-zero spacelike virtualities,  
$- q_1^2 = Q_1^2 \gg 0$ and/or $-q_2^2 = Q_2^2 \gg 0$,  is given by~\cite{Melic:2002ij}:
\bea
F_{\pi^0}(-Q_1^2, -Q_2^2) = \frac{2 f_\pi}{6} \int_0^1 dx \,\phi_\pi(x)\, T_H(x, Q_1^2, Q_2^2),
\label{eq:pi0pqcd}
\eea 
where $f_\pi = 0.0924$~GeV is the pion decay constant. Furthermore $\phi_\pi (x)$ is the pion distribution amplitude, which can be interpreted as the amplitude for finding a pion in a collinear quark/antiquark state, with quark (anti-quark) carrying a momentum fraction $x$ ($\bar x \equiv 1 - x$) of the meson respectively.  It satisfies the normalization condition:
\bea
\int_0^1 dx \,\phi_\pi(x) = 1.
\eea
The asymptotic distribution amplitude corresponds with  $\phi_\pi (x) = 6\, x \, \bar x$. In Eq.~(\ref{eq:pi0pqcd}), $T_H$ is the hard scattering kernel, which at leading order (LO) in the strong coupling $\alpha_s$ is given by
\bea
T_H^{LO}(x, Q_1^2, Q_2^2) = \frac{2}{(Q_1^2 + Q_2^2)} \left\{ \frac{1}{1 - \omega\, (2 x - 1)} + \frac{1}{1 + \omega\, (2 x - 1)} \right\},
\label{eq:hard}
\eea
where the kinematic variable $\omega$ is defined as:
\bea
\omega \equiv \frac{Q_1^2 - Q_2^2}{Q_1^2 + Q_2^2},
\label{eq:omega}
\eea
with $| \omega | \leq 1$. The corresponding expressions at NLO (order $\alpha_s$) and NNLO (order $\alpha_s^2$) can be found in 
Ref.~\cite{Melic:2002ij}. 

From Eq.~(\ref{eq:hard}), one deduces some special limits. 
For the case of the single virtual TFF, $Q_2^2 = 0$ and $Q_1^2 \gg 0$, one has $\omega = 1$, and 
\bea
T_H^{LO}(x, Q_1^2, 0) = \frac{1}{Q_1^2} \left\{ \frac{1}{\bar x} + \frac{1}{x} \right\},
\eea
which yields for an asymptotic $\pi^0$ distribution amplitude for $Q_1^2 \gg 0$: 
\bea
F_\pi(- Q_1^2, 0) \rightarrow \frac{2 f_\pi}{Q_1^2}.  \label{eq:bl} 
\eea
Another special limit is the double virtual symmetric limit, $Q_1^2 = Q_2^2$, corresponding with $\omega = 0$, and LO hard scattering kernel:
\bea
T_H^{LO}(x, Q_1^2, Q_1^2) = \frac{2}{Q_1^2}.
\eea
In this limit, one obtains for an asymptotic $\pi^0$ distribution amplitude for $Q_1^2 \gg 0$: 
\bea
F_{\pi^0}(- Q_1^2, - Q_1^2) \rightarrow \frac{2}{3}\frac{f_\pi}{Q_1^2}.   
\label{eq:dvsymm}
\eea
In the general double virtual asymmetric case, one can integrate Eq.~(\ref{eq:pi0pqcd}) for an asymptotic $\pi^0$ distribution amplitude and obtains:
\bea
F_{\pi^0}(- Q_1^2, - Q_2^2) \rightarrow \frac{2 f_\pi}{(Q_1^2 + Q_2^2)} f(\omega),
\label{eq:pi0dvpqcd}
\eea
with 
\bea
f(\omega) \equiv \frac{1}{\omega^2} \left\{ 1 - \frac{(1 - \omega^2)}{2 \,\omega} \ln \left( \frac{1 + \omega}{1 - \omega} \right) \right\} ,
\label{eq:fomega}
\eea
which is an even function of $\omega$. 
The above two special cases are obtained as limits by noting that $f(\omega = 1) = 1$ and $f(\omega = 0) = 2/3$. 

For the case of the $\pi^0$ single virtual TFF, the LO, NLO and NNLO pQCD predictions of Ref.~\cite{Melic:2002ij}, for an asymptotic $\pi^0$ distribution amplitude, are shown in \Figref{pi0pqcd}. Although the NNLO predictions with an asymptotic distribution amplitude show a good agreement with the BaBar and Belle data below $Q^2 = 10$~GeV$^2$, at larger $Q^2$ there is unfortunately a disagreement between both data sets, which leaves the onset towards the pQCD limit still as an open issue for the $\pi^0$ TFF.

\subsubsection{Phenomenological interpolating parameterization of TFFs}
\label{sec:phenoparam}

In this section we will discuss phenomenological parameterizations for the single and double virtual $\pi^0, \eta, \eta^\prime$ TFFs, with the aim of estimating the  $\pi^0, \eta, \eta^\prime$ pole contributions to $a_\mu$.

\begin{figure}
\centering{ \includegraphics[width=0.75\textwidth]{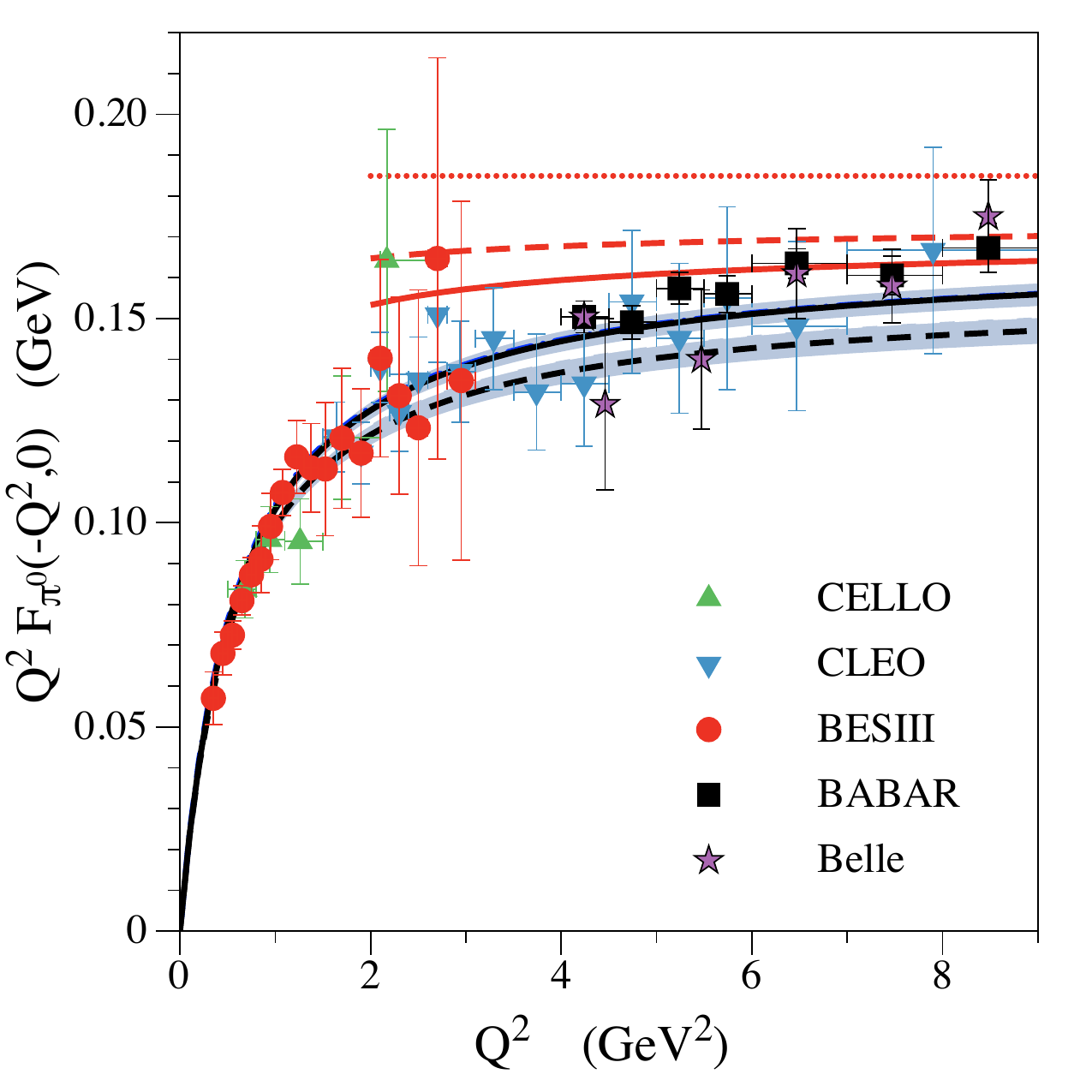}}
 \caption{Single virtual $\pi^0$ TFF up to 9~GeV$^2$ compared to the world data. The red curves are the pQCD predictions~\cite{Melic:2002ij} using an asymptotic $\pi^0$ distribution amplitude: LO (dotted), NLO (dashed), NNLO (solid). The black curves with corresponding error bands are monopole fits, using Eq.~(\ref{eq:monopole}), when fitting all data up to 9~GeV$^2$ (solid curve) and when fitting all data up to 4~GeV$^2$ (dashed curve). The resulting parameters are given in Table~\ref{tab:monopole}.}
 \label{fig:pi0pqcd}
\end{figure}

\begin{figure}
 \includegraphics[width=0.5\textwidth]{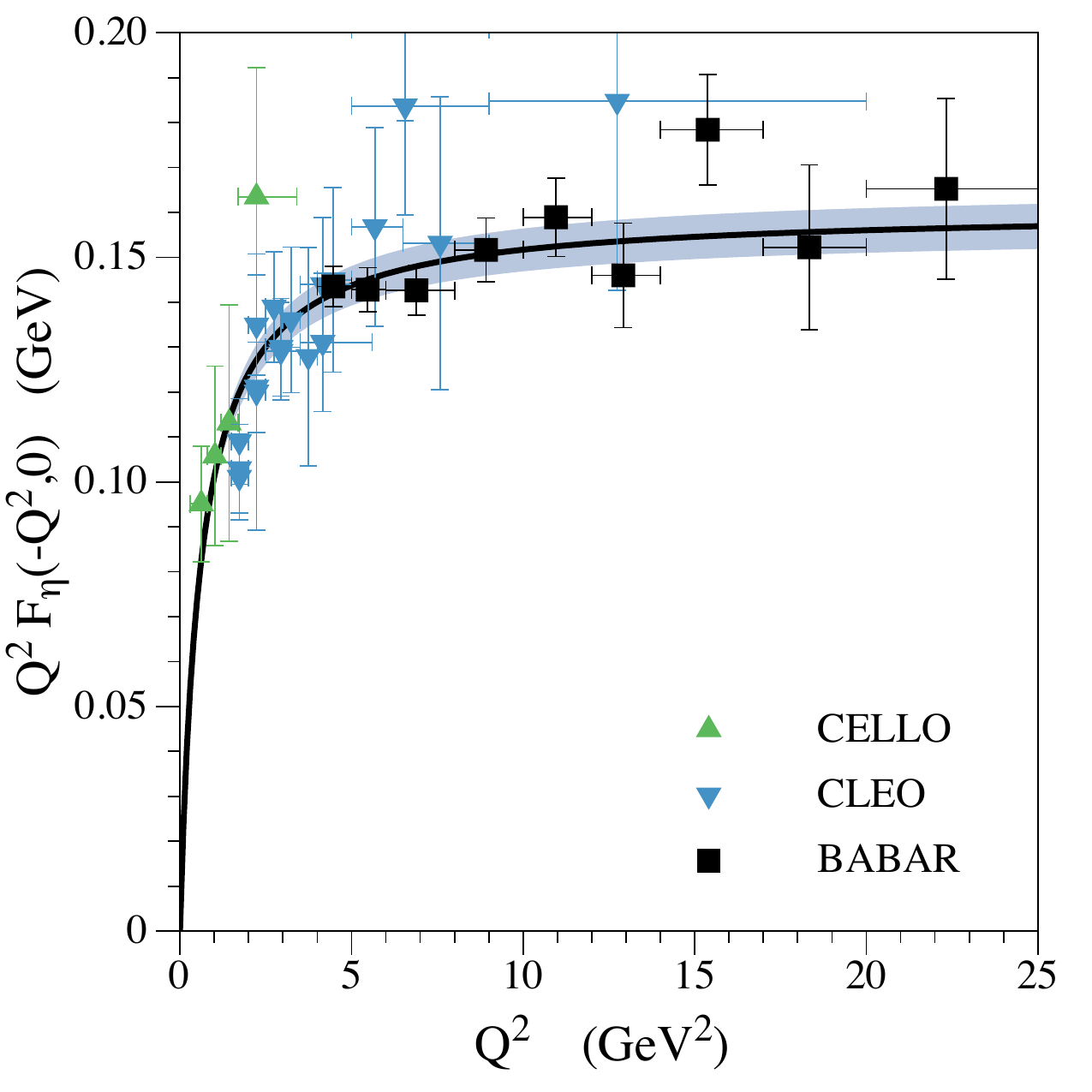}
 \includegraphics[width=0.5\textwidth]{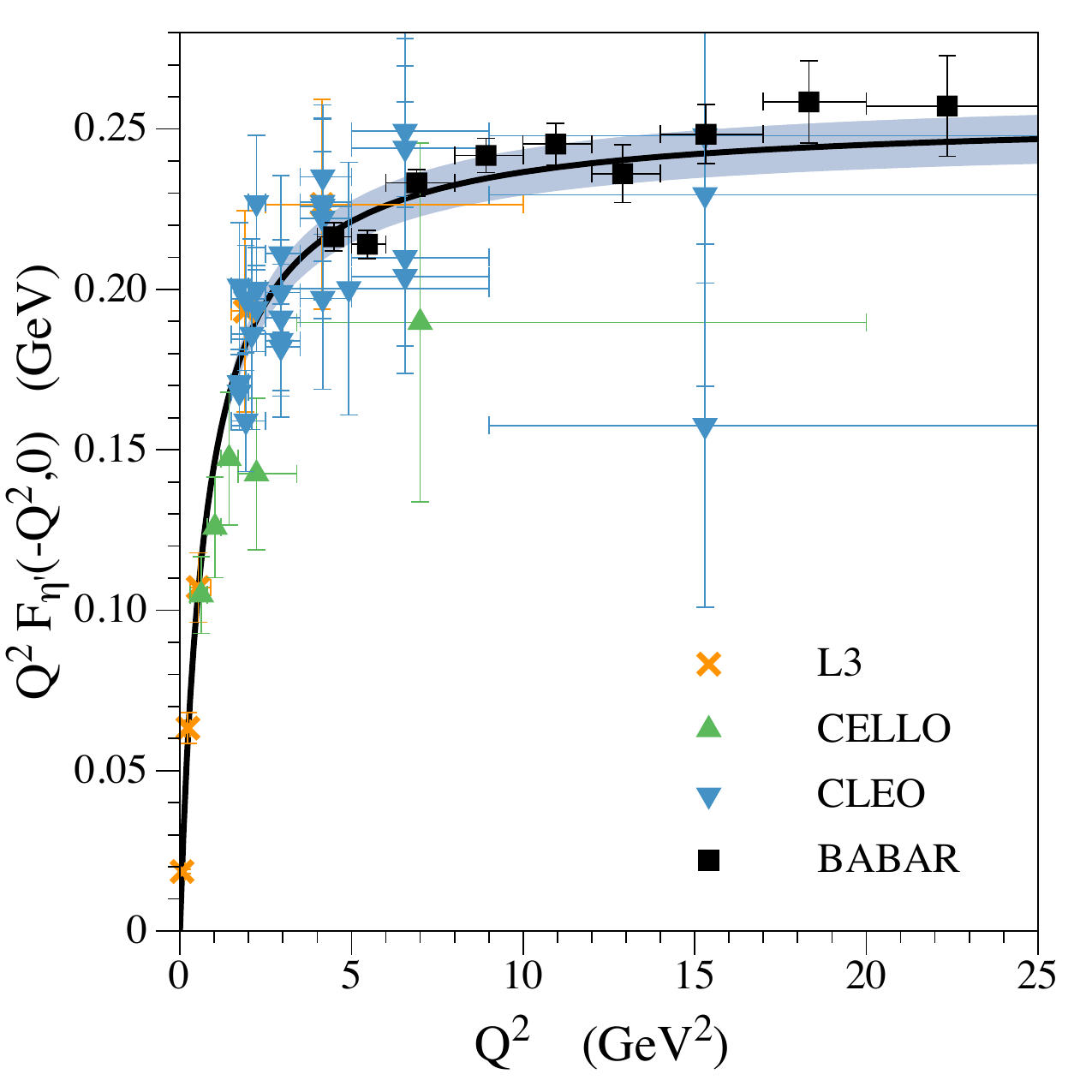}
 \caption{Single virtual $\eta$ TFF (left) and $\eta^\prime$ TFF (right) compared to the world data. The black curves with corresponding error bands are monopole fits, using Eq.~(\ref{eq:monopole}). The resulting parameters are given in Table~\ref{tab:monopole}.}
 \label{fig:etapqcd}
\end{figure}

We start with the single virtual TFF for $M = \pi^0, \eta, \eta^\prime$, which we will parameterize by a monopole form: 
\bea
F_M(-Q^2, 0) =  \frac{F_M(0,0)}{1 + Q^2/\Lambda^2}, 
\label{eq:monopole}
\eea
where the TFF values for real photons $F_M(0,0)$ are obtained from the experimental two-photon decays widths, as given by their PDG2018 values~\cite{Tanabashi:2018oca}.  
The mass parameters $\Lambda^2$ are obtained from a one-parameter fit to the single virtual TFF data shown in Figs.~\ref{fig:pi0pqcd} and \ref{fig:etapqcd}.  For the $\pi^0$ data set, which includes the new BESIII data with their high statistical accuracy in the lower $Q^2$ range, we perform two fits: one with all data up to 4 GeV$^2$ included, and another one which also includes the data up to 9 GeV$^2$. The corresponding $\chi^2$ values are calculated with only the statistical errors of the corresponding data sets. The corresponding value and its error for the mass parameters $\Lambda^2$ entering the $\pi^0, \eta$, and $\eta^\prime$ TFFs are given in Table~\ref{tab:monopole}.

\begin{table}[h]
\begin{tabular*}{\textwidth}{ccccc}
\hline\hline
$M$ & $\Gamma_{\gamma \gamma} $  & $F_M(0,0)$ & $\Lambda^2$  & $\chi^2/\mathrm{d.o.f.}$  \\
&  [keV] & [GeV$^{-1}$] & [GeV$^2$] & \\
\hline \hline
$\pi^0$ &  $\left(7.635 \pm 0.160 \right) \times 10^{-3}$ & $ 0.273 \pm 0.003$ &  $ 0.611 \pm 0.005$ & 2.1 ($< 9$~GeV$^2$)   \\
 &  & &  $ 0.574 \pm 0.007$ & 1.5 ($< 4$~GeV$^2$)  \\
$\eta$ &  $0.516 \pm 0.020$  & $0.274 \pm 0.005$  &   $ 0.587 \pm 0.009$ & 0.9   \\
$\eta^\prime$ &   $4.35 \pm 0.25$ & $0.341 \pm 0.008$  &  $ 0.745 \pm 0.007$  & 1.3  \\
\hline\hline
\end{tabular*}
\caption{Parameters in the monopole fits for the single virtual TFF of $\pi^0, \eta$, and $\eta^\prime$ according to Eq.~(\ref{eq:monopole}). The TFF values for real photons $F_M(0,0)$ are obtained from the corresponding two-photon decays widths 
$\Gamma_{\gamma \gamma}$, which are taken from PDG2018~\cite{Tanabashi:2018oca}. The mass parameters $\Lambda^2$ are obtained from a one-parameter fit to the single virtual TFF data, shown in Figs.~\ref{fig:pi0pqcd} and \ref{fig:etapqcd}. We emphasize that 
only the statistical errors were used in the fits.
For the $\pi^0$ two fits are shown: one with all data up to 4~GeV$^2$, and one with all data up to 9~GeV$^2$.}
\label{tab:monopole}
\end{table}

\begin{figure}
\centering{ \includegraphics[width=0.85\textwidth]{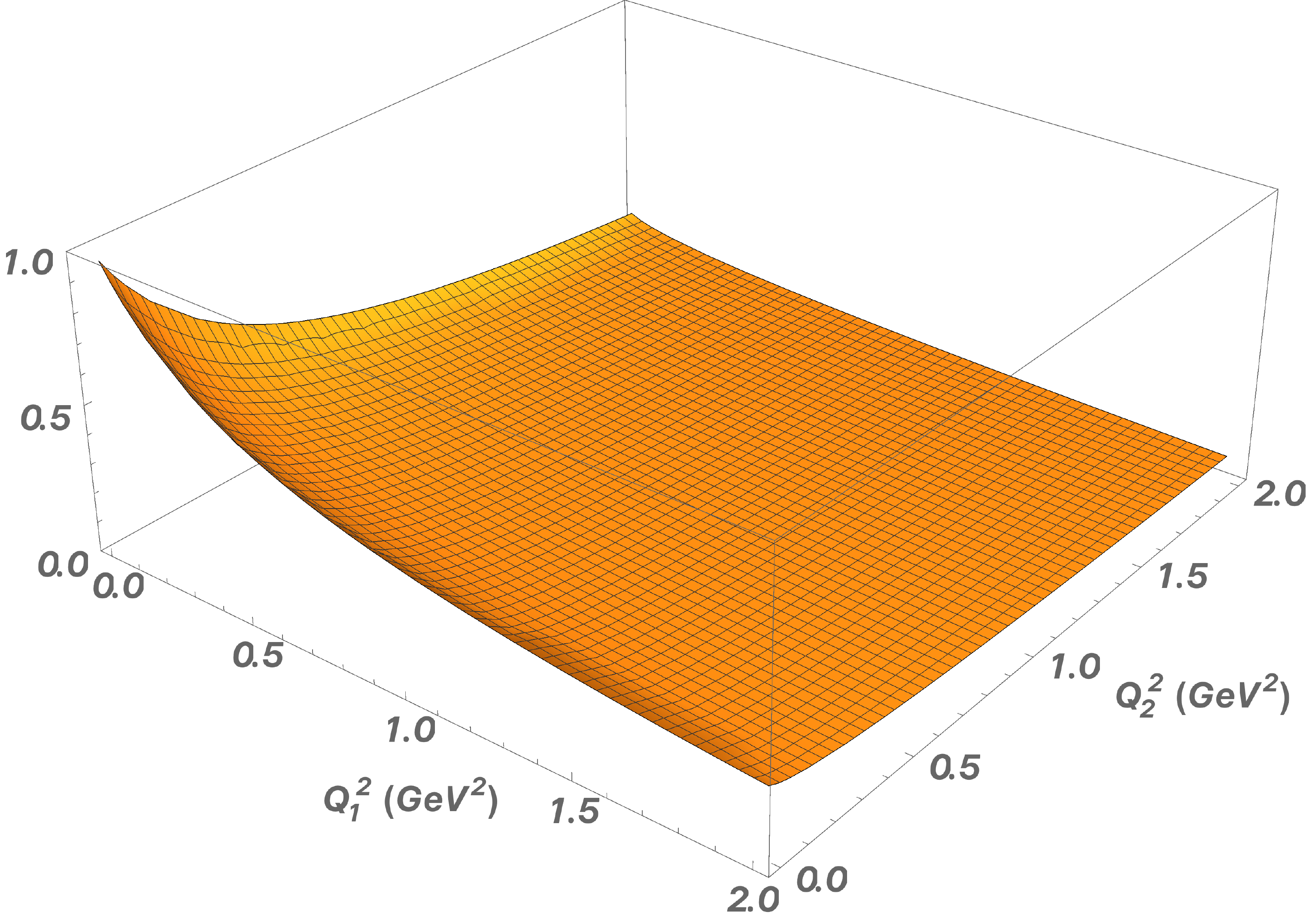} }
\centering{ \includegraphics[width=0.85\textwidth]{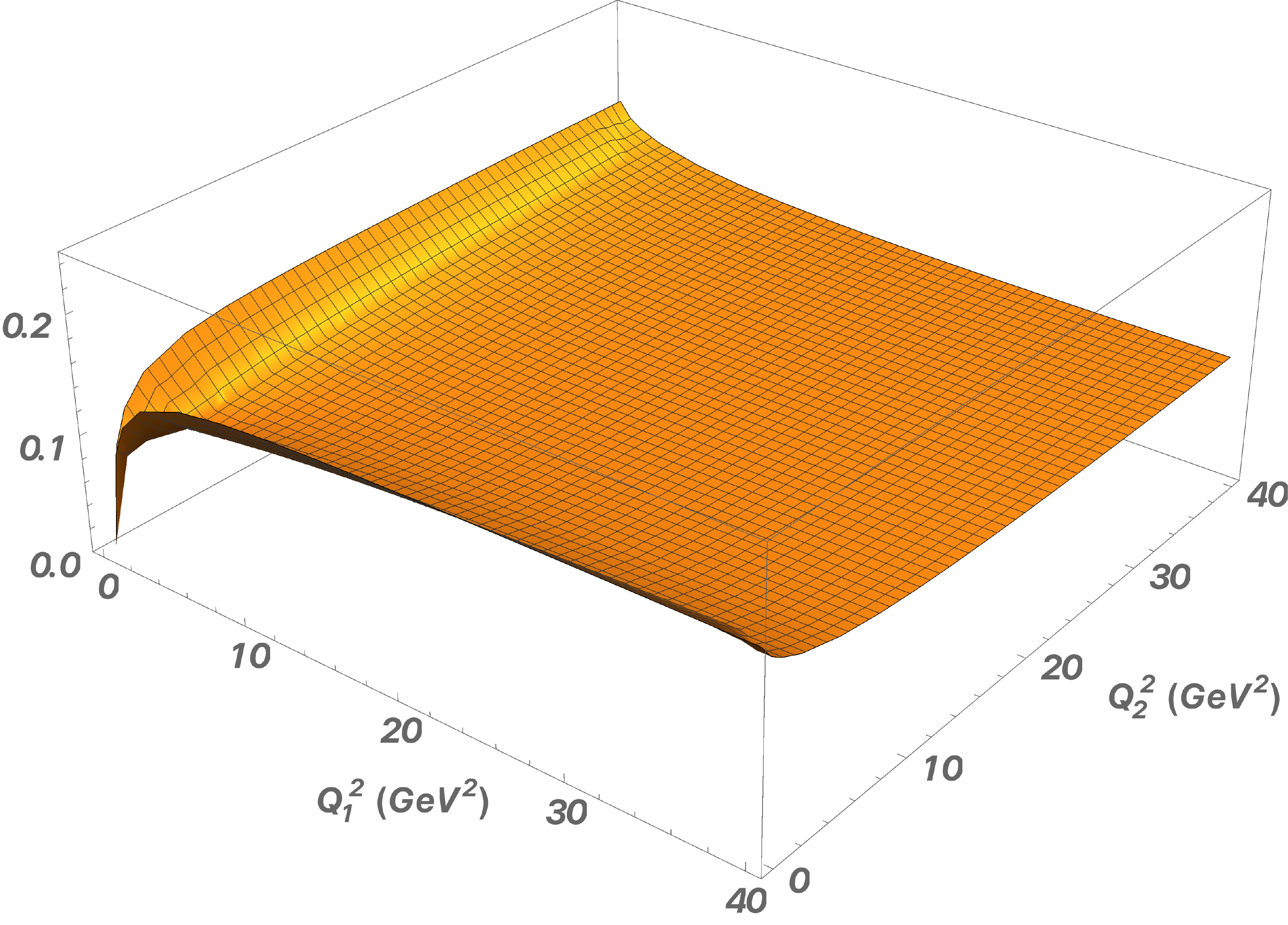} }
 \caption{Parameterization of Eq.~(\ref{eq:interpol}) for the $\pi^0$ double virtual TFF. Upper panel: $F_{\pi^0}(-Q_1^2, - Q_2^2) / F_{\pi^0}(0,0)$. 
 Lower panel: $ (Q_1^2 + Q_2^2) F_{\pi^0}(-Q_1^2, - Q_2^2)$ in GeV.}
 \label{fig:tff3d}
\end{figure}

In order to provide estimates for the pseudoscalar meson contribution to $a_\mu$, we also need the knowledge of the corresponding 
double virtual TFFs. For this purpose, we choose a parameterization with minimum number of free parameters, and which satisfies 
the following three criteria:
(i) the TFF is normalized at the real photon point to the empirical value obtained from the two-photon decay width, 
(ii) for the single virtual case, the TFF reduces to a monopole form, which provides a very efficient one-parameter fit of the corresponding TFF data over a wide range in $Q^2$, (iii) in the limit of large $Q_1^2$ and/or large $Q_2^2$, the TFF has to tend to the correct pQCD expression 
of Eq.~(\ref{eq:pi0dvpqcd}). The latter is obtained by noting that before the asymptotic regime is reached, the quarks are not entirely collinear, 
having a finite transverse momentum. This will introduce in the propagator denominators in Eq.~(\ref{eq:hard}) a scale corresponding with this average transverse momentum (which is expected to be in the order of a few hundred MeV/c). We can account for this in a phenomenological way by replacing the parameter $\omega$ of Eq.~(\ref{eq:omega}) by
\bea
\omega_\Lambda \equiv \left( \frac{(Q_1^2 - Q_2^2)^2 + \Lambda^4}{(Q_1^2 + Q_2^2)^2 + \Lambda^4} \right)^{1/2},
\label{eq:omegalam}
\eea
which we defined corresponding with $0 < \omega_\Lambda \leq 1$, 
where $\Lambda^2$ can be interpreted as proportional to a scale given by the average transverse momentum. 
We thus propose to parameterize the double virtual TFF by extending Eq.~(\ref{eq:pi0dvpqcd}) to the interpolating formula
\bea
F_{M}(- Q_1^2, - Q_2^2) = \frac{F_{M}(0,0) }{1 + (Q_1^2 + Q_2^2)/\Lambda^2} f(\omega_\Lambda).
\label{eq:interpol}
\eea
Note that the single virtual limit, $Q_2^2 = 0$, corresponds with $\omega_\Lambda = 1$, and because of $f(\omega_\Lambda = 1) = 1$, 
we exactly obtain the monopole form of Eq.~(\ref{eq:monopole}), which allows to fix the only fit parameter $\Lambda^2$ as given in Table~\ref{tab:monopole}. 
In the parameterization of Eq.~(\ref{eq:interpol}), the LO pQCD limit of the FF is not built in, and we make no assumptions about the pion distribution amplitude. 
It is however interesting to note from \Figref{pi0pqcd} that the empirical fit comes within 
5 \% of the NNLO pQCD result with asymptotic distribution amplitude around $Q^2 = 10$~GeV$^2$.  
Note also that the parameterization of Eq.~(\ref{eq:interpol}) will satisfy the pQCD limit in the general double virtual case, both for symmetric and asymmetric kinematics, as for large virtualities $\omega_\Lambda \to |\omega|$. We like to note that in common parameterizations for the $\pi^0$ TFF, which are used to estimate $a_\mu$,  either the correct pQCD limit is only satisfied for the single virtual case (e.g. for the VMD parameterization of Eq.~(\ref{eq:vmd})) or for the double virtual symmetric ($Q_1^2 = Q_2^2$) case (e.g. for the LMD+V parameterization of Eq.~(\ref{eq:lmdv})). 
The general double virtual case involves a logarithmic dependence as given by Eqs.~(\ref{eq:pi0dvpqcd},\ref{eq:fomega}). 
We also like to note that the parameterization of Eq.~(\ref{eq:interpol}) is entirely smooth, as displayed in \Figref{tff3d}. 

\begin{figure}
 \includegraphics[width=0.5\textwidth]{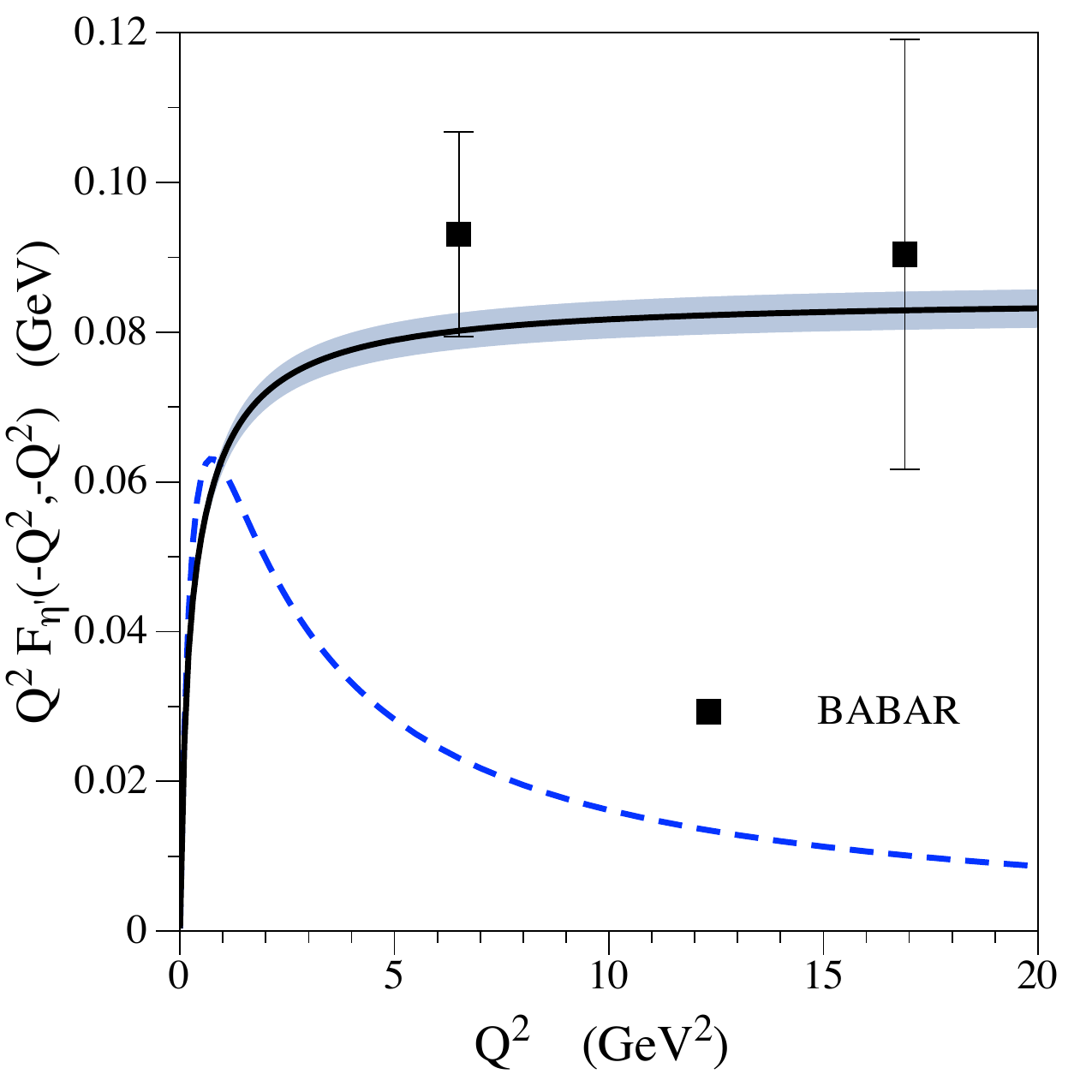}
 \includegraphics[width=0.5\textwidth]{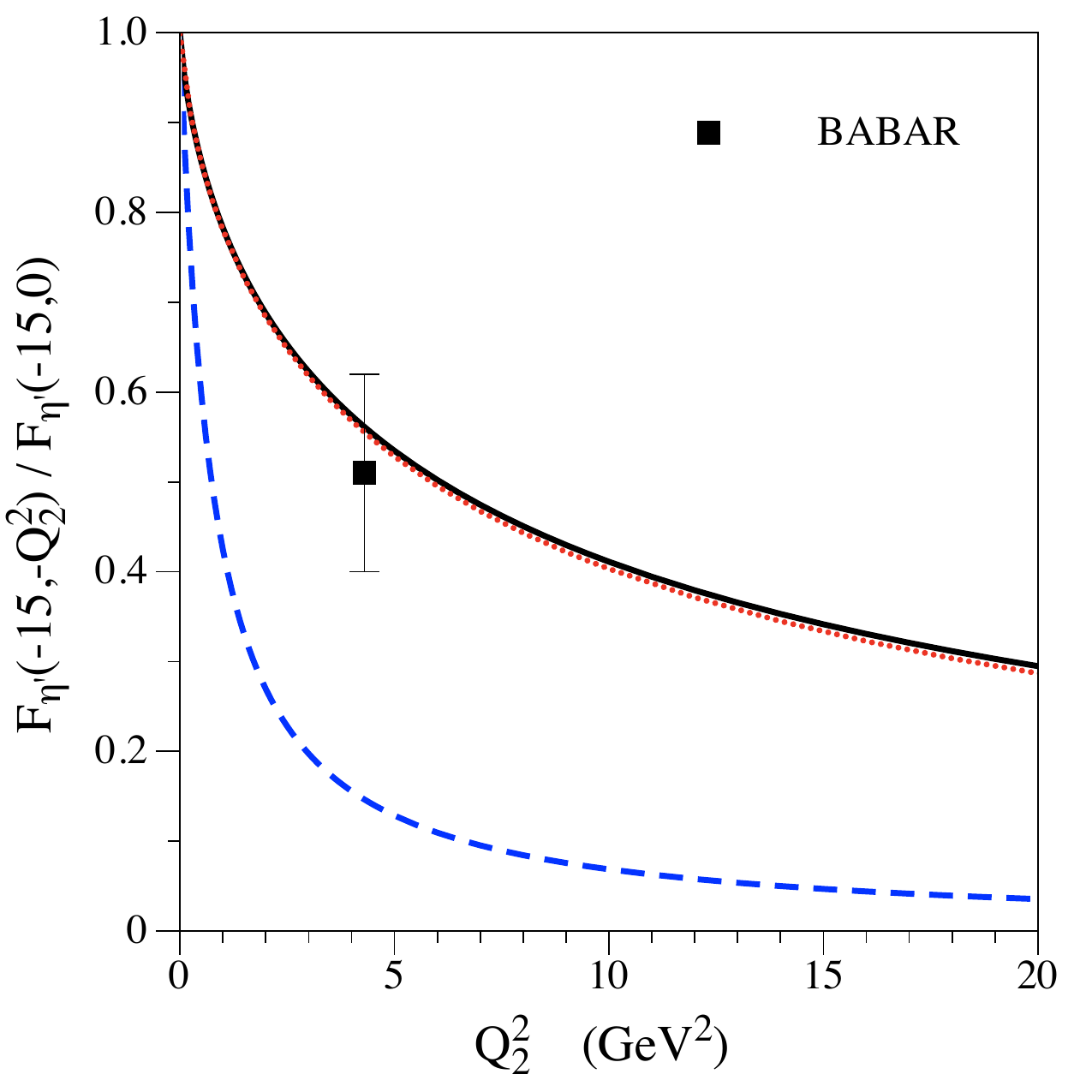}
 \caption{Double virtual $\eta^\prime$ TFF. Left panel:  double virtual symmetric case, $Q_1^2 = Q_2^2$. Right panel: double virtual asymmetric case, with $Q_1^2 = 15$~GeV$^2$ as function of $Q_2^2$, normalized to the single virtual TFF for the same $Q_1^2$. The data points are from 
 BaBar~\cite{BaBar:2018zpn}. The solid (black) curves with corresponding error bands are the predictions of the parameterization of Eq.~(\ref{eq:interpol}), with parameter $\Lambda^2$ fixed from the single virtual TFF as in Table~\ref{tab:monopole}. The dashed (blue) curves are the VMD prediction~\cite{Jegerlehner:2009ry,Nyffeler:2016gnb}. The red dotted curve on the right panel is the pQCD prediction.}
 \label{fig:etapdv}
\end{figure}

Recently, the BaBar Collaboration has released first data of the double virtual TFF for the $\eta^\prime$, both for symmetric and asymmetric kinematics at intermediate and large momentum transfers. As the parameterization of Eq.~(\ref{eq:interpol}) is fully determined from the fit of the scale parameter $\Lambda^2$ from the single virtual TFF data, such double virtual TFF data provide a strong test of the validity of the parameterization.   We show this comparison in \Figref{etapdv}, which shows that Eq.~(\ref{eq:interpol}) provides a very good description of the data both for the symmetric kinematics around 5 and 15 GeV$^2$ as well as for the asymmetric kinematics.  In contrast, around 5 GeV$^2$, the VMD prediction of Eqs.~(\ref{eq:vmd},\ref{eq:vmdparam}) is already around a factor of 3 below the data. 

\begin{figure}
 \includegraphics[width=0.5\textwidth]{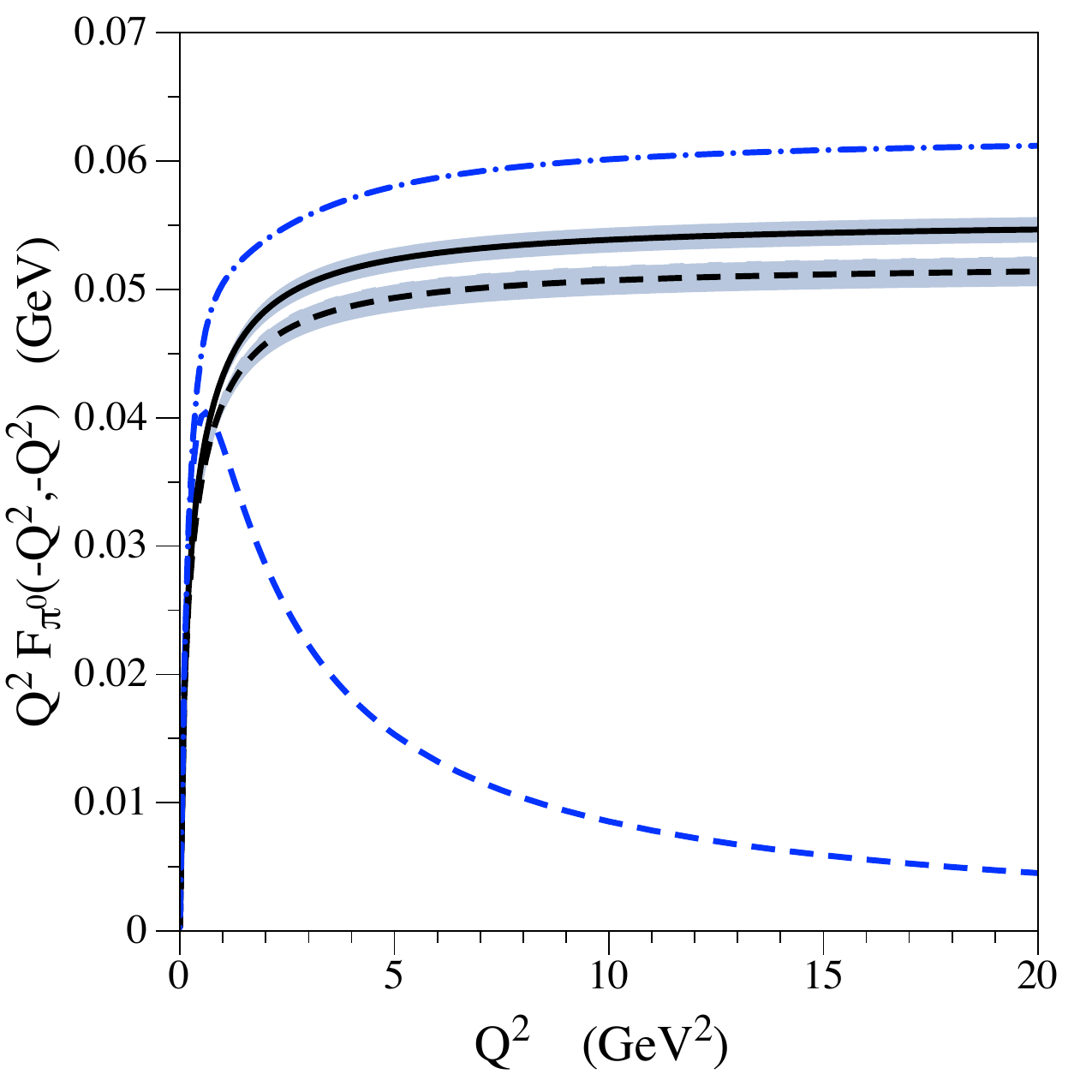}
 \includegraphics[width=0.5\textwidth]{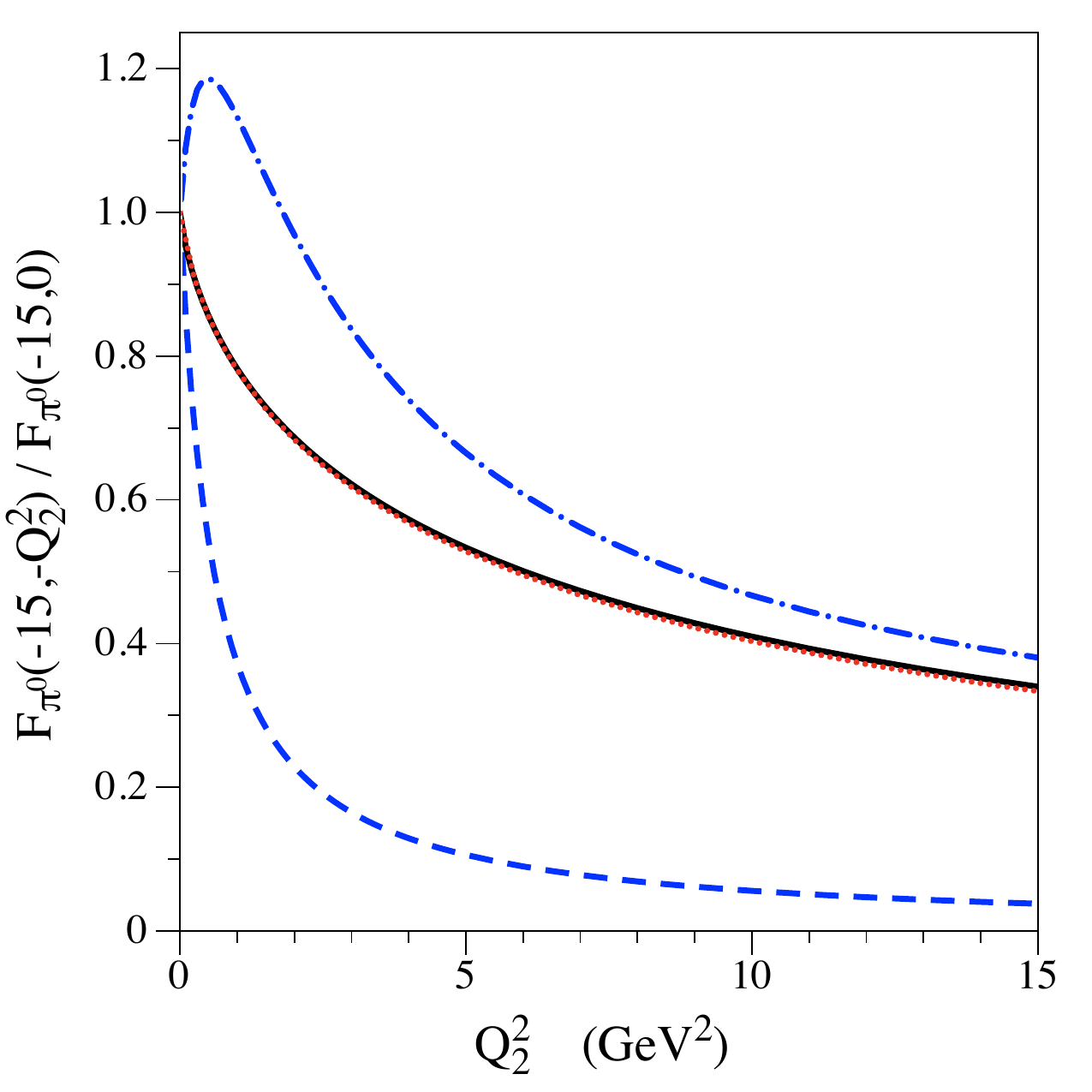}
 \caption{Double virtual $\pi^0$ TFF. Left panel: double virtual symmetric case, $Q_1^2 = Q_2^2$. Right panel: double virtual asymmetric case, with $Q_1^2 = 15$~GeV$^2$ as function of $Q_2^2$, normalized to the single virtual TFF for the same $Q_1^2$. The solid and dashed (black) curves with corresponding error bands are the predictions of the parameterization of Eq.~(\ref{eq:interpol}), with parameter $\Lambda^2$ fixed from the single virtual TFF as in Table~\ref{tab:monopole}. Dashed (blue) curves: VMD prediction~\cite{Jegerlehner:2009ry,Nyffeler:2016gnb}, dashed-dotted (blue) curves: LMD+V prediction~\cite{Jegerlehner:2009ry,Nyffeler:2016gnb}. The red dotted curve on the right panel is the pQCD prediction.}
 \label{fig:pi0dv}
\end{figure}

\begin{figure}
\centering{ \includegraphics[width=0.6\textwidth]{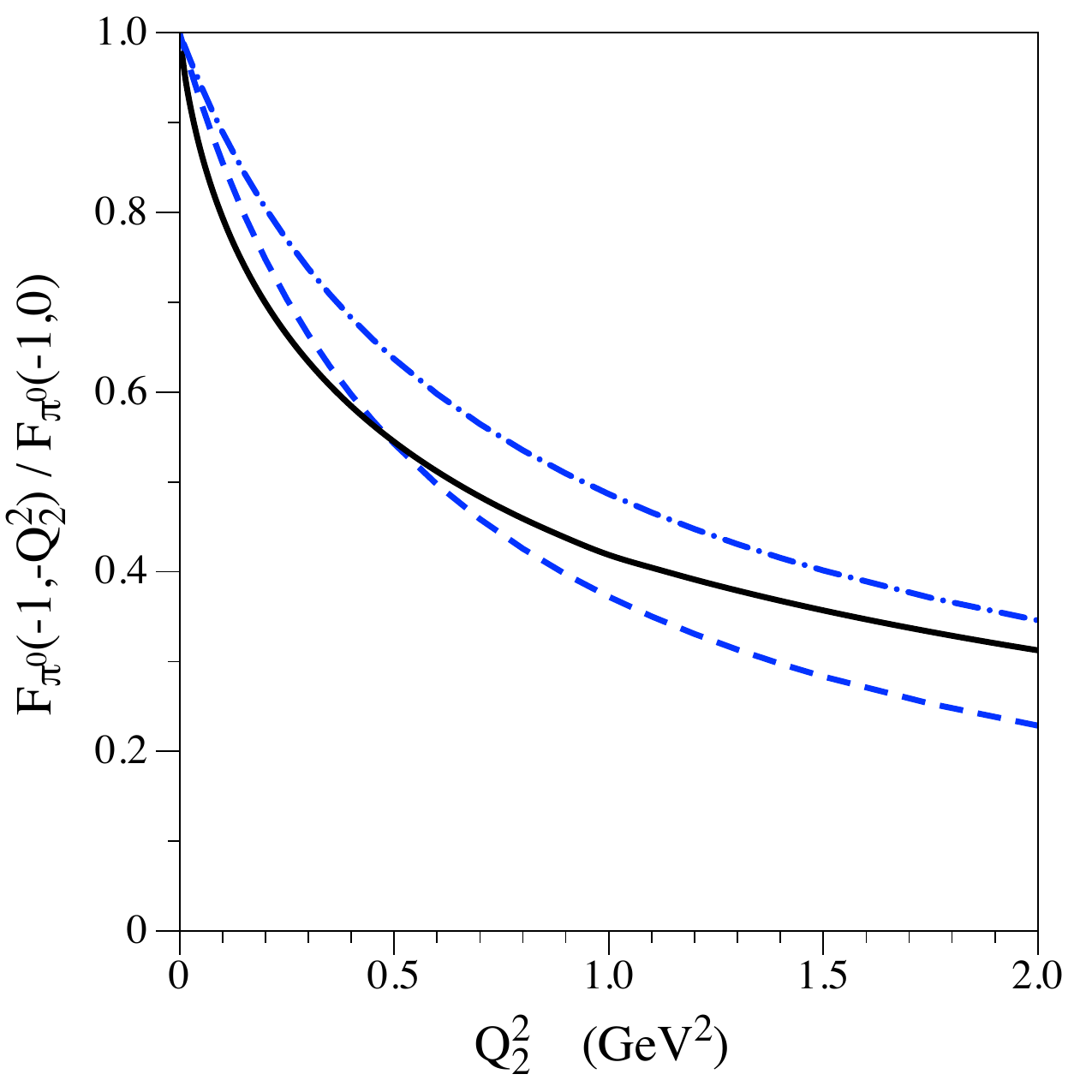} }
 \caption{$\pi^0$ TFF for the double virtual asymmetric case, with $Q_1^2 = 1$~GeV$^2$ as function of $Q_2^2$, normalized to the single virtual TFF for the same $Q_1^2$, in the kinematic range accessible at BESIII. The solid (black) curve is the predictions of the parameterization of Eq.~(\ref{eq:interpol}), with parameter $\Lambda^2$ fixed from the single virtual TFF as in Table~\ref{tab:monopole}. Dashed (blue) curve: VMD prediction~\cite{Jegerlehner:2009ry,Nyffeler:2016gnb}, dashed-dotted (blue) curve: LMD+V prediction~\cite{Jegerlehner:2009ry,Nyffeler:2016gnb}. }
 \label{fig:pi0dv2}
\end{figure}

The predictions for the $\pi^0$ double virtual TFF are shown in Fig.~\ref{fig:pi0dv}, where we compare the interpolating formula of Eq.~(\ref{eq:interpol}) to available VMD and LMD+V parameterizations~\cite{Jegerlehner:2009ry,Nyffeler:2016gnb} of Eqs.~(\ref{eq:vmd},\ref{eq:vmdparam}) and 
(\ref{eq:lmdv},\ref{eq:lmdvparam}) respectively. We again see that the VMD prediction drops very fast relative to the interpolating formula of Eq.~(\ref{eq:interpol}). We note that the LMD+V parameterization, which has the double virtual symmetric pQCD limit of Eq.~(\ref{eq:dvsymm}) built in, gives qualitatively similar results for the double virtual symmetric case, but shows an overshoot in the range of 20\% for the double virtual asymmetric case. Even at lower $Q^2$ this overshoot is visible, as shown in Fig.~\ref{fig:pi0dv2}, which displays the $\pi^0$ double virtual TFF in kinematics accessible at BESIII. As this region is crucial for an accurate estimate of $a_\mu$, a corresponding measurement with a 10\% accuracy, albeit challenging, will provide an important test.  

\subsection{Pseudoscalar meson pole contributions to $a_\mu$}

\subsubsection{Spacelike 3-dimensional integral representation}
\label{sec:3dint}

\begin{figure}[h]
\centering{ \includegraphics[width=0.6\textwidth]{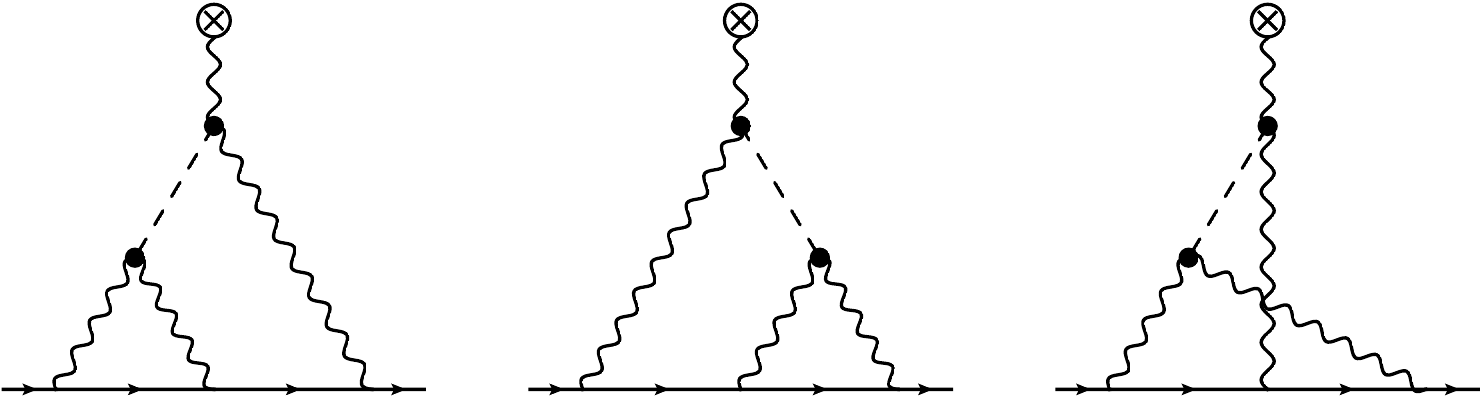} }
 \caption{Single meson pole HLbL contributions to $a_\mu$. The dashed lines correspond with the meson states. }
 \label{fig:hlblpole}
\end{figure}

The dominant hadronic contribution to $a_\mu$ arises from the lightest meson states coupling to two photons, shown in Fig.~\ref{fig:hlblpole}. As the lightest meson states with large two-photon couplings are the $\pi^0, \eta, \eta^\prime$ mesons, we will use the empirical information on the double virtual $\pi^0, \eta, \eta^\prime$ TFFs discussed above to estimate in this section the corresponding pseudoscalar pole contributions to $a_\mu$. The formalism has been pioneered in Ref.~\cite{Knecht:2001qf}, where the two-loop integral of Fig.~\ref{fig:hlblpole} is evaluated by a Wick rotation to Euclidean space, which allows to express the corresponding pole contributions to $a_\mu$ solely in terms of the single and double virtual TFFs for spacelike momentum transfers $Q_1^2$ and $Q_2^2$, with $Q_1$ and $Q_2$ denote the two independent Euclidean four-momenta in the two-loop process of Fig.~\ref{fig:hlblpole}.  
  
After an average over the muon momentum direction, and by using the hyperspherical approach (using Gegenbauer polynomials),  one can perform all angular integrals in the two-loop integral except for one over the hyperangle $\theta$ (with $\cos \theta \equiv \tau$) between the Euclidean four-momenta $Q_1$ and $Q_2$. In this way one obtains a three-dimensional integral representation, over $\tau$ as well as over both virtualities, which are denoted (for simplicity of notation) by $Q_i \equiv \sqrt{Q_i^2} (i = 1,2)$ as~\cite{Jegerlehner:2009ry}:
\bea
&&a_\mu^M = \left(\frac{\alpha}{\pi}\right)^3 \int_0^\infty dQ_1 \int_0^\infty dQ_2 \int_{-1}^{+1} d \tau  \\
&& \hspace{3.5cm} \times\left\{ w_1(Q_1, Q_2, \tau)\, F_M(- Q_1^2, - Q_3^2) \, F_M(- Q_2^2, 0) \right. \nonumber \\   
&& \hspace{3.75cm} \left. + w_2(Q_1, Q_2, \tau)\, F_M(- Q_1^2, - Q_2^2)\,  F_M(- Q_3^2, 0)\right\} , \quad \quad\nonumber
\label{eq:amups}
\eea
with $Q_3^2 \equiv Q_1^2 + Q_2^2 + 2\, Q_1 \,Q_2\,\m\, \tau$. 
In Eq.~(\ref{eq:amups}), the weighting functions $w_1$ and $w_2$ were derived in Refs.~\cite{Jegerlehner:2009ry,Nyffeler:2016gnb}, and can be expressed as:
\bea
&&\hspace{-0.5cm}w_1(Q_1, Q_2, \tau) = \frac{2 \pi}{3}  \frac{Q_1^2 \,Q_2^2\, \sqrt{1 - \tau^2}}{Q_3^2} \frac{1}{Q_2^2 + m_M^2} \\
&&\hspace{0.5cm} \times \left\{ -8 \,\sqrt{1 - \tau^2} \left( \frac{Q_2^2}{m^2} - 2 \right) \arctan\left( \frac{z\, \sqrt{1 - \tau^2} }{1 - z\, \tau} \right) + 4\, \tau \right. \nonumber \\
&&\hspace{0.5cm} - \left. \tau\, \frac{Q_1^2}{m^2}\, (1 - R_{m1})^2  - \frac{2}{m^2}  \left[ 2\, Q_1\, Q_2\, (1 - \tau^2) - Q_1^2\, \tau\right] (1 - R_{m1}) \right\}, 
\nonumber \\
\nonumber \\
&&\hspace{-0.5cm}w_2(Q_1, Q_2, \tau) = \frac{2 \pi}{3}  \frac{Q_1^2\, Q_2^2\, \sqrt{1 - \tau^2}}{Q_3^2} \frac{1}{Q_3^2 + m_M^2}  \\
&&\hspace{0.5cm} \times \left\{ \frac{8}{ \sqrt{1 - \tau^2}} \left[1 - \tau^2 + \frac{1}{m^2}\, (Q_1^2 + Q_1\, Q_2\, \tau) \right] 
\arctan\left( \frac{z\, \sqrt{1 - \tau^2} }{1 - z \tau} \right) + 4\, \tau \right. \nonumber \\
&&\hspace{0.5cm} - \left. \tau\, \frac{Q_2^2}{m^2}\, (1 - R_{m2})^2  
+ \frac{4}{m^2}  \left[ Q_1\, Q_2 + Q_2^2\, \tau\right] (1 - R_{m2}) 
+ 2\,\tau\, \frac{Q_1^2}{m^2} (1 - R_{m1})
\right\}, \nonumber
\eea
with $m$ the muon mass, $R_{mi} \equiv \sqrt{1 + 4\, m^2 / Q_i^2}$ (for $i = 1, 2$), and where $z$ is defined as:
\bea
z \equiv \frac{Q_1\, Q_2}{4\, m^2} (1 - R_{m1}) (1 - R_{m2}) .
\eea

In Table~\ref{tab:amupieta}, we estimate the pseudoscalar pole contributions by using different parameterizations for the double virtual TFFs 
for the $\pi^0, \eta$, and $\eta^\prime$ mesons. 
We show the early estimates within the HLS model~\cite{Hayakawa:1995ps,Hayakawa:1996ki} 
and ENJL model~\cite{Bijnens:1995xf}, as well as the non-pole estimate from Ref.~\cite{Melnikov:2003xd}, corresponding with a pointlike coupling at the external vertex, and an on-shell LMD+V TFF at the internal vertex. 
Furthermore, we show the results for the VMD parameterization of Eqs.~(\ref{eq:vmd}, \ref{eq:vmdparam}) for the $\eta$, and $\eta^\prime$ states~\cite{Jegerlehner:2009ry,Nyffeler:2016gnb}, the LMD + V parameterization of Eqs.~(\ref{eq:lmdv}, \ref{eq:lmdvparam}) for the $\pi^0$~\cite{Jegerlehner:2009ry,Nyffeler:2016gnb}, 
the Pad\'e estimate~\cite{Masjuan:2017tvw}, as well as the dispersive~\cite{Hoferichter:2018dmo,Hoferichter:2018kwz} and lattice~\cite{Gerardin:2016cqj,Gerardin:2019vio} 
estimates for the $\pi^0$ pole.
Lastly, we also show the interpolating parameterization of Eq.~(\ref{eq:interpol}) for $\pi^0, \eta$, and $\eta^\prime$ mesons, with parameter values given in Table~\ref{tab:monopole} (for the $\pi^0$ we use the fit to all single virtual TFF data below 9 GeV$^2$, including the recent BESIII data).

\begin{table}[h]
\centering
\begin{tabular*}{0.9 \textwidth}{cccc|c}
\hline\hline
model & $\pi^0$  & $\eta$  &  $\eta^\prime$ & sum  \\
\hline
HLS  \cite{Hayakawa:1995ps,Hayakawa:1996ki} & $5.7 \pm 0.4$ &  & & $8.3 \pm 0.6$ \\
ENJL  \cite{Bijnens:1995xf} & $5.9 \pm 0.9$ &  &  & $8.5 \pm 1.3$  \\
MV \cite{Melnikov:2003xd}  & $7.7$ & $1.8$ & $1.8$ & $11.4 \pm 1.0$  \\
N/JN09~\cite{Jegerlehner:2009ry} &  $7.2 \pm 1.2$ & $1.5 \pm 0.5$ & $1.3 \pm 0.4$ & $9.9 \pm 1.6$ \\ 
Pad\'e \cite{Masjuan:2017tvw} & $6.4 \pm 0.3$ & $1.6\pm0.1$  & $1.4\pm 0.2$ & $9.4\pm0.5$   \\
dispersive \cite{Hoferichter:2018dmo,Hoferichter:2018kwz} & $6.3 \pm 0.3$ & -  & - & -   \\
lattice \cite{Gerardin:2016cqj} &  $6.5 \pm 0.8$ & -  & - & -   \\
lattice \cite{Gerardin:2019vio} &  $6.0 \pm 0.4$ & -  & - & -   \\\hline 
Eq.~(\ref{eq:interpol}) this work &   $5.6 \pm 0.2$ & $1.5 \pm 0.1$  & $ 1.3 \pm 0.1$  & $8.4 \pm 0.4$ \\
\hline\hline
\end{tabular*}
\caption{Pseudoscalar meson pole contributions to $a_\mu$ (in units $10^{-10}$). The N/JN09 estimate~\cite{Jegerlehner:2009ry} employs the LMD+V parameterization $\pi^0$ and VMD parameterization for $\eta, \eta^\prime$. The MV estimate~\cite{Melnikov:2003xd} is a non-pole contribution, corresponding with a pointlike coupling at the external vertex, and an on-shell LMD+V TFF at the internal vertex. The last entry corresponds with the one-parameter interpolating fit  of Eq.~(\ref{eq:interpol}), with parameter values given in Table~\ref{tab:monopole}. }
\label{tab:amupieta}
\end{table}

For the $\pi^0$, the interpolating fit result is about 20\% smaller than the LMD+V result, which is a reflection from the corresponding double virtual TFFs shown in Figs.~\ref{fig:pi0dv} and \ref{fig:pi0dv2}, where it can be seen that the LMD+V result is larger than the interpolating parameterization of Eq.~(\ref{eq:interpol}) over the whole $Q^2$ range.  Note that around 85~\% of the $\pi^0$ contribution to $a_\mu$ arises from the region $Q_1^2, Q_2^2 < 1$~GeV$^2$.   

We note that for the $\eta$ and $\eta^\prime$ states, the comparison between the $a_\mu$ values using the VMD and interpolating parameterizations are consistent in their central values, although the VMD fits underestimate the double virtual TFF  strongly at larger $Q^2$, as can be seen from Fig.~\ref{fig:etapdv}. We checked that the similar central value between the VMD fit and the interpolating fit of Eq.~(\ref{eq:interpol}) arises due to the larger VMD results for the double virtual TFF for $Q^2 < 1$~GeV$^2$, which offsets the smaller results at larger $Q^2$ values, which contribute less to the integrals. 

We also note from Table~\ref{tab:amupieta}, that using the interpolating parameterization of Eq.~(\ref{eq:interpol}), which have only one fit parameter, which is obtained from a fit to the single virtual TFF data, the corresponding uncertainties are significantly reduced. At intermediate and larger $Q^2$ values, it was shown in Fig.~\ref{fig:etapdv} that the interpolating parameterization of Eq.~(\ref{eq:interpol}) is consistent with the recent BaBar data 
for the $\eta^\prime$ double virtual TFF. To validate the parameterization, it will be important to cross-check the double virtual TFF also at lower $Q^2$ values, around and below 1 GeV$^2$, which can be done at BESIII.

\subsubsection{Dispersive representation}

A data driven approach for calculating $a_\m$ based on the analytic properties of the muon's electromagnetic vertex function was presented in 
Ref.~\cite{Pauk:2014rfa}. In this approach,  $a_\m$ is expressed through a dispersive integral over the discontinuity of the muon's electromagnetic vertex function, which in turn can be related to observables. 
Defined as a static limit ($k^2=0$, with $k$ being the photon momentum) of the Pauli form factor $F_2(k^2)$, the anomalous magnetic moment can be extracted from the vertex function by a projection technique as was elaborated in~\cite{Barbieri:1972as, Roskies:1990ki}.  
When analytically continued to complex values of the external photon's virtuality $k^2$, the muon's electromagnetic vertex function possesses branch point singularities joining the physical production thresholds, as is dictated by unitarity. Using Cauchy's integral theorem, the Pauli form factor can be represented as an integral along a closed contour avoiding the cuts and extended to infinity. Assuming that the form factor vanishes uniformly when $k^2$ tends to infinity the contour integral reduces to an integral of the form factor's discontinuity $\mathrm{Disc}_{k^2}F_2(k^2)$ along the cut in the $k^2$-plane starting from the lowest branch point:
\bea
F_2(0)=\frac1{2 \pi i}\int\limits_0^\infty\frac{\mathrm{d} k^2}{k^2}\mathrm{Disc}_{k^2}\,F_2(k^2).
\label{eq:F2disp}
\eea

\begin{figure}[h]
\centering
  \includegraphics[width=5.1cm]{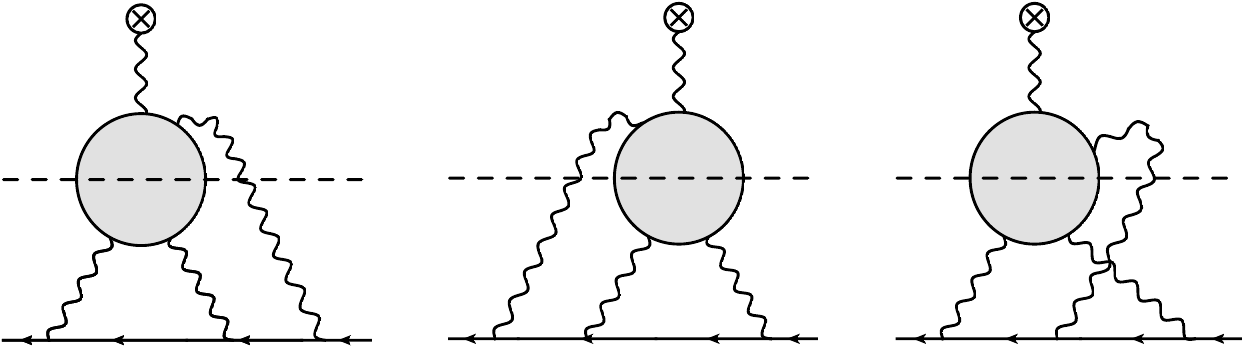}
  \vspace{.3cm}\\
  \includegraphics[width=5.1cm]{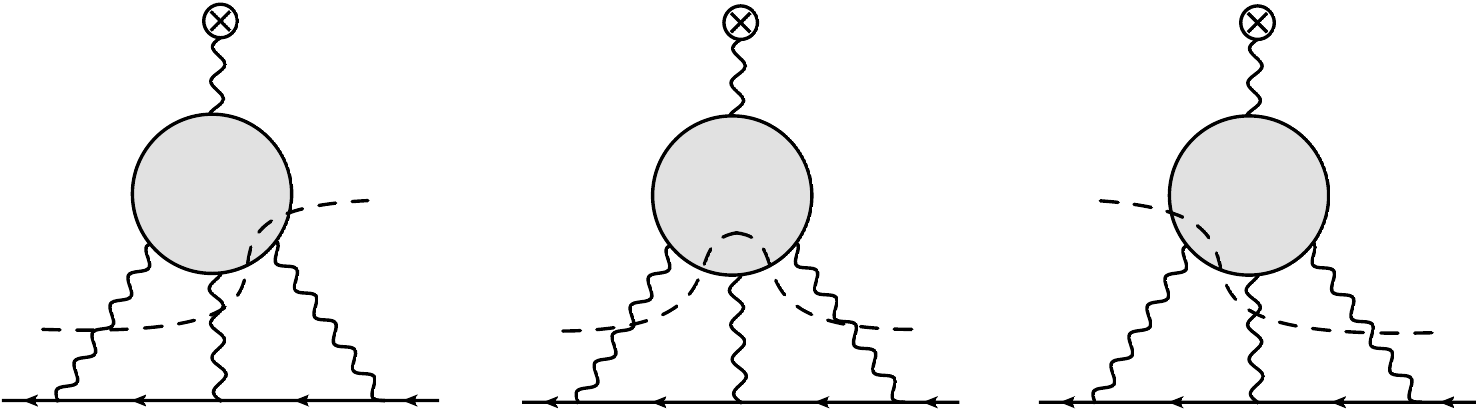}
  \vspace{.3cm}\\
  \includegraphics[width=1.65cm]{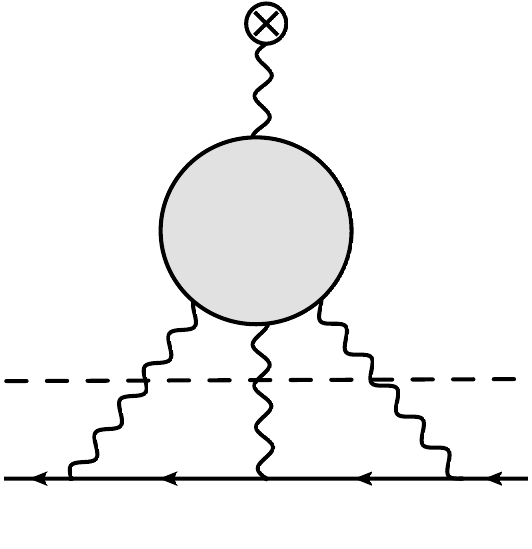}
  \hspace{.2cm}
    \includegraphics[width=1.65cm]{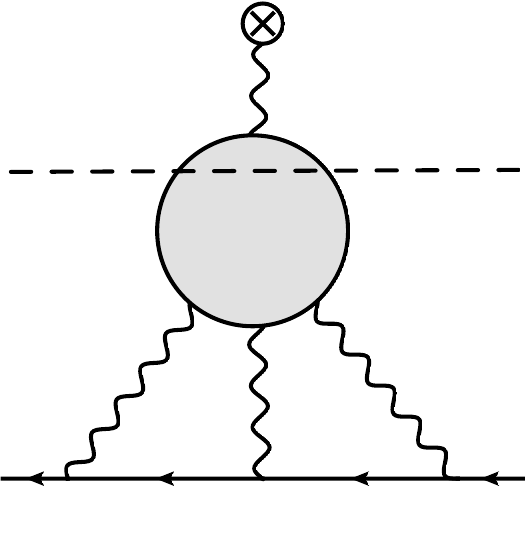}
  \hspace{.2cm}
  \includegraphics[width=1.65cm]{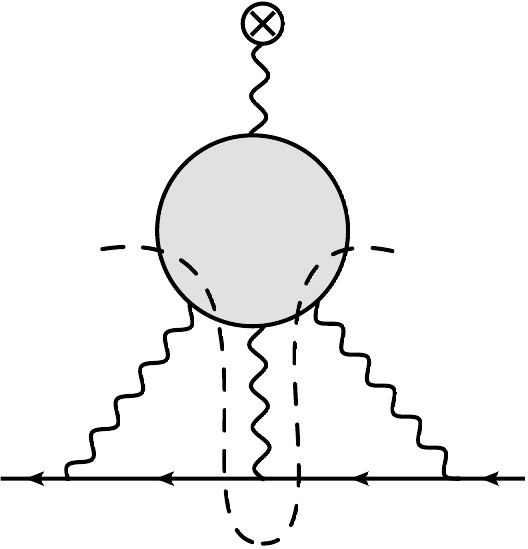}
  \caption{\label{fig:undiagr}
  Unitarity diagrams contributing to the imaginary part of the vertex function. The cut indicates the on-shell intermediate state. }
\end{figure}
 
The branch cuts of the Pauli form factor $F_2(k^2)$ are related to the propagators of virtual particles and non-analyticities of the HLbL tensor. 
The latter possesses two types of discontinuities, the corner (one-photon) and cross (two-photon) cuts. The corner cuts are related to a conversion of a photon to a hadronic state with negative $C$-parity, while the cross cuts are related to a two-photon production of a $C$-even hadronic state. As the dominant contributions originate from the lowest thresholds it is mainly governed by intermediate states including pions. In particular, the lowest threshold in the $C$-odd channel is related to a $\pi^+\pi^-$-pair production and in the $C$-even channel to a $\pi^0$ intermediate state. By virtue of unitarity, these discontinuities are related to amplitudes of two-photon and $e^+e^-$ hadron production processes, which are accessible experimentally.  
Taking into account the analytical structure of the HLbL tensor, the discontinuity in Eq.~(\ref{eq:F2disp}) is obtained as a sum of nine topologically different contributions, which are graphically represented by unitarity diagrams in Fig.~\ref{fig:undiagr}. On a practical level, the contribution of a particular unitarity diagram is obtained by replacing the cut virtual propagators in the two-loop integral by corresponding delta functions, and the cut vertices by their appropriate discontinuities. As an example for the first diagram in Fig.~\ref{fig:undiagr}, it implies:

\begin{eqnarray}
&&\hspace{-0.5cm}\mathrm{Disc}_{}F_2(k^2)=e^6\sum\limits_{\lambda_1,\lambda_2,\lambda_3,\lambda}(-1)^{\lambda+\lambda_1+\lambda_2+\lambda_3}\int\frac{\mathrm{d}^4 q_1}{(2\pi)^4}\int\frac{\mathrm{d}^4 q_2}{(2\pi)^4}\frac{(2 \pi i)\delta (q_2^2)  } {q_1^2 (k-q_1-q_2)^2} \nn \\
&&\times
\frac{L_{\lambda_1\lambda_2\lambda_3\lambda}(p,q_1,k-q_1-q_2,q_2) \mathrm{Disc}_{(k-q_2)^2}\Pi_{\lambda_1\lambda_2\lambda_3\lambda}(q_1,k-q_1-q_2,q_2,k)}{\left( (p+q_1)^2-m^2 \right) \left( (p+k-q_2)^2-m^2 \right)}, \nn \\
\label{eq:discpi}
\end{eqnarray}
where $L_{\lambda_1\lambda_2\lambda_3\lambda}$ denotes the leptonic helicity amplitude,  with $\lambda$ the helicity of the external photon, and $\lambda_i$ ($i = 1,2,3$) the helicities of the internal photons. Furthermore,  the non-perturbative discontinuity function $\mathrm{Disc}_{(k-q_2)^2}\Pi_{\lambda_1\lambda_2\lambda_3\lambda}$ in Eq.~(\ref{eq:discpi}) is directly related to amplitudes of processes $\gamma^\ast\gamma^\ast\to X$ and $\gamma^\ast\to\gamma X$, with $X$ denoting a $C$-even hadronic state, for details see Ref.~\cite{Pauk:2014rfa}. 
 
\begin{figure}[h]
\centering
  \includegraphics[width=6cm]{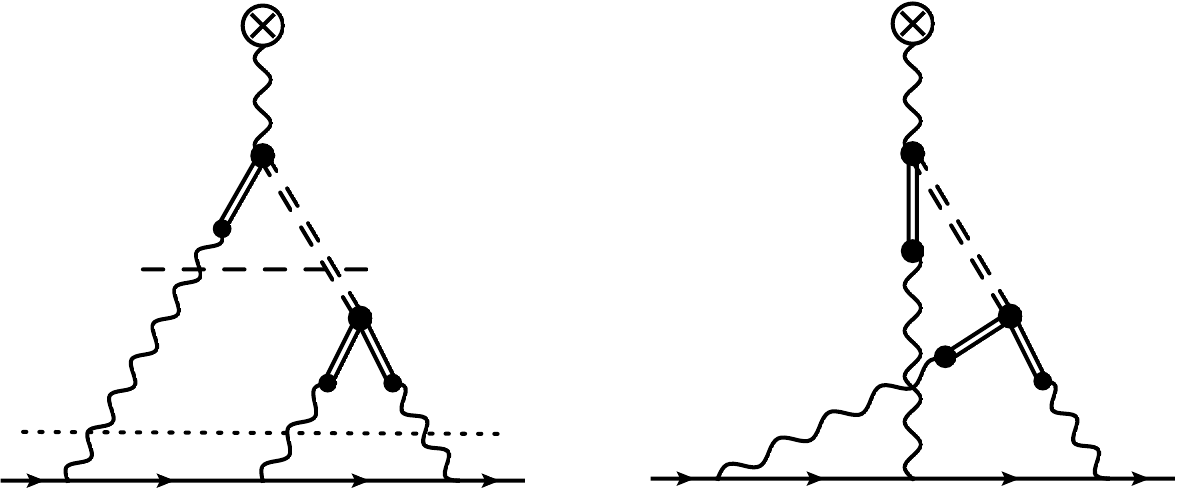}
  \caption{\label{fig:LbLpole}The two topologies of the HLbL contribution to $a_\m$ in the pole approximation and examples of the two-particle (dashed) and three-particle (dotted line) cuts for the first topology. The wavy lines stand for photons, whereas the double-dashed (double-solid) lines stand for pseudoscalar (vector) meson poles.}
\end{figure}

To set up and test the technique for evaluating the phase space and dispersion integrals, Ref.~\cite{Pauk:2014rfa} considered a well-studied approximation for the HLbL amplitude based on the large-$N_c$ limit~\cite{Knecht:2001qf}. In such approximation, the analytic structure of the HLbL amplitude is governed by simple poles. While  in the $C$-even channel it is defined by a pole due to an exchange of the pseudoscalar meson (corresponding to $\pi^0$, $\eta$ and $\eta^\prime$ exchanges), in the $C$-odd channel it is governed by a vector state exchange which can be confronted with the VMD model. As a result the HLbL amplitude is approximated by pole terms of the form:
\begin{eqnarray}
&&\Pi_{\rm pole}(q_1,k-q_1-q_2,q_2,k)=\frac{|F_M(0,0)|^2}{((q_1+q_2)^2-m_M^2)} \nonumber \\
&&\hspace{0.5cm}\times 
\frac{1}{(q_1^2-\Lambda^2)(q_2^2-\Lambda^2)((k-q_1-q_2)^2-\Lambda^2)(k^2-\Lambda^2)},
\end{eqnarray}
plus two additional terms obtained by crossing.
Here $m_M$ and $\Lambda$ denote masses of the pseudoscalar ($M$) and vector ($V$) mesons respectively and $F_M(0,0)$ stands for the pseudoscalar meson transition strength into real photons.
 The resulting analytic structure of the two distinct contributions to the muon's electromagnetic vertex function arising from such pole terms is equivalent to the structure of the two-loop diagrams shown in Fig.~\ref{fig:LbLpole}.
The process of computation is illustrated on the example of the first topology, illustrated by the diagram in the left panel of Fig.~\ref{fig:LbLpole}. The contribution of the second topology (right panel of Fig.~\ref{fig:LbLpole}) has a similar structure and is computed in an analogous way. We can consider the dispersive integral for $F_2(k^2)$ multiplied by $(k^2-\Lambda^2)$, which removes the pole in $k^2$ and its related discontinuity. The remaining discontinuities may be separated in two and three-particle cuts. The two-particle cuts include the $\gamma M$ and $V M$ intermediate states. The three-particle cuts include: 
$\gamma \gamma \gamma$, $\gamma \gamma V$, $\gamma V \gamma$, $V \gamma \gamma$, $\gamma VV$, $V \gamma V$, $VV \gamma$, $VVV$ intermediate states. Graphically they are represented by cuts shown in the left panel of Fig.~\ref{fig:LbLpole} for the case of 
$\gamma M$ (two-particle) and $\gamma \gamma \gamma$ (three-particle) intermediate states. 
 The lowest threshold for the two-particle cut is located at $k^2=m_M^2$ corresponding to $\gamma M$ intermediate state. For the three-particle discontinuity it is $k^2=0$ related to the $\gamma\gamma\gamma$ cut. Thus the dispersion integral has the form
\bea
F_2(0)=\frac{1}{2 \pi i} \int\limits_{m_M^2}^{\infty}\frac{{\rm{d}} k^2}{k^2} {\rm{Disc}}_2 F_2(k^2)
+\frac{1}{2\pi i} \int\limits_{0}^\infty\frac{{\rm{d}} k^2}{k^2} {\rm{Disc}}_3 F_2(k^2),
\eea
with ${\rm{Disc}}_2 F_2(k^2)$ and ${\rm{Disc}}_3 F_2(k^2)$ denoting the sum of two- and three-particle discontinuities. 
The resulting phase-space integrals and the one-loop insertions have been evaluated partially analytically with the subsequent numerical computation in Ref.~\cite{Pauk:2014rfa}, see~\cite{Pauk:2014jza} for some technical details in the case of scalar field theory.

\begin{figure}[h]
\centering
  \includegraphics[width=8.5cm]{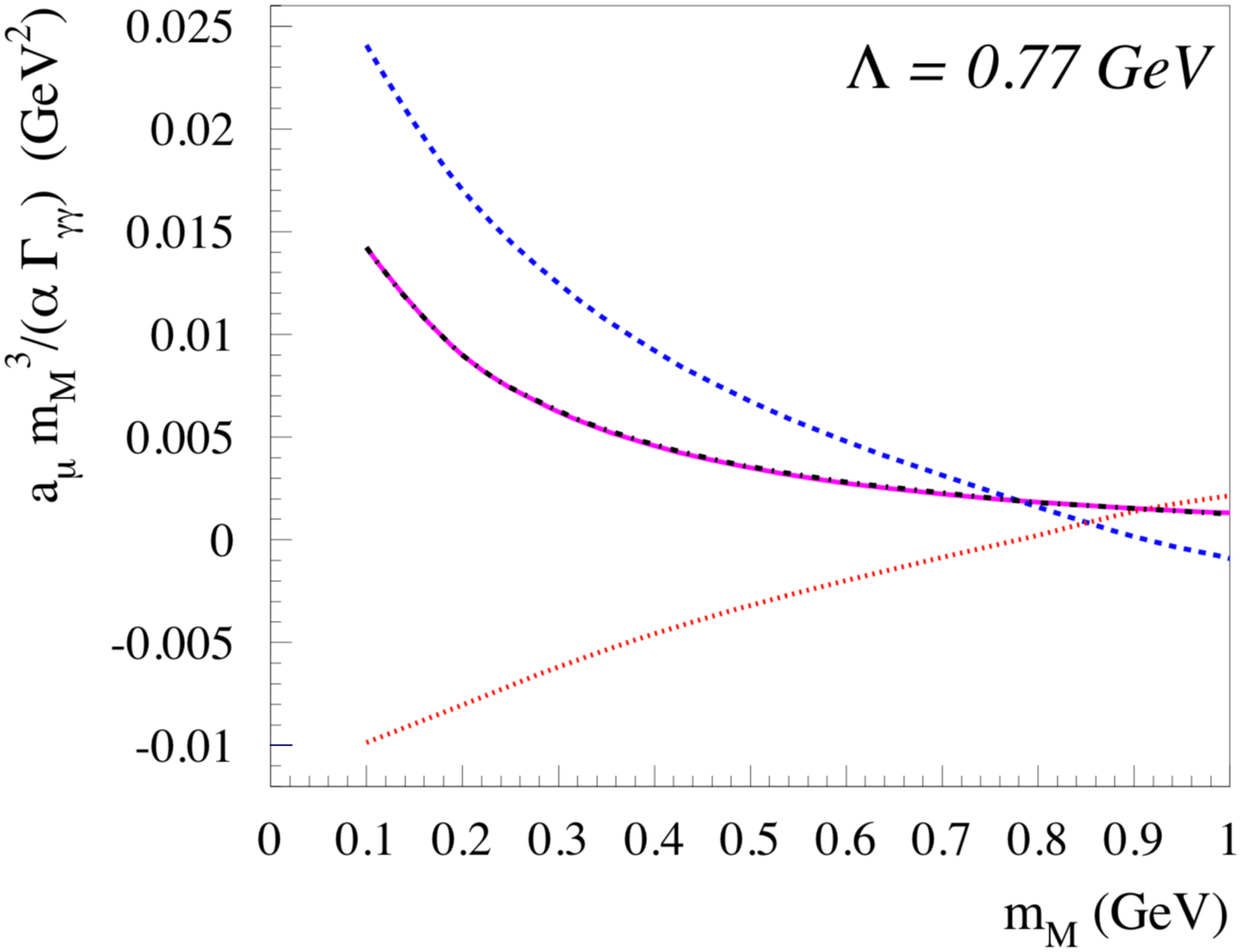}
  \caption{The value of the HLbL pseudoscalar pole contribution due to the diagram of topology (1) (left panel in Fig.~\ref{fig:LbLpole}) 
  to $a_\mu m_M^3/(\alpha \Gamma_{\gamma\gamma})$ depending on the mass of the pseudoscalar meson, with $\Gamma_{\gamma\gamma}$ the two-photon decay width of the pseudoscalar meson. The blue dashed (red dotted) curve represents the contribution of the two (three) particle cuts. Their sum is denoted by the black dashed-dotted curve. The result of the direct evaluation of the two-loop integral is illustrated by the pink solid curve.}
  \label{fig:impartscalar}
\end{figure}

Using the dispersive setup, Ref.~\cite{Pauk:2014rfa} analyzed the dependence of the HLbL contribution to $a_\m$ on the pseudoscalar meson mass $M$. For a test, the result is compared with the 3-dimensional integral representation by evaluating the two-loop integral in Euclidean space, according to the method presented in the section~\ref{sec:3dint}.
The contributions of the two types of discontinuities, their sum and the result of the conventional integration depending on the pseudoscalar meson mass are shown in Fig.~\ref{fig:impartscalar} for the quantity $a_\mu m_M^3/(\alpha \Gamma_{\gamma \gamma})$. When comparing the result obtained by the two different methods one finds an exact agreement confirming the consistency of the adopted procedure.
 
\subsubsection{Schwinger sum rule}

The above dispersive frameworks calculate the hadronic corrections to $a_\mu$ by summing over the individual hadronic channels. 
It was proposed in Ref.~\cite{Hagelstein:2017obr} that it is in principle possible to measure the hadronic contributions to $a_\mu$ directly through use of a sum rule, referred to as Schwinger sum rule~\cite{Schwinger:1975ti,HarunarRashid:1976qz}:
\bea
\label{schwinger}
a_\mu  &=& \frac{m^2}{\pi^2\alpha}
\int_{\nu_0}^\infty \! d \nu \, \left[\frac{\si_{LT}(\nu,Q^2)}{Q}\right]_{Q^2=0} \nonumber \\
&=& \lim_{Q^2\to 0} \frac{8m^2}{Q^2}
 \int_0^{x_0} d x \;  
 [g_1+g_2](x,Q^2)\,,
\eea
where $m$ is the muon mass, and 
where $\sigma_{LT}(\nu,Q^2)$ is the longitudinal-transverse photo-absorption cross section of a polarized virtual photon with energy $\nu$ and spacelike virtuality $Q^2$ on a muon target with transverse spin. The last line in Eq.~(\ref{schwinger}) is the equivalent expression in terms of the muon spin structure functions $g_1$ and $g_2$. A measurement involves the same type of experiment as to access the transverse spin structure of the proton. 

The Schwinger sum rule can be applied 
both for the HVP and HLbL contributions to $a_\mu$, and it was demonstrated in Ref.~\cite{Hagelstein:2017obr} that it exactly reproduces the standard expression for the HVP. 

Although a direct experiment to measure the HLbL contribution to $a_\mu$ using Schwinger's sum rule seems challenging, as it involves subtractions of large QED backgrounds in electron-muon collisions, as well as the measurement of a structure function over a large energy range, the Schwinger sum rule may be very useful as a theoretical tool, as it requires different input as the above dispersive techniques. As an example, it was discussed that the $\pi^0$ HLbL contribution involves the Primakoff mechanism as well as the meson-lepton-lepton coupling, which is fixed from the decay width of pseudoscalar mesons into leptons (i.e. $\pi^0 \to e^+ e^-$ and  $\pi^0 \to \mu^+ \mu^-$). It may thus e.g. provide a constraint on the double-virtual TFF entering these leptonic decays.

\section{Photon-photon fusion processes into two mesons}
\label{sec4}
The next important contribution beyond lightest pseudoscalar mesons ($\pi^0$, $\eta$ and $\eta'$) comes from  pairs of two pseudoscalar mesons, which we consider in this section.

\subsection{Observables for the $\gamma^\ast \gamma^\ast \to MM$ processes}

The two-photon fusion reaction $\gamma^*\gamma^*\to MM$ is a subprocess of the unpolarized doubled tagged process $e^+(k_1)e^{-}(k_2)$ $\to$ $e^{+}(k_1')e^{+}(k_2')$ $M(p_1)M(p_2)$ which is given (in Lorenz gauge) as
\begin{align}\label{ee->eepipi_1}
i\,{\cal M}&=\frac{i\,e^2}{q_1^2 q_2^2}\,[\bar{v}(k_1)\,\gamma_\mu\,v(k_1')]\,[\bar{u}(k_2')\,\gamma_\nu\,u(k_2)]\,H^{\mu\nu}\,,\\
H^{\mu\nu}&=  i\int d^4 x\,e^{-i\,q_1\cdot x} \langle M(p_1)M(p_2)|T(j^{\mu}_{em}(x)
\,j^{\nu}_{em}(0))|0\rangle\,, \nonumber
\end{align}
where the momenta of the leptons  $k^\prime_2$ and $k^\prime_1$ are detected. This corresponds with the kinematical situation where the photons with momenta $q_1$ and $q_2$ have finite spacelike virtualities $q_1^2=-Q_1^2$ and $q_2^2=-Q_2^2$. If only one lepton is detected, then one of the photons is quasi-real, \textit{i.e.} $q_1^2\simeq 0$. The hadron tensor $H^{\mu\nu}$ satisfies gauge invariance, \textit{i.e.} $q_{1\mu}\,H^{\mu\nu}=q_{2\nu}\,H^{\mu\nu}=0$. By contracting the hadronic tensor $H^{\mu\nu}$ with polarization vectors, one defines helicity amplitudes $H_{\lambda_1 \lambda_2}$,
\begin{align}
&\epsilon_\mu(q_1,\lambda_1)\,\epsilon_\nu(q_2,\lambda_2)\,H^{\mu\nu}\equiv e^{i\phi(\lambda_1-\lambda_2)}H_{\lambda_1 \lambda_2}\,.
\end{align}
From the helicity amplitudes it is straightforward to obtain the differential cross sections
\begin{align}\label{Eq:Cross_section1}
&\frac{d \sigma_{TT}}{d \cos\theta}=\frac{\beta_{\pi\pi}(s)}{128\,\pi\,\sqrt{X}}\left(|H_{++}|^2+|H_{+-}|^2\right)\,,\quad\\
&\frac{d \sigma_{TL}}{d \cos\theta}=\frac{\beta_{\pi\pi}(s)}{64\,\pi\,\sqrt{X}} |H_{+0}|^2\,,\quad
\frac{d \sigma_{LT}}{d \cos\theta}=\frac{\beta_{\pi\pi}(s)}{64\,\pi\,\sqrt{X}} |H_{0+}|^2\,,\nonumber\\
&\frac{d \sigma_{LL}}{d \cos\theta}=\frac{\beta_{\pi\pi}(s)}{64\,\pi\,\sqrt{X}} |H_{00}|^2\,,\quad \beta(s)=\frac{2p(s)}{\sqrt{s}}\nonumber
\end{align}
where $p(s)$ is the c.m. momentum of the final state  and $X$ is defined as in Eq.~(\ref{eq:defX}). 
The quantities $\sigma_{TT}$, $\sigma_{TL}$, $\sigma_{LT}$ and $\sigma_{LL}$
enter the cross section for the process $e^+e^{-}$ $\to$ $e^{+}e^{+}\mathrm{X}$ with $\mathrm{X}=MM$ given in Eq.(\ref{eq:gagacross}). The latter sets the convention for the longitudinal polarization vectors and the flux factor.

\subsection{Experimental situation of $\gamma^{\ast} \gamma^{\ast} \to MM$ processes}
\subsubsection{Untagged data}
The study of the two-photon fusion reactions entered a new era after the Belle Collaboration measured exclusive hadronic $\pi^+\pi^-$ \cite{Mori:2007bu}, $\pi^0\pi^0$ \cite{Uehara:2009cka}, $\pi^0\eta$ \cite{Uehara:2009cf}, $K^+K^-$ \cite{Abe:2003vn}, $K_S^0\bar K_S^0$ \cite{Uehara:2013mbo} and $\eta\eta$ \cite{Uehara:2010mq} productions with high statistics, more than a few orders of magnitude higher than any previous measurements \cite{Boyer:1990vu, Behrend:1992hy, Marsiske:1990hx, Antreasyan:1985wx, Aihara:1986qk, Albrecht:1989re, Behrend:1988hw, Althoff:1985yh}. Two-meson production is an ideal reaction for a systematic study of the properties of scalar, i.e. $f_0(500)$, $f_0(980)$, $a_0(980)$, and tensor $f_2(1270)$ and $a_2(1320)$ resonances and allows to extract the two-photon couplings of them.  Experimentally, the two-photon reactions are studied at $e^+e^-$ colliders.

\begin{figure}[!t]
\centering
\includegraphics[width =0.47\textwidth]{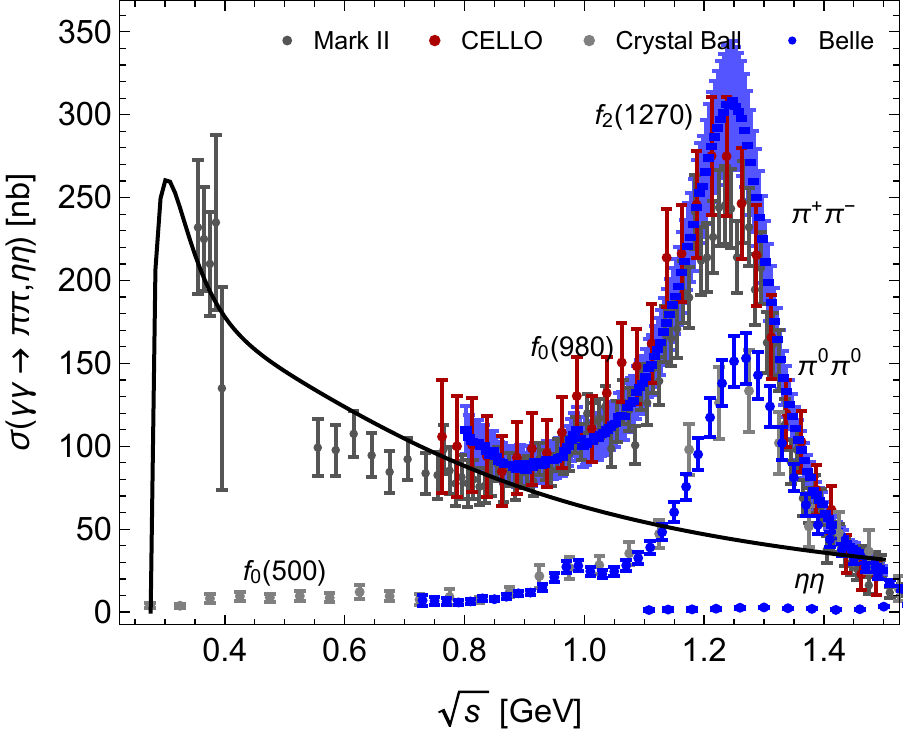}\quad
\includegraphics[width =0.47\textwidth]{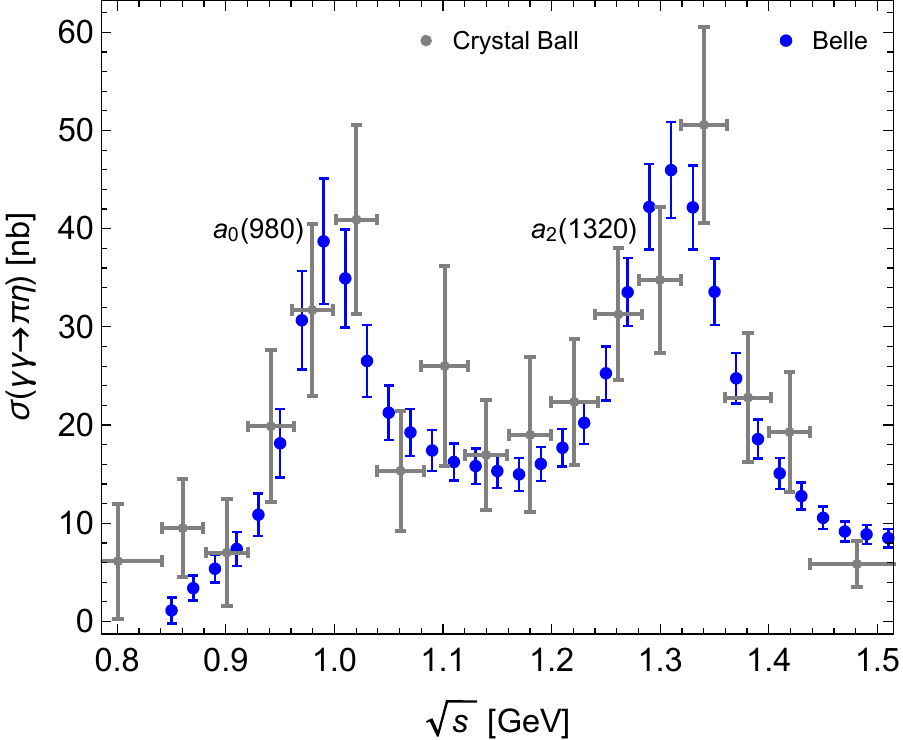}\\
\includegraphics[width =0.47\textwidth]{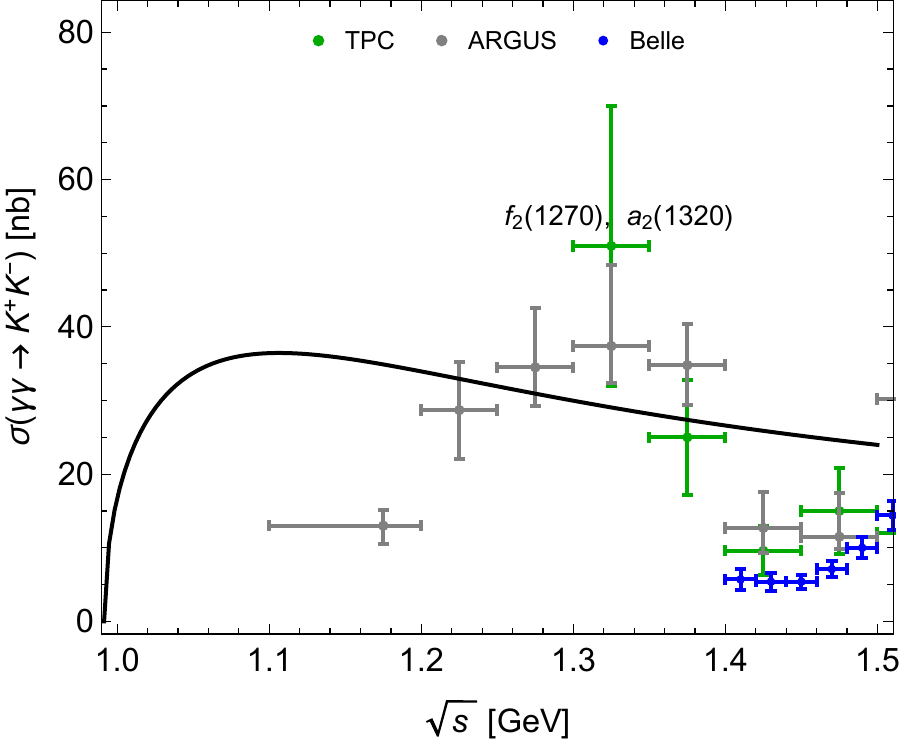}\quad
\includegraphics[width =0.47\textwidth]{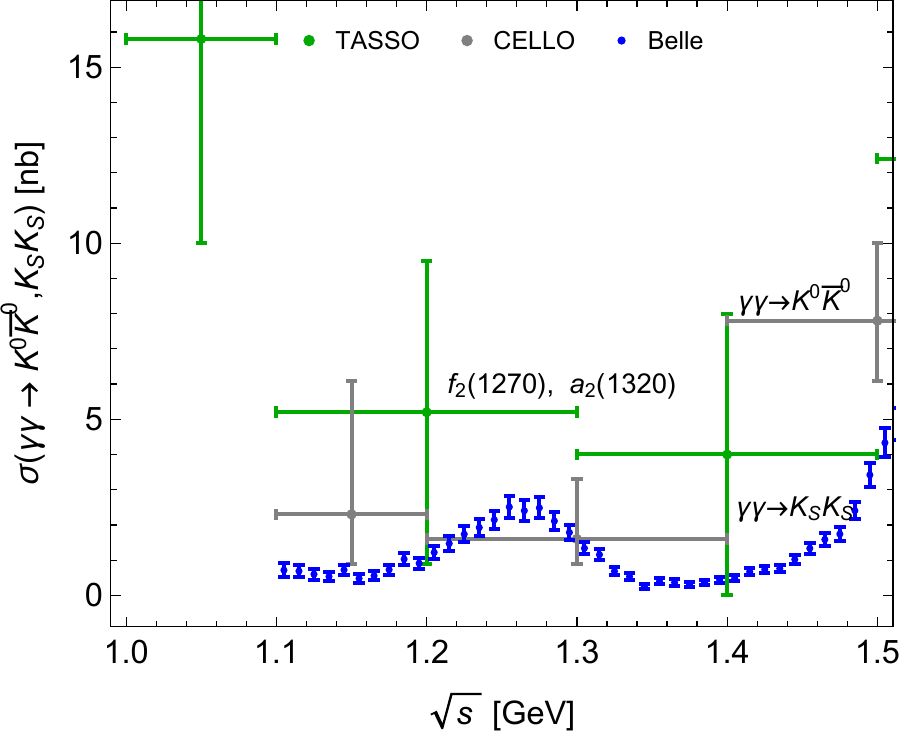}
\caption{Total cross sections for $\gamma\gamma \to \pi^+\pi^-\, (|\cos\theta|<0.6)$, $\gamma\gamma \to\pi^0\pi^0\,(|\cos\theta|<0.8)$, $\gamma\gamma \to \eta\eta\,(|\cos\theta|<1.0)$, $\gamma\gamma \to \pi^0\eta\,(|\cos\theta|<0.9)$,  $\gamma\gamma \to K^+K^-$ (ARGUS $|\cos\theta|<1.0$, TPC and Belle $|\cos\theta|<0.6$), $\gamma\gamma \to K^0\bar{K}^0$ (CELLO $|\cos\theta|<0.7$, TASSO $|\cos\theta|<0.87$) and $\gamma\gamma \to K_S K_S\, (|\cos\theta|<1.0)$.
The data are taken from \cite{Mori:2007bu, Boyer:1990vu, Behrend:1992hy, Marsiske:1990hx, Uehara:2009cka, Uehara:2009cf, Uehara:2010mq, Antreasyan:1985wx, Aihara:1986qk, Albrecht:1989re, Abe:2003vn, Behrend:1988hw, Althoff:1985yh, Uehara:2013mbo}. The Born result for $\gamma\gamma\to \pi^+\pi^-$ and $\gamma\gamma \to K^+K^-$ is shown by the black curves. 
\label{fig:Exp:Q2=0}}
\end{figure}

\begin{figure}[!t]
\centering
\includegraphics[width =0.45\textwidth]{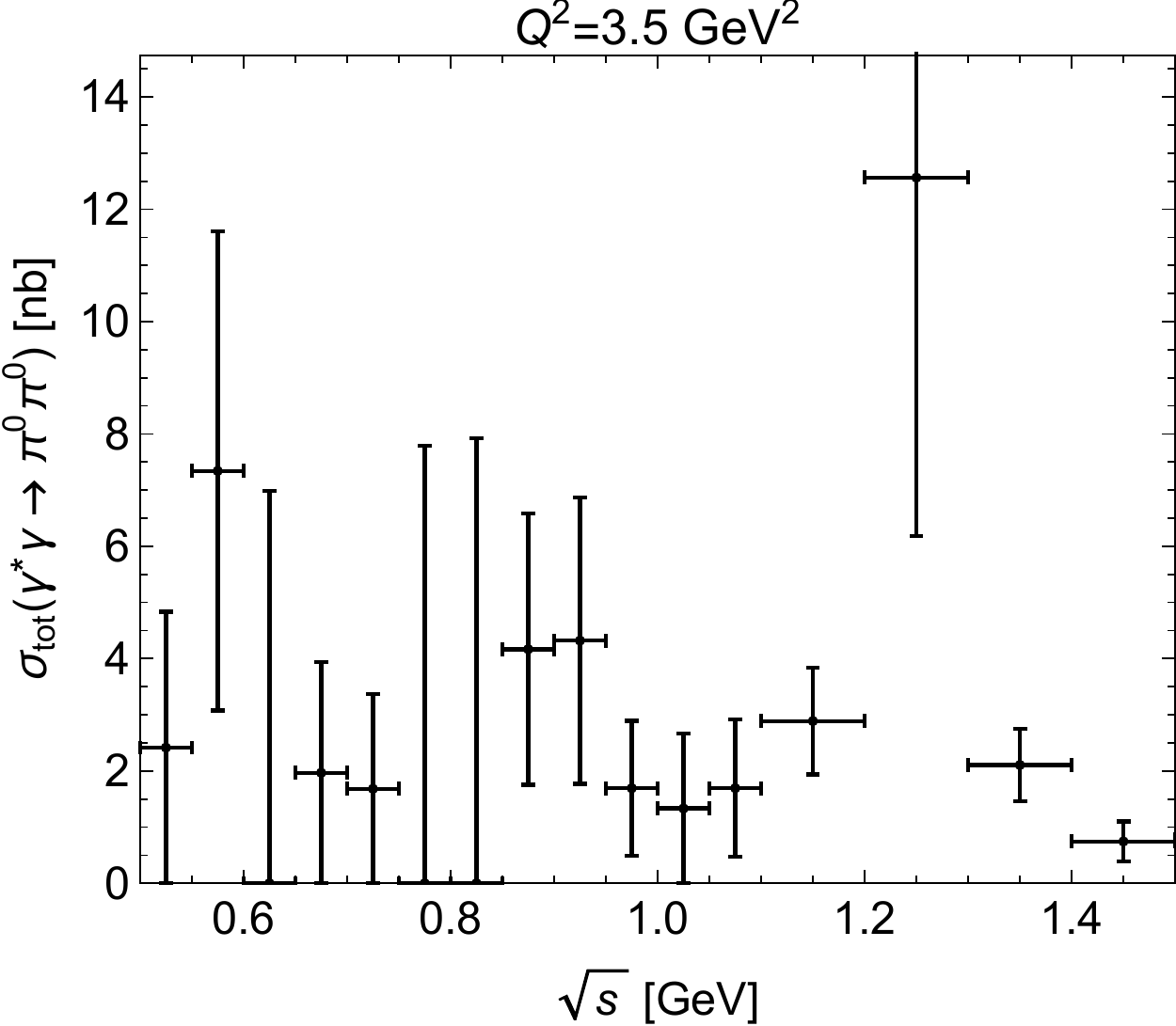}\quad
\includegraphics[width =0.45\textwidth]{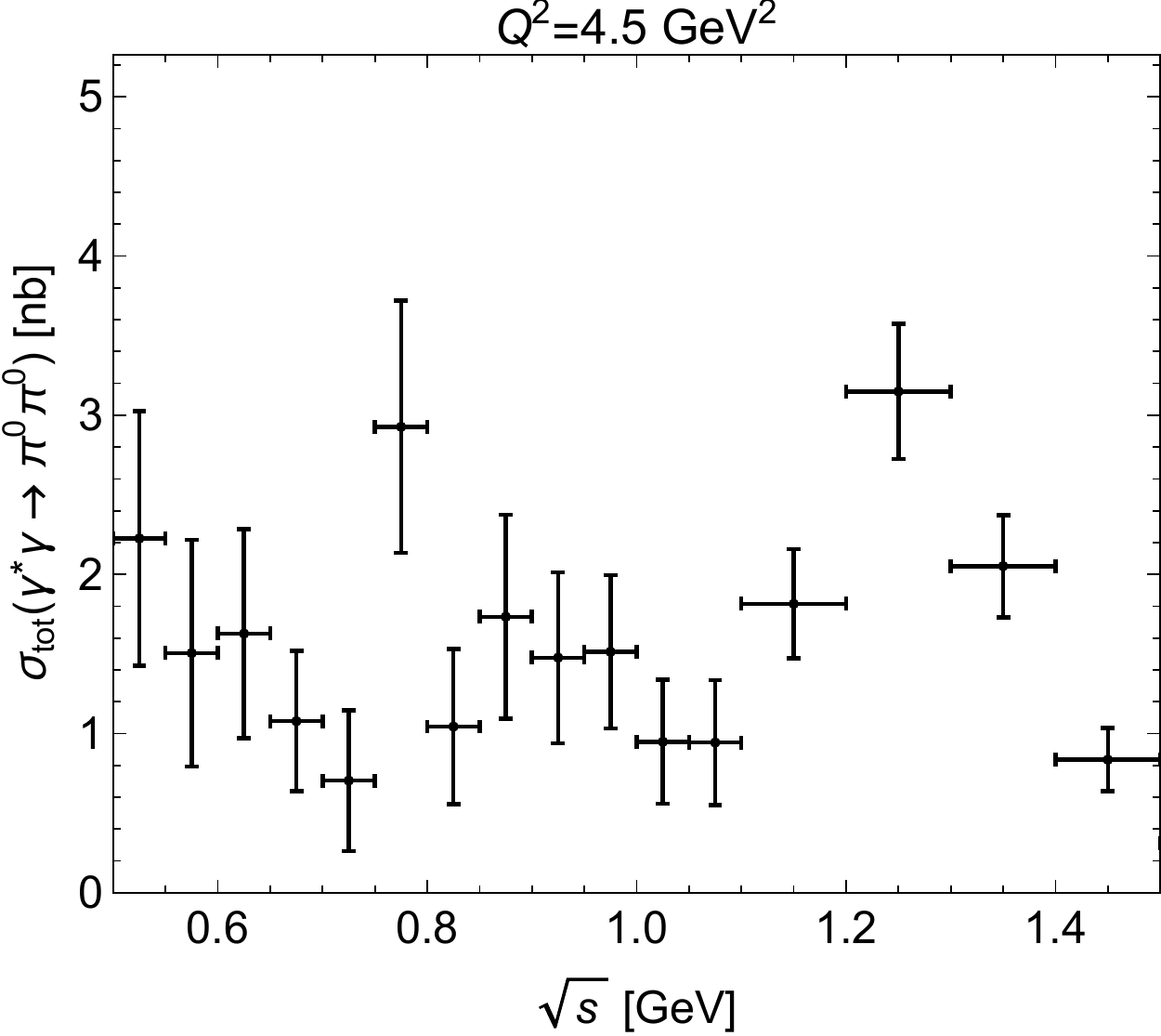}\\ \vspace*{0.25cm}
\includegraphics[width =0.49\textwidth]{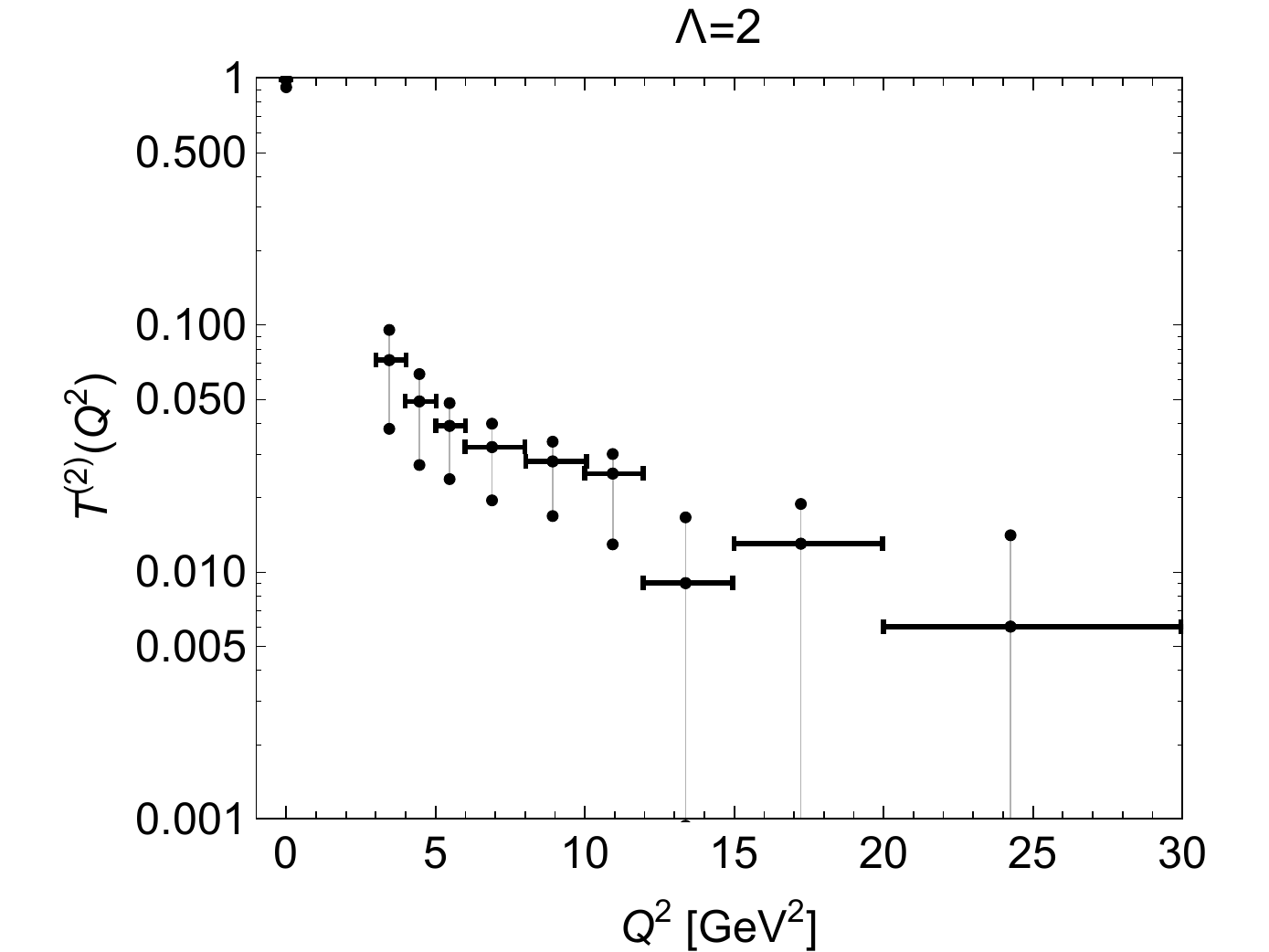}
\includegraphics[width =0.49\textwidth]{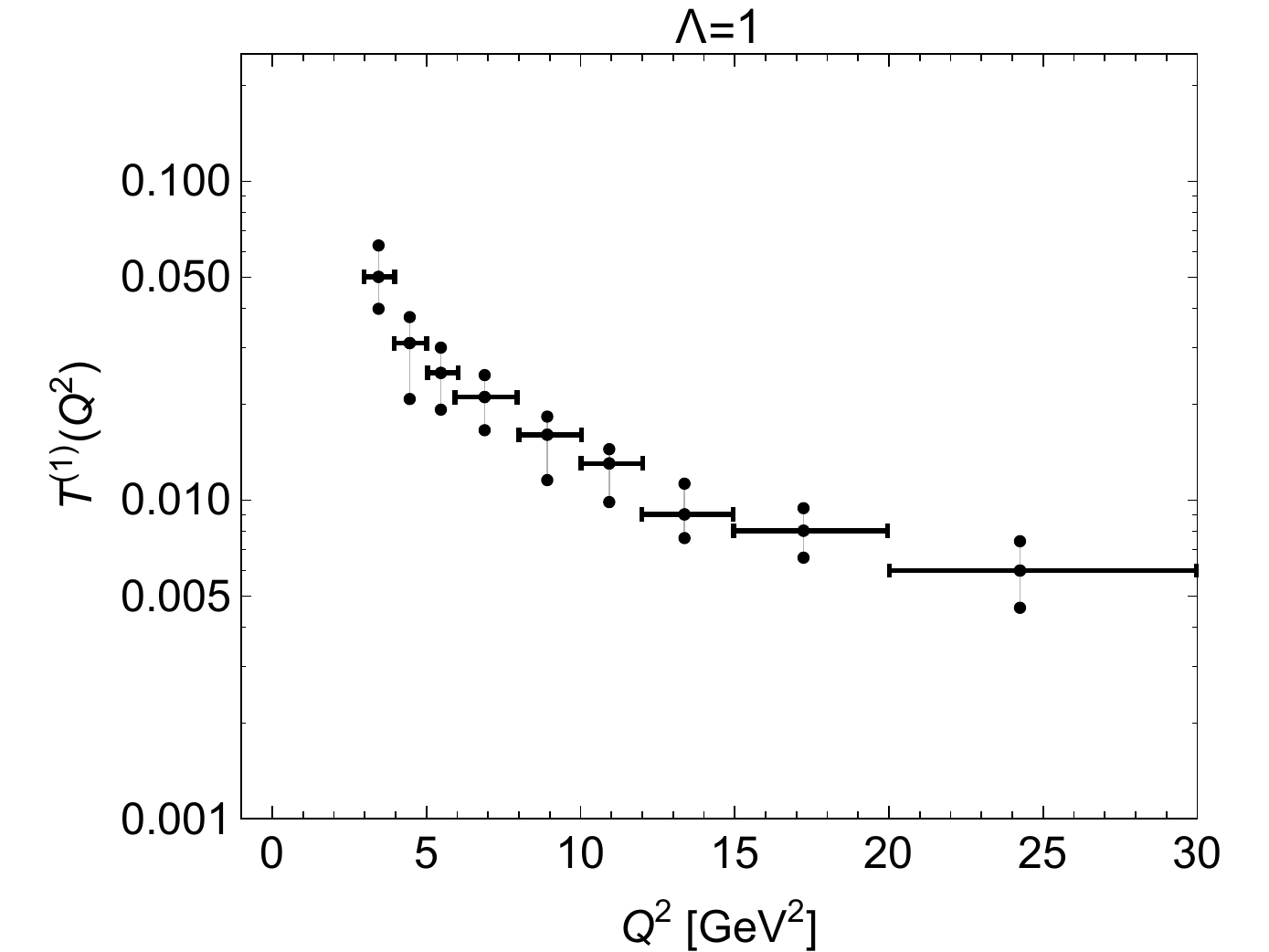}
\caption{Upper panels: Integrated total cross section for $\gamma\gamma^*\to \pi^0\pi^0$ for two bins $Q^2=3.5$ GeV$^2$ ($\epsilon=0.82$) and $Q^2=4.5$ GeV$^2$ ($\epsilon=0.88$) . Lower panels: $Q^2$ dependence of the TFFs of the $f_2(1270)$ resonance for helicity $\Lambda=2$ (left panel) and $\Lambda=1$ (right panel). The experimental data are taken from \cite{Masuda:2015yoh}.
\label{fig:Exp:Q2>0}}
\end{figure}

On Fig.\ref{fig:Exp:Q2=0} (left-panel) we show $\gamma \gamma \to MM$ ($M=\pi,K,\eta$) cross sections which indicate that the dominant contribution to $a_\mu$ calculation comes from the charged two-pion channel. The data from the Belle Collaboration with integrated luminosity 85.9 fb${}^{-1}$ shows a clear signal for $f_0(980)$ and $f_2(1270)$ resonances. The former was detected for the first time in the $\gamma\gamma$ reaction, compared to the previous data from the Mark II \cite{Boyer:1990vu} and CELLO \cite{Behrend:1992hy}. Unfortunately, the measured invariant mass range 0.8\,GeV $<\sqrt{s}<$\,1.5\,GeV did not cover the low and intermediate regions due to the difficulty in discriminating $\mu^{\pm}$ from $\pi^{\pm}$. In that region, there is only data from Mark~II Collaboration \cite{Boyer:1990vu} with large error bars. Measurements of the cross-section at low energy play a significant role for the pion polarizability determination \cite{Dai:2016ytz} and are of prime importance for the HLbL contribution to $a_\mu$. For the neutral channel, the Crystal Ball Collaboration with the integrated luminosity of 97\,pb${}^{-1}$ observed a broad bump around $500$\,MeV, which corresponds to $f_0(500)$, and a clear resonance at higher energies that corresponds to $f_2(1270)$ \cite{Marsiske:1990hx}. Besides, at intermediate energy, there is a hint of the $f_0(980)$ resonance, which was later confirmed by the high precision Belle data, where $f_0(980)$ shows up as a clear signal \cite{Uehara:2009cka}. The cross sections for $\gamma \gamma \to \eta \eta$ have been obtained in \cite{Uehara:2010mq}, which are very much suppressed ($\sim 1-4$\,nb) compared to the other channels. On Fig.\,\ref{fig:Exp:Q2=0} (right-panel) we show $\gamma \gamma \to \pi \eta$ cross-section, where data indicate two clear peaks due to $a_0(980)$ and $a_2(1320)$ resonances. The latest measurement was performed by the Belle Collaboration \cite{Uehara:2009cf}, with an integrated luminosity of 223\,fb${}^{-1}$. We expect this channel to be the second most important multi-meson contribution to $a_\mu$ after the two-pion channel. Regarding the two kaon channel shown in Fig.\,\ref{fig:Exp:Q2=0} (bottom panels), the data from Belle is available for $\gamma\gamma \to K_S K_S$ \cite{Uehara:2013mbo} and $\gamma\gamma \to K^+ K^-$ \cite{Abe:2003vn}. The latter is the dominant channel. However, the data do not cover the region around the $f_2(1270)$ and $a_2(1320)$ interference and starts only from 1.4\,GeV. The existing data from ARGUS \cite{Albrecht:1989re} indicate the drastic reduction of the Born term contribution due to re-scattering effects in the low energy region~\cite{Oller:1997yg,Danilkin:2012ua,Dai:2014zta} and become sizable only around 1.3\,GeV. High statistics data from the BESIII Collaboration for $\gamma\gamma \to K^+K^-$ in the region 1.0\,GeV $< \sqrt{s}< 2$\,GeV currently is under analysis.

\subsubsection{Single tagged data}
A single-tagged measurement of the two-photon production of $\pi^0$ pairs has been performed by the Belle Collaboration~\cite{Masuda:2015yoh} based on a data sample of 759\,fb${}^{-1}$. The kinematic range of the data is $0.5\,\text{GeV} < \sqrt{s} <2.1 \,\text{GeV}$ and $|cos\theta| < 1.0$ in the $\gamma\gamma^*$ center-of-mass system. The differential and integrated cross sections were measured in the $Q^2$ range $3\, \text{GeV}^2  < Q^2 < 30\,\text{GeV}^2$. The lowest two $Q^2$ bins with $3.5$\,GeV$^2$ and $4.5$\,GeV$^2$ are shown in Fig. \ref{fig:Exp:Q2>0}. Though the uncertainties are quite large, the peak corresponding to the $f_2(1270)$ resonance is evident. Based on these data the transition form factors (TFFs) of the $f_2(1270)$ resonance with the helicity-0, -1, and -2 were extracted separately by parametrizing partial-wave amplitudes in a Breit-Wigner form supplemented with a background. In Fig.\,\ref{fig:Exp:Q2>0} (lower panels) we show helicity -1, and -2 TFFs, which are supposed to be dominant at low $Q^2$. The effective parametrization of the $f_2(1270)$ TFFs allows for an independent look at the $f_2(1270)$ contribution into $a_\mu$ using a narrow resonance description. The single-tag two-photon measurement of $\pi^0\pi^0$ was followed by a measurement of the cross section for $K_S^0 K_S^0$ in the same  $Q^2$ range. Current statistics allowed to estimate TFFs for $f_2'(1525)$ for the first time.

Currently, single tagged BESIII data for $\pi^+\pi^-$ with the integrated luminosity of 7.51\,fb${}^{-1}$ in the range $0.2\,\text{GeV}^2 <Q^2<2.0\,\text{GeV}^2$ is under analysis~\cite{Redmer:2017fhg}. The kinematic range of the data will cover the threshold region up to 2\,GeV.

\subsection{Theoretical description of $\gamma^{\ast} \gamma^{\ast} \to MM$ processes}

For simplicity we discuss the case of two pions in the final state assuming that the generalization for the non-equal masses (like $\pi\eta$) can be performed straightforwardly. Following the work of of Jacob and Wick \cite{Jacob:1959at}, the hadronic tensor $H^{\mu\nu}$ contracted with polarization vectors can be decomposed into partial wave amplitudes characterized by the total angular momentum $J$ and helicities $\lambda_{1,2}$
\begin{align}\label{p.w.expansion}
H_{\lambda_1 \lambda_2}= \sum_{J}(2J+1)\,h^{(J)}_{\lambda_1\lambda_2}(s,Q_1^2,Q_2^2)\,d_{\Lambda,0}^{(J)}(\theta)\,,
\end{align}
where $\Lambda=\lambda_1-\lambda_2$, $d_{\Lambda,0}^{(J)}(\theta)$ is a Wigner rotation function and $\theta$ is the c.m. scattering angle. The two-photon initial state implies that the $C$-parity quantum number of the final particles should always be positive. For the case of two real photons, this implies that the partial-wave expansion involves only even $J$ and positive parity, independent on the two Goldstone bosons in the final state. However, when one or two photons are virtual, Bose symmetry of the two initial photons is broken and $C$-parity quantum number excludes odd partial waves only in the case of two pions due to their Bose symmetry. For the case of $\pi\eta$, however, one can have an odd partial wave. The p-wave would correspond to exotic quantum numbers $J^{PC}=1^{-+}$.

\subsubsection{Dispersion relations}
In order to write down dispersion relations (DRs) for the $\gamma^*\gamma^* \to \pi\pi$ process, one has to identify all the kinematic constraints of the p.w. helicity amplitudes.  While for the case of the on-shell photons, helicity amplitudes are not correlated at any kinematic point, and one can write the corresponding dispersion relations by just properly accounting for so-called barrier factors \cite{Morgan:1987gv} this is no longer the case for $Q_i^2 \neq 0$. For instance, for the S-wave it holds \cite{Morgan:1987gv,Moussallam:2013una, Colangelo:2017fiz}
\begin{align}\label{BarierFactors_Swave}
&\bar{h}^{(0)}_{++}(s) \simeq s\,,\nonumber \\
&\bar{h}^{(0)}_{++}(s,Q^2) \simeq s+Q^2\,,  \\
&\bar h^{(0)}_{++}(s,Q_1^2,Q_2^2)\pm \bar h^{(0)}_{00}(s,Q_1^2,Q_2^2) \simeq s+(Q_1\mp Q_2)^2\,, \nonumber
\end{align}
where $Q_i \equiv \sqrt{Q_i^2}$ ($i=1,2$) and
\begin{align}
\bar{h}^{(J)}_{\lambda_1\lambda_2}\equiv h^{(J)}_{\lambda_1\lambda_2}-h^{(J), Born}_{\lambda_1\lambda_2}\,,
\end{align}
stand for the Born subtracted amplitudes. While the first two constraints of Eq.(\ref{BarierFactors_Swave}) are required by the soft-photon theorem \cite{Low:1958sn}, the constraint for the double-virtual case is not that straightforward. Eq.(\ref{BarierFactors_Swave}) can be most easily seen by expanding the hadron tensor $H^{\mu\nu}$ in terms of a complete set of invariant amplitudes \cite{Colangelo:2015ama, Tarrach:1975tu, Drechsel:1997xv}. By construction, invariant amplitudes are free from kinematic singularities or constraints and expected to satisfy a Mandelstam’s dispersion representation. As a second step, one can express the invariant amplitudes in terms of the p.w. helicity amplitudes in order to pin down the kinematic correlations. As a result, for the S-wave the kinematically uncorrelated amplitudes can be written as
\begin{align}\label{new_pw_amplitudes}
&\bar h^{(0)}_{i=1,2}(s,Q_1^2,Q_2^2)=\frac{\bar h^{(0)}_{++}(s,Q_1^2,Q_2^2)\pm \bar h^{(0)}_{00}(s,Q_1^2,Q_2^2)}{s+(Q_1\mp Q_2)^2}\,.
\end{align}
In general, for $J \neq 0$, it is a non-trivial task to derive these transformations \cite{Lutz:2011xc, Gasparyan:2010xz}. In \cite{Danilkin:2018qfn} for the single virtual case, the kinematically unconstrained basis of the partial wave amplitudes were derived, which paves the way for the double-virtual case \cite{Danilkin:X}. After identifying all the kinematic constraints, one can write a dispersion relation, the solution of which will unitarize the p.w. amplitudes. It holds
\begin{align}\label{DR_1}
h^{(J)}_{I,i}(s)
&=h^{(J),Born}_{I,i}(s)+\int_L \frac{d s'}{\pi}\frac{\text{Disc}\,\bar{h}^{(J)}_{I,i}(s')}{s'-s}+\int_{4m_\pi^2}^{\infty} \frac{ds'}{\pi}\frac{\text{Disc}\,\bar{h}^{(J)}_{I,i}(s')}{s'-s}\nonumber\\
&\equiv \Delta^{(J)}_{I,i}(s)+\int_{4m_\pi^2}^{\infty} \frac{d s'}{\pi}\frac{\text{Disc}\,h^{(J)}_{I,i}(s')}{s'-s}\,,
\end{align}
where all left-hand cut singularities (including Born contribution) represented by $\Delta^{(J)}_{I,i}(s)$, $I$ is the isospin (working in the isospin limit). The unitarity relation for $s \geq 4\,m_\pi^2$ can be written as
\begin{align}\label{h_Unitarity}
&\text{Disc}\,h^{(J)}_{I,i}=t^{(J)*}_{I}\,\rho\,
h^{(J)}_{I,i}, \quad \rho=\frac{\beta_{\pi\pi}(s)}{16\,\pi}\,\theta(s-4m_\pi^2)\,,
\end{align}
where $\rho(s)$ is a two-body phase space factor and $t^{(J)}_{I}(s)$ is the hadronic scattering amplitude, which is normalized as $\text{Im }(t^{(J)}_{I})^{-1}=-\rho$. Since the intermediate states with two photons are proportional to $e^4$, they are suppressed and a solution for the $\gamma^*\gamma^* \to \pi\pi$ process is fully determined by the hadronic rescattering part. For the energy region above 1 GeV, it is necessary to take into account the inelasticity. The first relevant inelastic channel is $K\bar{K}$ which is required to capture the dynamics of the $f_0(980)$ scalar meson. For the coupled-channel case, the unitarity relation can be written in matrix form. See \cite{GarciaMartin:2010cw, Danilkin:2012ua, Danilkin:2017lyn} for more details.

The solution to Eq. (\ref{DR_1}) is given by the well known Muskhelishvili-Omn\`es (MO) method for treating the final-state interactions \cite{Omnes:1958hv}. As it was pointed out in \cite{GarciaMartin:2010cw}, the usual method \cite{Omnes:1958hv} is based on writing a dispersion relation for the function,
\begin{equation}\label{Sol1}
(\Omega^{(J)}_{I}(s))^{-1}(h^{(J)}_{I,i}(s)-\Delta^{(J)}_{I,i}(s))\,,
\end{equation}
which by construction, has only a right-hand cut. In Eq.(\ref{Sol1}), $\Omega^{(J)}_{I}$ is the Omn\`es function, which has only the right-hand cut
\begin{equation}\label{Omnes_CC_Unitarity}
\text{Disc}\,\Omega^{(J)}_{I}=t^{(J)}_{I}\rho\,\Omega^{(J)*}_{I}\,,
\end{equation}
and completely given in terms of the hadronic phase shifts. However, it is useful to treat differently the model independent left-hand cut associated with the QED Born term and the remaining part, which can be approximated by resonance exchanges \cite{GarciaMartin:2010cw}. Instead of (\ref{Sol1}), one can consider
\begin{equation}\label{Sol2}
(\Omega^{(J)}_{I}(s))^{-1}(h^{(J)}_{I,i}(s)-h^{(J), Born}_{I,i}(s))\,,
\end{equation}
which now contains both right- and left-hand cuts. This leads to the following dispersion relation,
\begin{align}\label{Rescattring:general}
&{h}^{(J)}_{I,i}(s)= {h}^{(J),Born}_{I,i}(s)+ 
\Omega^{(J)}_{I}(s) \bigg[
-\int_{4m_\pi^2}^{\infty}\frac{ds'}{\pi}\,\frac{\text{Disc}\,(\Omega^{(J)}_{I}(s'))^{-1}\,{h}^{(J),Born}_{I,i}(s')}{s'-s}
\nonumber \\
& \hspace{2cm}+
\int_{-\infty}^{s_L}
\frac{ds'}{\pi}\,\frac{(\Omega^{(J)}_{I}(s'))^{-1}\,\text{Disc}\,{\bar{h}}^{(J)}_{I,i}(s')}{s'-s}
\bigg]\,,
\end{align}
where $s_L$ defines the position of the left-hand singularity nearest to the physical region due to non-Born intermediate $t$ - and $u$- channel left-hand cuts. This particular separation is the most effective, because only $\text{Disc}\,{\bar{h}}^{(J)}_{I,i}(s)$ is required as input. If one uses the effective field theory Lagrangian to model the left-hand cuts, it will result in a bad high energy behavior and limit the range of applicability of the dispersion result. However, the discontinuity along the left-hand cut is typically asymptotically bounded at high energy and does not have any polynomial ambiguities \cite{GarciaMartin:2010cw}. Besides, for the finite virtuality $\text{Disc}\,{\bar{h}}^{(J)}_{I,i}(s)$ is unique for the vector pole contribution, which allows taking into account an off-shellness of the photon by multiplying the vertex by the TFF.

We want to emphasize that the gauge invariance property, which results in the constraints of Eq.(\ref{BarierFactors_Swave}), allows one to write an unsubtracted dispersion relation and have a predictive power. Consequently, one can predict pion (generalized) polarizabilities, which encode the two-pion rescattering at low energy and serve as a check of the low energy limit. It holds 
\begin{align}
\frac{1}{2\pi m_\pi}\,\bar{h}^{(0)}_{++}=(\alpha_1-\beta_1)\,(s+Q^2)+(\alpha_2-\beta_2)\,\frac{(s+Q^2)^2}{12}+...
\end{align}
where $(\alpha_1-\beta_1)$ and $(\alpha_2-\beta_2)$ are dipole and quadrupole polarizabilities, respectively.

\subsubsection{Hadronic input}
\begin{figure}[!t]
\centering
\includegraphics[width =0.50\textwidth]{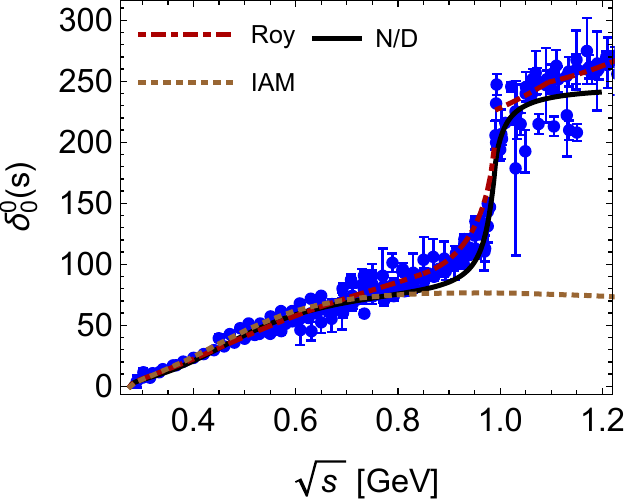}
\includegraphics[width =0.48\textwidth]{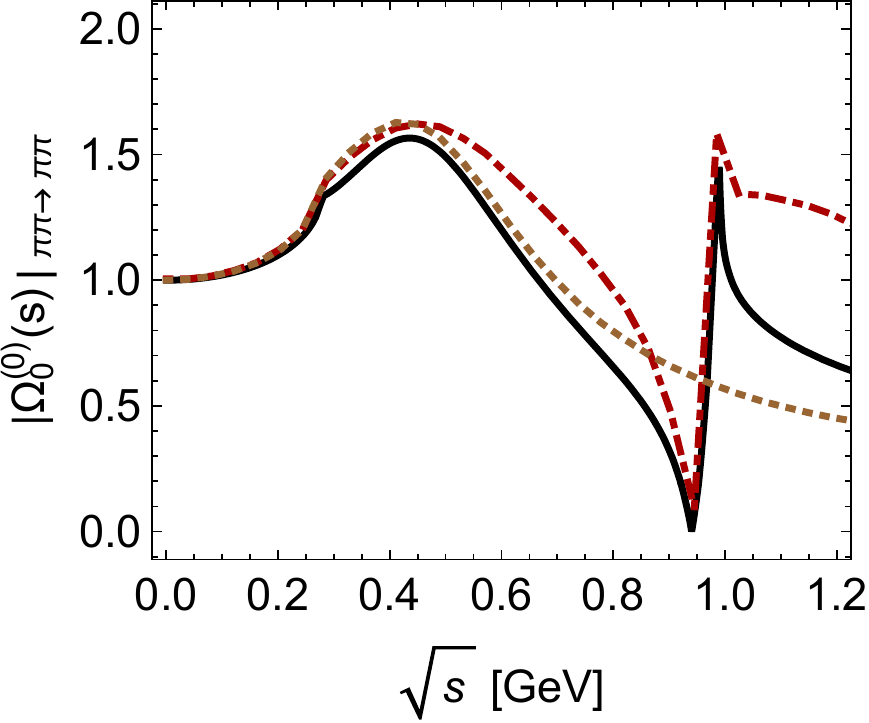}
\caption{Left panel: S-wave isoscalar $\pi\pi$ phase shifts from the single channel IAM (dotted brown line), Roy analysis (dot-dashed red line) and coupled channel $N/D$ approach (solid black line). Right panel: Modulus of the corresponding Omn\`es functions calculated using (\ref{OmenesPhaseShift}) and IAM phase shift (dotted brown line), the $\pi\pi \to \pi\pi$ element of the coupled-channel Omn\`es solution of Eq.(\ref{Omnes_CC_2}) with input from Roy analysis phase shift (dot-dashed red line) and Omn\`es function from the coupled-channel $N/D$ approach (solid black line) as described in the text.
\label{fig:Omnes}}
\end{figure}

For the single-channel (elastic) description of the hadronic rescattering one can use the Omn\`es functions given in terms of the corresponding experimental (or Roy analyses) phase shifts,
\begin{equation}\label{OmenesPhaseShift}
\Omega_I^{(J)}(s)=\exp\left(\frac{s}{\pi}\int_{4m_\pi^2}^{\infty} \frac{d s'}{s'}\frac{\delta_{I}^{(J)}(s')}{s'-s}\right)\,,
\end{equation}
with the plausible guesses concerning its high-energy behavior. For the S-wave isoscalar channel (which brings the main contribution to $a_\mu$), the single-channel description covers only the $f_0(500)$ resonance region. It can be produced using as input the phase shift from the single-channel inverse-amplitude method (IAM) \cite{GomezNicola:2007qj}, as it was done in \cite{Colangelo:2017fiz,Colangelo:2017qdm}. In Fig.\,\ref{fig:Omnes} we show $\pi\pi$ S-wave phase shifts from the IAM compared to Madrid/Krakow Roy-equation analyses and experimental data.

For a proper description of the $f_0(980)$ resonance one has to incorporate a couple-channel  $\{\pi\pi , K\bar{K}\}$ dynamics
\begin{equation}\label{Omnes_CC}
\Omega_0^{(0)}(s)=\left(
\begin{array}{cc}
\Omega(s)_{\pi\pi \to\pi\pi} & \Omega(s)_{\pi\pi \to K\bar{K}}\\
\Omega(s)_{K\bar{K} \to \pi\pi} & \Omega(s)_{K\bar{K} \to K\bar{K}}
\end{array}
\right)\,.
\end{equation}
Unlike the single-channel case, there is no analytic Omn\`es representation for two or more channels. However, one can obtain a numeric solution. There are several ways of accomplishing this. For two-dimensional matrices one can write a dispersion relation based on Eq.(\ref{Omnes_CC_Unitarity}) \cite{Donoghue:1990xh, Moussallam:1999aq}
\begin{equation}\label{Omnes_CC_2}
\Omega_0^{(0)}(s)=\int_{4m_\pi^2}^{\infty}\frac{ds'}{\pi}\frac{t^{(0)}_{0}(s')\rho(s')\,\Omega^{(0)*}_{0}(s')}{s'-s},\quad \Omega(0)=1,
\end{equation}
with the input entering the $t$-matrix: the $\pi\pi$ phase shift, the modulus and phase of the $\pi\pi \to K\bar{K}$ amplitude. They can be taken from the Roy (Roy-Steiner) analyses up to 1.3 GeV \cite{Ananthanarayan:2000ht, GarciaMartin:2011cn, Buettiker:2003pp, Pelaez:2018qny} and above that point the phase typically chosen to tend to $2\pi$ \cite{GarciaMartin:2010cw, Daub:2015xja}. The latter imposes $1/s$ asymptotic behavior of $\Omega_0^{(0)}(s)$. A complementary approach is to use a dispersive summation scheme \cite{Gasparyan:2010xz, Danilkin:2010xd} which is based on the $N/D$ ansatz \cite{Chew:1960iv}. With the input from the left-hand cuts which can be presented in a model-independent form as an expansion in a suitably constructed conformal mapping variable, the set of coupled-channel integral equations for the $N$-function can be solved numerically. After solving the linear integral equation for $N(s)$, the $D$-function (the inverse of the Omn\`es function) is restored. The unknown conformal mapping expansion coefficients can be determined from fitting to Roy analyses for $\pi\pi \to \pi\pi, K\bar{K}$ \cite{Ananthanarayan:2000ht, GarciaMartin:2011cn, Buettiker:2003pp, Pelaez:2018qny} and existing experimental data for these channels. Both approaches are data-driven in the low energy, but different in their high energy assumptions. In particularly, in the second approach $\text{Disc}\,(\Omega^{(0)}_{0}(s))^{-1}$ is asymptotically bounded at high energies making the dispersive integral (\ref{Rescattring:general}) numerically stable. The modulus of the two solutions for the $\pi\pi$ S-wave  Omn\`es function is shown in Fig.\,\ref{fig:Omnes} compared to a single-channel version of it. The partial waves beyond S- and D-waves are typically approximated by the Born terms.

\subsubsection{Left-hand cuts}
The most important left-hand cut contribution comes from the pion pole. For the case of virtual photons, the off-shellness of the photon can be accounted through the pion vector form factor, $f_{\pi}(Q^2)$, which is determined as a matrix element of the EM current between the two on-shell pions
\begin{equation}
\langle \pi^+(p') | j_{\mu }(0) | \pi^+(p)\rangle =e \left(p+p'\right)_{\mu}f_{\pi }\left((p'-p)^2\right). \nonumber
\end{equation}
It was shown in \cite{Colangelo:2015ama} (see also \cite{Fearing:1996gs}) using the fixed-$s$ Mandelstam representation, that the pion-pole contribution coincides exactly with the scalar QED Born contribution multiplied by the electromagnetic pion form factors. Experimentally, precise data are available from both time-like and space-like \cite{Ackermann:1977rp,Dally:1981ur,Amendolia:1986wj} regions. For the space-like region, one can adopt a simple monopole form inspired by the vector-meson dominance representation and fit the unknown mass parameter directly to the data \cite{Danilkin:2018qfn}. Another way is to use a description that can profit from time-like high-statistics data as well. Both methods show that the pion vector factor in the space-like regions can be calculated with almost negligible uncertainties.

The next left-hand cuts should come from the multi-pion exchanges. In practice, they can be approximated by the resonance exchanges. For the vector-meson exchange, it can be obtained by the effective Lagrangian
\begin{equation}\label{LVPg}
{\cal  L}_{VP\gamma}=e\,C_V\,\epsilon^{\mu\nu\alpha\beta}\,F_{\mu\nu}\,\partial_\alpha P\,V_\beta\,,
\end{equation}
where $F_{\mu\nu}=\partial_\mu\,A_\nu-\partial_\nu\,A_\mu$ and the couplings can be obtained from the PDG values \cite{Tanabashi:2018oca} for the partial decay widths
\begin{equation}
\Gamma _{V\to P\gamma}=\frac{e^2\,C_{V}^2\left(m_V^2-m_P^2\right)^3 }{24\,\pi\,m_V^3}\,.
\end{equation}
The off-shellness of the photon can be taken into account by the vector TFF which is defined as
\begin{align}\label{def:FVi}
\langle V(k, \lambda)| j_{\mu}(0) | \pi(p)\rangle = 2e\,C_Vf_{V,\pi}(Q^2)\,\epsilon _{\mu \alpha \beta \gamma}k^{\alpha}\,p^{\beta }\epsilon^{\gamma *}(k, \lambda).
\end{align}
As for the pion pole, one can show that for vector pole contribution $\text{Disc}\, h^{(J),Vexch}_{\lambda_1\lambda_2}(s)$ is uniquely defined and does not have any polynomial ambiguities. For the electromagnetic transition form factor of the $\omega$ one uses the dispersive analysis from \cite{Schneider:2012ez,Danilkin:2014cra}. For the sub-dominant $\rho$ contribution, one can use the VMD model \cite{Sakurai1969}. Another way that would capture the finite $\rho$ meson width is to write a spectral representation with the $\gamma^* \pi \to \pi\pi$ amplitude as input \cite{Colangelo:2014pva}. Both form factors, however, suffer from the unphysical high $Q^2$ behavior, which is not encoded in the dispersive analysis or in the VMD model. A simple modification has been proposed in \cite{Khodjamirian:1997tk}
\begin{align}\label{ModifiedVMD}
f_{V,\pi}(Q^2)=\frac{1}{1+Q^2/\Lambda^2+Q^4/\tilde{\Lambda}^4}\,.
\end{align}
We will discuss this modification in the next section.

\subsubsection{Practical implementation}

We start with an overview of the recent dispersive analyses listed in Table\,\ref{tab_DispersiveggtoMM}, which follow the philosophy of \cite{Gourdin1960, Goble:1972rz, Goble:1988cg, Morgan:1987gv, Morgan:1990kw}. In \cite{Pennington:2006dg} a once-subtracted dispersion relation was employed for the S-wave in order to extract the two-photon coupling of the $f_0(500)$ meson. The unknown subtraction constants were fixed from the Adler zero requirement in the $\gamma\gamma \to \pi^0\pi^0$ S-wave amplitude and an assumption about $\pi^+$ dipole polarizability. The results obtained in \cite{Pennington:2006dg} show that the $\gamma \gamma \to \pi^0 \pi^0$ cross section is very sensitive to the choice of the $\gamma\gamma\to\pi\pi$ phase above the two kaon threshold. Indeed, Watson's theorem ensures that the phase $\phi$ of $\gamma\gamma\to\pi\pi$ equals the hadronic phase shift $\delta$ in the elastic region. This is relatively good approximation for $I=2$ channel, however in the isoscalar channel the elastic region extends roughly till $K\bar{K}$ threshold. Above that energy, the phase of the $\gamma\gamma\to\pi\pi$ can continue to behave like the elastic scattering phase shifts $\delta_{0}^{(0)}$, or exhibit a sharp dip. The cross section is sensitive to that choice already at energies of $\sqrt{s}=0.6$ GeV. A possible way out was suggested in \cite{Oller:2007sh, Oller:2008kf} based on the experimental information of the smallness of the $f_0(980)$ peak. It was proposed to use in the dispersive analyses a pure Omn\`es function (calculated with a hadronic phase shift) multiplied by a first order polynomial when the phase shift is larger than $\pi$. It is equivalent to constructing the Omn\`es function with a phase that satisfies $\phi_{0}^{(0)}=\{\delta_{0}^{(0)}~\text{for}~ s<s_\pi;~\delta_{0}^{(0)}-\pi ~\text{for}~  s>s_\pi\}$, where $\delta_{0}^{(0)}(s_\pi)=\pi$. This allows to extend the validity of the single channel MO representation. The same prescription was used in \cite{Moussallam:2013una} for the single virtual case. In \cite{Hoferichter:2011wk} Roy-Steiner equations were derived and solved using MO representation with a finite matching point \cite{Buettiker:2003pp, Hoferichter:2015hva}. This implies a cutoff in the Omn\`es function at $s_m=1$ GeV which also generates a zero at $s=s_m$. Recently, it was suggested in \cite{ Colangelo:2017fiz,Colangelo:2017qdm} to disentangle the effects of $f_0(980)$ in the single channel Omn\`es function by approximating the $\gamma\gamma\to \pi\pi$ phase by a hadronic phase shift calculated using the single-channel IAM \cite{GomezNicola:2007qj}. The obtained phase shift takes the $f_0(500)$ resonance into account, but never crosses $\pi$ (see Fig.\ref{fig:Omnes}). As a result, the Omn\`es function (and the inverse of Omn\`es function) is smooth around the two-kaon threshold.

An amplitude analysis was performed in \cite{Dai:2014zta, Dai:2014lza}, where the $K\bar{K}$ channel was included. In the analysis, a dispersive way of calculating amplitudes was only built in the low energy region $\sqrt{s}<0.6$ GeV in a single channel manner, similar to \cite{Pennington:2006dg}. Above that energy a partial-wave amplitude analysis was performed with the aim to determine the $s$-channel amplitudes from a simultaneous fit to all the available data. Given the significant amount of fitted parameters such method is currently not able to predict the data in other channels or to be extended to the single (double) virtual case.

\begin{table}
\centering
\begin{tabular*}{\textwidth}{@{\extracolsep{\fill}}lllll@{}}
\hline\hline 
$Q^2=0$ & Approach & Inelast. &N & Range [GeV] \\
\hline 
Pennington 2006 \cite{Pennington:2006dg}	      & Disp, Omn\`es  	&	$\pi\pi, J=0 $	&   $0$	        &	$\sqrt{s}<0.6$\\

Oller et al. 2007 \cite{Oller:2007sh}	      & Disp, Omn\`es  	&	$\pi\pi, J=0$	&   $0$	        &	$\sqrt{s}<0.8$\\

Hoferichter et al.  2011 \cite{Hoferichter:2011wk} &	Roy-Steiner	  & 	$\pi\pi $	&	   $1$	&	$\sqrt{s}<1.0$ \\

Garcia-Martin et al. 2010 \cite{GarciaMartin:2010cw}&   Disp, Omn\`es	&	$\pi\pi, K\bar{K}$	&	  $ 6$	&	$\sqrt{s}<1.3$ \\

Danilkin et al. 2017 \cite{Danilkin:2017lyn}&   Disp, Omn\`es	&	$\pi\eta, K\bar{K}$	&	  $ (1)$	&	$\sqrt{s}<1.4$ \\
\hline
$Q_1^2= 0$, $Q_2^2\neq 0$ &&&&\\
\hline 

Moussallam 2013 \cite{Moussallam:2013una}       &   Disp, Omn\`es	&	$\pi\pi, J=0 $		&   $0$	&	$\sqrt{s}<0.8$ \\

Danilkin et al. 2018 \cite{Danilkin:2018qfn}&   Disp, Omn\`es	&	$\pi\pi, K\bar{K}$	&	  $1$	&	$\sqrt{s}<1.4$ \\

Deineka et al. 2018 \cite{Deineka:2018nuh}&   Disp, Omn\`es	&	$\pi\eta, K\bar{K}$	&	  $ 1$	&	$\sqrt{s}<1.4$ \\

\hline
$Q_1^2\neq 0$, $Q_2^2\neq 0$ &&&&\\
\hline
Colangelo et al. 2017 \cite{Colangelo:2017qdm}	&          Roy-Steiner	&	$\pi\pi, J=0 $		&	   $0$	&	$\sqrt{s}<0.8$ \\ 
\hline\hline 
\end{tabular*}
\caption{Recent dispersive analyses of $\gamma^*\gamma^*\to \pi\pi, \pi\eta$ and $K\bar{K}$.  The third column shows the number of parameter fitted to the real photon data, while the fourth column indicates the range of the validity of the approach.
\label{tab_DispersiveggtoMM}}
\end{table}

The coupled-channel MO equations were considered for a first time in \cite{GarciaMartin:2010cw, Moussallam:2011zg}. The dispersive integrals were over-subtracted and the unknown subtraction constants were determined from the fit to the cross section data. As a result, the pion polarizabilities were extracted. It was also shown that in order to construct the Omn\`es representation for D-waves the contribution from the Born left-hand cut should be supplemented by higher-mass intermediate state exchanges. Extending such dispersive technique to the partial-wave helicity amplitudes of the single virtual $\gamma\gamma^* \to\pi\pi$ process is not straightforward, as in addition to the well-known low-energy constraints, partial-wave amplitudes exhibit kinematic constraints. Therefore, the first dispersive analyses of $\gamma\gamma^* \to\pi\pi$ \cite{Moussallam:2013una} and $\gamma^{*}\gamma^{*} \to\pi\pi$ \cite{Colangelo:2017qdm, Colangelo:2017fiz} have been limited to the S-wave and single-channel description which only covers the $f_0(500)$ resonance region. Recently, these ideas were extended for the coupled-channel MO equations for $\gamma\gamma^* \to\pi\pi$ by including $K\bar{K}$ intermediate states and D-waves \cite{Danilkin:2018qfn}, which allow for a full dispersive formalism through the prominent $f_2(1270)$ tensor meson region. These results will be shown below.

As for the other channels, a coupled-channel dispersive analysis at low energies of the $\gamma\gamma \to K\bar{K}$, $\gamma\gamma \to \eta\eta$ and $\gamma\gamma\to\pi\eta$ processes has been performed in \cite{Danilkin:2012ua}. The latter was revisited in \cite{Danilkin:2017lyn} by significantly reducing the number of fitted parameters and adding the D-wave. In \cite{Deineka:2018nuh} the first result of the single virtual process  $\gamma\gamma^{*} \to \pi\eta $ has been presented. An important ingredient in these calculations is a coupled-channel $\{\pi\eta, K\bar{K}\}$ Omn\`es function which was constructed in \cite{Danilkin:2011fz,Danilkin:2012ap}. An alternative form was suggested in \cite{Albaladejo:2015aca} which yet remains to be tested against $\gamma\gamma \to \pi\eta$ data. 

\begin{figure}[!t]
\centering
\centering
\includegraphics[width =0.47\textwidth]{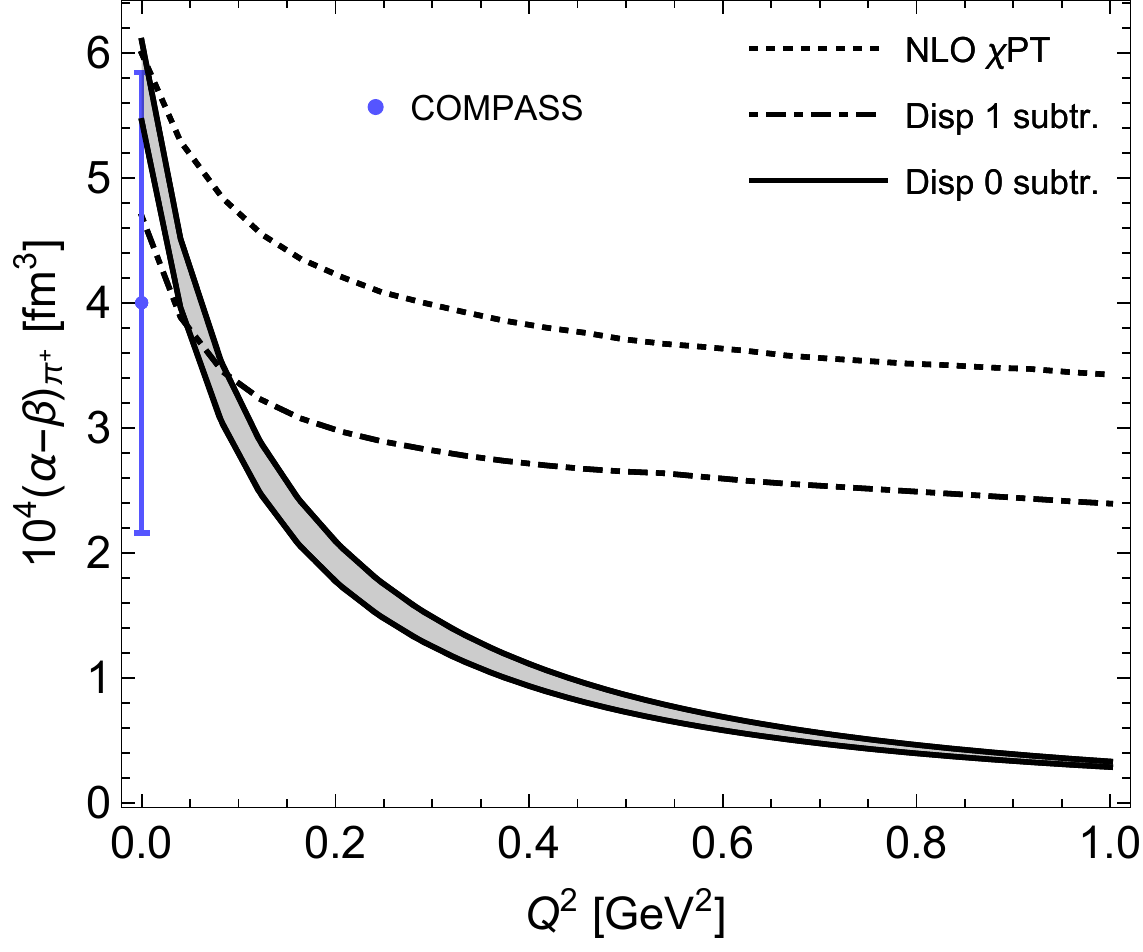}\quad \includegraphics[width =0.48\textwidth]{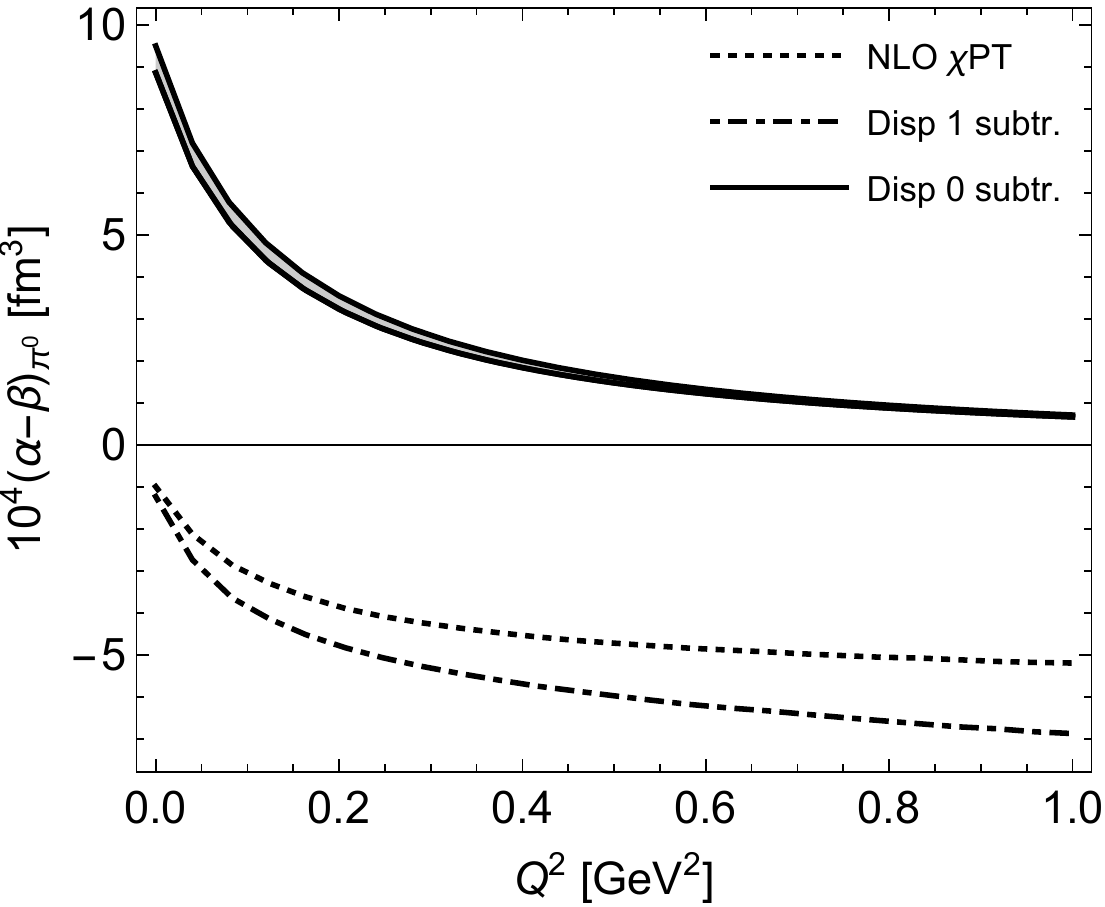}\\
\caption{Generalized dipole polarizability difference for $\pi^+$ and $\pi^0$. Unsubtracted dispersive prediction from \cite{Danilkin:2018qfn} shown as a band between coupled- and single-channel analyses. The latter is consistent with \cite{Colangelo:2017qdm}. We also show the result from once subtracted dispersive study \cite{Moussallam:2013una} (dot-dashed curve) and chiral NLO result (dotted curve) \cite{Fuchs:2000pn, Lvov:2001zdg}. The COMPASS data point is from \cite{Adolph:2014kgj}.
\label{fig:Polar}}
\end{figure}

\subsubsection{Numerical results}
\textbf{$\pi\pi$ channel: S-wave contribution}\\
We start the discussion of the results with the S-wave contribution to $\gamma\gamma \to \pi\pi$ cross sections. As it was found in \cite{Colangelo:2017qdm, Colangelo:2017fiz, Danilkin:2018qfn}, the rescattering of the Born terms alone can be taken into account using the unsubtracted DR given in Eq.(\ref{Rescattring:general}). In this way one can predict the pion dipole polarizabilities as a check of the low-energy limit
\begin{align}
(\alpha_1-\beta_1)_{\text{single channel}}^{\pi^\pm, \pi\text{-pole LHC}}&= 5.4-5.8\times 10^{-4}\,\text{fm}^3\, \text{\cite{Colangelo:2017fiz}}\,,\nonumber\\
(\alpha_1-\beta_1)_{\text{coupled channel}}^{\pi^\pm, \pi\text{-pole LHC}}&= 6.1\times 10^{-4}\,\text{fm}^3\,\text{\cite{Danilkin:2018qfn}}\,.
\end{align}
This result is consistent with NNLO $\chi$PT~\cite{Gasser:2006qa}:
\begin{equation}
(\alpha_1-\beta_1)^{\chi PT}_{\pi^\pm}= 5.7(1.0)\times 10^{-4} \text{fm}^3,
\end{equation}
and with the recent COMPASS measurement~\cite{Adolph:2014kgj}:
\begin{equation}
(\alpha_1-\beta_1)^{exp}_{\pi^\pm}=4.0(1.2)_{stat}(1.4)_{syst}\times 10^{-4} \text{fm}^3. 
\end{equation}
The dipole polarizability for the neutral pion comes out as 
\begin{align}
(\alpha_1-\beta_1)_{\text{single channel}}^{\pi^0, \pi\text{-pole LHC}}&= 11.2...8.9\times 10^{-4}\,\text{fm}^3\,  \text{\cite{Colangelo:2017fiz}}, \nonumber\\
(\alpha_1-\beta_1)_{\text{coupled channel}}^{\pi^0, \pi\text{-pole LHC}}&= 9.5\times 10^{-4}\,\text{fm}^3\,\text{\cite{Danilkin:2018qfn}}, 
\end{align}
which is far away from the NNLO $\chi$PT value~\cite{Gasser:2005ud}:
\bea
(\alpha_1-\beta_1)^{\chi PT}_{\pi^0}= -1.9(0.2)\times 10^{-4} \text{fm}^3.
\eea
The large value of the $\pi^0$ dipole polarizability is reflected in the absence of the Adler zero. Nevertheless, this mismatch to $\chi$PT is hardly visible on the $\gamma\gamma \to \pi^0\pi^0$ cross section, since its main contribution comes from the rescattering process $\gamma\gamma \to \pi^+\pi^-\to\pi^0\pi^0$. We note that the polarizabilities are saturated by $90\%$ from the dispersion integral over the low energies $<1.4\,\text{GeV}$. This is no longer the case for the generalized polarizabilities. For instance, for $Q^2=0.5\,\text{GeV}^2$, the generalized polarizbilities $(\alpha_1-\beta_1)_{\text{coupled channel}}^{\pi^\pm, \pi\text{-pole LHC}}= 0.86\times 10^{-4}$ fm$^3$ and $(\alpha_1-\beta_1)_{\text{coupled channel}}^{\pi^0, \pi\text{-pole LHC}}= 1.62\times 10^{-4}$ fm$^3$ are saturated by $70\%$ from the region $<1.4\,\text{GeV}$, indicating the importance of higher energies. 

\begin{figure}[!t]
\centering
\includegraphics[width =0.47\textwidth]{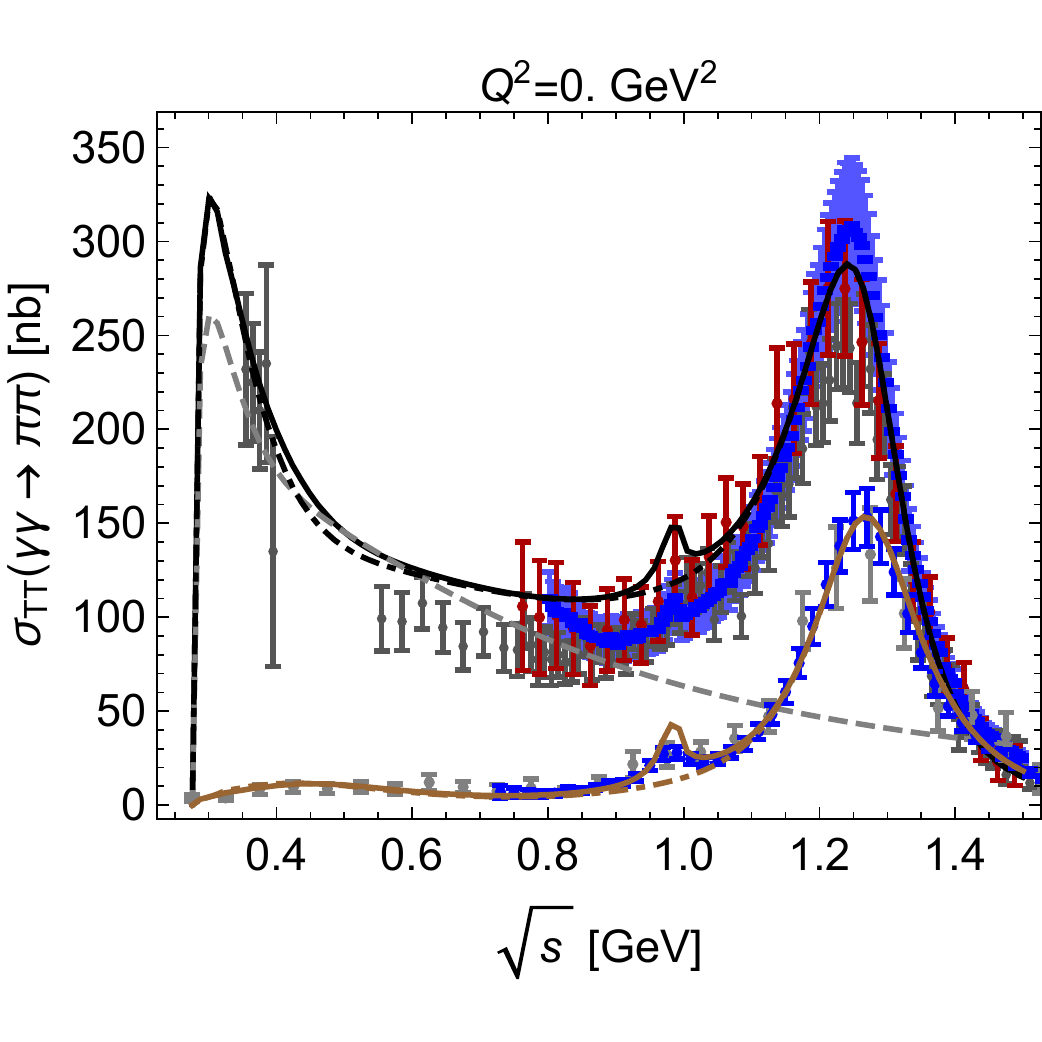}\quad \includegraphics[width =0.47\textwidth]{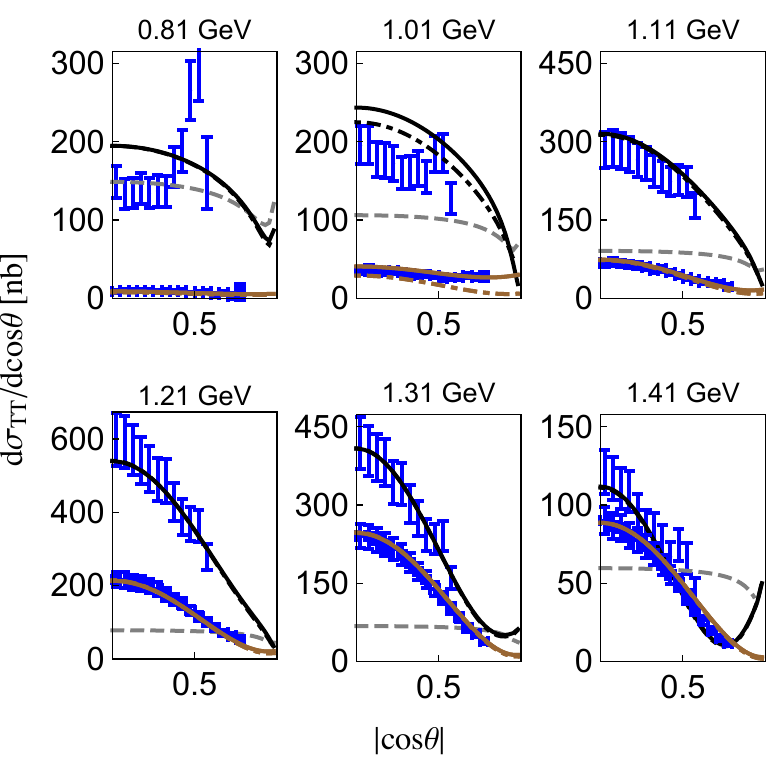}\\
\caption{Total and differential cross sections for $\gamma\gamma \to \pi^+\pi^-$ (upper black curves) and $\gamma\gamma \to \pi^0\pi^0$ (lower brown curves) taken from \cite{Danilkin:2018qfn}. The coupled-channel (single-channel) results are shown by the solid (dashed-dotted) curves. The Born result is shown by dashed gray curves. 
The data are taken from
\cite{Mori:2007bu,Uehara:2009cka,Boyer:1990vu,Behrend:1992hy,Marsiske:1990hx}.
\label{fig:Q2=0}}
\end{figure}

\begin{figure}[!t]
\centering
\includegraphics[width =0.47\textwidth]{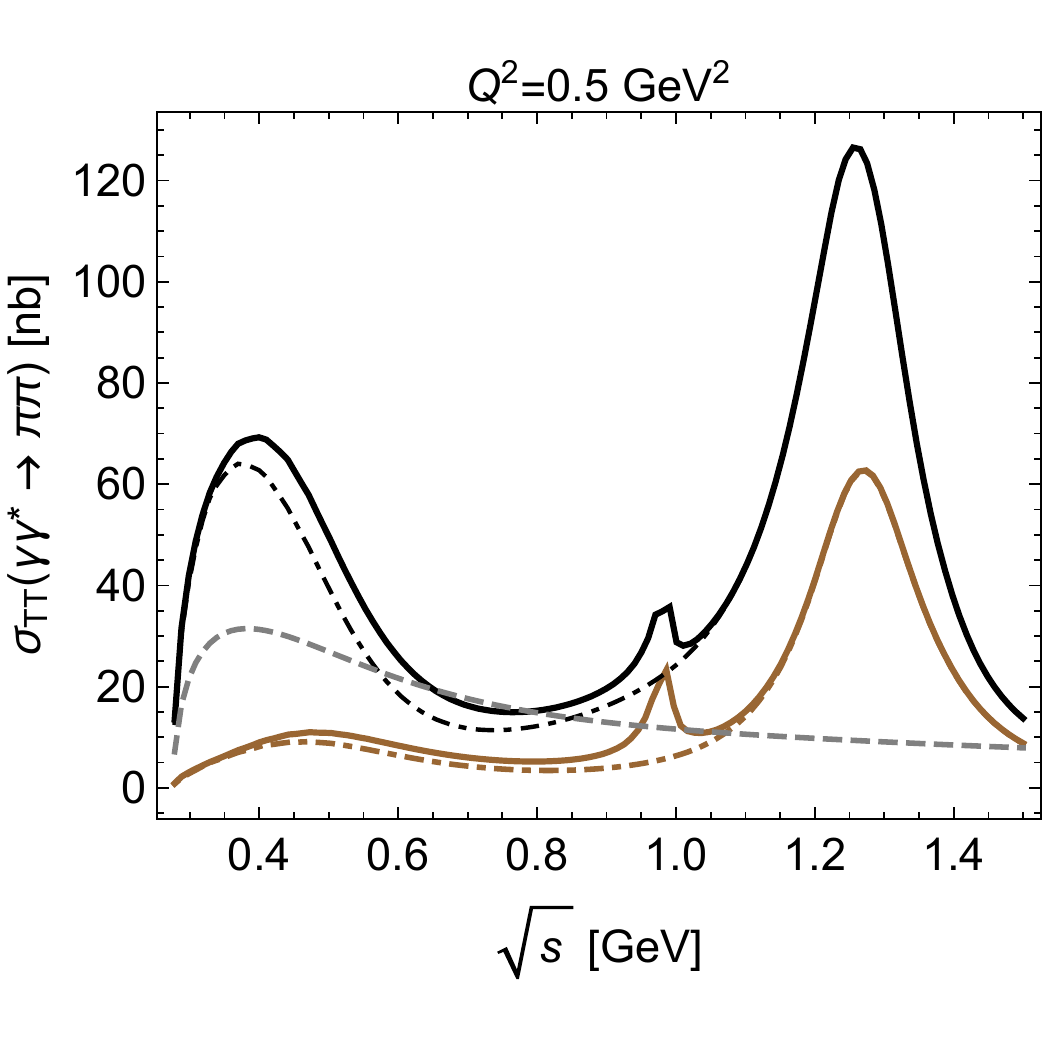}\quad \includegraphics[width =0.47\textwidth]{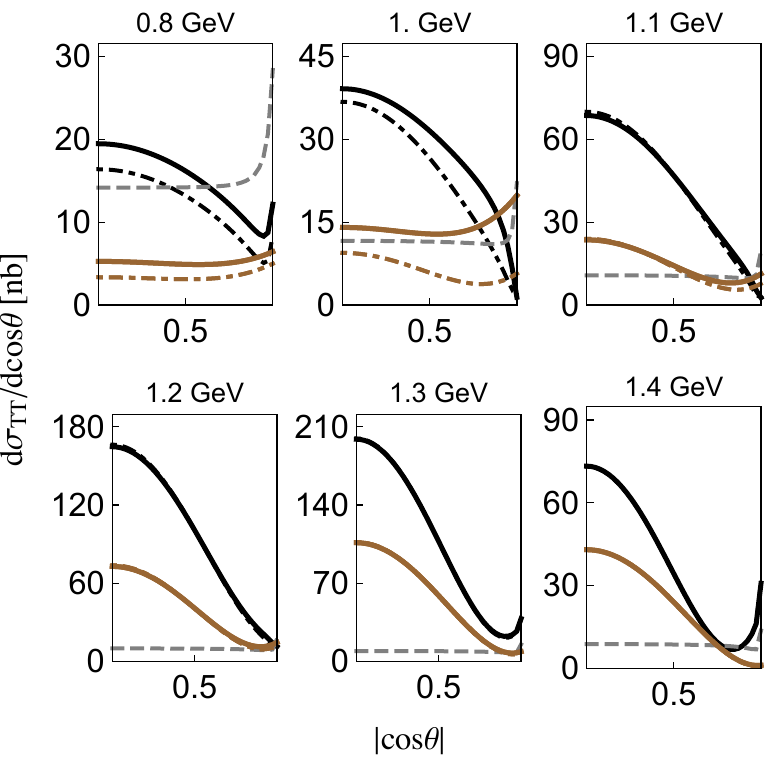}\\
\includegraphics[width =0.47\textwidth]{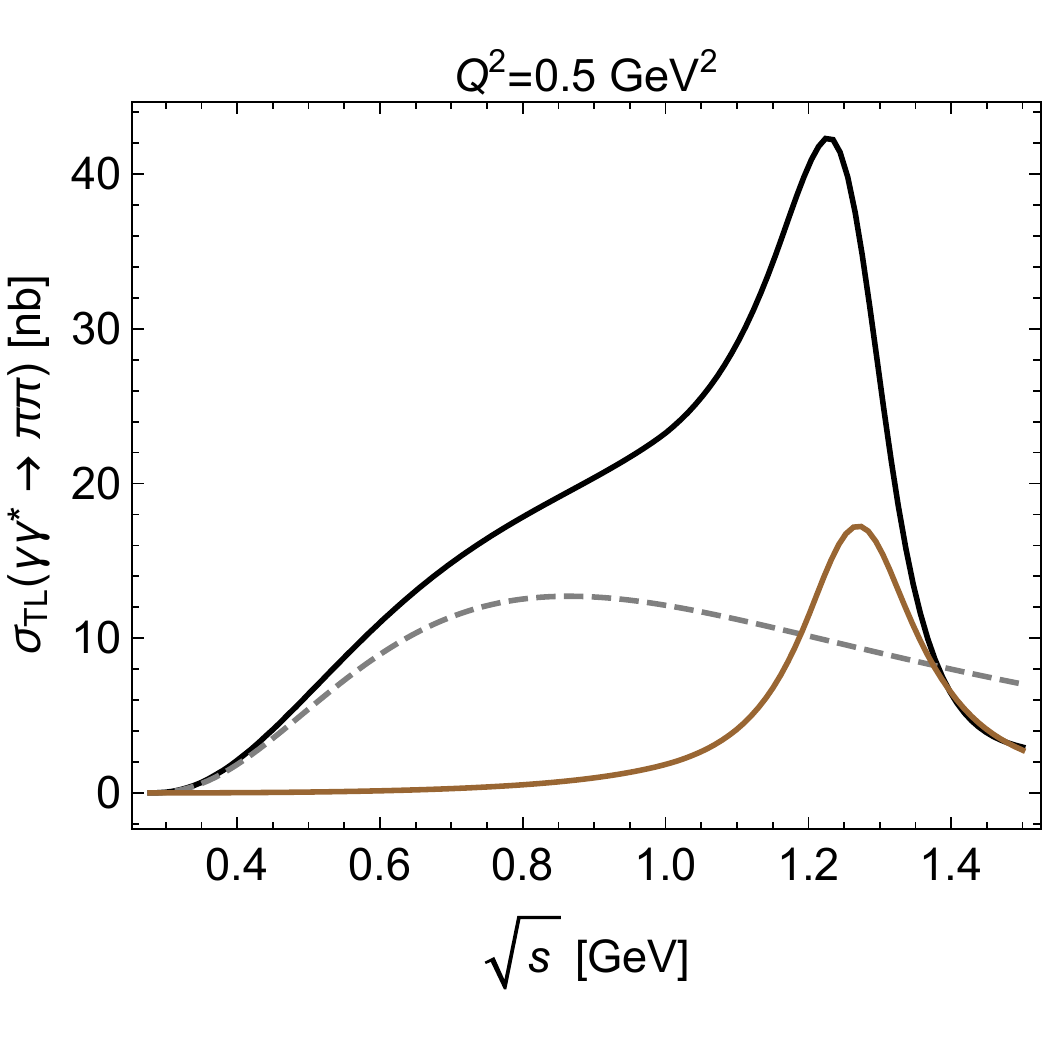}\quad \includegraphics[width =0.47\textwidth]{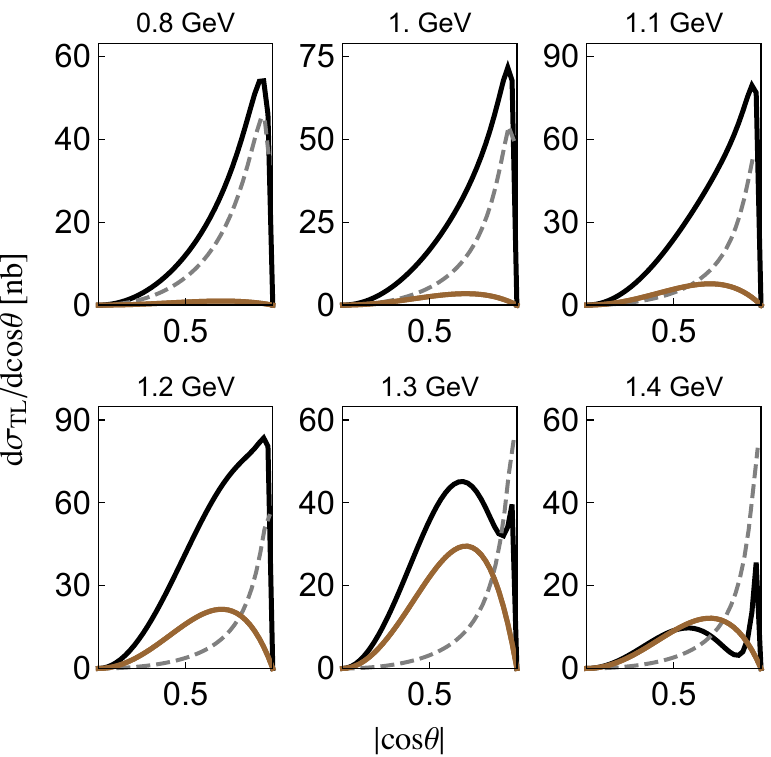}\\
\caption{Predictions for total and differential cross sections for $\gamma\gamma^* \to \pi^+\pi^-$ (upper black curves) and $\gamma\gamma^* \to \pi^0\pi^0$ (lower brown curves) with $Q^2=0.5$ GeV$^2$ and full angular coverage $|\cos\theta| \leq 1$ taken from \cite{Danilkin:2018qfn}.
The coupled-channel (single-channel) results are shown by the solid (dashed-dotted) curves. The Born result is shown by dashed gray curves. 
\label{fig:pipiQ2=0.5}}
\end{figure}

\begin{figure}[!t]
\centering
\includegraphics[width =0.47\textwidth]{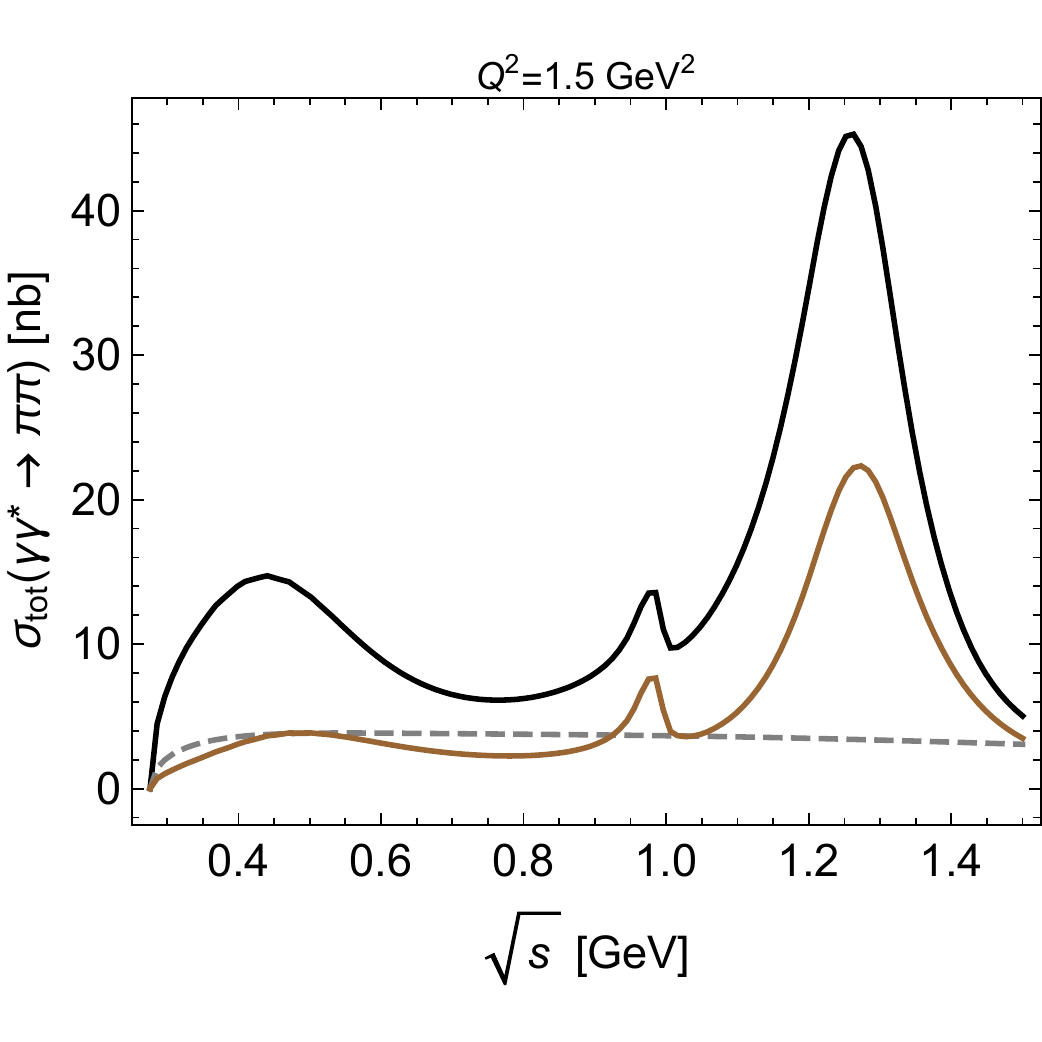}\quad 
\includegraphics[width =0.47\textwidth]{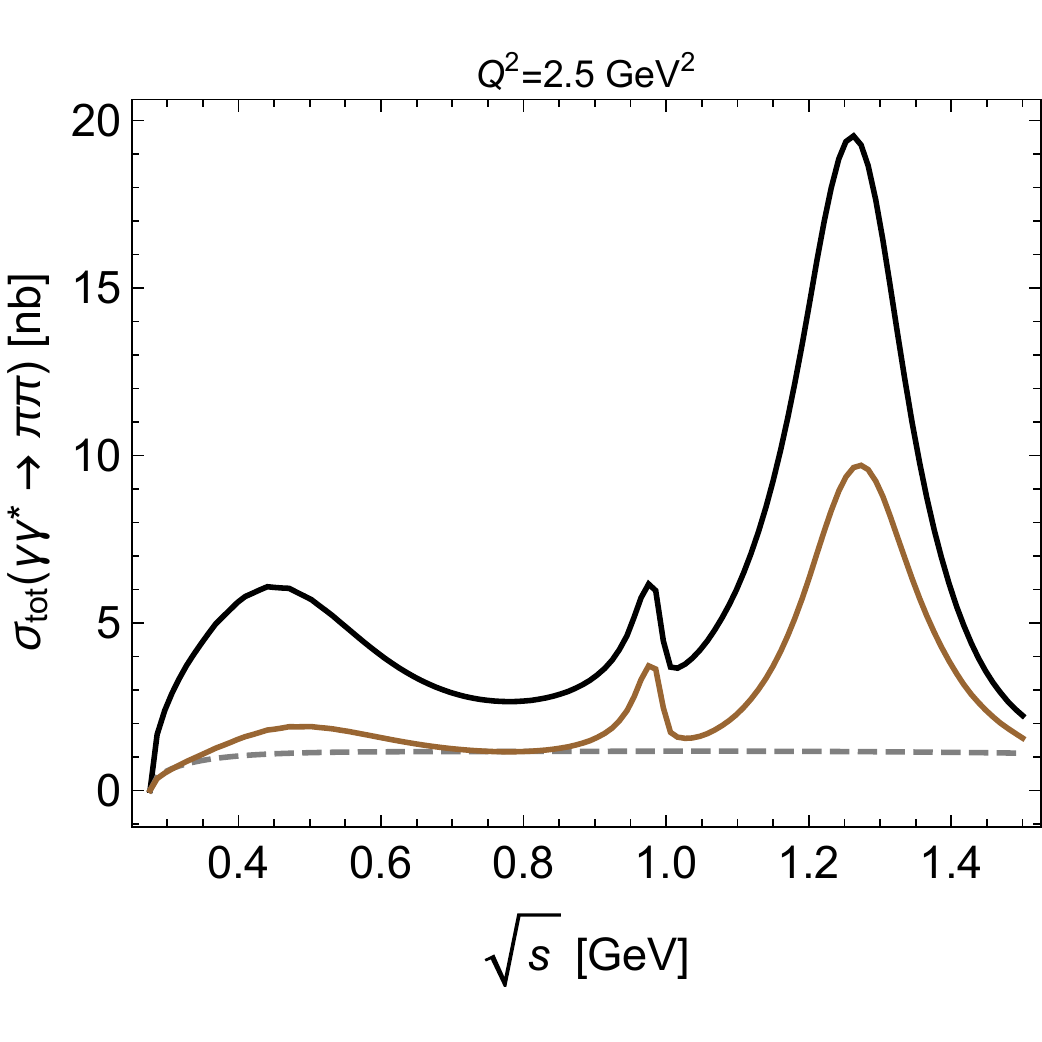}\\
\includegraphics[width =0.47\textwidth]{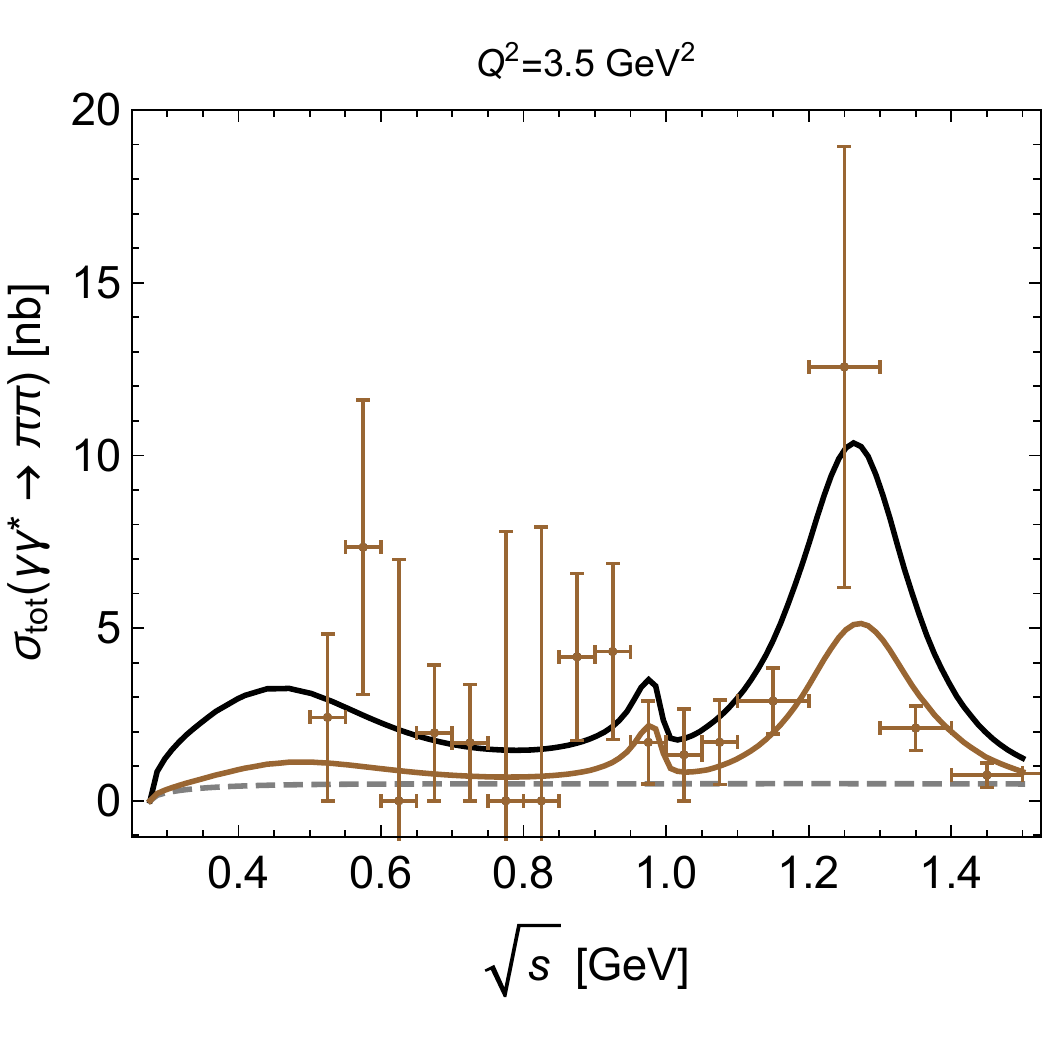}\quad 
\includegraphics[width =0.47\textwidth]{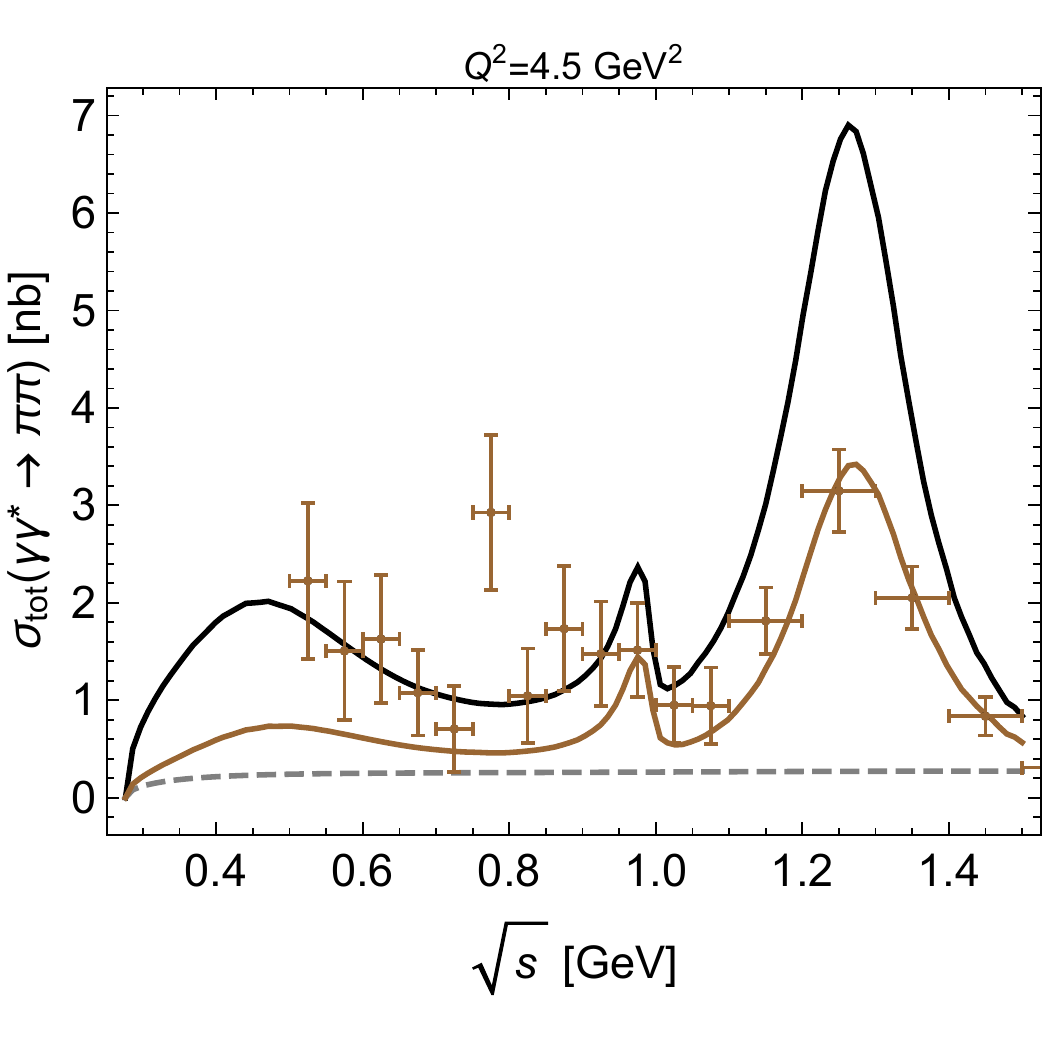}\\
\caption{Total cross sections for $\gamma\gamma^* \to \pi^+\pi^-$ (upper black curves) and $\gamma\gamma^* \to \pi^0\pi^0$ (lower brown curves) defined as $\sigma_{tot}=\sigma_{TT}+\epsilon\,\sigma_{TL}$ (see Eq.(\ref{eq:cross1tag3})) with $Q^2=1.5,\,2.5,\,3.5,$ and $4.5$ GeV$^2$ and full angular coverage $|\cos\theta| \leq 1$. For $Q^2=3.5$ GeV$^2$ and $Q^2=4.5$ GeV$^2$ bins, $\epsilon$ is taken as in the experimental analysis, i.e. $\epsilon(Q^2=3.5)=0.82$ and $\epsilon(Q^2=4.5)=0.88$, while for the other bins $\epsilon=0.9$ was assumed. The Born result is shown by dashed gray curves. The experimental data are taken from \cite{Masuda:2015yoh}.
\label{fig:sigma_tot}}
\end{figure}

\begin{figure}[!t]
\includegraphics[width =0.47\textwidth]{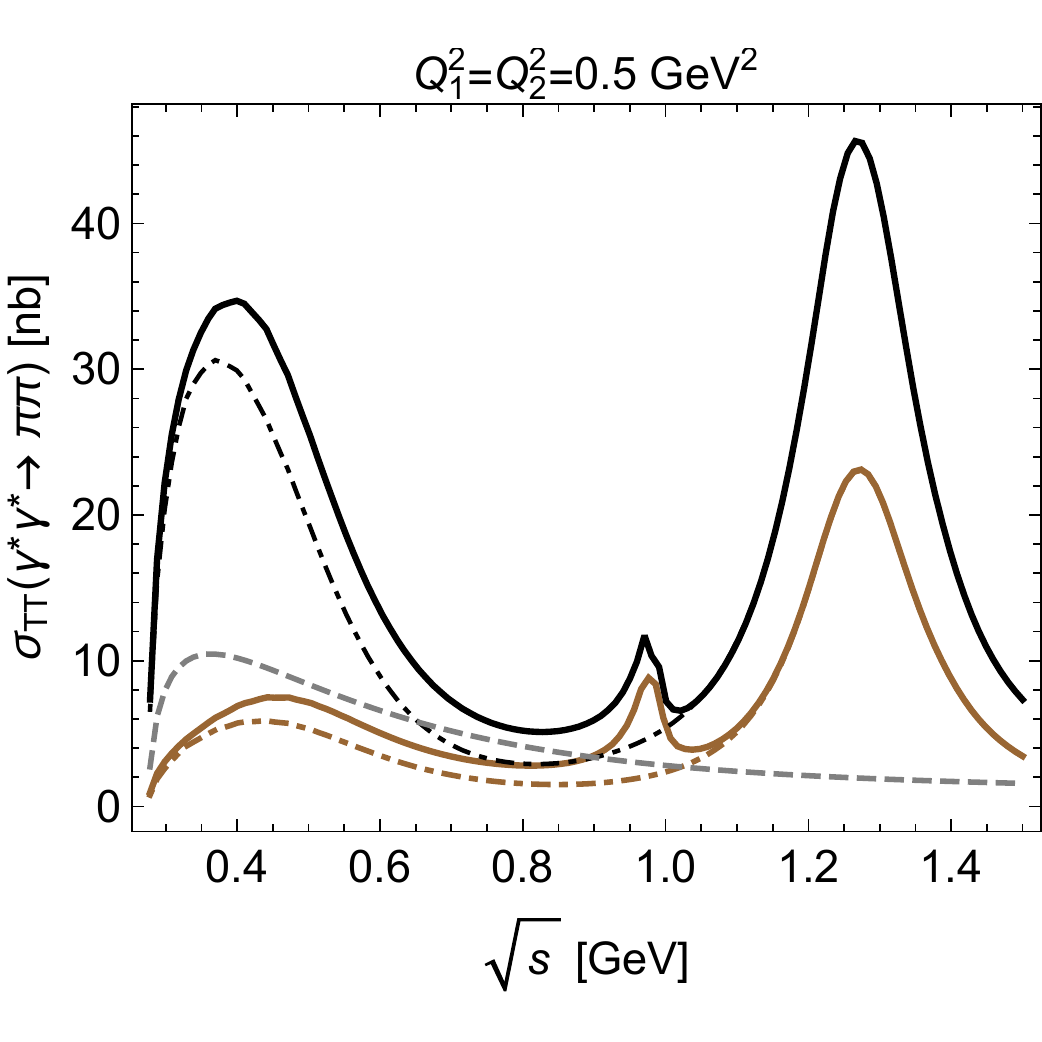}\quad 
\includegraphics[width =0.47\textwidth]{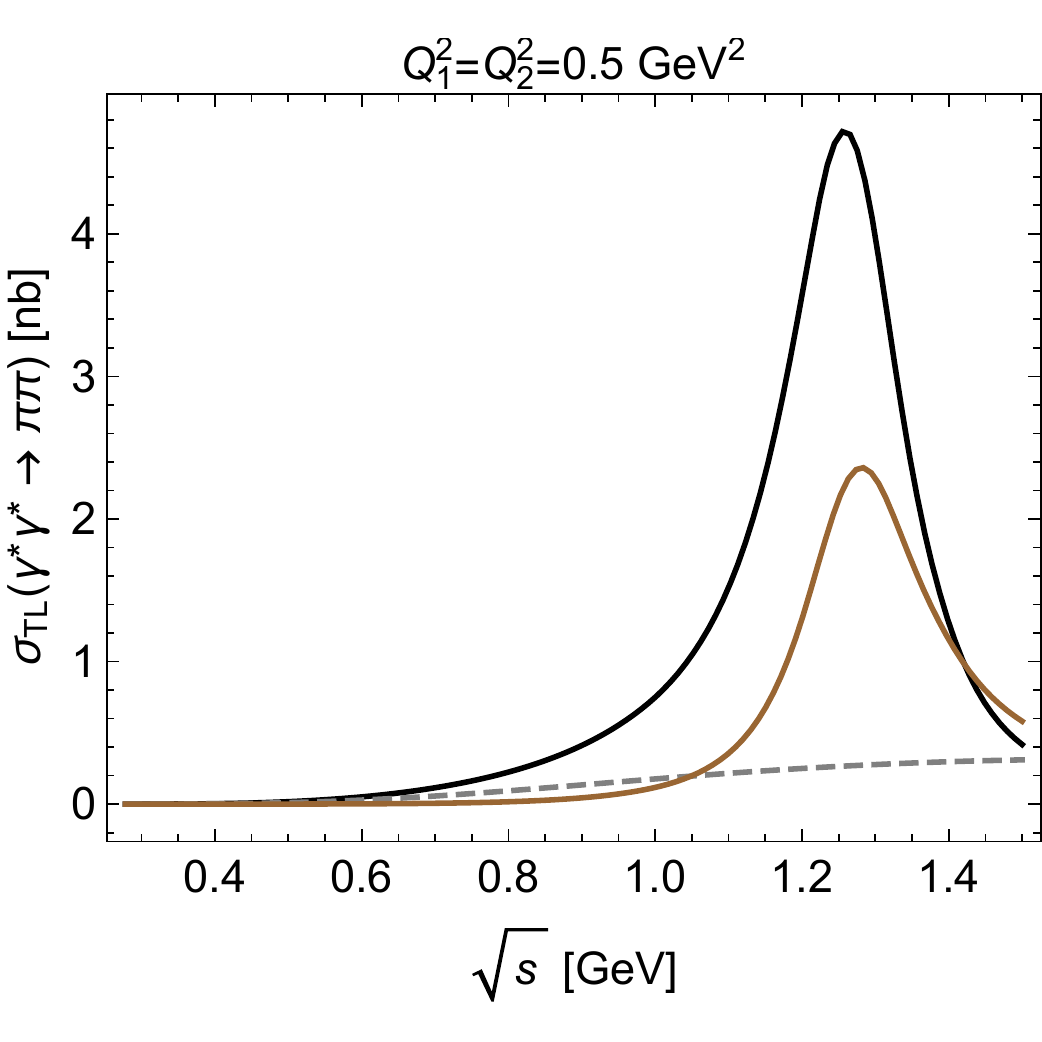}\\
\includegraphics[width =0.47\textwidth]{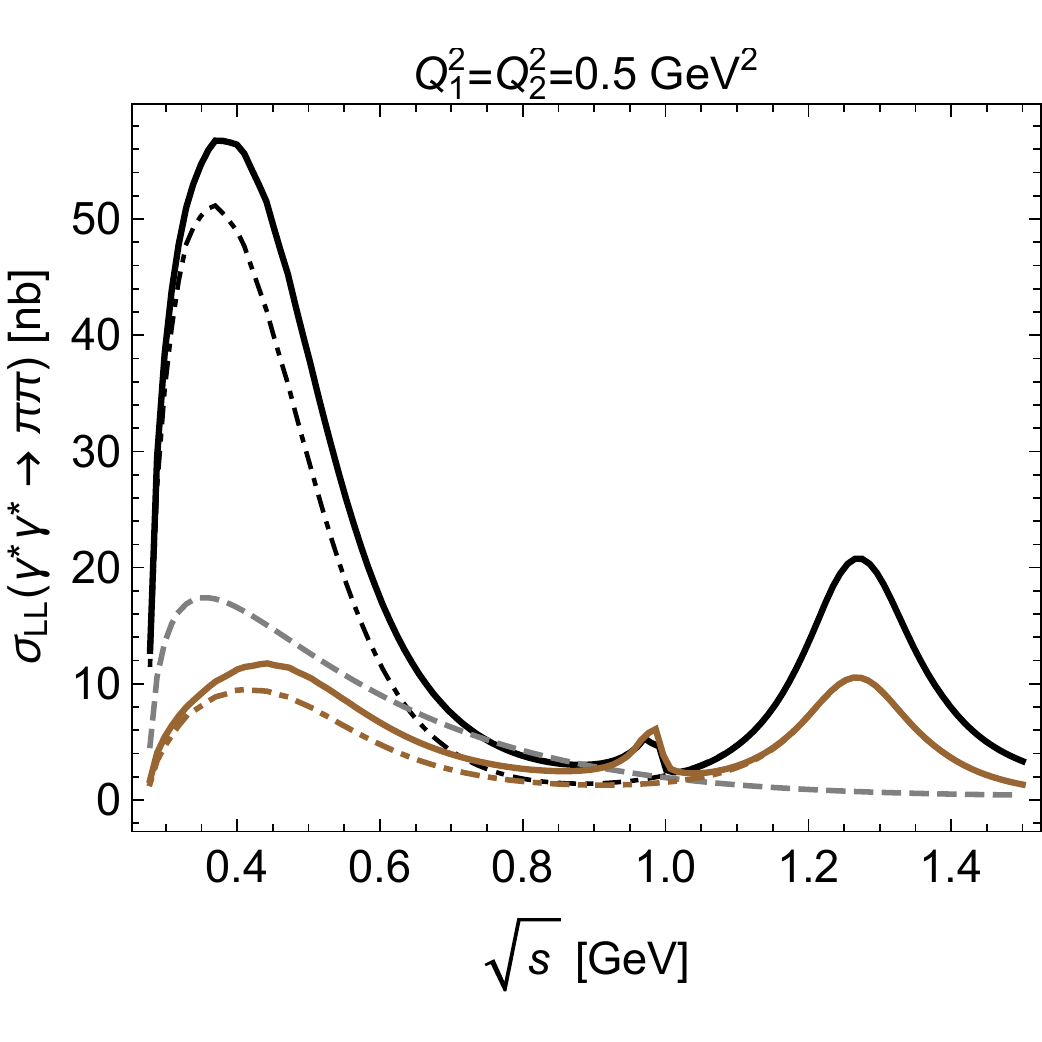}\quad 
\caption{Predictions for $\sigma_{TT}$, $\sigma_{TL}$ and $\sigma_{LL}$ cross sections for $\gamma^*\gamma^* \to \pi^+\pi^-$ (upper black curves) and $\gamma^*\gamma^* \to \pi^0\pi^0$ (lower brown curves) with $Q_1^2=Q_2^2=0.5$ GeV$^2$ and full angular coverage $|\cos\theta| \leq 1$ taken from \cite{Danilkin:X}. The coupled-channel (single-channel) results are shown by the solid (dashed-dotted) curves. The Born result is shown by dashed gray curves.\label{fig:pipiQ1Q2=0.5}}
\end{figure}

\begin{figure}[!t]
\centering
\includegraphics[width =0.47\textwidth]{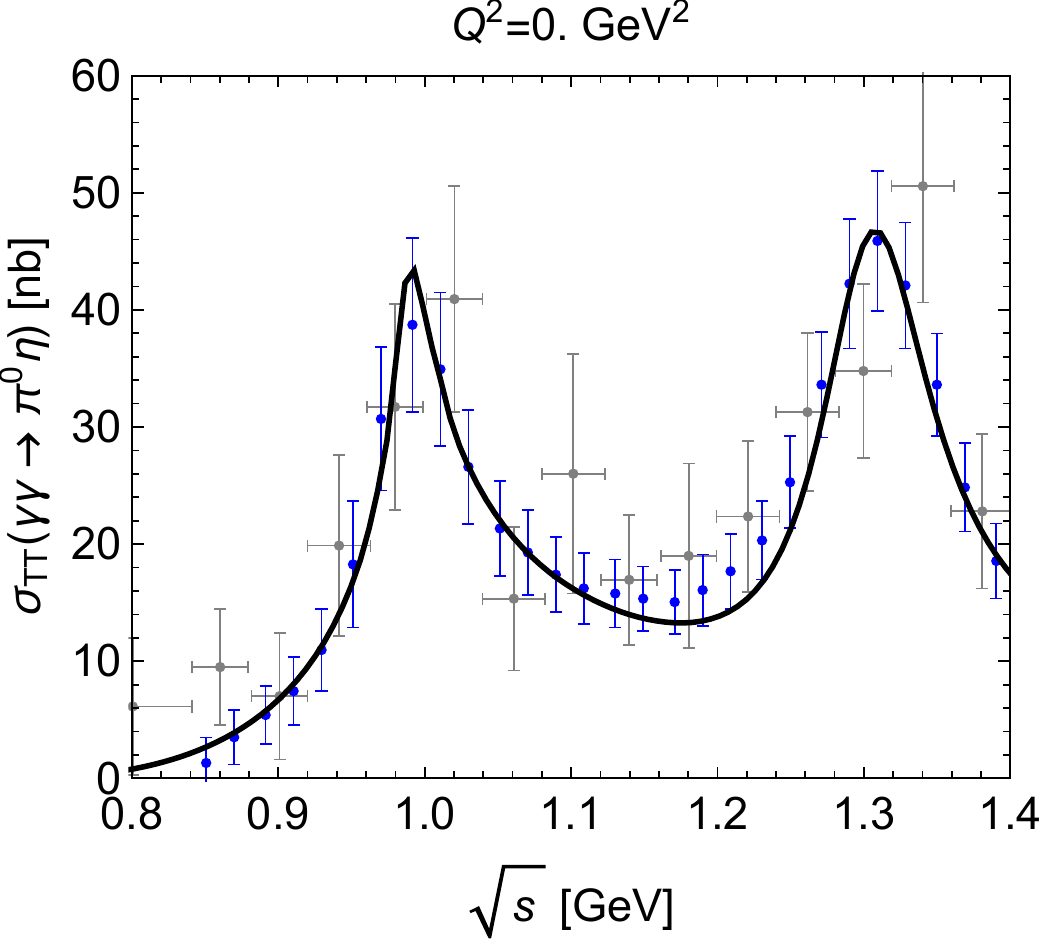}\quad \includegraphics[width =0.47\textwidth]{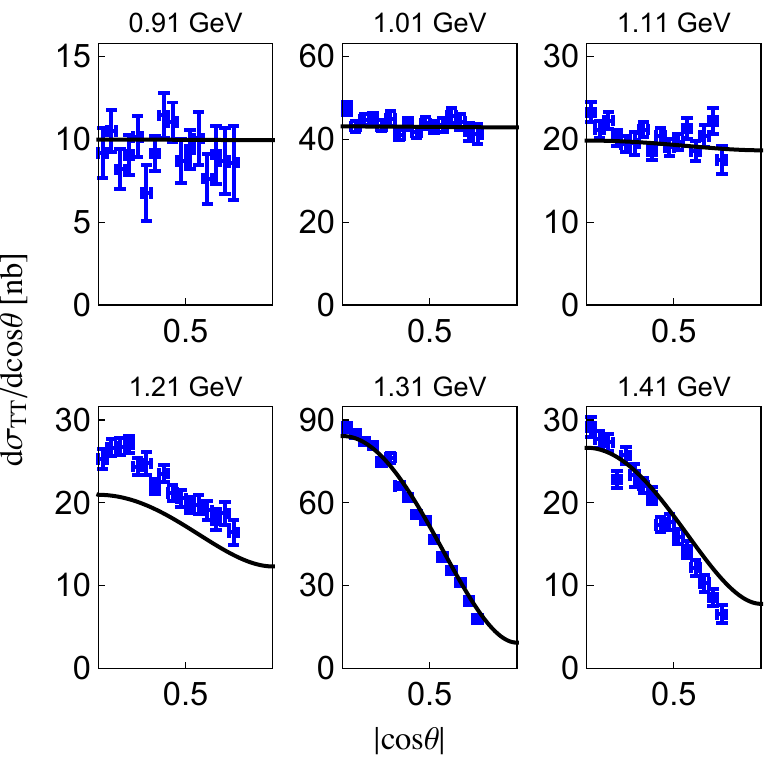}\\
\includegraphics[width =0.47\textwidth]{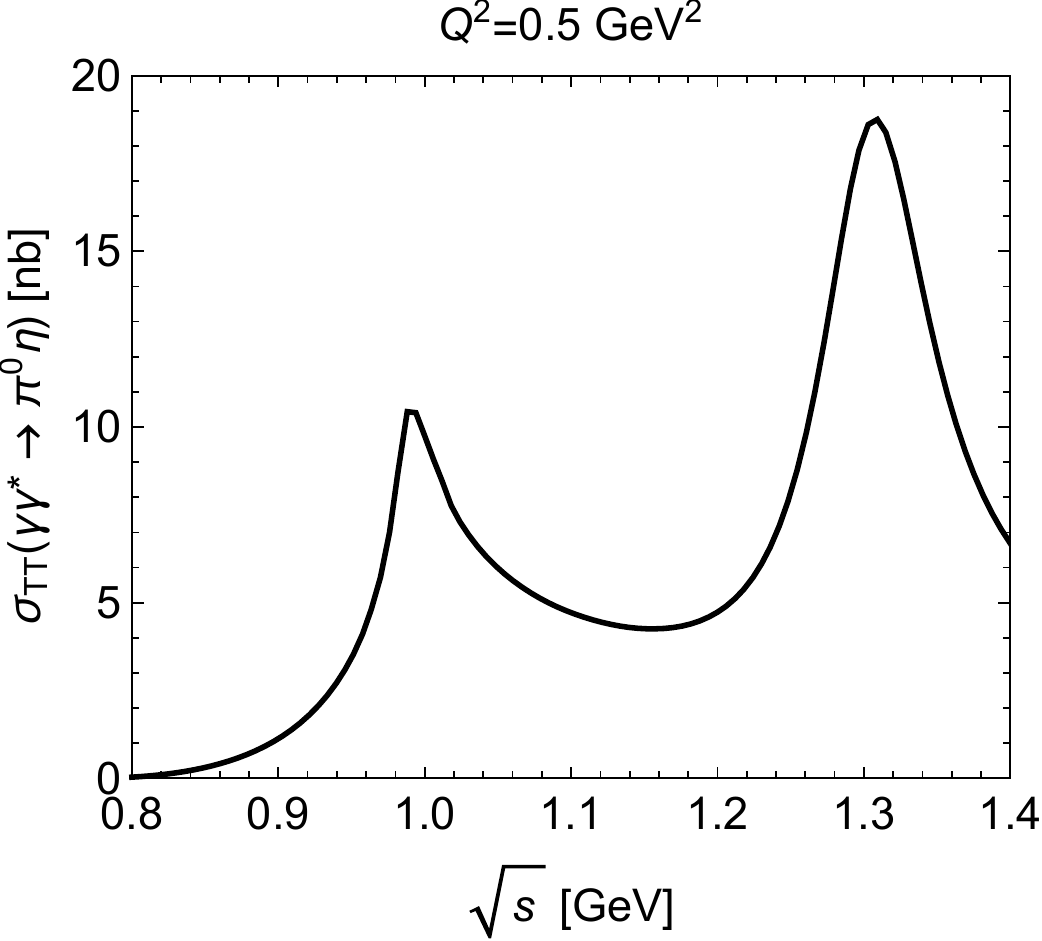}\quad \includegraphics[width =0.47\textwidth]{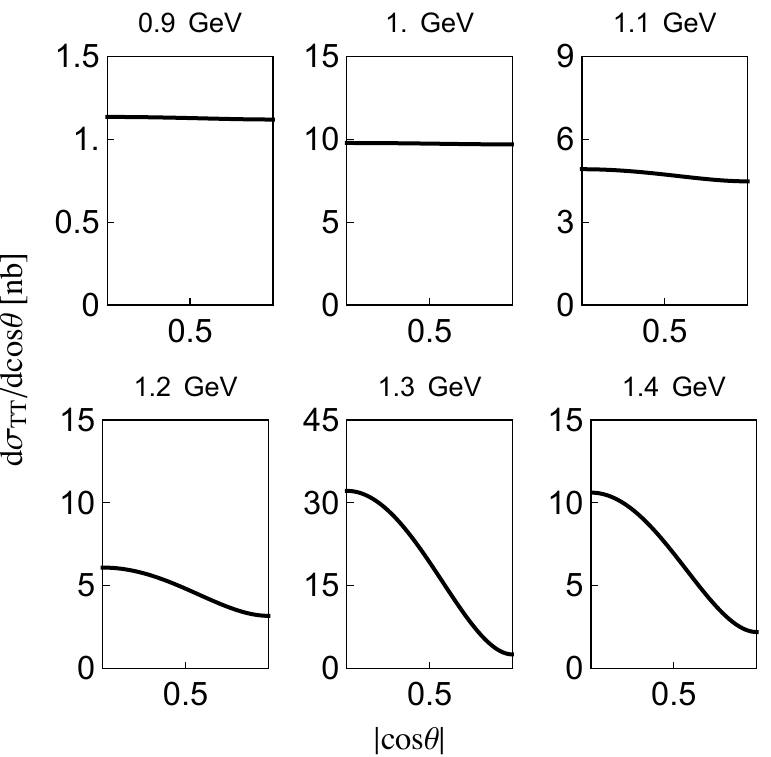}\\
\caption{Total and differential cross sections for $\gamma\gamma \to \pi^0\eta$ ($|\cos\theta| \leq 0.9$) and $\gamma\gamma^* \to \pi^0\eta$ ($|\cos\theta| \leq1.0$) with $Q^2=0.5$ GeV$^2$ taken from \cite{Danilkin:2017lyn,Deineka:2018nuh}. The data are taken from \cite{Uehara:2009cf,Antreasyan:1985wx}.
\label{fig:pieta}}
\end{figure}

The dipole polarizabilities for $\pi^0$ are expected to get large corrections once vector-meson left-hand cuts are added since they are much stronger for the neutral channel due to $\omega$-exchange. However, any Lagrangian-based field theory result has a bad high energy behavior and requires adding at least one subtraction in the dispersion relation to cure it. This reduces the predictive power of the dispersion relations. With light vector mesons as additional left-hand cuts, the once-subtracted result can be fixed to the COMPASS result for the $\pi^{\pm}$ and NLO $\chi$PT for the $\pi^0$ and $K$. As it was shown \cite{Danilkin:2018qfn} the comparison between unsubtracted and once-subtracted cases indicates a very similar description up to about 1.1 GeV for $Q_i^2=0$. However, since the finite $Q^2$ prediction from $\chi$PT for the generalized polarizabilities are expected to be valid only in a very small $Q^2$ region it is safer to stay with the unsubtracted dispersion relation. Results for the generalized polarizabilities predicted from unsubtracted dispersion relation and NLO $\chi$PT are shown in Fig.\,\ref{fig:Polar}. In addition we show the curve from the once-subtracted dispersive analyses given in \cite{Moussallam:2013una}. In Ref.\,\cite{Moussallam:2013una}, vector-meson left-hand cuts were added in the form of Eq.(\ref{Sol1}), which as we discussed above do not correspond to the vector pole contributions. The unknown subtraction functions were parametrized by a simple two parameter representation based on the chiral limit arguments and a fit to the experimental data on $e^+e^-\to \gamma \pi^0\pi^0$. One can see that the unsubtracted dispersion result predicts significant reduction of the generalized polarizability of $\pi^+$ compared to the NLO $\chi$PT. In turn, the result from dispersive analyses given in \cite{Moussallam:2013una} lies in between, but shows an increasingly larger deviation with the unsubtracted dispersion result with increasing $Q^2$. Once the data from BESIII will be available, one can extract generalized polarizabilities for each $Q^2$ by performing a fit of the differential and total cross sections.

\textbf{$\pi\pi$ channel: D-wave contribution}\\
While the contribution from the Born left-hand cut should be dominant at low energies (due to small pion mass in the t-channel), a description the $f_2(1270)$ region requires adding higher-mass intermediate states in the left-hand cuts \cite{GarciaMartin:2010cw}. If one approximates them only with vector-pole contributions, then the radiative decay coupling in SU(3) limit 
\begin{equation}\label{gVP}
g_{V}\simeq C_{\rho^{\pm,0}}\simeq \frac{C_{\omega}}{3} \simeq \frac{1}{2} C_{K^{*0}} \simeq C_{K^{*\pm}}
\end{equation}
can be fixed at the $f_2(1270)$ resonance position from the $\gamma\gamma\to \pi^0\pi^0$ cross section \cite{Danilkin:2018qfn}. Its value yields $g_{V} =0.33$ GeV$^{-1}$, which is well in agreement with the PDG spread $g_{V}= 0.4(1) \,\text{GeV}^{-1}$ \cite{Tanabashi:2018oca}. We emphasize that this is the only parameter that was adjusted to the real photon data in \cite{Danilkin:2018qfn}, within its expected range. The results for the differential and total cross section are shown in Fig.\,\ref{fig:Q2=0}. One can see that one can achieve a reasonable description of the charged and neutral channels with the exception of the intermediate region in $\gamma\gamma \to \pi^+\pi^-$ and a slightly stronger $f_0(980)$. This in principle can be fixed by over-subtracting the dispersion relations and fitting this unknown subtraction constant to the data, as it was done in \cite{GarciaMartin:2010cw}. However, this procedure lacks a predictive power for the single or double virtual processes.

The prediction for the spacelike single virtual case using the unsubtracted dispersion relation formalism from \cite{Danilkin:2018qfn} is shown in Fig.\,\ref{fig:pipiQ2=0.5} for $\sigma_{TT}$ and $\sigma_{TL}$. The latter is fully determined by the helicity-1 contributions and increases with increasing $Q^2$ in the low $Q^2$ regime. For the $\sigma_{TT}$ we emphasize the importance of the unitarization, which increases the pure Born prediction at low energy by approximately a factor of two. Coupled-channel effects are important not only in the $f_0(980)$ region, increasing its importance of the future $a_\mu$ extraction. For $\sigma_{TL}$ we notice that the angular distribution is forward peaked due to the Born contribution. For higher $Q^2$ one has to incorporate constraints from perturbative QCD at least for the vector transition from factors $f_{V,\pi}(Q^2)$ which is the driving force of the $Q^2$ dependence of the $f_2(1270)$ resonance. As it was shown in \cite{Khodjamirian:1997tk}, the correct asymptotic behavior at large $Q^2$ can be implemented in a simple form (\ref{ModifiedVMD}). The numerical values of the $\Lambda$ and $\tilde \Lambda$ can be adjusted directly to the $\gamma\gamma^* \to \pi^0\pi^0$ cross sections and an assumption that at low $Q^2$ VMD for $\pi\rho$ and dispersive analyses for $\pi\omega$ TFF give reasonable estimates. The results are given in Fig.~\ref{fig:sigma_tot} for $Q^2=1.5$, $2.5$, $3.5$ and $4.5$ GeV$^2$. 

For the double virtual process $\gamma^*\gamma^* \to \pi\pi$, in addition to $\sigma_{TT}$ and $\sigma_{TL}$, there is a contribution coming from $\sigma_{LL}$, i.e. when both photons longitudinally polarized. The latter is numerically significant already for $Q_1^2=Q_2^2=0.5$ GeV$^2$, as it can be seen from Fig.~\ref{fig:pipiQ1Q2=0.5}, where the predictions from \cite{Danilkin:X} are shown. It is based on a dispersive treatment of the S-wave and D-waves.

\textbf{$\pi\eta$ channel}\\
We now turn to the second most important multimeson channel, $\gamma^*\gamma^*\to \pi\eta$. Since both experimental and theoretical results serve in favor of $a_0(980)$ being a coupled-channel $\{\pi\eta, K\bar{K}\}$ system, the inclusion of $K\bar{K}$ intermediate state appears to be necessary in order to describe $\gamma\gamma \to \pi\eta$ data \cite{Oller:1997yg,Danilkin:2012ua,Danilkin:2017lyn}.  
In the recent dispersive analysis \cite{Danilkin:2017lyn} the left-had cuts coming from the $t$- and $u$ - channel vector-meson exchanges were tested against the data in the crossed process, the $\eta \to \pi^0 \gamma \gamma$ decay. The $a_2(1320)$ resonance was taken into account explicitly within the assumption that it is predominantly produced by the helicity-2 state (similar to $f_2(1270)$ in \cite{Drechsel:1999rf}). Together with the proposed dispersive method for $a_0(980)$ this yielded a parameter-free description of the $\gamma\gamma\to\pi^0\eta$ cross section \cite{Danilkin:2017lyn}, which was in the reasonable agreement with the data from the Belle Collaboration \cite{Uehara:2009cf}. Then the uncertainty coming from the hadronic final state interactions were narrowed down using this data. The results for the total and differential cross sections are shown in Fig.~\ref{fig:pieta}. For the single virtual process of $\gamma\gamma^* \to \pi\eta$ we show the results of \cite{Deineka:2018nuh}
for $Q^2=0.5$ GeV$^2$ and the case of $\sigma_{TT}$, i.e. when both photons are transversal. In contrast to the real case, there is also a non-zero P-wave amplitude $h^1_{1,++}(s)$, which gives a negligible contribution to the cross section compared to the combined result of the S- and D-waves shown on Fig.~\ref{fig:pieta}. The resulting cross section $\sigma_{TT}$ should be further confronted with the future data from BESIII \cite{Redmer:2018gah}.

\subsection{Two-meson channel contributions to $a_\mu$}
\label{sec:2mesonamu}

In order to estimate the two-meson (and eventually multi-meson) channel contributions to $a_\mu$ in a systematic way, a dispersive framework was developed in a pioneering set of papers~\cite{Colangelo:2014dfa,Colangelo:2015ama}. In particular, a Lorentz decomposition of the HLbL tensor was derived that manifestly implements crossing symmetry and gauge invariance, with scalar coefficient functions free of kinematic singularities and zeros that fulfill the Mandelstam double-spectral representation. As a generalization of the integral representation for the single pseudoscalar pole contributions to $a_\mu$, given by Eq.~(\ref{eq:amups}), 
a master formula has been presented in Ref.~\cite{Colangelo:2015ama} which provides a three-dimensional integral representation of the general HLbL contribution to $a_\mu$ as an integral over spacelike momentum transfers:
\bea
&&a_\mu^{HLbL} = \frac{2 \alpha^3}{3 \pi^2} \int_0^\infty dQ_1 \int_0^\infty dQ_2 \int_{-1}^{+1} d \tau \sqrt{1 - \tau^2} Q_1^3 Q_2^3 \nonumber \\
&& \hspace{5cm} \times \sum_{i = 1}^{12} T_i(Q_1, Q_2, \tau) \bar \Pi_i(Q_1, Q_2, \tau), \quad \quad
\label{eq:amuhlbl}
\eea
where $T_i$ are the integral kernels which have been worked out in Ref.~\cite{Colangelo:2015ama} for the general light-by-light process, 
and where the scalar functions $\bar \Pi_i$ parameterize the hadronic light-by-light scattering input. 

\subsubsection{Pion-loop contribution}  

The next important contribution to $a_\mu$, besides the pseudoscalar meson poles, arises from a charged pion loop. This contribution has been calculated in early works \cite{Bijnens:2016hgx,Hayakawa:1995ps,Hayakawa:1996ki} using scalar QED and including pion vector form factors. For the latter, two approaches were used: a full VMD model was used in Ref.~\cite{Bijnens:2016hgx} yielding $a_\mu^{\pi-loop} = - 1.9 \times 10^{-10}$, whereas a hidden local  symmetry (HLS) model was used in Refs.~\cite{Hayakawa:1995ps,Hayakawa:1996ki}, yielding $a_\mu^{\pi-loop} = - 0.45 \times 10^{-10}$. The reason for the relatively large difference was found~\cite{Bijnens:2016hgx} to arise from the short distance behavior, which in the HLS model yields a relatively large contribution of opposite sign above photon virtualities of 1 GeV$^2$. The full VMD  model on the other hand has the correct short-distance behavior. Within the full VMD model,  an updated value for the $\pi$-loop contribution, containing effects of the pion polarizability, was obtained as~\cite{Bijnens:2016hgx}: 
\bea
a_\mu^{\pi-loop} = - (2.0 \pm 0.5) \times 10^{-10}.
\label{eq:piloop}
\eea

Obtaining a fully data-driven estimate of the $\pi \pi$ contribution is a test and first application of the general dispersive formalism underlying Eq.~(\ref{eq:amuhlbl}). Within this formalism, the charged pion-box contribution has been calculated in 
Refs.~\cite{Colangelo:2015ama,Colangelo:2017fiz}. It corresponds with the two-pion intermediate state with a pion left-hand cut, and is fully determined by specifying the pion vector form factors, which are known with very good accuracy. Therefore, this contribution has been pinned down to very high precision as~\cite{Colangelo:2017fiz}:
\bea
a_\mu^{\pi-box} = - \left( 1.59 \pm 0.02 \right) \times 10^{-10}. 
\label{eq:pibox}
\eea

\subsubsection{Scalar meson contributions}

To quantify the HLbL contribution due to two-meson intermediate states beyond the charged pion loop, requires a dispersive formalism to evaluate the partial-wave amplitudes for the $\gamma^\ast \gamma^\ast \to \pi \pi, \pi \eta, ....$ channels. Such dispersive formalisms are at present under development,  as discussed in Section~\ref{sec4}, and will be validated by a dedicated experimental program underway at BESIII and Belle/Belle-II.  
As the leading partial waves in the two-photon fusion processes at low energies are the $J = 0$ and $J = 2$ partial waves, we will subsequently discuss the contribution from these channels. 

Besides the Born contribution to the $\gamma^\ast \gamma^\ast \to \pi^+ \pi^-$ channel, yielding the pion-box contribution to $a_\mu$,  the next important feature in the description of the 
$\gamma^\ast \gamma^\ast \to \pi \pi$ process 
is the $\pi \pi$ rescattering in the S-wave, which defines the $\sigma/f_0(500)$ region. 
To obtain a first estimate of such rescattering effects, Ref.~\cite{Colangelo:2017qdm,Colangelo:2017fiz} has evaluated the S-wave rescattering by using a pion left-hand cut (LHC)  only, and by using the $\pi \pi$ phase-shifts from the inverse-amplitude method, which reproduces the $f_0(500)$ properties at low energies. Such approach is expected to give a reasonable description of the $\gamma^\ast \gamma^\ast \to \pi \pi$ channels up to about $K \bar K$ threshold. Within such approach, Ref.~\cite{Colangelo:2017qdm,Colangelo:2017fiz} obtained as estimate for the low-energy $\pi\pi$ S-wave rescattering contribution: 
\bea
a_\mu^{\pi\pi, \pi-pole LHC} = - \left( 0.8 \pm 0.1 \right) \times 10^{-10}. 
\label{eq:pipiswave}
\eea
Note that the sum of the $\pi$-box contribution of Eq.~(\ref{eq:pibox}) and the S-wave $\pi\pi$ rescattering contribution of Eq.~(\ref{eq:pipiswave}) 
yields: $a_\mu^{\pi-box} + a_\mu^{\pi\pi, \pi-pole LHC} = -(2.4 \pm 0.1) \times 10^{-10}$, which is in good agreement with the $\pi$-loop model estimate of Eq.~(\ref{eq:piloop}), which includes the pion polarizability.  

The low-energy $\pi \pi$ rescattering contributions to $a_\mu$ are encoded in the pion (generalized) polarizabilities, as  shown in Fig.~\ref{fig:Polar}. Forthcoming single and double tagged measurements of the $\gamma^\ast \gamma^\ast \to \pi^+ \pi^-, \pi^0 \pi^0$  channels will allow to extract such generalized polarizabilities and thus quantify the full S-wave $\pi \pi$ rescattering contribution to $a_\mu$.

For the scalar states with masses $\gtrsim$~1 GeV, the dominant contributions are expected to arise from the $f_0(980)$ and $a_0(980)$ states which show up at the $K \bar K$ threshold in the $\pi \pi$ and $\pi^0 \eta$ channels respectively. Ideally, once the data for the $\gamma^\ast \gamma^\ast \to \pi \pi, \pi^0 \eta$ channels will be available, one can estimate those from an extraction of the TFFs using the coupled-channel dispersive formalism presented in Section~\ref{sec4}. In the meantime, several model estimates have been performed for the contribution of these scalar mesons to $a_\mu$, 
which are shown in Table~\ref{tab:amuscalar}. 

\begin{table}[h]
\centering
\begin{tabular*}{0.95 \textwidth}{c|c|c}
\hline\hline
model & Ref. & value  \\
&& \\
\hline
ENJL  & \cite{Bijnens:1995xf} & $-0.7 \pm 0.2$  \\
ENJL  & \cite{Prades:2009tw} & $-0.7 \pm 0.7$ \\
narrow scalars $\gtrsim$~1 GeV  & \cite{Pauk:2014rfa} & $[-(0.3 \pm 0.1), -(0.09 \pm 0.02)]$ \\
LMD $a_0, f_0, f_0^\prime $  & \cite{Jegerlehner:2017gek} & $-0.6 \pm 0.1$ \\
S-wave $\pi \pi$ rescatt., $\pi$-LHC  & \cite{Colangelo:2017qdm,Colangelo:2017fiz} & $- 0.8 \pm 0.1$ \\
$\sigma/f_0(500)$  & \cite{Knecht:2018sci} & $[- (0.3 \pm 0.1),  -(0.03^{+0.04}_{-0.08})]$ \\
narrow scalars $\gtrsim$~1 GeV  & \cite{Knecht:2018sci} & $[- (0.2^{+0.3}_{-0.07}),  -(0.1^{+0.2}_{-0.04})]$ \\
\hline
\hline
\end{tabular*}
\caption{Different scalar meson HLbL contributions to $a_\mu$ (in units $10^{-10}$). }
\label{tab:amuscalar}
\end{table}

An initial estimate for scalar mesons was performed within an extended Nambu-Jona-Lasinio (ENJL) model in Refs.~\cite{Bijnens:1995xf,Prades:2009tw}. 
In Ref.~\cite{Pauk:2014rfa}, an estimate of scalar mesons $f_0(980)$, $a_0(980)$, and $f_0(1370)$ was performed in the narrow resonance approximation by using the empirical information from the two-photon decay widths and using VMD type parameterizations for the transverse TFF (the longitudinal TFF was neglected in that estimate). Such estimate was further improved in  Ref.~\cite{Jegerlehner:2017gek} by including short-distance constraints. Recently, a new estimate was performed for scalar mesons in Ref.~\cite{Knecht:2018sci}, again in the narrow resonance approximation and including short-distance constraints, by including the effect of both transverse and longitudinal TFFs. Ref.~\cite{Knecht:2018sci} has estimated both the 
$\sigma/f_0(500)$ as well as the narrow scalar meson contributions with masses $\gtrsim$~1 GeV. For the $\sigma/f_0(500)$ contribution, the narrow resonance description of Ref.~\cite{Knecht:2018sci} found a value around 2 - 3 times smaller in absolute value than the $\pi \pi$ S-wave rescattering estimate with $\pi$-LHC only from Ref.~\cite{Colangelo:2017qdm,Colangelo:2017fiz}, given by Eq.~(\ref{eq:pipiswave}). 
For the scalar mesons with masses $\gtrsim$~1 GeV, all  estimates in Table~\ref{tab:amuscalar} give negative contributions, with $a_\mu$ values in the range $[-0.7, 0]$ in units $10^{-10}$. 

\subsubsection{Tensor meson contributions}  

Besides the S-wave $\pi\pi$ rescattering, the next prominent feature in the $\gamma \gamma \to \pi \pi$ and $\gamma \gamma \to \pi \eta$ channels are the prominent tensor resonance contributions, in particular $f_2(1270)$ and $a_2(1320)$.  
A first estimate of the contributions of the tensor mesons $f_2(1270), f_2(1565), a_2(1320)$ and $a_2(1700)$ to $a_\mu$ was performed in Ref.~\cite{Pauk:2014rfa} treating them as narrow resonances and assuming that the tensor mesons are dominantly produced in the helicity-2 states. The TFFs of the tensor mesons were constrained from forward light-by-light sum rules. This estimate was updated in Ref.~\cite{Danilkin:2016hnh} in light of new data on the $f_2(1270)$ TFFs from the Belle Collaboration, by allowing the production of tensor states in helicity 0, 1, and 2 states. 
The resulting contribution to $a_\mu$ was found as~\cite{Danilkin:2016hnh}:
\bea
a_\mu^{tensors} = (0.09 \pm 0.01) \times 10^{-10}. 
\label{eq:tensors}
\eea

\section{Two-photon couplings to higher mass multi-meson states}
\label{sec5}

\subsection{Axial-vector mesons}

Although the production of an axial-vector meson by two real photons is forbidden by the Landau-Yang theorem~\cite{Landau:1948kw,Yang:1950rg}, an axial-vector meson can be produced when one or both photons are virtual, and thus contributes to the HLbL contribution to $a_\mu$. Early estimates in the ENJL model~\cite{Bijnens:1995xf} and HLS model~\cite{Hayakawa:1995ps,Hayakawa:1996ki} found contributions to $a_\mu$ in the range 
$(0.2 - 0.3) \times 10^{-10}$. A more phenomenological evaluation of the $a_1(1260), f_1(1285)$, and $f_1(1420)$ was performed in Ref.~\cite{Melnikov:2003xd}, with TFFs consistent with perturbative QCD constraints, which resulted in a ten times larger estimate as shown in Table~\ref{tab:amuaxial}. Such value corresponds with about 20\% of the total HLbL contribution. This large contribution was subsequently scrutinized in Refs.~\cite{Pauk:2014rfa,Jegerlehner:2017gek}.  A general discussion of the $A \gamma^\ast \gamma^\ast$ vertex has to allow for three independent Lorentz structures and to satisfy the Landau-Yang result. This imposes some antisymmetry constraint on the form factors, which was not accounted for in Ref.~\cite{Melnikov:2003xd}. Both Refs.~\cite{Pauk:2014rfa,Jegerlehner:2017gek} have included these constraints. The L3 Collaboration measured the two-photon fusion to $f_1(1285)$ and $f_1(1420)$ at LEP using the decays to $\pi^+\pi^-\eta$~\cite{Achard:2001uu} and $K^0_S K^\pm \pi^\mp$~\cite{Achard:2007hm}. The $Q^2$ dependence is studied assuming the squared transverse momentum of the reconstructed final state to be equivalent to $Q^2$. In Ref.~\cite{Pauk:2014rfa}, this empirical information was used for  the TFFs of the $f_1(1285)$, and $f_1(1420)$, whereas Ref.~\cite{Jegerlehner:2017gek}  has also estimated the $a_1(1260)$ contribution and imposed short-distance constraints. As can be seen from Table~\ref{tab:amuaxial} both estimates are  consistent with each other, yielding contributions to $a_\mu$ in the range $(0.6 - 0.8) \times 10^{-10}$, around 3 times smaller than the estimate of Ref.\cite{Melnikov:2003xd}.   

\begin{table}[h]
\centering
\begin{tabular*}{0.7 \textwidth}{c|c|c}
\hline
\hline
model & Ref. & value  \\
&& \\
\hline
ENJL  $a_1$ & \cite{Bijnens:1995xf} & $0.25 \pm 0.1$  \\
HLS  & \cite{Hayakawa:1995ps,Hayakawa:1996ki} & $0.2 \pm 0.1$ \\
MV & \cite{Melnikov:2003xd}  & $2.2 \pm 0.5$ \\
$f_1(1285), f_1(1420)$  & \cite{Pauk:2014rfa} & $0.64 \pm 0.20$ \\
$a_1, f_1, f_1^\prime$  & \cite{Jegerlehner:2017gek} & $0.76 \pm 0.20$ \\
\hline
\hline
\end{tabular*}
\caption{Different axial-vector meson HLbL contributions to $a_\mu$ (in units $10^{-10}$). }
\label{tab:amuaxial}
\end{table}

\subsection{Two-photon fusion into $3 \pi$, $4 \pi$ processes}
Three or four pion final states have only been studied in two-photon fusion of quasi-real photons. The main aim for the three pion case is the investigation of the tensor state $a_2(1320)$ and the pseudotensor $\pi_2^0(1670)$. The Crystal Ball Collaboration investigated the neutral three pion final state in $257\,\textrm{pb}^{-1}$ of data taken at the DORIS~II $e^+e^-$ storage ring of DESY~\cite{Antreasyan:1985wx}. The radiative width of $\pi_2^0(1670)$ is determined reconstructing $\pi_2^0(1670) \to \pi^0 f_2(1270)$, which is found to be the dominating contribution to the neutral three pion mode.

The charged three pion final state has been measured by the CELLO and ARGUS Collaborations at DESY. While the CELLO Collaboration also observes the $\pi_2^0(1670) \to \pi^0 f_2(1270)$ in a $86\,\textrm{pb}^{-1}$ sample acquired at $\sqrt{s}=35\,\textrm{GeV}$ at the PETRA rings, the ARGUS Collaboration performed a partial wave analysis on $456\,\textrm{pb}^{-1}$ of data taken at $\sqrt{s}=10.4\,\textrm{GeV}$ at the DORIS~II rings, and does not observe a contribution of $\pi_2(1670)$. In addition, the ARGUS Collaboration was able to determine the spin alignment of the $a_2(1320)$ comparing the helicity states 0 and 2, confirming the dominance of the helicity-2 state.

With the BESIII detector at the BEPCII storage ring there is a possibility to continue these studies. Moreover, it is also possible to investigate the three pion final states in a single-tagged analysis. The virtual photon allows for the production of axial states, {e.g.} the $a_1(1320)$, which is suppressed in quasi-real photon fusion due to the Landau-Yang theorem. The feasibility of these investigations at BESIII is indicated by the ongoing studies of the spacelike TFF of the $\eta$ meson, which is reconstructed from its three pion decay mode~\cite{Redmer:2018uew}. The meson production is tagged by applying a mass window around the nominal $\eta$ mass, rejecting events at higher invariant masses.

The four-pion final state has been studied extensively in the past in view of double vector meson production. Both, the fully charged as well as the $\pi^+\pi^-\pi^0\pi^0$ final state have been investigated. The most recent data are provided by the L3 Collaboration at LEP. The charged and the neutral $\rho\rho$ systems are investigated for invariant masses between $1.1\,\textrm{GeV}\leq W \leq3\,\textrm{GeV}$ and for momentum transfers between $0.2\,\textrm{GeV}^2\leq Q^2 \leq 30\,\textrm{GeV}^2$~\cite{Achard:2003qa,Achard:2004ux,Achard:2004us,Achard:2005pb}. Despite the investigations of the four pion system at large virtualities, which allows the formation of states beyond the restrictions of the Landau-Yang theorem, axial mesons like the $f_1(1285)$ were so far not considered.

\subsection{Quark-loop contribution}  

After including the low-energy contributions to $a_\mu$ through pseudo-scalar meson poles, as well as two- and multi-meson channels, the remaining question is how to quantify the residual short-distance contribution, which perturbatively is given by a quark loop. The analytical expression of the quark loop contribution with a fixed value of the quark mass is well known. What is ambiguous is which part of it should be added to $a_\mu$ when including the low-energy region through empirical information on meson channels. This question has been addressed in models such as the ENJL model, where it was found that for a constituent quark mass of 300 MeV, and a cut-off of 1 (2) GeV, the quark-loop diagram only saturates 50\%(75\%) of the total contribution. This means that even if one replaces the low energy region (say $< 2$~GeV) by empirical information on meson channels, one can expect a significant fraction from the high-energy region. Within the ENJL model~\cite{Bijnens:1995xf} the long-distance (mesonic) contributions were calculated up to a cut-off $\Lambda$ and the remaining part was estimated by a short-distance quark-loop with a quark mass $m_Q = \Lambda$. This resulted in the estimate~\cite{Bijnens:1995xf}
\bea
a_\mu^{quark-loop} = (2.0 \pm 0.4) \times 10^{-10}.
\eea  
It is interesting to observe that within the ENJL model there is a large cancellation between the pion-loop contribution, Eq.~(\ref{eq:piloop}), and the quark-loop contribution. It shows the importance of performing the matching between long-distance and short-distance contributions properly in order to obtain a reliable estimate for the sum of both.
It remains an open question how to perform such matching with the  short-distance quark-loop contribution properly 
when using an empirical estimate for the region up to 1 - 2~GeV.

\section{Conclusion and Outlook}
\label{sec6}
The current 3 - 4~$\sigma$ deviation between theoretical and experimental values  for the muon's anomalous magnetic moment has triggered a worldwide effort on both theoretical and experimental fronts to clarify the situation. Ongoing experimental programs at FERMILAB and J-PARC will soon reach a fourfold increase in precision in the direct measurement of $a_\mu$. On the theoretical side, a dedicated effort to reach a similar improvement in the dominant hadronic contribution to $a_\mu$, which arises from the hadronic vacuum polarization and from the hadronic light-by-light scattering,  is underway. To further constrain the hadronic corrections, both lattice QCD estimates as well as dispersive techniques combined with measurements of the required hadronic input are at present being pursued. In the present work, we reviewed the ongoing efforts in constraining the hadronic light-by-light contribution to $a_\mu$ by using dispersive techniques combined with a dedicated experimental program to obtain the required hadronic input. 

We started this work by reviewing the model independent relations between the forward light-by-light scattering amplitude and the experimentally accessible structure functions which fully characterize the light-by-light fusion process. In particular, we discussed applications of three superconvergence sum rules, which provide model independent relations between the TFFs of mesons with different quantum numbers. Such relations have recently allowed to constrain e.g. the TFFs of tensor mesons from Belle data.  

We have subsequently discussed the status of the TFFs of the $\pi^0, \eta, \eta^\prime$ 
pseudoscalar mesons. As these are the lightest mesons with large couplings to two-photons, they provide by far the dominant contribution to $a_\mu$. 
Furthermore, as about 85\% of their contribution to $a_\mu$ comes from the region of virtualities below 1\,GeV$^2$, data in this region is of particular importance to improve on the error estimate. 
We have presented new data from the BESIII Collaboration for the $\pi^0$ spacelike single-virtual TFF  in the region $0.3$~GeV$^2< Q^2 < 3.1$~GeV$^2$. While in the overlap region with the CLEO data at $Q^2 \geq 1.5$~GeV$^2$ both results show good agreement, the BESIII result extends the CLEO and CELLO measurements towards lower $Q^2$, where it exceeds their accuracy. Furthermore, the BaBar Collaboration has recently provided the first double-virtual TFF measurement for the $\eta^\prime$ both for symmetric and asymmetric kinematics of both photon virtualities, which provides strong constraints on current TFF parameterizations. We have also reviewed the complementary information available on the timelike TFFs and from the $\pi^0, \eta, \eta^\prime$ Dalitz decays. 
The new data for the pseudoscalar meson TFFs provide new challenges for a theoretical understanding. We have discussed recent progress on the $\pi^0, \eta, \eta^\prime$ TFFs in dispersion theory, lattice QCD, as well as the status of the perturbative QCD limit. For the purpose of a new estimate of the $\pi^0, \eta, \eta^\prime$ pole contributions to $a_\mu$, we have provided a new parameterization of these TFFs, given by Eq.~(\ref{eq:interpol}) which satisfies the following three criteria: (i) its value at the real photon point is normalized to the empirical value obtained from the two-photon decay width, (ii) for the single-virtual case, the TFF reduces to a monopole form which was found to provide a very efficient parameterization of the world data up to around 9 GeV$^2$ for $\pi^0, \eta, \eta^\prime$, in particular including the new BESIII data for the $\pi^0$, (iii) the double-virtual TFF has to tend to the correct pQCD expression {\it both} for the symmetric ($Q_1^2 = Q_2^2$) and asymmetric ($Q_1^2 \neq Q_2^2$) kinematics, satisfying the short-distance constraints. The new parameterization has only a single free parameter/scale, which is fixed from the fit to the single-virtual TFF data. The resulting prediction for the double-virtual TFF was found to be in perfect agreement with the new BaBar data for $\eta^\prime$. 
As a result of having only a single free parameter, which is well constrained by the single-virtual TFF data, the error estimate for $a_\mu$ was significantly reduced compared to previous evaluations, especially for the $\pi^0$, for which the new BESIII data were included in the fit, as can be seen from the corresponding entry in Table~\ref{tab:amuconclusion}. 

We next discussed the hadronic information required to constrain the two-meson and multi-meson HLbL contributions to $a_\mu$. We discussed the ongoing experimental program to access the 
$\gamma^\ast \gamma \to \pi^+ \pi^-, \pi^0 \pi^0$ channels, as well as the dispersive frameworks which are being developed to describe the $\gamma^\ast \gamma^\ast \to \pi^+ \pi^-, \pi^0 \pi^0$, and $\pi \eta$ channels, which correspond with the largest cross sections. A new dispersive formalism for the HLbL contribution to $a_\mu$ will allow to fully quantify the two-meson contributions, once the partial-wave amplitudes for the $\gamma^\ast \gamma^\ast \to \pi^+ \pi^-, \pi^0 \pi^0, \pi \eta$ processes are known. We reviewed the initial estimates to $a_\mu$, resulting from a charged pion left-hand cut in the dispersive framework, which have been given by the $\pi$-box contribution and the $\pi \pi$ S-wave rescattering. We have furthermore reviewed the current estimates for $a_\mu$ resulting from scalar mesons with masses $\gtrsim$~1 GeV, from tensor mesons, as well as from axial-vector mesons. 

As a summary, we provide an improved HLbL estimate   
for $a_\mu$, which is shown in Table~\ref{tab:amuconclusion}. For the pseudoscalar meson pole contributions to $a_\mu$, we use the fits for the $\pi^0, \eta, \eta^\prime$ TFFs, including the new BESIII data for $\pi^0$ and using the new TFF parameterization of Eq.~(\ref{eq:interpol}).
For the low-energy $\pi^+ \pi^-$ channel, we use the 
$\pi$-box and S-wave $\pi \pi$ rescattering contribution of Ref.~\cite{Colangelo:2017qdm,Colangelo:2017fiz}. 
For the narrow scalars $f_0, a_0$ with masses $\gtrsim$~1 GeV, we use the recent estimate of Ref.~ \cite{Knecht:2018sci}, and for the tensor mesons 
$f_2$, $a_2$ the recent estimate of Ref.~\cite{Danilkin:2016hnh}.
For the axial-vector mesons, $a_1, f_1, f_1^\prime$, we use the estimate of 
Ref.~\cite{Jegerlehner:2017gek} which has the constraint from the Landau-Yang theorem included.
For the short-distance part of the quark-loop, we use the estimate of Ref.~\cite{Bijnens:1995xf}. 
As a result, upon adding all errors linearly, we obtain the new HLbL estimate for $a_\mu$:
\bea
a_\mu^{HLbL} = (8.7 \pm 1.3) \times 10^{-10}.
\eea  

\begin{table}[h]
\centering
\begin{tabular*}{0.9 \textwidth}{c|c|c}
\hline
\hline
contribution & Ref. & value  \\
&& \\
\hline
$\pi^0, \eta, \eta^\prime$ poles &  this work, Eq.~(\ref{eq:interpol})  &  $8.4 \pm 0.4$   \\ 
charged $\pi$-loop ($\pi$-box) & \cite{Colangelo:2017fiz} & $ - 1.59 \pm 0.02$ \\
S-wave $\pi \pi$ rescatt., $\pi$-LHC & \cite{Colangelo:2017qdm,Colangelo:2017fiz} & $- 0.8 \pm 0.1$  \\
narrow scalars $f_0, a_0$ $\gtrsim$~1 GeV & \cite{Knecht:2018sci} & $-0.2 \pm 0.2$ \\
axial-vectors $a_1, f_1, f_1^\prime$ & \cite{Jegerlehner:2017gek} & $0.8 \pm 0.2$  \\
tensors $f_2$, $a_2$ & \cite{Danilkin:2016hnh} & $0.09 \pm 0.01$ \\
quark-loop & \cite{Bijnens:1995xf}  & $2.0 \pm 0.4$ \\
\hline
sum & & $8.7 \pm 1.3$ \\
\hline
\hline
\end{tabular*}
\caption{Different hadronic light-by-light contributions to $a_\mu$ (in units $10^{-10}$). }
\label{tab:amuconclusion}
\end{table}

We like to end this review by spelling out a few open issues and challenges in this field. 
\begin{enumerate}
\item
{\bf Single-virtual spacelike TFF for $\pi^0, \eta$ and $\eta^\prime$}

The preliminary result of the BESIII Collaboration for the single-virtual spacelike $\pi^0$ TFF illustrates the potential of the experiment to provide high accuracy data in the range of momentum transfer relevant for $a_\mu$. The final result is expected to be available soon, taking into account the full radiative corrections. At the same time the investigations are extended to $\eta$ and $\eta^\prime$ mesons. Combining the most abundant decay modes of each meson, a similar statistical accuracy can be expected as found in case of the $\pi^0$ measurement. Also the range of momentum transfer, for which the TFFs can be determined is similar. It is limited at small values of $Q^2$ by the boost acting on the mesons. At larger momentum transfers the measurement is limited by statistics. So far only a fraction of the total data set of the BESIII collaboration has been considered. A more than four times larger data set is currently available, and more data is to expected in the upcoming years, allowing for high accuracy studies of the single-virtual TFF of $\pi^0, \eta$ and $\eta^\prime$ at small momentum transfers in the spacelike regime. 

At large momentum transfers new data can be provided by the Belle~II Collaboration. These will be of interest to test the pQCD predictions, but also to shed more light on the BaBar-Belle-puzzle.

\item
{\bf Double-virtual spacelike TFF for $\pi^0, \eta$ and $\eta^\prime$}

After the pioneering double-virtual spacelike TFF data from the BaBar Collaboration for $\eta^\prime$, obtaining such a data set for $\eta$ and in particular for $\pi^0$ will be important to validate the existing parameterizations for the double virtual TFF.
Especially the region for virtualities below $Q^2 = 1$\,GeV$^2$ is of importance, as around 85~\% of the pseudoscalar meson contribution to $a_\mu$ originates from  this region.
 
Double-tag measurements are planned at BESIII for all three of the light pseudoscalar mesons. First feasibility studies indicate that it is possible to map out the TFF in a significant region around $Q_1^2=Q_2^2=1\,\textrm{GeV}^2$. In order to perform a different kind of double-tag measurements, the BESIII collaboration plans to install additional tagging detectors at small angles. These will allow to collect high statistics samples of double-tag events at small momentum transfers. In comparison to the single-tag measurements, the exact measurement of the momentum of the second lepton allows to study possible polarization effects in the two-photon production of pseudoscalar mesons. These result in forward-backward asymmetries in the distribution of the dihedral angle $\Tilde{\phi}$ between the planes of the incoming and outgoing leptons in the rest frame of the two virtual photons.

From the theory side, all the existing approaches need to be checked at low $Q^2$. In addition, they need to comply with high-energy constraints, not only for the single virtual, but also for the double virtual asymmetric cases, which can be validated by the current data from BaBar for double virtual $\eta'$ TFF, and future BESIII measurements for $\pi^0$, $\eta$ and $\eta'$ around and below 1 GeV$^2$. 

\item 
{\bf Timelike TFFs for $\pi^0, \eta$ and $\eta^\prime$}

Studying the timelike TFF in radiative meson production at $e^+e^-$ colliders can serve as a cross-check of the measurement in the spacelike regime. Also here results from the BESIII Collaboration around 16 GeV$^2$ are expected. In principle data can be provided in the momentum range $4\,\textrm{GeV}^2\leq q^2\leq 21.16\,\textrm{GeV}^2$, which corresponds exactly to the momenta where the spacelike results for $\pi^0$ of BaBar and Belle start to differ.

\item 
{\bf Dalitz decays of $\pi^0, \eta$ and $\eta^\prime$}

Measurements of Dalitz decays of $\pi^0$, $\eta$, and $\eta^\prime$ into electrons are still the most viable approach to provide precision data on TFFs at smallest values of momentum transfer. At the moment a new measurement of the $\pi^0$ Dalitz decay is performed by the A2 Collaboration. The aim is increase the statistics by a factor a two with respect to the recent NA60 result, providing the most precise directs measurement of the timelike TFF slope. Also in case of the Dalitz decay of $\eta^\prime$ data has been announced by the A2 Collaboration, which will be competitive with the existing BESIII data. Further results with similar accuracy have been announced also by other meson factories. In addition, a new data taking is ongoing at BESIII, aiming to improve the statistics of the $J/\psi$ data set by an order of magnitude, which will allow for a similarly improved determination of the $\eta^\prime$ TFF.

 \item 
 {\bf $ \gamma^\ast \gamma^\ast \to \pi \pi, \pi \eta $ channels}
 
The BESIII Collaboration is currently analyzing both $\gamma\gamma^* \to \pi^+ \pi^-$ and $\gamma\gamma^*\to \pi^0 \pi^0$ reactions in the $0.2~{\rm {GeV}}^2 \lesssim Q^2 \lesssim 2~{\rm {GeV}}^2$ range, corresponding with the most relevant kinematic region for quantifying the HLbL contribution to $a_\mu$. An important aspect of the analysis concerns background subtraction coming from $e^+e^-\to e^+e^-\mu^+\mu^-$ and $\rho$ meson (i.e. $e^+e^-\to e^+e^-\gamma^*$, where $\gamma^*\to \rho \to \pi^+\pi^-$). To accomplish this program, a new version of the \textsc{Ekhara} Monte Carlo generator is being developed. In general, as the data will be available both for integrated and differential cross sections, one can contemplate a partial wave analysis, similar to what has been done for the real photon case. From the low pion invariant mass measurements, it is planned to extract the generalized polarizability, which plays a significant role in the $a_\mu$ estimate. The intermediate energy region will constrain the $f_0(500)$, $f_0(980)$ and $f_2(1270)$ resonance contributions. Ultimately, the ongoing analysis is planned to be extended for $\pi\eta$ channel, which is the second most important two-meson contribution to $a_\mu$. 

From the theoretical side, within the dispersive approach, the dominant uncertainties lie in the treatment of the left-hand cuts beyond the pion (kaon) poles. Once the singly-virtual measurements validate them, an extension to the double-virtual $\gamma^*\gamma^* \to \pi\pi,\pi\eta$ would contain several non-trivial, but somewhat technical challenges. In particular, it is related to a more complicated structure of kinematic constraints and behavior of the left-hand cuts at large virtualities.

\item 
{\bf Axial-vector meson and multi-meson channel contributions}

Experimental studies of multi-meson channels in two-photon fusion experiments were so far focused on spectroscopy using quasi real photons. The three pion final state has been predominantly used to establish the resonance parameters of the tensor $a_2(1230)$, while the production of significant amount of the pseudotensor $\pi_2^0(1670)$ seems to be ruled out by the high statistics partial wave analysis of the ARGUS collaboration. In view of the HLbL contributions of axial-vector mesons, a study of the three pion system at large virtualities is of interest, which allows to investigate the $a_1(1260)$. The ongoing studies of the $\eta$ TFF at BESIII, where the three pion decay modes are investigated, indicate that the extension of the analysis towards higher invariant masses is feasible.  

A similar situation is found for the four pion channels. Existing studies focused on the double vector meson production in quasi-real two-photon fusion. The latest detailed investigations of the L3 collaboration also determined the $Q^2$ dependence of the $\rho\rho$ system, however, the possibility to produce axial-vectors due to the large virtuality is not considered. Here, the BESIII and Belle~II collaborations can provide additional data.

\end{enumerate}

\section*{Acknowledgments}
The authors like to thank Henryk Czyz, Oleksandra Deineka, Achim Denig, Yuping Guo, Andreas Nyffeler, Vladimir Pascalutsa, and Vladyslav Pauk for many discussions and collaborations. 
This work was funded by the Deutsche Forschungsgemeinschaft (DFG, German Research Foundation), 
in part through the Collaborative Research Center [The Low-Energy Frontier of the Standard Model, Projektnummer 204404729 - SFB 1044], 
and in part through the 
Cluster of Excellence “Precision Physics, Fundamental Interactions, and Structure of Matter” (PRISMA$^+$ EXC 2118/1) within the German Excellence Strategy (Project ID 39083149).

\section*{References}
\bibliographystyle{elsarticle-num}
\bibliography{Bib/biblio_hlblppnp}

\end{document}